\documentclass[a4paper]{article}
\usepackage{appendix}
\usepackage{mhchem}
\usepackage{placeins}
\usepackage[english]{babel}
\usepackage[T1]{fontenc}
\usepackage{bm}
\usepackage{amsfonts,amsmath,amssymb}
\usepackage[square,numbers]{natbib}
\usepackage{rotating}
\usepackage{subcaption}
\usepackage{upgreek}
\usepackage{csquotes}

\usepackage[a4paper,top=3cm,bottom=2cm,left=3cm,right=3cm,marginparwidth=1.75cm]{geometry}

\newcommand{\Bas}{\#PlasmaBasic \\}
\newcommand{\Voc}{\#PlasmaVocab \\}
\newcommand{\Mod}{\#PlasmaModel \\}
\newcommand{\Adv}{\#PlasmaAdvan \\}
\newcommand{\partder}[2]{\dfrac{\partial  #1}{\partial  #2}} 
\newcommand{\der}[2]{\dfrac{d #1}{d  #2}}

\newcommand{\TBD}{}

 \usepackage{color}
 \definecolor{red}{rgb}{1,0,0}
 \definecolor{gre}{rgb}{0,1,0}
 \definecolor{blu}{rgb}{0,0,1}

\usepackage{amsmath}
\usepackage{graphicx}
\usepackage[colorinlistoftodos]{todonotes}
\usepackage[colorlinks=true]{hyperref}
\usepackage{soul}
\setstcolor{blue}

\title{
An Introduction to Stellarators \\\large From magnetic fields to symmetries and optimization
}
\author{Lise-Marie Imbert-G\'erard, Elizabeth J. Paul, Adelle M. Wright}
\date{}

\begin{document}
\maketitle

\tableofcontents

\newpage

\section*{Preface}

In this self-contained document, we aim 
to present the basic theoretical building blocks to understand modeling of stellarator magnetic fields, some of the challenges associated with modeling, and optimization for designing stellarators.
As often as possible, the ideas will be presented using equations and pictures, and references to other relevant introductory material will be included. This document is accessible to those who may not have a physics background but are interested in applications of mathematical and computational tools to stellarator research. Readers are simply expected to have basic knowledge on classical physics, Partial Differential Equations (PDEs), and variational calculus, but prior knowledge of plasma physics is not required. We present the relevant models and their derivation, when it is not too involved. We aim to provide enough details for a reader without any background while using language close enough to the plasma physics literature. 


This document arose out of the Simons Collaboration on Hidden Symmetries and Fusion Energy, a collaboration between experts in different fields of mathematics and plasma physics, to tackle fundamental problems in stellarator theory and design the next generation of optimized stellarators. Given the diverse backgrounds of the participants, establishing a common language was a first challenge to tackle. We hope this document, beyond making these topics accessible to a broader audience of mathematicians and physicists, stimulates new contributions to the field of stellarator research.

\section{Introduction}

\TBD 


Harnessing fusion reactions holds promise for a clean, safe source of energy. Under the conditions necessary to achieve sustained fusion reactions, matter exists in a plasma state and can be confined by magnetic fields.
Magnetic confinement fusion techniques have been explored by experimental and theoretical research for several decades.
Much of the research effort in magnetic confinement of plasmas has focused on two toroidal configurations: the tokamak and the stellarator. The tokamak, due to its symmetry with respect to the toroidal coordinate, features a much simpler geometry. 
While the tokamak relies on plasma current for confinement, the stellarator relies on breaking of toroidal symmetry. 
Although stellarators tend to be more difficult to design because of their lack of inherent symmetry, they provide several advantages over tokamak configurations as they do not require a large current in the confinement region. Even though the first stellarator experiments predated the first tokamak, the tokamak concept soon took precedence in the 1960s, as early stellarators had poor confinement properties. With the increase in available computing power, modern stellarators have been carefully designed with numerical optimization techniques. Thus stellarator physics has seen a resurgence since the 1980s. Although stellarators do not have continuous toroidal symmetry, other \enquote{hidden symmetries} have been leveraged to achieve confinement properties similar to tokamaks. 

\begin{figure}
\begin{center}
\begin{tikzpicture}
\node[draw,align=center](z) at (0,0) {Electromagnetic\\ fields\\
$\bm{E}$, $\bm{B}$};
\node[draw,align=center](y) at (7,0) {Plasma \\(Charged particles)\\
$q$, $\bm{v}$};
\draw[->] (z) .. controls +(up:2cm) and +(up:2cm) .. node[above] {Lorentz force} (y);
\draw[->] (z) .. controls +(up:2cm) and +(up:2cm) .. node[below] {$\bm{F}=q(\bm{v}\times \bm{B}+\bm{E})$} (y);
\draw[->] (y) .. controls +(down:2cm) and +(down:2cm) .. node[below] {Currents/charge density} (z);
\draw[->] (y) .. controls +(down:2cm) and +(down:2cm) .. node[above] {$\bm{J}$, $\rho$} (z);
\end{tikzpicture}
\end{center}
\caption{Diagram of the coupling between fields (left box) and particles (right box). 
Particles are affected by the fields through the action of the Lorentz force, while fields are affected by the currents and charge density through Maxwell's equations.}\label{diag}
\end{figure}
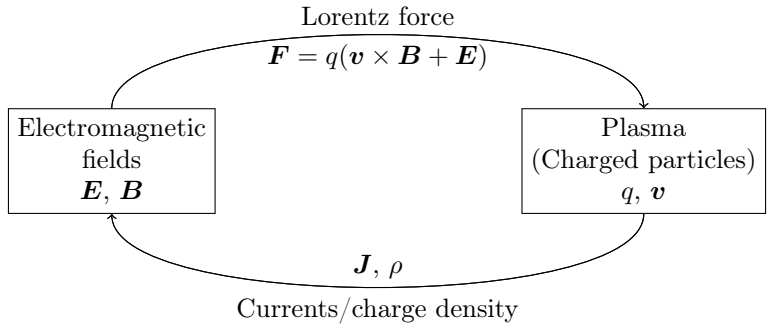

The general setting of plasma modeling is represented in Figure \ref{diag}, which describes the interaction between electromagnetic fields and charged particle motion. Charged particle motion depends on  electromagnetic fields through the Lorentz force, and electromagnetic fields depend on charged particle motion through the resulting current and density which appear in Maxwell's equations. While a set of equations describing the motion of individual particles coupled to Maxwell's equations provides a complete picture of how a coupled system evolves, it is hopeless to solve these equations in practice for any physical system of interest, as all of the particles are coupled through the fields. To glean physical understanding and for computational tractability, it is therefore necessary to make approximations.
The choice of approximation depends on the problem at hand or physical regime of interest, leading to a hierarchy of models which are used to describe the physics of plasmas. In this way, there is a wealth of interesting open problems related to the mathematical properties and numerical approximations for models of magnetic confinement plasmas. 

 The design of any magnetic confinement device must take into account properties of the plasma to ensure good confinement in an experiment, or good energy efficiency in a reactor. In order for the magnetic field to hold the plasma in place for a sufficient period of time for fusion reactions to occur, the plasma pressure must be balanced by the magnetic pressure. The magnetic field in such devices is provided by some combination of currents in the plasma and external currents due to electromagnetic coils. In tokamaks, the large plasma current necessary for confinement must be treated carefully, which poses a challenge in the design process. Axisymmetry, however, is advantageous in some regards, as certain confinement properties are guaranteed. Because the plasma current of a stellarator is relatively small in magnitude, the magnetic field is largely provided by external currents. For this reason the performance of a stellarator is much more sensitive to the coil configuration. From a theoretical point of view, the design of a stellarator device refers to the search for both a desired confining magnetic field and the electromagnetic coils producing this field. In particular, the magnetic field must be consistent with a time-independent state such that the plasma is in equilibrium with respect to the magnetic field. Thus the concept of equilibrium magnetic fields is central to the stellarator design process.

%


We will leverage a series of increasingly complex models to study desirable properties of the fields. For simplicity, it is standard to initially ignore any coupling between the plasma and fields. We will first discuss particle confinement in a steady background magnetic field and motivate the desire for magnetic fields which lie on nested toroidal surfaces.
Next we will introduce coupling between the plasma and fields. Under simplifying assumptions valid for magnetic confinement fusion, we will model the plasma by a single fluid coupled to the fields through the ideal magnetohydrodynamic (MHD) equations. 
We will describe challenges associated with the ideal MHD model in stellarators and introduce several approaches to calculating 3D equilibrium fields. While axisymmetry of tokamaks ensures the existence of continuously nested magnetic surfaces and single particle confinement, stellarators lack this symmetry. We will discuss how magnetic fields in a stellarator can provide some nested magnetic surfaces. Furthermore, we will define several stellarator design concepts which can provide comparable confinement properties to those of a tokamak. We will conclude with a discussion on optimization of the equilibrium magnetic field.

Section \ref{sec:big_picture} provides basic terminology for plasmas and their relation to magnetic confinement fusion. Section \ref{sec:maxwell} presents the set of equations that 
govern the evolution of electromagnetic fields, namely Maxwell's equations. Section \ref{sec:classical_mechanics} reviews the equations of motion that 
describe the trajectories of charged particles in electromagnetic fields and their relation to the variational principles associated with the Lagrangian and Hamiltonian. Under the assumption that electromagnetic fields can be imposed without feedback from the particles on the fields, Section \ref{sec:magnetic_confinement} discusses the motion of charged particles within these given fields. Section \ref{sec:coordinates} introduces convenient coordinate systems to describe toroidal confinement devices. In Section \ref{sec:magnetic_confinement_devices}, these ideas are applied to discuss concepts related to magnetic confinement. Here we provide motivation for the stellarator concept. Section \ref{coupling} discusses the coupling of the electromagnetic fields with particles through the magneto-hydrodynamic (MHD) model, including the set of equations used to describe the equilibrium state. Section \ref{sec:magnetic_coordinates} introduces coordinate systems relying on the additional assumption of MHD equilibrium. Section \ref{sec:3D_difficulties} focuses on the existence of surfaces in stellarator devices and other challenges associated with 3D ideal MHD equilibria. Several models for MHD equilibria in stellarators are presented in Section \ref{sec:equilibrium_fields}. These models provide the equations which determine the time-independent fields, from which the particle trajectories and other physical quantities of interest can be computed. Finally, several symmetries common to stellarator configurations and their consequences are described in Section \ref{sec:symmetry}. The symmetries can be approximately realized in a configuration with optimization of the equilibrium magnetic field and coil shapes using techniques described in Section \ref{sec:optimization}. We conclude in Section \ref{sec:new_frontiers} with an overview of several current research problems in stellarator optimization.

The Sections on classical mechanic, non-orthogonal coordinate systems, and magnetic coordinate systems are provided with the objective of presenting a self-contained document. They appear only when they become necessary to carry on the main discussion.
Readers familiar with these topics can skip the corresponding Sections as they are limited to standard material. 

The Sections on 3D equilibrium fields, symmetry, and optimization are the most important as they are fundamental to an understanding of stellarators. The rest of the document is constructed to enable discussion of these three central topics.








\section{Background}
\label{sec:big_picture}

In this Section we will briefly define and discuss central ideas to the field of plasma physics and magnetic confinement. This will not include many mathematical details, but will simply provide some background and motivation for what will follow. 




\subsection{Plasma}
A plasma is a partially or fully ionized gas that exhibits collective behavior due to long-range electromagnetic forces, in contrast to a neutral gas where the particle dynamics are determined essentially by collisions between neutral particles. 
As the behavior of a plasma is rather different from that of a gas, plasma is often called the fourth state of matter. In an ionized gas, some electrons are stripped off of the nuclei, and ions and those electrons move independently rather than bound together as atoms. 


Ionization can occur due to the collision of an atom with an electron or the absorption of a photon with sufficient energy, causing an electron to be removed from the atom.
The inverse processes can also occur, in which an atom and two electrons collide to form an atom and one electron, or an electron and ion can combine, releasing a photon. In equilibrium, these processes balance each other to determine the degree of ionization. The fraction of ionized atoms depends on the temperature and density of the gas. At room temperature, the ionization fraction of a typical gas is negligibly small. At the typical densities and temperatures relevant for fusion experiments, the ionization fraction is large enough that collisions between charged particles dominate over those between charged and neutral particles. For this reason, we will focus our attention on charged particle motion when discussing plasmas in this document, ignoring any interaction with neutrals. See Chapter 10 in \cite{Goldston1995} for a further discussion on ionization. 

On earth, naturally occurring plasmas are not especially common but can be found in
lightning and auroras
. Laboratory created plasmas are widely used in many industrial processing applications \cite{lieberman2005} such as
the deposition of thin layers of metal on surfaces like in solar panels or watches, or for processing of materials including the etching of superconductors. In medicine, plasma is used to treat certain cells \cite{kong2009}. On the other hand, plasmas are ubiquitous in space and astrophysical environments.
For example, the earth's upper atmosphere (the ionosphere) is ionized and plays a critical role in shielding the planet from potentially harmful radiation from the sun.
More generally, the magnetospheres of magnetized astronomical objects are important for determining the interaction of the body with the surrounding medium.
Plasma thrusters have also been explored for use in satellite propulsion \cite{levchenko2018}.

A hot, fusion-relevant plasma is typically fully ionized, with the ions and electrons in thermal equilibrium. If we consider the temperature, $T$, of each species, we often find that $T_e \approx T_i$ if the plasma is confined long enough for the temperatures to equilibrate. Temperature can be thought of as being a measure of the energy per particle; thus temperature equilibration corresponds to the equal partition of energy among particles.
A classical plasma must be hot enough that the electrons have the necessary energy to overcome the potential of the nucleus to ionize, and diffuse enough that the plasma appears neutral on length scales larger than the Debye length (see Appendix \ref{debye}). This is known as quasineutrality,
\begin{gather}
\sum_s n_s q_s = 0,
\label{eq:quasineutrality}
\end{gather}
where $n_s$ and $q_s$ are the number density and charge of species $s$, respectively, and the sum is taken over species. We call this quasineutrality rather than neutrality, since if one considers short enough length scales this assumption is violated. 

\subsection{Fusion reactions and power source}
\label{sec:fusion_reactions}


Stars, including the sun, are giant balls of plasma bound by large gravitational forces. Stars are fuelled by nuclear fusion reactions, when two atomic nuclei combine to form atomic nuclei in addition to other products. The energy associated with nuclear fusion reactions is due to the strong force, the attractive force that binds nuclei together. In order for two particles to undergo a fusion reaction, they must be brought close enough for the strong force, which acts on the scale length of protons and neutrons, to act. This requires overcoming the repulsive Coulomb force, which makes like charges repel and opposite charges attract, and acts over much larger length scales.

The nuclear fusion process leads to a slight decrease in mass, resulting in the release of energy according to the famous $E=mc^2$ equation, where $E$ is the energy released, $m$ is the mass, and $c$ is the speed of light (a physical constant). Therefore, during a nuclear fusion reaction, the difference in mass $m$ between the reactants and the products determines the change in energy, $E$. 


In stars, the combination of high temperature and strong gravitational field ensures the probability per interaction, or cross-section, of a fusion reaction occurring is sufficient to power the star. At standard conditions on earth, the probability of a fusion reaction is negligibly small. To realize terrestrial fusion power, we therefore need to create conditions which sufficiently increase the fusion reaction cross-section. In practice, this is achieved by increasing temperatures, up to ten times hotter than the center of the sun. The candidate reaction for terrestrial fusion with the largest cross-section is the fusion of deuterium, $\ce{^2_1D}$, and tritium, $\ce{^3_1T}$, to produce helium, $\ce{^4_2He}$, and an energetic neutron, $\ce{n^0}$, which can be represented as,
\begin{align}
    \ce{^2_1D + ^3_1T -> ^4_2He (3.5 MeV) + n^0 (14.1 MeV)},
    \label{eq:DT}
\end{align}
known as the D-T reaction.
At the requisite temperatures for the D-T reaction to occur, matter exists in a plasma state. 
Therefore, we need to consider how to confine hot plasmas on earth. We will discuss magnetic confinement with the stellarator concept in detail in this document.



Nuclear fusion is one of the most energetic reactions known in nature. The D-T reaction produces $3.4 \times 10^8$ MJ of energy for every kg of fuel, in comparison with the combustion of gasoline, which produces $40$ MJ.
If we could harness this mechanism as a power source on earth, it would yield numerous benefits.
Fusion as a power source requires very little fuel, produces no greenhouse gasses, is safe, and produces no long-lived radioactive waste. In other words, it could provide a source of clean, virtually limitless energy. Both of the fuels necessary for the D-T reaction are readily available on earth: deuterium is found abundantly in the earth's oceans, and
tritium can be produced with irradiation of lithium by an energetic neutron. There are now schemes 
proposed to produce tritium as a by-product of fusion reactions (see Section 5.5 in \cite{Freidberg2008}). 

Nuclear fission and fusion yield a similar amount of energy per kg of fuel. However, nuclear fusion has several advantages over nuclear fission. While the fusion reaction itself does not produce any radioactive by-products, isotopes result from the reaction of the fusion-produced neutrons with material structures in the plasma vessel. Isotopes associated with the production of fusion power have half-lives of around 10 years, compared to fission by-products which can have half-lives of over $10^4$ years, eliminating challenges associated with long-lived radioactive waste.  Unlike plutonium and uranium, the fuels of fission power, the fuel required for fusion power also have zero potential for enrichment or weaponization.
Finally, compared to fission, fusion power is inherently safer as it does not depend on a chain reaction, is very sensitive to its conditions, and the amount of fuel present in a device at any given time is fairly small meaning there is no potential for runaway reactions.

\subsection{Magnetic confinement for fusion}
\label{sec:magnetic_confinement_fusion}
In fusion experiments, 
the center of the device where fusion reactions take place is very hot and must be kept well away from the wall of the experimental vessel to avoid damage.
The interaction of charged particles, which constitute a plasma, with electromagnetic fields is one important property which can be exploited to confine plasmas. This is known as magnetic confinement (see Section \ref{sec:magnetic_confinement_devices}), the focus of this document. Magnetic confinement is one method to thermally insulate plasmas at temperatures and densities necessary for fusion to occur.
Other principles for fusion plasma confinement include inertial confinement, as in the laser facility at the National Ignition Facility \cite{NIF}.
Regardless of the method, the ultimate goal of all fusion power devices is to produce more energy than what is required to initiate the reactions.
This is measured by $Q$, the ratio of the fusion output power to the power used to heat the plasma.
The break-even point corresponds to $Q=1$ and ignition, a self-sustaining reaction, occurs as $Q\to\infty$.
The goal of fusion research is to reach $Q\gg1$.

The first laboratory magnetic confinement fusion device was built in the late 1940s, a toroidal device known as a Z-pinch \cite{Bishop1958}. Many have been built around the world since, and there continues to be active experimental research at magnetic confinement fusion devices today. The largest magnetic confinement device in operation is the Joint European Torus (JET), a tokamak which has set the record for the largest value of $Q = 0.6$ \cite{wesson2011}. 

The D-T fusion reaction \eqref{eq:DT} results in a helium nucleus (alpha particle) and a neutron. The neutron, being charge-neutral, is not confined by the magnetic fields and can leave the device. The alpha particle, on the other hand, may be confined by the magnetic fields and through collisions with the bulk plasma can deposit energy to the plasma.
Ignition ($Q\to\infty$) is achieved if the energy deposited by the alpha particles is sufficient to sustain fusion plasma conditions without external heating.

The plasma conditions necessary for ignition, a self-sustaining fusion reaction, to occur is described by the Lawson criterion, a lower bound on the fusion triple product,
\begin{gather}
nT \tau_E > 3 \times 10^{21} \text{ m}^{-3}\text{keVs},
\end{gather}
here $n$ is the number density in m$^{-3}$, $T$ is the temperature in keV \footnote{The plasma temperature is typically measured in units of energy. If measured in Kelvin, we multiply by Boltzmann's constant $k_B = 1.38\times 10^{-23}$J/K)}, and $\tau_E$ is the energy confinement time in seconds (the timescale of energy loss from the plasma). While magnetic confinement experiments can reach the conditions necessary for fusion to occur, the Lawson criteria has not been met in any experiments to date, though some have come close. 

Ultimately, for electricity production, fusion power aims to achieve a burning plasma ($Q\ge5$) operating regime where most of the energy required to heat the plasma is obtained from fusion reactions.
The ITER project is a multi-national (35 countries) collaboration to demonstrate the scientific viability of fusion power. The experimental device is currently under construction in France and aims to demonstrate $Q>10$ \cite{ITER}.

From the fusion triple product, it is clear that, for a given temperature, the Lawson criterion can be achieved by increasing confinement time.
Increasing plasma temperature involves heating which, if external, is expensive.
One of the main challenges of magnetic confinement fusion is thus to achieve ignition by maximizing the energy confinement time which requires confining a hot, turbulent plasma for a sufficient period of time.

See Chapter 1 in \cite{wesson2011} or Chapter 6 in \cite{Freidberg2008} for a more complete overview of magnetic confinement for fusion. 

\subsection{Charged particles and trajectories}

While not all plasmas are fully ionized, in a fusion plasma the interactions between charged particles dominate over those between charged and neutral particles. Thus for many applications, it suffices to only consider the behavior of charged particles, ions and electrons, rather than neutral atoms. 
In the plasma physics literature, the term ion refers to a positively charged nucleus while an electron is negatively charged. The bulk of fusion plasmas consist of hydrogen or one of its isotopes, deuterium and tritium. Other ions may exist in fusion plasmas, such as alpha particles, which are a product of a fusion reaction, or heavier impurities which enter the plasma through interaction with material structures of the the device. Charged particles are accelerated by electric fields and, when moving, interact with magnetic fields. 

Trajectories, or orbits, refer to the motion of charged particles as the equations of motion that describe a particle's position and velocity are evolved in time. For example, in a straight magnetic field, particles exhibit helical trajectories, as will be discussed in Section \ref{sec:cylindrical}. In Section \ref{ref:gyroaveraged_lagrangian} we will consider the trajectories of charged particles in the presence of strong magnetic fields. This provides a basis for confinement in the tokamak and stellarator configurations.

\subsection{Separation of length and time scales}
\label{sec:LTscales}

As demonstrated in Figure \ref{diag}, modeling of physical processes in a plasma is complex, as each particle is coupled to every other particle through the electromagnetic fields. In practice, accounting for each of these interactions is both impractical and unnecessary. In fusion plasmas, there is a good separation between length and time scale that has lead to the development of \textit{reduced models}, the asymptotic reductions of a set of model equations based on the smallness of a physical parameter. Table \ref{LTscales} provides a few typical length and time scales that exist in magnetic confinement fusion plasmas.
\begin{table}
\begin{center}
\begin{tabular}{|c|c||c|c|c|}
\hline
  Name & Parameter  & W7-AS \cite{Mccormick2002} & LHD \cite{Komori2006} & W7-X \cite{Sunn2017} \\ 
  \hline \hline
 Electron Debye length  & $\lambda_{D,e}$ [cm] & $3 \times 10^{-3}$ & $2 \times 10^{-3}$ & $9 \times 10^{-3}$ \\ 
  Ion gyroradius &  $\rho_i$ [cm] & $2 \times 10^{-1}$ & $3 \times 10^{-1}$ & $2\times 10^{-1}$ \\ 
   Device minor radius &  $a$ [cm] & 20 & 60 & 50 \\ \hline
    Ion gyrofrequency & $\Omega_i$ [s$^{-1}$] & $9\times 10^7$ & $1 \times 10^8$ & $2\times 10^8$ \\
  Electron-Electron collision frequency &  $\nu_{ee}$ [s$^{-1}$] & $1\times 10^5$ & $2\times 10^5$ & $4\times 10^3$ \\ 
   Energy confinement time &  $\tau_E^{-1}$ [s$^{-1}$] & 2 & 3 & 10 \\ \hline
\end{tabular}
\caption{This chart displays the typical of length and time scales for several stellarator experiments. Here $\lambda_{D,e}$ is the electron Debye length (see Appendix \ref{debye}), $\rho_i$ is the ion gyroradius (see Section \ref{sec:cylindrical}), $a$ is the average minor radius of the device (see Section \ref{sec:toroidal_geom}), $\nu_{ee}$ is the electron-electron collision frequency, $\Omega_i$ is the ion gyrofrequency (see Section \ref{sec:cylindrical}), and $\tau_E$ is the energy confinement time. Due to the large range of scales, it is intractable to model all of these scales simultaneously. This provides motivation for the application of reduced models. }
\label{LTscales}
\end{center}
\end{table}
%
%
The Debye length, $\lambda_D$, is much smaller than typical length scales for fusion devices, so the plasma can be assumed to be quasineutral, see Appendix \ref{debye}. The ion gyrofrequency, $\Omega_i$, is much larger than typical frequencies, so motion can be averaged over the fast gyrofrequency to consider guiding center models, see Section \ref{ref:gyroaveraged_lagrangian}. 

The modeling of plasmas can be largely grouped into several categories, each with its own assumptions of length and time scales. 
\begin{enumerate}
    \item The \textit{single particle approach} studies single particle motion in stationary background fields. Thus feedback of particles on the electromagnetic fields or collisions between particles will be neglected. This approach is useful for discussing the confinement properties of several magnetic field configurations in the absence of plasma waves. Single particle trajectories are also considered when modeling a small population of particles whose feedback on the fields can be neglected. This approach will be discussed in Section \ref{sec:magnetic_confinement}.
    \item The \textit{kinetic approach} studies the evolution of distributions of particles in velocity and position space. Particles can interact with each other through collisions and are coupled to electromagnetic fields. Rather than consider the equations of motion of individual particles, the distribution of the positions and velocities of particles are computed.
    This approach will not be discussed in this document. For an introduction, see Chapters 3-5 in \cite{Nicholson1983} or Chapters 4-5 in \cite{Hazeltine2003}. One kinetic model of relevance for fusion plasmas, known as gyrokinetics, is used to study phenomena on length scales comparable to the gyroradius. 
    \item The \textit{fluid approach} studies the plasma at a macroscopic scale, as one or several fluids. The fluid is described by scalar density and pressure fields and vector flow velocity field rather than a distribution of all particles in velocity and position space. As one averages over all particle velocities, a certain distribution of velocities must be assumed, such as an equilibrium distribution. If the velocity distribution is not in equilibrium, then this approach may not capture the necessary physics. The fluid approach is useful for studying large-scale effects (lengths much larger than the gyroradius) on long time scales (much longer than the gyroperiod). It provides a set of equations which is simpler than the kinetic approach. Some aspects of this approach will be discussed in Section \ref{coupling}. 
\end{enumerate}

\FloatBarrier
\section{Electric and magnetic fields: Maxwell's equations}
\label{sec:maxwell}

Here we describe models for the electromagnetic fields, respectively denoted $\bm{E}$ and $\bm{B}$. In this Section, the plasma density $\rho$ and current density $\bm{J}$ are considered to be prescribed by another model, and the electromagnetic fields consistent with these sources will be obtained. In the following Sections we will discuss several such models. In Section \ref{sec:magnetic_confinement}, we will discuss the trajectories of particles in static electromagnetic fields, without considering the feedback of $\rho$ and $\bm{J}$ on the fields. A more realistic model would include this coupling: the Lorentz force describes how the electric and magnetic fields act on a charged particle motion, while Maxwell's equations describe how electric and magnetic fields evolve in the presence of  $\rho$ and $\bm{J}$. See Figure \ref{diag} and Section \ref{coupling}. 

In the remainder of the text, we will use the SI system of units. Maxwell's equations are sometimes presented in Gaussian units such that the physical constants differ. See \cite{Huba2006} for a comparison between the SI and Gaussian systems.

\subsection{Electromagnetics}
Maxwell's equations describe how electric $\bm{E}$ and magnetic $\bm{B}$ fields propagate and interact together, as well as with currents and charges. Maxwell's equations refer to the four following equations.

Gauss's law is
\begin{gather}\label{GL}
\nabla \cdot \bm{E} = \frac{\rho}{\epsilon_0},
\end{gather}
Ampere's law is
\begin{align}
\nabla \times \bm{B} = \mu_0 \bm{J} + \frac{1}{c^2} \partder{\bm{E}}{t},
\label{eq:Amp}
\end{align}
Faraday's law is
\begin{gather}
\nabla \times \bm{E} = - \partder{\bm{B}}{t},
\label{eq:faraday}
\end{gather}
and magnetic fields must be divergence-free,
\begin{gather}
\nabla \cdot \bm{B} = 0. 
\label{eq:gauss_mag}
\end{gather}
Here $\mu_0$ is the permeability of free space, $\epsilon_0$ is the permitivity of free space, and $c$ is the speed of light. This set of PDEs describe how the electric and magnetic fields evolve in time, $t$, in response to charge density $\rho$ and current density $\bm{J}$. In a plasma, this set of equations forms describing the interactions of the electric and magnetic fields with the particles. Under further assumptions on time scales of interest, Maxwell's equations can be further reduced as we will see in the following Sections.


Often the electric and magnetic fields are expressed in terms of scalar and vector potentials. As $\bm{B}$ is divergence-free \eqref{eq:gauss_mag}, it can always be written in terms of a vector potential, 
\begin{align}
    \bm{B} &= \nabla \times \bm{A}. \label{eq:vector_potential} 
\end{align}
We note that there is some non-uniqueness in the choice of the vector potential, as the gradient of any scalar function can be added to $\bm{A}$ without altering the magnetic field. A standard by which the vector potential is chosen in order to remove this freedom is often referred to as the gauge in the physics literature. 



Inserting \eqref{eq:vector_potential} into \eqref{eq:faraday}, we see that $\bm{E}$ can be written as $- \partial \bm{A}/\partial t$ with the addition of a curl-free vector field,
\begin{align}
    \bm{E} &= - \nabla \Phi - \partder{\bm{A}}{t}.
    \label{eq:scalar_potential}
\end{align}
Often $\Phi$ is referred to as the scalar potential and $\bm{A}$ as the vector potential. 

\subsection{Electrostatics}
\label{sec:electrostatics}
In the static case ($\partial/\partial t = 0$), the equations satisfied by $\bm{E}$ and $\bm{B}$ decouple. As $\nabla \times \bm{E} = 0$ under this assumption, the electric field can be written in terms of only a scalar potential
\begin{gather}
\bm{E} = -\nabla \Phi.
\end{gather}
From Gauss's law \eqref{GL} the electrostatic potential satisfies Poisson's equation,
\begin{gather}
\Delta \Phi = - \frac{\rho}{\epsilon_0}.
\end{gather}
Gauss's law can be written in an equivalent integral form by integrating over a volume $\Omega$,
\begin{gather}
     \text{for all volume } \Omega \subset \mathbb{R}^3, \int_{\partial \Omega} \bm{E} \cdot \hat{\bm{n}} \, d A = \frac{1}{\epsilon_0} \int_{\Omega} \rho \, d^3 x,
\end{gather}
where $\hat{\bm{n}}$ is the outward unit normal on $\partial \Omega$.

\subsection{Magnetostatics}
\label{sec:magnetostatics}
Often the displacement current term, $\partial \bm{E}/\partial t$ in \eqref{eq:Amp}, can be neglected if the typical velocities of a system, $v$, are non-relativistic ($v/c \ll 1$). This is the assumption that the model does not need to include light waves, which are associated with very short time scales. Under this assumption, Ampere's law becomes
\begin{gather}
    \nabla \times \bm{B} = \mu_0 \bm{J}.
    \label{eq:ampere_magnetostatic}
\end{gather}
This can be written in an equivalent integral form,
\begin{gather}
 \text{for all surface }S \subset \mathbb R^3 ,
\oint_{\partial S} \bm{B} \cdot d \bm{l} = \mu_0 \int_S\hat{\bm{n}} \cdot \bm{J} \, d^2x ,
\label{eq:ampere_surface_int}
\end{gather}
where for an open surface the line integral is taken along a closed curve forming the boundary of the surface $S$, while for a closed surface the left hand side is zero. Another integral form that is consistent with the magnetostatic equations is the Biot-Savart law,
\begin{gather}
\bm{B}(\bm{r}) = \frac{\mu_0}{4\pi} \int_{\mathbb R^3} \frac{\bm{J}(\bm{r}') \times (\bm{r}-\bm{r}')}{\rvert{\bm{r}-\bm{r}'\rvert^3}} \, d\bm{r}'.
\label{eq:biot_savart}
\end{gather}
This follows from application of a Green's function approach to \eqref{eq:ampere_magnetostatic}. For example, in a magnetic confinement fusion device, the integral is taken throughout the plasma volume and along the electromagnetic coils. 

\subsection{Vacuum fields}
\label{sec:vacuum}

The term vacuum field is used to describe the magnetic field in a region $\Omega$ without currents under the magnetostatics assumptions. In magnetic confinement fusion, this could be the region outside the plasma, not including the electromagnetic coils. In this case, we have $\nabla \times \bm{B} = 0$, so a scalar potential can be used to describe the magnetic field,
\begin{gather}
\bm{B} = \nabla \Phi_B \text{ in }\Omega. 
\label{eq:phi_B}
\end{gather}
The magnetic field must also satisfy (\ref{eq:gauss_mag}), 
so the scalar potential must satisfy Laplace's equation,
\begin{gather}
\Delta \Phi_B = 0 \text{ in }\Omega.
\label{eq:laplace}
\end{gather}
Laplace's equation is often solved with a Neumann boundary condition such that $\bm{B} \cdot \hat{\bm{n}} = \hat{\bm{n}} \cdot \nabla \Phi_B $ is prescribed on $\partial \Omega$. Depending on the topology of $\Omega$, an additional scalar quantity must be specified to obtain a unique solution. The solution in a toroidal domain will be discussed in Section \ref{sec:vacuum_toroidal}. 

Instead of solving Laplace's equation, the magnetic field in the vacuum region can be determined using the Biot-Savart law \eqref{eq:biot_savart}, integrating over all currents outside of the vacuum region, that is to say over $\mathbb R^3\backslash \Omega$. 

\subsection{Summary}
Here we describe Maxwell's equations in the presence of some charge density $\rho$ and current density $\bm{J}$. A realistic model would include coupling to a set of equations which describes the evolution of the charge and current density. Under various sets of hypotheses, Maxwell's equations can be reduced to simpler models. Common reduced models are gathered in the following Table. For each reduced model, the Table provides the hypotheses (Hyp.), the PDE model, a different formulation of the model, and the model data. Computational domains, as well as boundary conditions, will be addressed later (see Section \ref{sec:vacuum_toroidal}).

\begin{center}
{\renewcommand{\arraystretch}{2.0}%
\begin{tabular}{|c|c|c|c|c|}
\hline
&Maxwell & Electrostatics& Magnetostatics  & Vacuum fields \\\hline
Hyp. & 
& $\partial \bm{E}/\partial t=0$
& $\partial \bm{E}/\partial t=0$
& $\partial \bm{E}/\partial t=0$
\\
 & 
& $\partial \bm{B}/\partial t=0$
& $\partial \bm{J}/\partial t=0$
& $\bm{J} = 0\text{ in }\Omega$\\\hline
&  $\nabla \cdot \bm{E} = \frac{\rho}{\epsilon_0}$
& $\nabla \cdot \bm{E} = \frac{\rho}{\epsilon_0}$ 
& 
&
\\ PDE&
  $\nabla \times \bm{B} = \mu_0 \bm{J} + \frac{1}{c^2} \partder{\bm{E}}{t}$
& 
& $\nabla \times \bm{B} = \mu_0 \bm{J} $
& $\nabla \times \bm{B} = 0 \text{ in }\Omega$
\\model&
  $\nabla \times \bm{E} = - \partder{\bm{B}}{t}$
& $\nabla \times \bm{E} = 0$
&
& 
\\&
  $\nabla \cdot \bm{B} = 0$
&
& $\nabla \cdot \bm{B} = 0$
& $\nabla \cdot \bm{B} = 0 \text{ in }\Omega$
\\\hline Model 
&
&$\bm{E} = -\nabla \Phi$
& $\bm{B}(\bm{r}) = \frac{\mu_0}{4\pi} \times \qquad\qquad $
&$\bm{B}(\bm{r}) = \frac{\mu_0}{4\pi} \times\ \qquad\qquad $
\\$\Leftrightarrow$
&
&$\Delta \Phi = - \frac{\rho}{\epsilon_0}$
& $\int_{\mathbb R^3} \frac{\bm{J}(\bm{r}') \times (\bm{r}-\bm{r}')}{\rvert{\bm{r}-\bm{r}'\rvert^3}} \, d \bm{r}'$
&$\int_{\mathbb R^3\backslash \Omega} \frac{\bm{J}(\bm{r}') \times (\bm{r}-\bm{r}')}{\rvert{\bm{r}-\bm{r}'\rvert^3}} \, d \bm{r}'$\\
& & & & $\Leftrightarrow
\left\{\begin{array}{l}
\bm{B} = \nabla \Phi_B\\\Delta \Phi_B = 0
\end{array}\right.\text{ in }\Omega
$\\ \hline
Given
& $\bm{J},\rho$ & $\rho$ & $\bm{J}$ & $\bm{J}$ outside of $\Omega$\\ 
&&&& 
or $\bm{B}\cdot\hat{\bm{n}}$ on $\partial \Omega$
\\ \hline
\end{tabular}
}
\end{center}

\FloatBarrier
\section{Classical mechanics}
\label{sec:classical_mechanics}

Classical mechanics refers to the study of phenomena occurring at velocities much smaller than the speed of light, and at length scales much larger than the atomic size. We briefly review concepts from classical mechanics, related to three formulations to study the evolution of a mechanical system.
Newton’s approach relies on balancing forces acting on a particle and the rate of change of its momentum, while variational approaches  
rely on the principle of least action. Instead of focusing on forces, variational methods focus on other physical quantities such as kinetic and potential energy. They facilitate the identification of conserved quantities, moreover, restating a variational formulation in a different coordinate system is not complicated unlike it would be for a Newtonian formulation. The Lagrangian approach will enable us to more easily derive the guiding center model under the assumption of large magnetic field strength in Section  \ref{ref:gyroaveraged_lagrangian}, as scalar functions are simpler to manipulate than a set of ODEs. However, non-conservative forces, such as frictional forces, are more easily represented in the Newtonian approach. For further reading on classical mechanics see \cite{1998classical,2015Tong,Goldstein2002}.

We begin in Section \ref{sec:Newton} with Newton's law, a set of ODEs which describe the motion of a particle in the presence of a force. We will then show that these ODEs can be expressed in terms of two variational principles involving the Lagrangian and Hamiltonian functionals, respectively in Sections \ref{sec:lagrangian} and \ref{sec:Hamiltonian}. These concepts will later be relevant to study single particle motion in electromagnetic fields in Section \ref{sec:magnetic_confinement} as well as  magnetic field lines in Section \ref{sec:3D_difficulties}.

\subsection{Newton's law}
\label{sec:Newton}

Newton's law is a set of ODEs which describe the trajectory of a point particle of mass $m$,
\begin{gather}
m\der{^2\bm{r}(t)}{t^2} = \bm{F}(\bm{r}(t),\dot{\bm{r}}(t)).
\label{eq:Newton}
\end{gather}
 On the right hand side, $\bm{F}$ is a force, which can in general depend on both the position, $\bm{r}$, and its time derivative, the velocity $\dot{\bm{r}}$. In words, Newton's law states that the force is equal to the mass of a particle multiplied by its acceleration. To study single particle motion in electric and magnetic fields, we will consider the Lorentz force,
\begin{gather}
\bm{F}(\bm{r},\dot{\bm{r}}) = q (\dot{\bm{r}} \times \bm{B}(\bm{r}) + \bm{E}(\bm{r}) ),
\label{eq:lorentz}
\end{gather}
which depends explicitly on $\dot{\bm{r}}$ and on $\bm{r}$ through $\bm{B}$ and $\bm{E}$. Newton's law can be solved to obtain the trajectory, $\bm{r}(t)$ and $\dot{\bm{r}}(t)$, of a particle given initial conditions $\bm{r}_{\text{init}}$ and $\dot{\bm{r}}_{\text{init}}$.

\subsection{Lagrangian mechanics}
\label{sec:lagrangian}


While Newton's law, in principle, provides us with all of the information we need to obtain trajectories of charged particles, another more general approach is desirable. Specifically, we will discuss a variational formalism from which Newton's law can be obtained. The Lagrangian approach will provide several advantages as we consider reduced models in the limit of a strong magnetic field in Section \ref{ref:gyroaveraged_lagrangian}. Further, calculations are simplified when manipulating the scalar Lagrangian as opposed to the vectorial equations of motion. Working within the Lagrangian formalism will provide insight into certain conserved quantities as will be seen in Sections \ref{sec:adiabatic_invariant} and \ref{sec:energy_conservation}. For further reading on Lagrangian mechanics, see the notes of David Tong \cite{2015Tong} and Chapter 2 in \cite{Goldstein2002}.

The methods of Lagrangian mechanics involve a scalar functional $L$, from which \eqref{eq:Newton} is naturally derived. The Lagrangian functional depends on the position ($\bm{r}$), velocity ($\dot{\bm{r}}$), and time ($t$),
\begin{align}
    L:(\bm{r},\dot{\bm{r}},t)\in\mathbb R^3\times\mathbb R^3\times \mathbb R.
\end{align}
Although the Lagrangian will be evaluated along a particle trajectory, $\bm{r}_T\in\mathcal C^1(\mathbb R, \mathbb R^3)$ parametrized by $t$, along which a particle's position and velocity are related through $\dot{\bm{r}}_T(t) = d\bm{r}_T(t)/dt$, it is important that $\bm{r}$ and $\dot{\bm{r}}$ are treated as independent variables in the Lagrangian. 
The Lagrangian along a trajectory is then expressed as $L(\bm{r}_T(t),\dot{\bm{r}}_T(t),t)$. We will consistently write ($\bm{r},\dot{\bm{r}}$) for independent variables, as opposed to ($\bm{r}_T,\dot{\bm{r}}_T$) along a trajectory. In \eqref{eq:lagrangian} we will consider a specific Lagrangian for a particle in the presence of electro-magnetic fields.

Consider the following functional of a trajectory $\bm{r}_T$ through $L$,
\begin{gather}
\text{ for all } \bm{r}_T\in\mathcal C^1(\mathbb R, \mathbb R^3), \quad 
    S[\bm{r}_T] := \int_{t_{\text{init}}}^{t_{\text{final}}} L (\bm{r}_T(t),\dot{\bm{r}}_T(t),t)\, dt, 
\end{gather}
called the action integral. We will show that trajectories which are stationary points of $S$ correspond with those which satisfy Newton's law. This is known as the principle of stationary action. We compute the first variation of $S$ with respect to $\bm{r}_T(t)$, keeping the end points of the trajectory fixed. Consider the perturbation to $S$ which results from perturbing the trajectory by a vector field of displacements of the trajectory, $\delta \bm{r}_T(t)$, 
\begin{gather}
    \delta S[\bm{r}_T;\delta \bm{r}_T] = \int_{t_{\text{init}}}^{t_{\text{final}}} \left( \partder{L(\bm{r}_T(t),\dot{\bm{r}}_T(t),t)}{\bm{r}} \cdot \delta \bm{r}_T(t) + \partder{ L (\bm{r}_T(t),\dot{\bm{r}}_T(t),t)}{ \dot{\bm{r}}} \cdot \delta \dot{\bm{r}}_T(t) \right) \, dt,
\end{gather}
where $\delta \dot{\bm{r}}_T(t) = d(\delta \bm{r}_T(t))/dt$.
The second term on the right hand side can be integrated by parts, using the condition $\delta \bm{r}_T(t_{\text{init}}) = \delta \bm{r}_T(t_{\text{final}}) = 0$, 
\begin{gather}
    \delta S[\bm{r}_T;\delta \bm{r}_T] = \int_{t_{\text{init}}}^{t_{\text{final}}} \left( \partder{ L (\bm{r}_T(t),\dot{\bm{r}}_T(t),t)}{ \bm{r}} - \der{}{t} \left(\partder{ L(\bm{r}_T(t),\dot{\bm{r}}_T(t),t)}{\dot{\bm{r}}} \right) \right) \cdot \delta \bm{r}_T(t) \, d t.
\end{gather}
In order for $S$ to be stationary, it is necessary for $\delta S[\bm{r}_T;\delta \bm{r}_T]$ to vanish; thus, the above integrand must vanish for all $\delta \bm{r}_T$. So we arrive at the following condition,
\begin{gather}
    \der{}{t} \left( \partder{L(\bm{r}_T(t),\dot{\bm{r}}_T(t),t)}{\dot{\bm{r}}} \right) = \partder{ L(\bm{r}_T(t),\dot{\bm{r}}_T(t),t)}{\bm{r}}. 
    \label{eq:euler_lagrange}
\end{gather}
The above is referred to as the Euler-Lagrange equations. 

To obtain the Lagrangian for motion in electromagnetic fields, we first write the electric and magnetic fields in terms of vector and scalar potentials, 
\begin{subequations}
\begin{align}
    \bm{B} &= \nabla \times \bm{A} \label{eq:potentials_lagrangian_B} \\
    \bm{E} &= - \nabla \Phi - \partder{\bm{A}}{t}.
    \label{eq:potentials_lagrangian}
\end{align}
\label{eq:potentials_lagrangian_subequations}
\end{subequations}
The Lagrangian for charged particles in electromagnetic fields is,
\begin{gather}
    L(\bm{r},\dot{\bm{r}},t) =\frac{m|\dot{\bm{r}}|^2}{2} + q \bm{A}(\bm{r},t) \cdot \dot{\bm{r}} - q \Phi(\bm{r},t).
    \label{eq:lagrangian}
\end{gather}

We will next see that \eqref{eq:euler_lagrange} is equivalent to Newton's law. By computing $\partial L/\partial \bm{r}$ and $\partial L/\partial \dot{\bm{r}}$ from \eqref{eq:lagrangian}, we find
\begin{gather}
     \der{}{t} \left( m \dot{\bm{r}}_T(t) + q \bm{A}(\bm{r}_T(t)) \right) = q \nabla \left( \dot{\bm{r}}_T(t) \cdot \bm{A}(\bm{r}_T(t)) \right) - q \nabla \Phi(\bm{r}_T(t)).
\end{gather}
We use the notation $\nabla = \partial/\partial \bm{r}$, noting that $\nabla \dot{\bm{r}}_T = 0$ as $\bm{r}$ and $\dot{\bm{r}}$ are independent coordinates in the Lagrangian. To evaluate $\nabla(\dot{\bm{r}}_T\cdot \bm{A})$ we use the vector identity $\nabla \left( \bm{a} \cdot \bm{b} \right) = \bm{a} \times \nabla \times \bm{b} + \bm{b} \times \nabla \times \bm{a} + \bm{a} \cdot \nabla \bm{b} + \bm{b} \cdot \nabla \bm{a}$. We can also note that the total time derivative of $\bm{A}$ along the trajectory is $d \bm{A}/dt = \partial{\bm{A}}/\partial t + \dot{\bm{r}}_T \cdot \nabla \bm{A}$. Thus we obtain the following equation of motion from the Euler-Lagrange equation \eqref{eq:euler_lagrange},
\begin{gather}
    m\der{^2\bm{r}_T(t)}{t^2}  = q \dot{\bm{r}}_T(t) \times (\nabla \times \bm{A}(\bm{r}_T(t))) - q \nabla \Phi(\bm{r}_T(t)) - q \partder{\bm{A}(\bm{r}_T(t))}{t}. 
    \label{eq:euler_lagrange_lorentz}
\end{gather}
Here we can use the expressions for the electromagnetic fields in terms of vector and scalar potentials \eqref{eq:potentials_lagrangian_subequations} to obtain the familiar Lorentz force \eqref{eq:lorentz}. Therefore, we find that the Lagrangian reproduces Newton's law \eqref{eq:Newton} for electromagnetic fields with a force given by \eqref{eq:lorentz}. Thus, we have shown that the Lagrangian given by \eqref{eq:lagrangian} reduces to Newton's law \eqref{eq:Newton} on the path that minimizes the action. 

In Section \ref{sec:cylindrical} we will study trajectories within a uniform, straight magnetic field using the above ODEs directly. We will then consider motion in a strong magnetic field which is not necessarily straight or uniform in Section \ref{ref:gyroaveraged_lagrangian}.

\subsection{Hamiltonian mechanics}
\label{sec:Hamiltonian}
Newton's law can also be obtained from a variational principle involving the Hamiltonian functional. Rather than treating $\bm{r}$ and $\dot{\bm{r}}$ as independent coordinates, in the Hamiltonian formalism we take the coordinate $\bm{q} := \bm{r}$ and the momentum, $\bm{p}$, which satisfies
\begin{gather}
    \bm{p} = \partder{L(\bm{q},\dot{\bm{q}},t)}{\dot{\bm{q}}},
    \label{eq:p_Hamiltonian}
\end{gather}
as independent coordinates. The correspondence between the Hamiltonian and the Lagrangian reads
\begin{align}
    H(\bm{q},\bm{p},t) = \bm{p} \cdot \dot{\bm{q}}(\bm{q},\bm{p}) - L(\bm{q},\dot{\bm{q}}(\bm{q},\bm{p}),t), 
    \label{eq:Hamiltonian_Lagrangian}
\end{align}
where we have denoted the transformation between Lagrangian $( \bm{q},\dot{\bm{q}})$ and Hamiltonian $( \bm{q},\bm{p} )$ variables through \eqref{eq:p_Hamiltonian}.

Given a trajectory $\{\bm{q}_T(t),\bm{p}_T(t)\}$, consider the following functional,
\begin{gather}
    \mathcal{W}[\bm{q}_T,\bm{p}_T] := \int_{t_{\text{init}}}^{t_{\text{final}}} \, \left(\bm{p}_T(t) \cdot \der{\bm{q}_T(t)}{t} - H(\bm{q}_T(t),\bm{p}_T(t),t) \right) dt, 
    \label{eq:hamiltonian_functional}
\end{gather}
where the integral is taken along the trajectory. 

We consider the perturbation to $\mathcal{W}$ due to the perturbation of the trajectory, using the boundary condition $\delta \bm{p}_T(t_{\text{init}}) = \delta \bm{q}_T(t_{\text{init}}) = \delta \bm{p}_T(t_{\text{final}}) = \delta \bm{q}_T(t_{\text{final}}) = 0$,
\begin{multline}
    \delta \mathcal{W}[\bm{q}_T,\bm{p}_T;\delta \bm{q}_T,\delta \bm{p}_T] = \int_{t_{\text{init}}}^{t_{\text{final}}} \Bigg( \delta \bm{p}_T(t) \cdot \der{\bm{q}_T(t)}{t} + \bm{p}_T(t) \cdot \der{\delta \bm{q}_T(t)}{t} - \partder{H(\bm{q}_T(t),\bm{p}_T(t),t)}{\bm{q}} \cdot \delta \bm{q}_T(t) \\ - \partder{H(\bm{q}_T(t),\bm{p}_T(t),t)}{\bm{p}} \cdot \delta \bm{p}_T(t) \Bigg) dt. 
\end{multline}
Upon integration by parts with respect to $t$ we obtain
\begin{multline}
    \delta \mathcal{W}[\bm{q}_T,\bm{p}_T;\delta \bm{q}_T,\delta \bm{p}_T] = \int_{t_{\text{init}}}^{t_{\text{final}}} \Bigg( \delta \bm{p}_T(t) \cdot \der{\bm{q}_T(t)}{t} - \der{\bm{p}_T(t)}{t} \cdot \delta \bm{q}_T(t) - \partder{H(\bm{q}_T(t),\bm{p}_T(t),t)}{\bm{q}} \cdot \delta \bm{q}_T(t) \\
    - \partder{H(\bm{q}_T(t),\bm{p}_T(t),t)}{\bm{p}} \cdot \delta \bm{p}_T(t) \Bigg) dt. 
\end{multline}
For $\delta \mathcal{W}$ to vanish for every $\delta \bm{p}_T$ and $\delta \bm{q}_T$, the following equations must be satisfied
\begin{subequations}
\begin{align}
    \der{\bm{q}_T(t)}{t} &= \partder{H(\bm{q}_T(t),\bm{p}_T(t),t)}{\bm{p}} \label{eq:Hamiltons_eqns1} \\
    \der{\bm{p}_T(t)}{t} &= -\partder{H(\bm{q}_T(t),\bm{p}_T(t),t)}{\bm{q}}.
    \label{eq:Hamiltons_eqns}
\end{align}
\label{eq:Hamiltons_subequations}
\end{subequations}
These are known as Hamilton's equations. For an $N$-dimensional system such that $\bm{q} \in \mathbb{R}^N$, the Lagrangian provides a set of $N$ second order ODEs while the Hamiltonian provides a set of $2N$ first order ODEs. 

Using \eqref{eq:Hamiltons_subequations}, the total time derivative of $H$ along the trajectory is given by,
\begin{align}
    &\der{H(\bm{q}_T(t),\bm{p}_T(t),t)}{t} \nonumber\\&= \partder{H(\bm{q}_T(t),\bm{p}_T(t),t)}{t} + \partder{H(\bm{q}_T(t),\bm{p}_T(t),t)}{\bm{q}} \cdot \der{\bm{q}_T(t)}{t} + \partder{H(\bm{q}_T(t),\bm{p}_T(t),t)}{\bm{p}} \cdot \der{\bm{p}_T(t)}{t} \nonumber\\&= \partder{H(\bm{q}_T(t),\bm{p}_T(t),t)}{t},
    \label{eq:Hamiltonian_time}
\end{align}
from which we note that if $H$ does not depend explicitly on time, or is autonomous, then $H$ is a constant of the motion.
In many physical systems, $H$ corresponds to total energy; a physical system which can be described by an autonomous Hamiltonian thus conserves energy.

Equivalence of the Hamiltonian and Lagrangian variational principles can be seen by showing that they lead to the same Euler-Lagrange equations. We will demonstrate by considering the Hamiltonian for charged particles in electromagnetic fields. We first compute the momentum from \eqref{eq:p_Hamiltonian} to be
\begin{align}
    \bm{p} = m \dot{\bm{q}} + q \bm{A}(\bm{q}), 
\end{align}
from which we can note that $\dot{\bm{q}} = (\bm{p}-q\bm{A}(\bm{q}))/m$.
The Hamiltonian is computed from \eqref{eq:Hamiltonian_Lagrangian},
\begin{align}
    H(\bm{q},\bm{p},t) &= \frac{|\bm{p}-q\bm{A}(\bm{q})|^2}{2m} + q \Phi(\bm{q}).
\end{align}
We now apply Hamilton's equations \eqref{eq:Hamiltons_subequations} to obtain
\begin{subequations}
\begin{align}
    \der{\bm{q}_T(t)}{t} &= \frac{(\bm{p}_T(t)-q \bm{A}(\bm{q}(t)))}{m}  \\
    \der{\bm{p}_T(t)}{t} &=  \frac{q}{m}\left(\bm{p}_T(t)-q\bm{A}(\bm{q}_T(t))\right)\times \bm{B}(\bm{q}_T(t)) + \frac{q}{m} \left(\bm{p}_T(t)-q\bm{A}(\bm{q}_T(t))\right) \cdot \nabla \bm{A}(\bm{q}_T(t)) -q \nabla \Phi(\bm{q}_T(t)),
\end{align}
\end{subequations}
where we use the notation $\nabla = \partial/\partial \bm{q}$. Using the transformation from $(\bm{q},\dot{\bm{q}})$ to $(\bm{q},\bm{p})$ and the expression of the fields in terms of the potentials \eqref{eq:potentials_lagrangian_subequations}, we therefore recover the same Euler-Lagrange equations \eqref{eq:euler_lagrange_lorentz}.

For further reading on Hamiltonian mechanics, see \cite{2015Tongb} and Chapter 8 of \cite{Goldstein2002}.

\FloatBarrier
\section{Single particle motion in electromagnetic fields}
\label{sec:magnetic_confinement}

Magnetic fields can be designed to confine the orbits of charged particles.
Examples of magnetic confinement configurations include the tokamak, see Section \ref{sec:tokamak}, stellarator, see Section \ref{sec:stellarator}, and mirror machines. 
In order to introduce the ideas behind these confinement concepts, we first study the motion of charged particles in the presence of electric and magnetic fields. 
We describe here models for individual charged particle motion with stationary electromagnetic fields considered to be inputs. A particle in these fields undergoes complex orbits which are solutions of the equations of classical mechanics introduced in Section \ref{sec:classical_mechanics}. In practice, the electric and magnetic fields also depend on the charged particle motion, so a more realistic model would include the coupling between charged particle motion and electromagnetic fields, see Section \ref{coupling}. We will also ignore the effects of collisions between particles. 

We begin in Section \ref{sec:cylindrical} by considering the equations of motion a simple setting: a straight, uniform magnetic field. In Section \ref{ref:gyroaveraged_lagrangian}, we discuss a reduction of the Lagrangian in the presence of a strong magnetic field which can vary in space, while consequences are discussed in Section \ref{gradBdrift}. This model will become important as we discuss confinement of charged particles in toroidal magnetic confinement devices in Section \ref{sec:magnetic_confinement_devices}. Finally  Section \ref{confiningMagneitcField} introduces the basic ideas of toroidal confinement.

\subsection{Motion in a straight and uniform magnetic field}
\label{sec:cylindrical}

\begin{figure}
\begin{center}
\includegraphics[trim=12cm 6cm 12cm 6cm,clip,width=\textwidth]{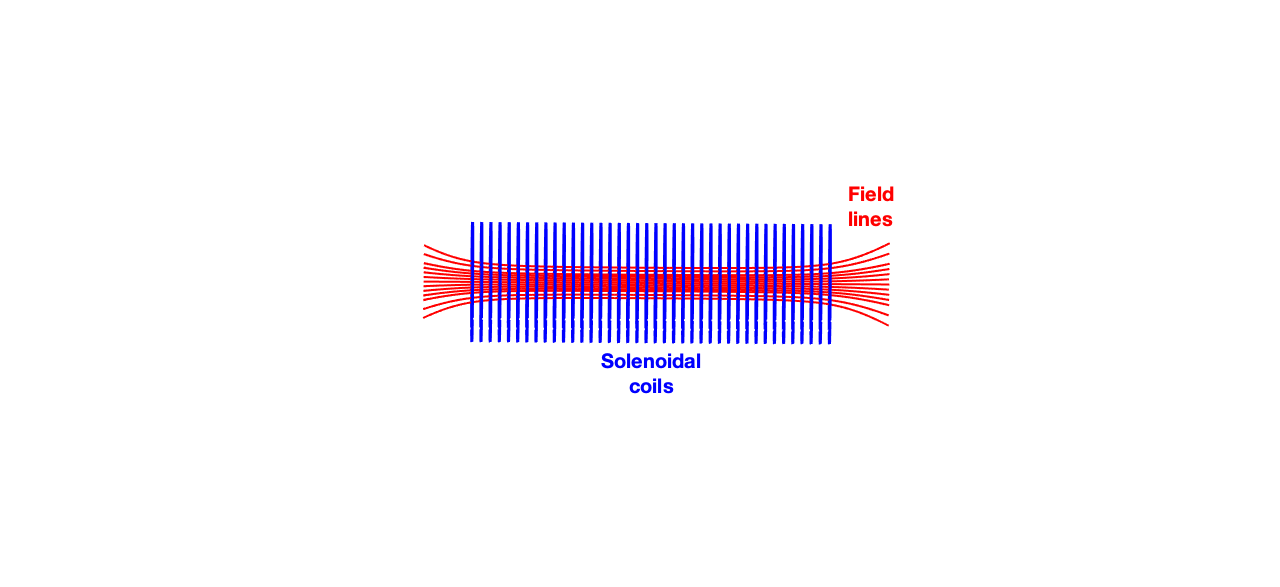}
\caption{A solenoid is use to produce an (approximately) straight, uniform magnetic field.}
\label{fig:solenoid}
\end{center}
\end{figure}

The concept of magnetic confinement can be illustrated by studying the trajectory of a particle in a straight, uniform magnetic field $\bm{B} = B \hat{\bm{e}}_1$. In practice, a solenoid,  a cylindrical coil with several turns, can be used in order to generate a set of straight field lines in a given volume of space (see Figure \ref{fig:solenoid}). We will use the orthonormal coordinate system $(\hat{\bm{e}}_1, \hat{\bm{e}}_2, \hat{\bm{e}}_3)$ such that $\hat{\bm{e}}_1 \times \hat{\bm{e}}_2 =\hat{\bm{e}}_3$. 

The Lorentz force \eqref{eq:lorentz} on a particle of charge $q$ and velocity $\bm{v}$ is given by $\bm{F} = qB\left( \bm{v} \times \hat{\bm{e}}_1\right)$, and the resulting trajectory obeys the following equations of motion from \eqref{eq:Newton},
\begin{subequations}
\begin{align}
\der{\bm{r}(t)}{t} &= \bm{v}(t) \\
m\der{\bm{v}(t)}{t} &= qB\left( \bm{v}(t) \times \hat{\bm{e}}_1\right),
\label{eq:lorentz_cylindrical}
\end{align}
\end{subequations}
where we have taken the electric field to vanish. If a particle has initial velocity, $\bm{v}_{\text{init}} = v_{||} \hat{\bm{e}}_1  + v_{\perp} \left(-\sin(\varphi_{\text{init}}) \hat{\bm{e}}_3 + \cos(\varphi_{\text{init}}) \hat{\bm{e}}_2\right)$, it will spiral about the magnetic field in a helical orbit, 
\begin{gather}
\bm{v} (t)= v_{||} \hat{\bm{e}}_1 + v_{\perp} \left( -\sin \left(\Omega t+\varphi_{\text{init}}\right)\hat{\bm{e}}_3 +  \cos \left(\Omega t+\varphi_{\text{init}}\right)\hat{\bm{e}}_2\right).
\label{eq:v_t_uniform}
\end{gather}
Here $\Omega = qB/m$ is the gyrofrequency. The particle position is given by,
\begin{multline}
\bm{r} (t)= (z_{\text{init}} + v_{||} t) \hat{\bm{e}}_1 + \left(x_{\text{init}} - \frac{v_{\perp}}{\Omega} \cos(\varphi_{\text{init}}) + \frac{v_{\perp}}{\Omega}\cos \left( \Omega t+
\varphi_{\text{init}} \right) \right) \hat{\bm{e}}_3 \\ + \left(y_{\text{init}} - \frac{v_{\perp}}{\Omega}\sin(\varphi_{\text{init}}) +  \frac{v_{\perp}}{\Omega} \sin \left(\Omega t +\varphi_{\text{init}}\right) \right) \hat{\bm{e}}_2,
\label{eq:r_t_uniform}
\end{multline}
for initial position $\bm{r}_{\text{init}} = (x_{\text{init}},y_{\text{init}},z_{\text{init}})$. We see that the motion in the direction of the magnetic field is constant, while the perpendicular motion is periodic with a frequency given by $\Omega$. If $\Omega$ is positive (for an ion species), the particle will rotate clockwise in the $\hat{\bm{e}}_2- \hat{\bm{e}}_3$ plane, while if $\Omega$ is negative the particle will rotate counter-clockwise.

\begin{figure}
    \centering
    \includegraphics[trim = 14cm 4cm 14cm 3cm,clip,width=0.7\textwidth]{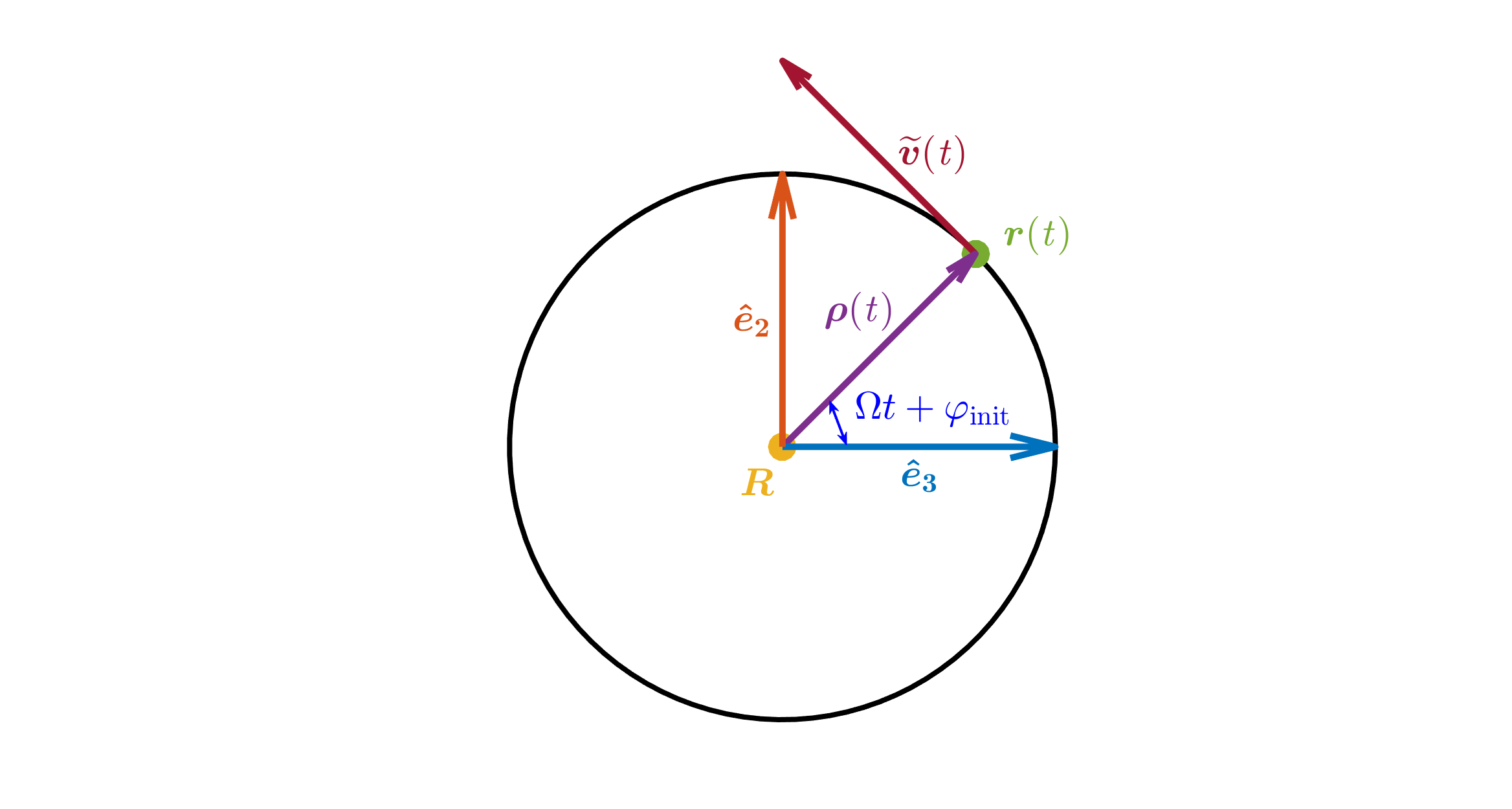}
    \caption{The motion in the plane perpendicular to the magnetic field is described by the orthonormal $\hat{\bm{e}}_2$-$\hat{\bm{e}}_3$ basis. The quantity $\widetilde{\bm{v}}$ describes the periodic velocity in the perpendicular plane. The particle position is decomposed into periodic and non-periodic pieces as $\bm{r} = \bm{R} + \bm{\rho}$, where $\bm{\rho}$ is the gyroradius and $\bm{R}$ is the guiding center. }
    \label{fig:gyromotion}
\end{figure}

Typically in experiments (see Table \ref{LTscales}) the ion gyrofrequency is $\Omega_i\approx 10^{8}\text{s}^{-1}$, and the electron gyrofrequency is $\Omega_e = \left( m_i/m_e \right)$, where the mass ratio $m_i/m_e \sim 10^3$. When considering time scales $t\gg \Omega^{-1}$, it is useful to separate the secular motion along a field line from the periodic motion. We will define a new set of variables to allow us to isolate the gyromotion. The velocity associated with the periodic motion is,
\begin{gather}
    \widetilde{\bm{v}}(t) = v_{\perp} \left( -\sin(\Omega t+\varphi_{\text{init}})\hat{\bm{e}}_3 +\cos(\Omega t+\varphi_{\text{init}})\hat{\bm{e}}_2\right).
    \label{eq:periodic_velocity}
\end{gather}
The position associated with this periodic motion, $\bm{\rho}(t)$, can be obtained by integrating $\widetilde{\bm{v}}(t)$ with respect to time. The integration constants are chosen such that $\bm{\rho}(t) \cdot \hat{\bm{e}}_1 = 0$, as the gyromotion occurs in the plane perpendicular to the magnetic field, and $\widetilde{\bm{v}}(t) \cdot \bm{\rho}(t) = 0$ for all $t$, as should be true for uniform periodic motion,
\begin{gather}
    \bm{\rho}(t) = \frac{v_{\perp}}{\Omega} \left(\cos(\Omega t+\varphi_{\text{init}})\hat{\bm{e}}_3 + \sin(\Omega t+\varphi_{\text{init}})\hat{\bm{e}}_2\right). 
    \label{eq:gyroradius}
\end{gather}
As we can be seen from Figure \ref{fig:gyromotion}, $\bm{\rho}(t)$ points from the center of the circular orbit in the $\hat{\bm{e}}_2-\hat{\bm{e}}_3$ plane to the particle position. For this reason, the term \textit{gyroradius} is often used to refer to $\bm{\rho}(t)$. The \textit{guiding center} velocity, $\bm{V} = \bm{v}(t)-\widetilde{\bm{v}}(t)$, accounts for the secular motion along field lines,
\begin{gather}
    \bm{V} = v_{||} \hat{\bm{e}}_1. 
    \label{eq:guiding_center_cylindrical}
\end{gather}
The guiding center position, $\bm{R}(t)$, can be determined such that $\bm{r}(t) = \bm{R}(t) + \bm{\rho}(t)$,
\begin{gather}
   \bm{R}(t) = \left(z_{\text{init}} +v_{||}t\right) \hat{\bm{e}}_1 + \left( x_{\text{init}} - \frac{v_{\perp}}{\Omega} \cos(\varphi_{\text{init}}) \right) \hat{\bm{e}}_3 + \left(y_{\text{init}} - \frac{v_{\perp}}{\Omega} \sin(\varphi_{\text{init}}) \right) \hat{\bm{e}}_2. 
    \label{eq:guiding_center}
\end{gather}
The quantity $\bm{R}(t)$ is referred to as the \textit{guiding center} position as it is the center about which the particle is said to gyrate. The guiding center moves purely along the field line at the center of the helical motion. 
As $|\bm{\rho}| = v_{\perp}/\Omega$ is typically much smaller than most length scales of interest for a magnetic confinement device (see Section \ref{sec:LTscales}), considering the motion of the guiding center is a very good approximation. In Section \ref{ref:gyroaveraged_lagrangian} we will use this assumption to explore the trajectories of particles in the presence of more general electromagnetic fields.

\begin{figure}
\begin{center}
\includegraphics[trim=37cm 7cm 32cm 6cm,clip,width=0.7\textwidth]{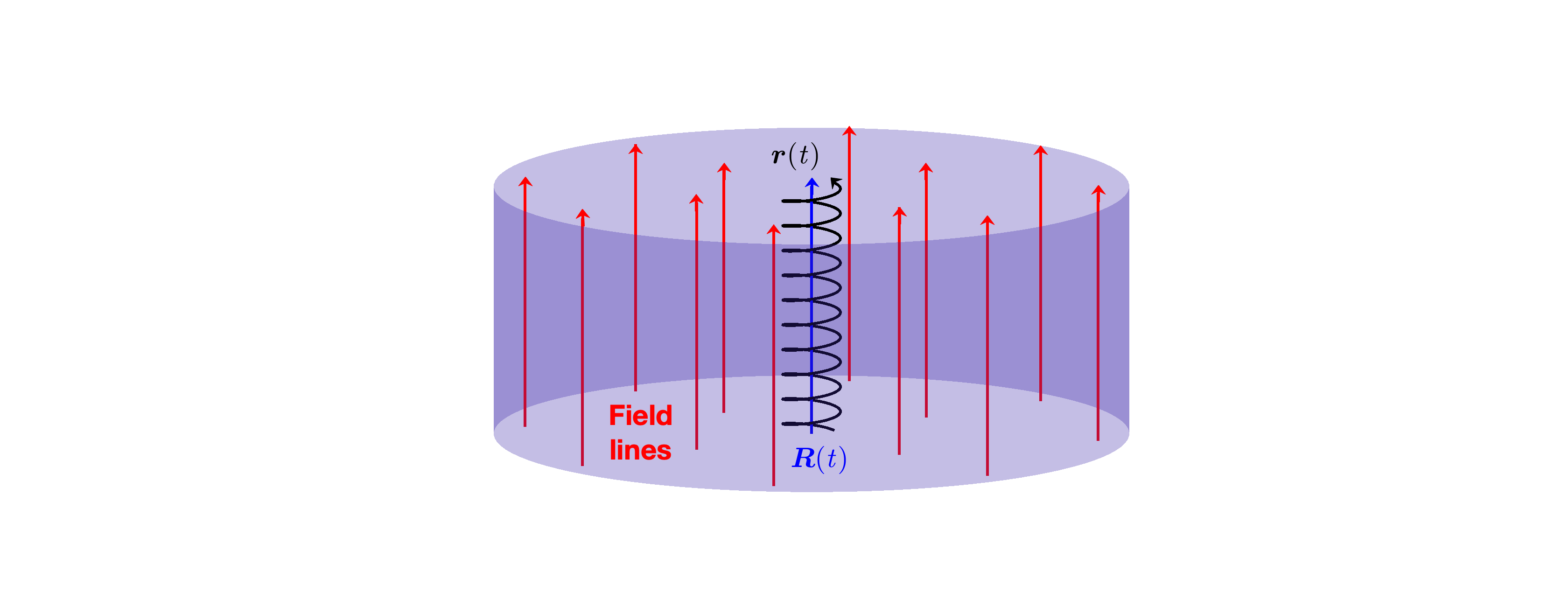}
\caption{In a straight, uniform magnetic field, charged particles exhibit fast helical motion about field lines (red). All charged particles are confined in the direction perpendicular to the magnetic field, but are free to move in the direction parallel to the magnetic field. The guiding center trajectory (blue) describes the particle's motion (black) after averaging over the fast gyration.}
\label{cylindrical_Figure}
\end{center}
\end{figure}

In a straight, uniform field, the motion of the guiding center is purely along a field line. In this way, the particle is confined in the direction perpendicular to the magnetic field (in an experiment it would stay away from the walls of the cylinder) but unconfined in the direction parallel to the magnetic field (particles can escape out the ends). See Figure \ref{cylindrical_Figure}. An additional confining mechanism is needed to avoid losses of particles along the field lines, such as toroidal confinement or the mirror force. This will be discussed in Section \ref{confiningMagneitcField}.

\subsection{Gyroaveraged Lagrangian}
\label{ref:gyroaveraged_lagrangian}

In this Section we consider the motion of charged particles in static electric and magnetic fields within the Lagrangian framework. Applying knowledge of the length and time scales involved, we will reduce the Lagrangian in order to gain insight about motion in a strong magnetic field. The results will be used to arrive at the well-known expressions for the drifts of particles across field lines in Section \ref{gradBdrift}. While many sources simply average the Lorentz force in order to obtain the cross-field drifts, working within the Lagrangian framework will provide additional insight into the conserved quantities of the system. 

We consider motion in a field $\bm{B}(\bm{r}) = B(\bm{r}) \hat{\bm{e}}_1(\bm{r})$ that is no longer assumed to be straight and uniform. Here $\hat{\bm{e}}_1 = \hat{\bm{b}}$ is a local unit vector in the direction of the magnetic field at $\bm{r}$. The unit vectors $\hat{\bm{e}}_2(\bm{r})$ and $\hat{\bm{e}}_3(\bm{r})$ form a basis of the plane perpendicular to $\bm{B}(\bm{r})$. At each point in $\bm{r}\in\mathbb{R}^3$, $(\hat{\bm{e}}_1,\hat{\bm{e}}_2,\hat{\bm{e}}_3)$ forms a local orthonormal basis, independent of the motion.

Motivated by the calculation in Section \ref{sec:cylindrical}, we would like to perform a similar coordinate transformation to study the motion of guiding centers in the Lagrangian framework. We separate the position coordinate into the guiding center position, $\bm{R}$, and gyroradius, $\bm{\rho}$,
\begin{gather}
    \bm{r} = \bm{R} + \bm{\rho}.
    \label{eq:coordinate_transformation}
\end{gather}
The gyroradius lies in the plane perpendicular to $\bm B$ and is parameterized by $(\rho,\varphi)\in\mathbb R^2$ as follows
\begin{gather}
    \bm{\rho} = \rho \left(-\hat{\bm{e}}_2(\bm{R}) \cos(\varphi) + \hat{\bm{e}}_3(\bm{R}) \sin(\varphi) \right),
\end{gather}
where $\varphi$ is called the gyroangle and describes the angle of the gyroradius in the plane perpendicular to the magnetic field, while $\rho$ describes the magnitude of the gyroradius in this plane. We have replaced one vector describing the position $\bm{r}\in\mathbb R^3$, for a vector coordinate $\bm{R} \in\mathbb R^3$
and two scalar coordinates $(\rho,\varphi)\in\mathbb R^2$ constrained by \eqref{eq:coordinate_transformation}. 
We perform the coordinate transformation
$(\bm{r}) \rightarrow (\bm{R},\rho,\varphi)$,
\begin{gather}
    L(\bm{r},\dot{\bm{r}}) = \widetilde{L}(\bm{R},\rho,\varphi,\dot{\bm{R}},\dot{\rho},\dot{\varphi}).
\end{gather}
The Lagrangian in this new set of coordinates can now be expressed as
\begin{multline}
        \widetilde{L}(\bm{R},\rho,\varphi,\dot{\bm{R}},\dot{\rho},\dot{\varphi}) = \frac{m}{2} \left( \dot{\bm{R}} + \dot{\bm{\rho}}(\bm{R},\rho,\varphi,\dot{\bm{R}},\dot{\rho},\dot{\varphi}) \right)^2  \\ + q \bm{A}(\bm{R} + \bm{\rho}(\bm{R},\rho,\varphi)) \cdot \left(\dot{\bm{R}} + \dot{\bm{\rho}}(\bm{R},\rho,\varphi,\dot{\bm{R}},\dot{\rho},\dot{\varphi}) \right) - q \Phi(\bm{R} + \bm{\rho}(\bm{R},\rho,\varphi)),
    \label{eq:guiding_center_lagrangian}
\end{multline}
where we have dropped any time dependence of the fields. 

We will be able to simplify this Lagrangian based on the following assumptions:
\begin{itemize}
    \item The gyroradius is small compared with typical length scales of our system. This implies that $\rho/L_B\ll 1$, where $L_B^{-1} = |\nabla B|/B$ is the typical length scale for the magnetic field variation. This means that  This is justified considering Table \ref{LTscales}. 
    \item The gyrofrequency is much larger than other frequencies of our system. This implies that $\omega_B/\dot{\varphi}\ll 1$, where $\omega_B \sim v_t/L_B$ is the frequency associated with the variation of the magnetic field strength and $v_t = \sqrt{2T/m}$ is the thermal velocity associated with the temperature $T$. The thermal speed characterizes the typical speed of a particle in any direction. This is also justified considering physical scales given in Table \ref{LTscales}.
    \item All of the drifts across the field lines are small compared with the velocity along the field lines, which is of similar order to the thermal speed, $v_{||} \sim v_t$. This implies that $q \Phi/T \ll 1$.
\end{itemize}
Based on these assumptions, we will perform an asymptotic expansion in the small parameter,
\begin{gather}
    \epsilon \sim \frac{\rho}{L_B} \sim \frac{\omega_B}{\dot{\varphi}} \sim \frac{q\Phi}{T} \ll 1.
    \label{eq:gyro_ordering}
\end{gather}

Based on the solution in a straight, uniform magnetic field, we assume the following:
\begin{itemize}
    \item  The parallel guiding center velocity $ V_{||} =\hat{\bm{b}}\cdot \dot{\bm{R}}$ satisfies $V_{||} \sim v_t$ due to \eqref{eq:guiding_center_cylindrical}. We found in the straight field line case that the guiding center velocity is constant along field lines, $\hat{\bm{b}} \cdot \bm{V} = v_{||}$, and we can assume $v_{||} \sim v_t$. As the guiding center parallel velocity is the same as the particle parallel velocity, our assumption is sensible.
    \item The perpendicular guiding center velocity $\bm{V}_{\perp} = \dot{\bm{R}} - V_{||}\hat{\bm{b}}$ satisfies $|\bm{V}_{\perp}| \sim \epsilon v_t$, as we assume that the guiding center drifts are slow with respect to  the velocity of the gyromotion or the motion along field lines. In a strong magnetic field, we expect the largest contribution to the guiding center motion to be in the parallel direction.
    \item The gyrofrequency $\dot{\varphi} \sim qB/m$ which arises from \eqref{eq:periodic_velocity}, as we found $\dot{\varphi} = \Omega = qB/m$ in the case of cylindrical confinement. 
    \item The time variation of the gyroradius scales as, $\dot{\rho} \sim \omega_B \rho$, which arises from \eqref{eq:gyroradius}, as we found $\rho = v_{\perp}/\Omega$ in the case of cylindrical confinement. Therefore, the time-variation of $\rho$ arises due to the variation in $B$, which has characteristic frequency $\omega_B$.
\end{itemize}

 We wish to average the Lagrangian over the fast gyromotion under the assumption that $\omega_B/\dot{\varphi}\ll1$, by performing the gyroaveraging operation, defined as,
\begin{gather}
    \langle F \rangle_{\varphi} = \frac{1}{2\pi} \int_0^{2\pi} F(\bm{R},\rho,\varphi,\dot{\bm{R}},\dot{\rho},\dot{\varphi}) \, d \varphi,
\end{gather}
for a function $F$. Performing this operation will allow us to effectively remove frequencies $\omega \sim \dot{\varphi}$ from our system in accordance with the assumptions made in \eqref{eq:gyro_ordering}, enabling us to more effectively study phenomena that occur on a slower time scale.

As the gyroaverage is performed, the other coordinates are considered fixed $(\bm{R},\rho,\dot{\bm{R}},\dot{\rho},\dot{\varphi})$. At this point we want to identify the leading terms in \eqref{eq:guiding_center_lagrangian} with respect to $\epsilon$ after performing the gyroaverage. Note that any term with a single factor of $\bm{\rho}$ does not contribute after gyroaveraging, as it is periodic in $\varphi$.

Gyroaveraging the first term in \eqref{eq:guiding_center_lagrangian} we obtain,
\begin{multline}
    \left\langle \frac{m}{2} \left( \dot{\bm{R}} + \dot{\bm{\rho}} \right)^2 \right\rangle_{\varphi} = 
    \frac{m}{2} \left(V_{||}^2 + V_{\perp}^2 +  \dot{\rho}^2 + \rho^2 \dot{\varphi}^2 + \frac{\rho^2}{2} \left(|\dot{\hat{\bm{e}}}_2|^2 + |\dot{\hat{\bm{e}}}_3|^2 \right) + 2\rho^2 \dot{\varphi}\dot{\hat{\bm{e}}}_3 \cdot \hat{\bm{e}}_2\right),
\end{multline}
where $V_{||} = \hat{\bm{b}}(\bm{R}) \cdot \dot{\bm{R}}$ and $V_{\perp} = |\dot{\bm{R}}-V_{||}\hat{\bm{b}}(\bm{R})|$ and the unit vectors, $\hat{\bm{e}}_2$ and $\hat{\bm{e}}_3$, are evaluated at the guiding center position.
The time derivatives of the unit vectors can be evaluated using $\dot{\hat{\bm{e}}}_{2,3} = \dot{\bm{R}} \cdot \nabla \hat{\bm{e}}_{2,3}$. Throughout we will use the notation $\nabla = \partial/\partial \bm{R}$. The quantity $\langle \dot{\bm{\rho}} \cdot \dot{\bm{\rho}} \rangle_{\varphi}$ is evaluated thanks to the identity
\begin{gather}    \dot{\bm{\rho}}  = \left(\dot{\rho}/\rho\right) \bm{\rho}  - \dot{\varphi}  \left(\hat{\bm{b}} \times \bm{\rho}\right) + \left(\dot{\bm{R}} \cdot \nabla \right) \bm{\rho}.
    \label{eq:dot_rho}
\end{gather}
From our set of assumptions, we can summarize the ordering in $\epsilon$ of terms in the following Table.
{\renewcommand{\arraystretch}{1.5}
$$
\begin{array}{|c||c|c|c|c|c|c|}
\hline
    \text{Terms} & \frac{m}{2}V_{||}^2 &
    \frac{m}{2} V_{\perp}^2 &
    \frac{m}{2}\dot{\rho}^2 &
    \frac{m}{2}\rho^2 \dot{\varphi}^2 
    &
    \frac{m}{2}\frac{\rho^2}{2} \left(|\dot{\hat{\bm{e}}}_2|^2 + |\dot{\hat{\bm{e}}}_3|^2 \right)
    &
    m \rho^2 \dot{\varphi} \dot{\hat{\bm{e}}}_2 \cdot \hat{\bm{e}}_3
    \\\hline 
    \text{Order} & 
    \sim m v_t^2 & \sim\epsilon^2 m v_t^2 & \sim\epsilon^2 m  v_t^2 &\sim m v_t^2 & \sim \epsilon^2 m v_t^2 & \sim \epsilon m v_t^2
    \\\hline
\end{array}  
$$
}

So we obtain
\begin{gather}\label{eq:firstT}
    \left \langle \frac{m}{2} \left( \dot{\bm{R}} + \dot{\bm{\rho}} \right)^2 \right \rangle_{\varphi} = \frac{m}{2} \left( V_{||}^2 + \rho^2 \dot{\varphi}^2 \right) + \mathcal{O}(\epsilon m v_t^2).
\end{gather}

To compute the gyroaverage of the second term in \eqref{eq:guiding_center_lagrangian}, we Taylor expand $\bm{A}(\bm{r})$ about $\bm{R}$,
\begin{gather}
    \bm{A}(\bm{R} + \bm{\rho}) = \bm{A}(\bm{R}) + (\bm{\rho} \cdot \nabla) \bm{A}(\bm{R}) + \mathcal{O}(\epsilon^2 \bm{A}(\bm{R})),
\end{gather}
so by gyroaveraging we obtain
\begin{align}
    \left \langle q \bm{A}(\bm{R} + \bm{\rho}) \cdot \left( \dot{\bm{R}} + \dot{\bm{\rho}} \right) \right \rangle_{\varphi} &= \left \langle q \left(\bm{A}(\bm{R}) + (\bm{\rho} \cdot \nabla) \bm{A}(\bm{R}) +\mathcal{O}(\epsilon^2 \bm{A}(\bm{R}))  \right) \cdot \left( \dot{\bm{R}} + \dot{\bm{\rho}} \right) \right \rangle_{\varphi} \\    &
    = q \bm{A}(\bm{R}) \cdot  \bm{V}_{||}+ q \bm{A}(\bm{R}) \cdot \bm{V}_{\perp} + q\frac{\dot{\rho}}{\rho}\langle \bm{\rho} \bm{\rho} \rangle_{\varphi} : \nabla \bm{A}(\bm{R}) \nonumber
    \\ & \hspace{0.25cm} -     q\dot{\varphi}  \left\langle \left( \hat{\bm{b}} \times \bm{\rho} \right)\bm{\rho} \right \rangle_{\varphi} : \nabla \bm{A}(\bm{R}) + q \langle (\dot{\bm{R}} \cdot \nabla) \bm{\rho} \bm{\rho}  \rangle_{\varphi} : \nabla \bm{A}(\bm{R})
    + \mathcal{O}(\epsilon^2 m v_t^2). \nonumber
\end{align}
Here \eqref{eq:dot_rho} is used to express $\dot{\bm{\rho}}$. The double-dot ($:$) indicates contraction between two tensors, $\overleftrightarrow{\bm{A}}$ and $\overleftrightarrow{\bm{B}}$, in the following way, $\overleftrightarrow{\bm{A}}:\overleftrightarrow{\bm{B}} = \sum_i \sum_j A_{ij}B_{ji}$.
Again, from our set of assumptions, we can summarize the ordering in $\epsilon$ of terms in the following Table.
{\renewcommand{\arraystretch}{1.5}
$$
\begin{array}{|c||c|c|c|c|c|}
\hline
    \text{Terms} & q \bm{A} \cdot \bm{V}_{||} &
     q \bm{A} \cdot  \bm{V}_{\perp}&
     \frac{q\dot{\rho}}{\rho} \langle \bm{\rho} \bm{\rho} \rangle_{\varphi} :\nabla \bm{A}& q
    \dot{\varphi} \langle (\hat{\bm{b}}\times \bm{\rho}) \bm{\rho} \rangle_{\varphi} : \nabla \bm{A} & q \langle (\dot{\bm{R}} \cdot \nabla) \bm{\rho} \bm{\rho}  \rangle_{\varphi} : \nabla \bm{A}
    \\\hline 
    \text{Order} & \sim \epsilon^{-1} m  v_t^2 & \sim m v_t^2 & \sim \epsilon m v_t^2 &\sim  m v_t^2 & \sim \epsilon m v_t^2 
    \\ \hline
\end{array}  
$$
}

The fourth term can be evaluated using $\langle (\hat{\bm{b}} \times \bm{\rho}) \bm{\rho}  \rangle_{\varphi} = (\rho^2/2) \left( \hat{\bm{e}}_3 \hat{\bm{e}}_2 - \hat{\bm{e}}_2 \hat{\bm{e}}_3\right) = (\rho^2/2) \hat{\bm{b}} \times \overleftrightarrow{\bm{I}}$, where $\overleftrightarrow{\bm{I}} = \hat{\bm{e}}_1 \hat{\bm{e}}_1 + \hat{\bm{e}}_2 \hat{\bm{e}}_2 + \hat{\bm{e}}_3 \hat{\bm{e}}_3$ is the identity tensor, and $\hat{\bm{b}} \times \overleftrightarrow{\bm{I}} : \nabla \bm{A} = \hat{\bm{b}} \cdot \nabla \times \bm{A} = B$. So we obtain
\begin{align}\label{eq:secondT}
    \left \langle q \bm{A}(\bm{R} + \bm{\rho}) \cdot \left( \dot{\bm{R}} + \dot{\bm{\rho}} \right) \right \rangle_{\varphi} &= q \bm{A}(\bm{R}) \cdot \dot{\bm{R}} - q\frac{\dot{\varphi}\rho^2}{2} B(\bm{R}) + \mathcal{O}(\epsilon m v_t^2).
\end{align}
We similarly expand $\Phi(\bm{r})$ about $\bm{R}$ to evaluate the third term in \eqref{eq:guiding_center_lagrangian},
\begin{gather}\label{eq:thirdT}
    \langle \Phi(\bm{R} + \bm{\rho}) \rangle_{\varphi}  = \Phi(\bm{R}) + \mathcal{O}(\epsilon^2 \Phi(\bm{R})).
\end{gather}
Inserting \eqref{eq:firstT}, \eqref{eq:secondT}, and \eqref{eq:thirdT} into \eqref{eq:guiding_center_lagrangian}, we thus obtain the following gyroaveraged Lagrangian to $\mathcal{O}(\epsilon m v_t^2)$,
\begin{multline}
    \mathcal{L}(\bm{R},\rho,\dot{\bm{R}},\dot{\varphi}):=\langle L(\bm{R},\bm{\rho},\varphi,\dot{\bm{R}},\dot{\rho},\dot{\varphi}) \rangle_{\varphi} = \frac{m}{2} \left(\left(\dot{\bm{R}} \cdot \hat{\bm{b}}(\bm{R})\right)^2 + \rho^2 \dot{\varphi}^2 \right) + q \bm{A}(\bm{R}) \cdot \dot{\bm{R}} \\ - q\frac{\dot{\varphi} \rho^2}{2} B(\bm{R}) - q \Phi(\bm{R}).
    \label{eq:gyroaveraged_L}
\end{multline}
By construction, our gyroaveraged Lagrangian, $\mathcal{L}$, no longer depends on $\varphi$.

Further discussion of the gyroaveraged Lagrangian can be found in \cite{Littlejohn1983}, Chapter 6 of \cite{Helander2005}, and \cite{Jorge2017}.

\subsubsection{Adiabatic invariant for gyromotion}
\label{sec:adiabatic_invariant}

An adiabatic invariant is an approximately conserved quantity associated with nearly periodic motion. We will obtain a conserved quantity consistent with the assumption of fast gyromotion, $\epsilon \ll 1$, by considering the Euler-Lagrange equations for $\mathcal{L}$.

We evaluate the Euler-Lagrange equation from \eqref{eq:gyroaveraged_L} corresponding to the gyroradius, noting that $\partial \mathcal{L}/\partial \dot{\rho} = 0$,
\begin{gather}
    \partder{ \mathcal{L}}{\rho} = \der{}{t} \left( \partder{ \mathcal{L}}{\dot{\rho}} \right)
\Rightarrow
    \dot{\varphi}= \frac{q B(\bm{R})}{m}.
\end{gather}
We see that $\dot{\varphi}$ corresponds to the gyrofrequency $\Omega $ found in Section \ref{sec:cylindrical} but for a space dependent magnetic field $B(\bm{R})$, so we define $\Omega (\bm{R}) = q B(\bm{R})/m$. 

Now we evaluate the Euler-Lagrange equation for the gyroangle, noting that $\partial \mathcal{L}/\partial \varphi = 0$,
\begin{gather}
    \partder{\mathcal{L}}{\varphi} = \der{}{t} \left( \partder{\mathcal{L}}{\dot{\varphi}} \right)
\Rightarrow    
    \der{}{t} \left( m \rho^2 \dot{\varphi} -  \frac{\rho^2 q B(\bm{R})}{2} \right) = 0.
\end{gather}
This implies the conservation of the quantity $2m\rho^2 \dot{\varphi} -  \rho^2 q B(\bm{R}) = m\rho^2 \Omega (\bm{R})$ along trajectories. Any multiple of this quantity is conserved along trajectories, but the one that is most often used in the literature is defined in terms of the perpendicular velocity $v_\perp(\rho,\bm{R}):=\rho\Omega(\bm{R})$ as follows,
\begin{gather}
    \mu = \frac{v_{\perp}^2(\rho,\bm{R})}{2B(\bm{R})},
    \label{eq:mu}
\end{gather}
and is often referred to as the magnetic moment. Here $v_{\perp}$ is the velocity associated with the gyromotion, as opposed to $V_{\perp}$ which is associated with the guiding center motion.

\subsubsection{Energy conservation}
\label{sec:energy_conservation}

As we have made the assumption of time-independence of the fields ($\partial \Phi/\partial t =0$,  $\partial \bm{A}/\partial t = 0$), the Lagrangian does not depend explicitly on time. This results in a conserved quantity. Consider the total time derivative of $\mathcal{L}$, 
\begin{gather}
    \der{\mathcal{L}}{t} =
    \dot{\varphi} \partder{\mathcal{L}}{\varphi}  + 
    \dot{\rho} \partder{\mathcal{L}}{\rho}  + \dot{\bm{R}} \cdot \partder{\mathcal{L}}{\bm{R}}  +\ddot{\varphi} \partder{\mathcal{L}}{\dot{\varphi}}  + \ddot{\rho} \partder{\mathcal{L}}{\dot{\rho}}  +\ddot{\bm{R}} \cdot \partder{\mathcal{L}}{\dot{\bm{R}}}.
\end{gather}
Applying the Euler-Lagrange equations, we obtain
\begin{gather}
    \der{\mathcal{L}}{t} =\der{}{t}\left(\dot{\varphi}\partder{\mathcal{L}}{\dot{\varphi}} +\dot{\rho} \partder{\mathcal{L}}{\dot{\rho}}  +  \dot{\bm{R}} \cdot \partder{\mathcal{L}}{\dot{\bm{R}}} \right).
\end{gather}
If the fields are assumed to be time-independent, then $\mathcal{L}$ has no explicit time dependence, and therefore the total time derivative of the following quantity vanishes,
\begin{gather}
    E = \dot{\varphi} \partder{\mathcal{L}}{\dot{\varphi}}  + \dot{\rho}\partder{\mathcal{L}}{\dot{\rho}}  + \dot{\bm{R}} \cdot \partder{\mathcal{L}}{\dot{\bm{R}}} - \mathcal{L}.
\end{gather}
Thus $E$ is conserved along a trajectory. From the gyroaveraged Lagrangian  $\mathcal{L}$ \eqref{eq:gyroaveraged_L}, we then obtain
\begin{gather}
    E = \frac{m \left(v_{\perp}^2(\rho,\bm{R})+ V_{||}^2(\dot{\bm{R}},{\bm{R}})\right)}{2} + q \Phi(\bm{R}),
    \label{eq:lagrangian_energy}
\end{gather}
where the parallel guiding center velocity is defined as $V_{||} (\dot{\bm{R}},{\bm{R}}) := \dot{\bm{R}} \cdot \hat{\bm{b}}({\bm{R}})$ and the perpendicular velocity is $v_{\perp}(\rho,\bm{R}) = \rho \Omega(\bm{R})$. The conserved quantity $E$ represents the total energy of a particle: the first term in \eqref{eq:lagrangian_energy} accounts for the kinetic energy, the energy due to the motion of the particle, while the second accounts for the potential energy, the energy due to the fields.

\subsubsection{Particle trapping}
\label{sec:particle_trapping}

Combining the expression of the adiabatic invariant \eqref{eq:mu} with the energy invariant \eqref{eq:lagrangian_energy}, we obtain the following expression for the parallel velocity in terms of the conserved quantities
\begin{gather}
    \left(V_{||}(\dot{\bm{R}},{\bm{R}})\right)^2 = \frac{2(E - q\Phi(\bm{R}))}{m} - 2\mu B(\bm{R}).
\end{gather} 
Often the electrostatic potential can be assumed to be constant in a given region. Therefore, only the second term in the above depends on space, and $V_{||}$ vanishes at points where $B(\bm{R}) = (E-q\Phi))/(\mu m) =: B_{\text{crit}}$. In particular, along a given trajectory, the values of the invariants $(E,\mu)$ are given so $B_{\text{crit}}$ is fixed, and necessarily  $B(\bm{R}) \le B_{\text{crit}}$. Therefore the particle cannot access regions where $B(\bm{R}) > B_{\text{crit}}$.

The conservation of $E$ and $\mu$ leads to the result that particles cannot access regions of sufficiently large magnetic field. This effect is known as mirroring, as a particle will be reflected away from high field regions. For a given magnetic field, particles with sufficiently large values of $\mu/E$ may become trapped in regions of low field strength, often referred to as trapped particles. Particles with sufficiently small values $\mu/E$ will not become trapped and are referred to as passing particles. 

Particle trapping is an important concept for confinement. One of the first magnetic confinement devices, known as the mirror machine, relies on a strong magnetic field to confine a large fraction of particles. Trapped and passing particles tend to have very different confinement properties in magnetic confinement devices due to their distinct trajectories. In particular, a major challenge of designing a stellarator is obtaining confinement of trapped particles. See Section 8.9 in \cite{Freidberg2008} for additional details.

\subsection{Guiding center motion}
\label{gradBdrift}
In a uniform and straight magnetic field, guiding centers exhibit a constant velocity along field lines (see Section \ref{sec:cylindrical}). In the presence of a non-uniform magnetic field and curvature in the field, guiding centers have an additional slow drift across field lines. 

We will obtain these drifts by considering the Euler-Lagrange equation for the guiding center position from the gyroaveraged Lagrangian expression \eqref{eq:gyroaveraged_L},
\begin{multline}
\label{eq:EulLagGyroAv}
    \der{}{t} \left( m \left( \dot{\bm{R}} \cdot \hat{\bm{b}}(\bm{R}) \right) \hat{\bm{b}}(\bm{R}) + q \bm{A}(\bm{R}) \right) = m \nabla \left(\dot{\bm{R}} \cdot \hat{\bm{b}}(\bm{R}) \right) \left( \dot{\bm{R}} \cdot \hat{\bm{b}}\right)
    \\
    + q \nabla \left( \bm{A}(\bm{R})  \cdot \dot{\bm{R}} \right) -  \frac{q\dot{\varphi} \rho^2}{2} \nabla B(\bm{R}) - q \nabla \Phi(\bm{R}).
\end{multline}
The time derivative on the left hand side can be written as $d/dt = \dot{\bm{R}} \cdot \nabla + \ddot{\bm{R}} \cdot \partial/\dot{\bm{R}}$, as $\rho$ and $\dot{\rho}$ do not appear and the Lagrangian does not have explicit time dependence.

Using the vector identity $\nabla (\bm{a} \cdot \bm{b}) = \bm{a} \times (\nabla \times \bm{b}) + \bm{b} \times (\nabla \times \bm{a}) + (\bm{a} \cdot \nabla) \bm{b} + (\bm{b} \cdot \nabla)\bm{a}$ as well as the definitions of the fields in terms of the vector and scalar potentials \eqref{eq:potentials_lagrangian_subequations} we obtain
\begin{gather}
    m\dot{V}_{||}\hat{\bm{b}} +  m V_{||} \left( \dot{\bm{R}} \cdot \nabla \hat{\bm{b}}(\bm{R}) \right) = q \dot{\bm{R}} \times \bm{B}(\bm{R}) - \frac{q\dot{\varphi} \rho^2}{2} \nabla B(\bm{R}) + q \bm{E}(\bm{R}),
    \label{eq:gc_euler_lagrange}
\end{gather}
where the parallel guiding center velocity and acceleration are $V_{||}(\dot{\bm{R}},\bm{R}) = \dot{\bm{R}} \cdot \hat{\bm{b}}(\bm{R})$ and $\dot{V}_{||}(\ddot{\bm{R}},\bm{R}) = \ddot{\bm{R}} \cdot \hat{\bm{b}}(\bm{R})$, respectively. 

We now can check that we have not introduced any terms of higher order in $\epsilon$ while computing the Euler-Lagrange equations. We can see that the term on the left hand side of \eqref{eq:gc_euler_lagrange} involving the perpendicular guiding center velocity, $\bm{V}_{\perp} = \dot{\bm{R}}-V_{||}\hat{\bm{b}}$, are  $\mathcal{O}(\epsilon \omega_B m v_t)$
while the other terms are $\mathcal{O}(\omega_B m v_t)$ or larger; thus we need only retain the component of these terms involving $V_{||}$.
The resulting expression can also be written in terms of the magnetic curvature, $\bm{\kappa} = \hat{\bm{b}} \cdot \nabla \hat{\bm{b}}$, and the magnetic moment \eqref{eq:mu},
\begin{gather}
    m\dot{V}_{||}\hat{\bm{b}}(\bm{R})= - m V_{||}^2 \bm{\kappa}(\bm{R}) + q \dot{\bm{R}}_{\perp} \times \bm{B}(\bm{R}) - m\mu \nabla B(\bm{R}) + q \bm{E}(\bm{R}).
    \label{eq:guiding_center_EOM}
\end{gather}

\subsubsection{Parallel guiding center motion}

We can take the dot product of \eqref{eq:guiding_center_EOM} with $\hat{\bm{b}}$ to obtain the parallel guiding center acceleration,
\begin{gather}
\dot{V}_{||} = - \mu \hat{\bm{b}}(\bm{R}) \cdot \nabla B(\bm{R}) + \frac{q}{m} \bm{E}(\bm{R}) \cdot \hat{\bm{b}}(\bm{R}).
\end{gather}
\begin{itemize}
\item The first term in the above expression expresses the fact that particles are repelled from regions of large field strength, as discussed in Section \ref{sec:particle_trapping}. 
\item The second term accounts for acceleration by electric fields parallel to the magnetic field.
\end{itemize}

\subsubsection{Perpendicular guiding center motion}

\begin{figure}
\begin{center}
\includegraphics[trim=18cm 7cm 31cm 5cm,clip,width=0.7\textwidth]{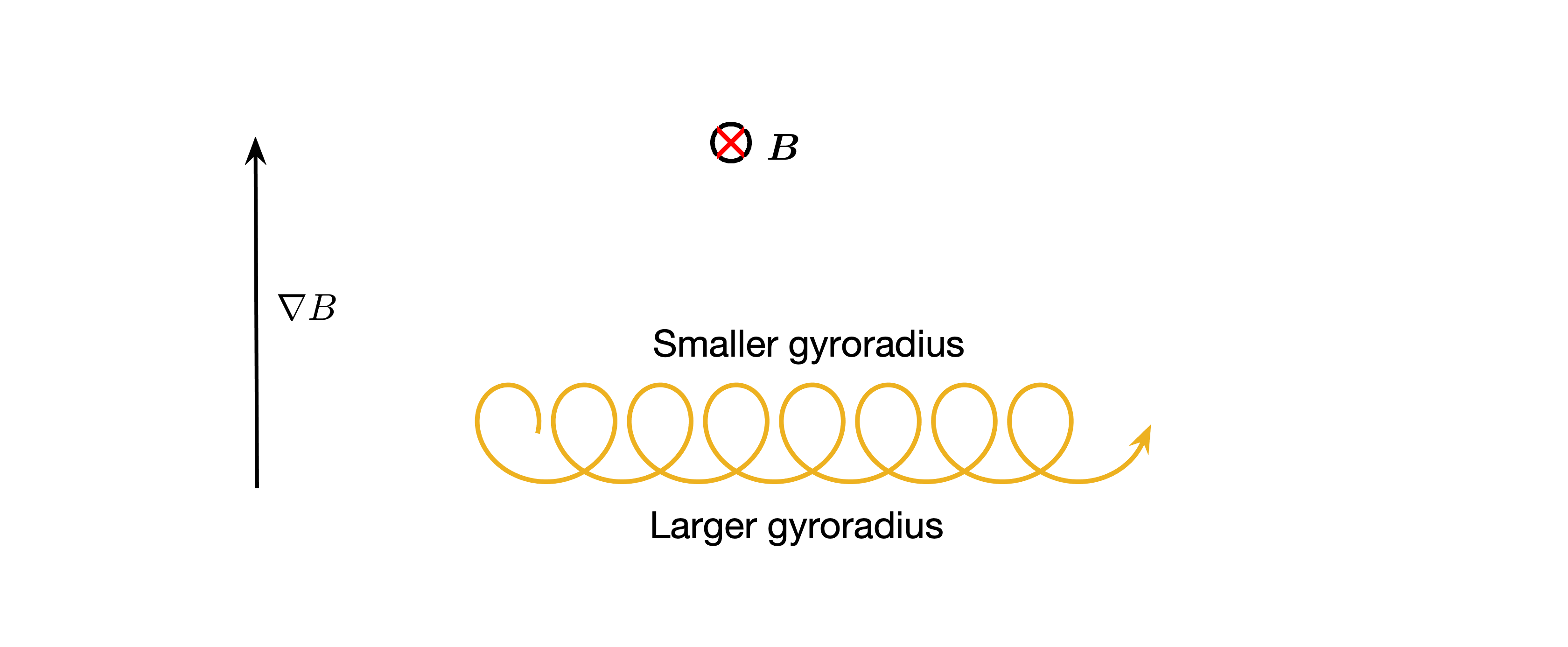}
\caption{We illustrate a particle orbit in a magnetic field pointing into the page with a gradient in the field strength pointing up. The trajectory of an ion is shown, with all motion projected into the plane perpendicular to $\bm{B}$. The ion will gyrate counter-clockwise, with a smaller gyroradius in the region of stronger field. This results in what is called a grad-$B$ drift to the right.}
\label{gradB_diagram}
\end{center}
\end{figure}

We now take the cross product of \eqref{eq:guiding_center_EOM} with $\hat{\bm{b}}$ to obtain the following expression for the perpendicular guiding center acceleration,
\begin{gather}
    \dot{\bm{R}}_{\perp} = \frac{V_{||}^2(\dot{\bm{R}},{\bm{R}})}{\Omega(\bm{R})} \hat{\bm{b}}(\bm{R}) \times \bm{\kappa}(\bm{R}) + \frac{\mu}{\Omega(\bm{R})} \hat{\bm{b}}(\bm{R}) \times \nabla B(\bm{R}) + \frac{\bm{E}(\bm{R}) \times \bm{B}(\bm{R})}{B(\bm{R})^2}.
    \label{eq:drifts}
\end{gather}
The right hand side has three separate terms:
\begin{itemize}
\item The first term is known as the curvature drift, denoted by $\bm{v}_{\text{curv}}$, resulting from curvature in the field lines. It depends on the mass, field strength, and charge through $\Omega$. The sign of this drift is different for ions and electrons as the species gyrate in opposite directions.
\item The second term is the grad-$B$ drift, denoted by $\bm{v}_{\nabla B}$. This drift also depends on the mass, field strength, and charge through the gyrofrequency. A physical picture for the grad-$B$ drift can be found in Figure \ref{gradB_diagram}.
\item The third term is the $\bm{E} \times \bm{B}$ drift, denoted by $\bm{v}_E$. This drift does not depend on the charge or mass, so is the same for all species.
\end{itemize}

Further discussion of guiding center drifts can be found in Chapter 2 of \cite{Nicholson1983}, Section 2.4 of \cite{Hazeltine2003}, Chapter 8 of \cite{Freidberg2008}, Chapter 6 of \cite{Helander2005}.

\FloatBarrier

\subsection{Introduction to toroidal confinement} 
\label{confiningMagneitcField}

In Section \ref{sec:cylindrical} we found that in a straight, uniform magnetic field, particles are confined in the direction perpendicular to the magnetic field lines but unconfined in the parallel direction. In order to avoid losses of particles along straight field lines, one can consider a modification of the field lines. A natural idea is to bend a set of straight field lines into closed field lines; this results in a toroidal shape. In this way the magnetic field points in the toroidal direction, the long way around the torus. 

\begin{figure}
\begin{center}
\includegraphics[width=.9\textwidth,trim=1cm 1cm 1cm 2cm,clip]{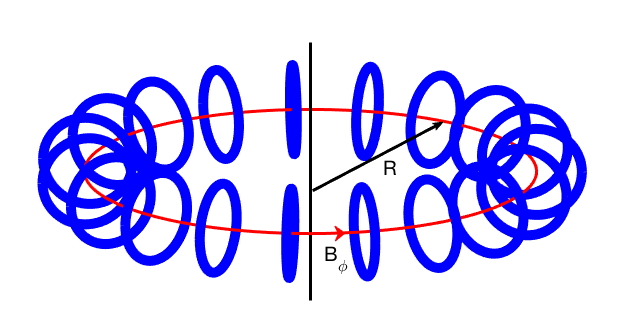}
\caption{In a toroidal vacuum magnetic field, the magnitude of the toroidal field varies as $B_{\phi} \propto 1/R$ where $R$ is the major radius, as can be seen by performing a line integral along the toroidal loop (red) which encloses coil current $I$, including contributions from the electromagnetic coils (blue).}
\label{fig:ampere_loop}
\end{center}
\end{figure}

One could imagine generating a set of toroidally closed field lines by bending a long solenoid to join its two open ends, forming a circle. Nearly toroidal field lines can be generated thanks to several individual coils placed along a common circular axis (see Figure \ref{fig:ampere_loop}). This produces a magnetic field whose magnitude is nearly axisymmetric (does not depend on the toroidal angle about the axis of symmetry) and points in the toroidal direction. We assume that such an axisymmetric magnetic field which points in the toroidal direction can be produced.

To analyze such a configuration we will use the canonical cylindrical coordinates $(R,\phi,Z)$, see Section \ref{sec:toroidal_geom} for a reminder. We will find that the magnitude of the toroidal field generated by such coils is a non-constant function of the position; the field is stronger inside (closer to the $\hat{\bm{z}}$ axis of symmetry) of the toroidal shape, where the coils are closer to each other, and decreases as a function of the major radius $R$. This can be seen by computing the current passing through a surface, $S$, lying in the $Z = 0$ plane whose boundary is a circle with major radius $R$. From Ampere's law \eqref{eq:ampere_magnetostatic} we find
\begin{gather}
I = \int_{S} \bm{J} \cdot \hat{\bm{n}} \, d^2 x = \frac{1}{\mu_0} \oint_{\partial S} \bm{B}\cdot d \bm{l} = \frac{2\pi R B_{\phi}}{\mu_0}. 
\end{gather}
Here we have used $\phi$ to parameterize the line integral such that $\bm{B} \cdot d \bm{l} = R B_{\phi} d \phi$. Due to the assumption of axisymmetry, $B_{\phi}$ is constant along the line integral. If moreover the loop is taken to go through the electromagnetic coils that link the plasma poloidally, the total current enclosed by the loop is the sum of the currents in each coil. We furthermore assume that there are no other sources of current such that $I$ does not depend on the radius $R$ of the circle for any curve that encloses the coils (see Figure \ref{fig:ampere_loop}). Thus the toroidal field strength varies as $B_{\phi} \propto 1/R$ due to the toroidal geometry.  

As discussed in Section \ref{sec:cylindrical}, in a straight, uniform field, particles exhibit gyromotion about field lines. If the magnetic field is non-uniform or curved or if an electric field is introduced, a particle will drift off of a field line on average, as discussed in Section \ref{gradBdrift}. Because the toroidal field varies as $1/R$, it is impossible to have good confinement with only a toroidal field.

Consider the grad-B drift of a particle, the second term in \eqref{eq:drifts},
\begin{gather}
\bm{v}_{\nabla B} = \frac{v_{\perp}^2}{2\Omega} \frac{\bm{B} \times \nabla B}{B^2}.
\label{eq:gradB_drift}
\end{gather}
The gyrofrequency is $\Omega = qB/m$ and $v_{\perp}$ is the magnitude of the velocity perpendicular to the magnetic field. The quantity $\bm{v}_{\nabla B}$ is the velocity at which guiding centers drift off of field lines in the presence of a gradient in the field strength. If the field is purely toroidal, a particle will drift in the $-q\hat{\bm{\phi}} \times \hat{\bm{R}} = q\hat{\bm{Z}}$ direction, either up or down depending on the sign of $q$. 

As ions and electrons move in opposite directions, an electric field will be set up in the $-\hat{\bm{Z}}$ direction to try to restore charge neutrality. This results in an additional $\bm{E} \times \bm{B}$ drift,
\begin{gather}
\bm{v}_E = \frac{\bm{E} \times \bm{B}}{B^2}.
\end{gather}
We see that this drift is in the $-\hat{\bm{Z}} \times \hat{\bm{\phi}} = \hat{\bm{R}}$ direction. As the direction of the $\bm{E}\times \bm{B}$ drift is the same for both electrons and ions, both species will drift radially out of the device. For this reason a purely toroidal field cannot provide sufficient confinement.

\begin{figure}
    \centering
    \begin{subfigure}{0.49\textwidth}
    \includegraphics[trim=42cm 13cm 39cm 12cm,clip,width=1.0\textwidth]{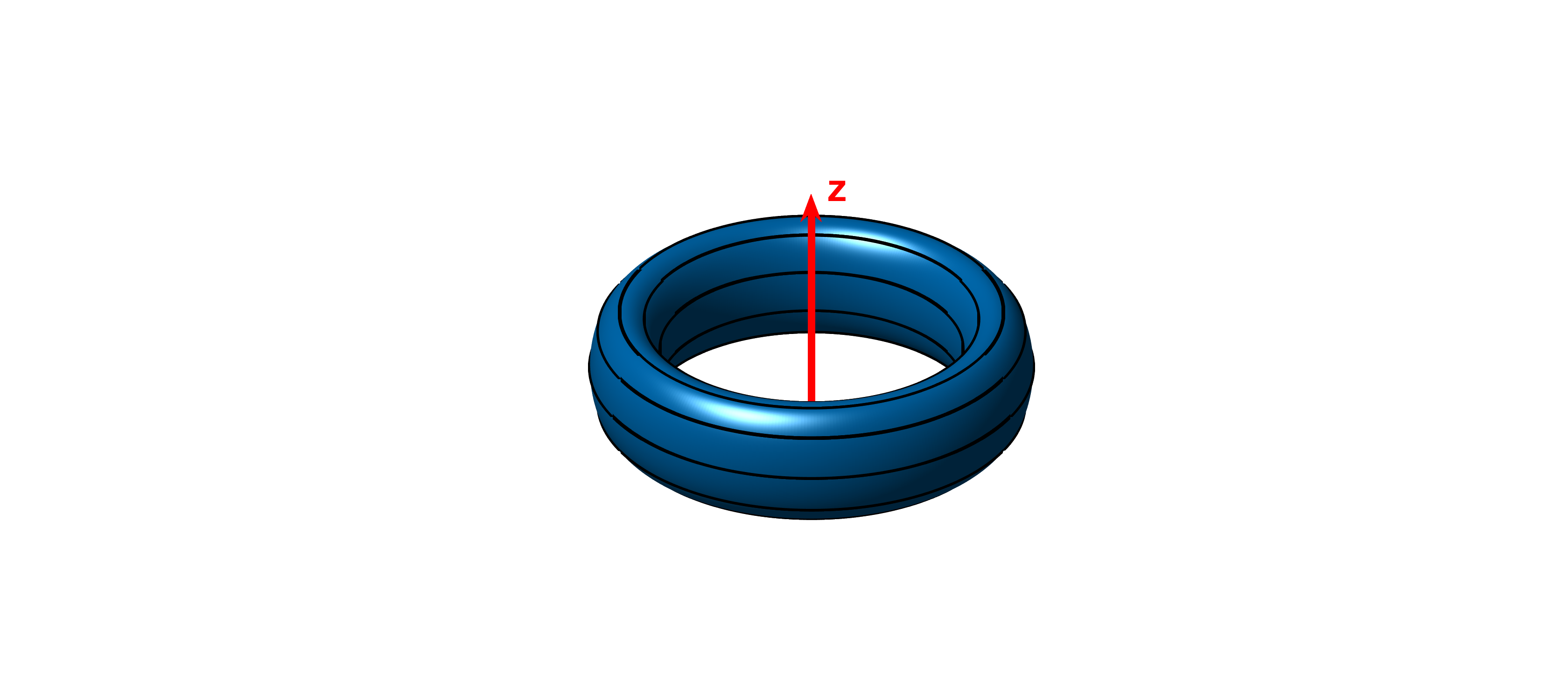}
    \caption{}
    \label{fig:surf_field_lines_toroidal}
    \end{subfigure}
    \begin{subfigure}{0.49\textwidth}
    \includegraphics[trim=42cm 13cm 39cm 12cm,clip,width=1.0\textwidth]{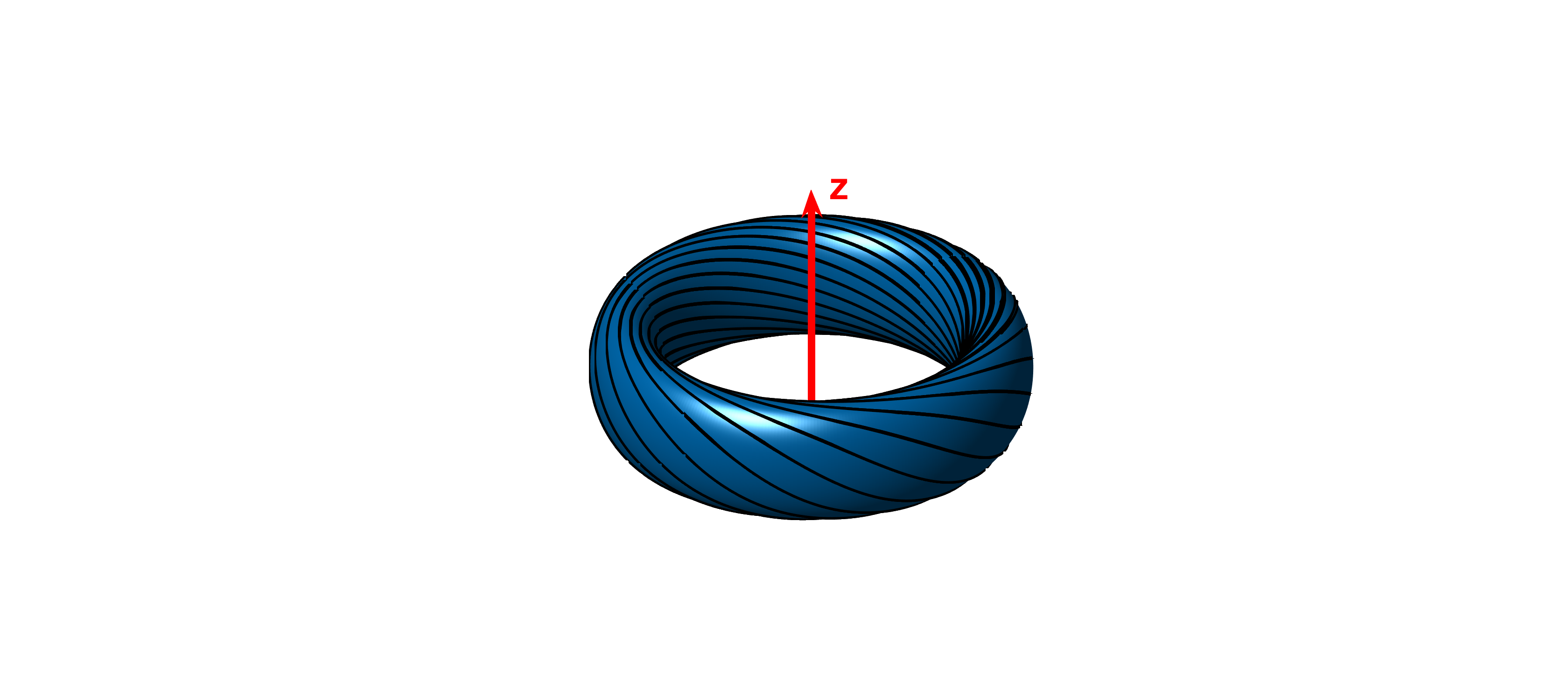}
    \caption{}
    \label{fig:surf_field_lines_poloidal}
    \end{subfigure}
    \caption{A purely toroidal field (a) cannot provide confinement due to the guiding center drifts. Therefore, we are interested in a magnetic field with both toroidal and poloidal components such that field lines twist to cover magnetic surfaces (b).}
    \label{fig:surf_field_lines}
\end{figure}

Thanks to a poloidal magnetic field, pointing the short way around the torus, these losses can be avoided. As we will discuss in Section \ref{sec:field_line_flow}, the existence of a poloidal magnetic field in axisymmetry ensures the existence of nested, toroidal magnetic surfaces. Consider field lines that twist to lie on a toroidal surface, having both a poloidal and toroidal component (see Figure \ref{fig:surf_field_lines_poloidal}). As particles move along field lines, they will move above and below the $Z = 0$ plane. When an electron is above the $Z = 0$ plane, it will grad-$B$ drift in the $-\hat{\bm{Z}}$ direction away from a given surface, and when it is below the $Z = 0$ plane it will drift in the $-\hat{\bm{Z}}$ direction back toward the magnetic surface. In this way, on an average, charged particles stay close to a magnetic surface. The poloidal magnetic field is used for the magnetic confinement of tokamaks and stellarators. A more quantitative explanation for the necessity of a poloidal magnetic field is provided in Section \ref{sec:confinement_axisymmetry}.

An analogy can be made with the motion of honey on a rotating honey dipper. As gravity always pulls the fluid down, the honey will fall off the dipper if it is stationary. However, if the dipper is rotated, the honey will fall away from the dipper while it is on the bottom half and toward the dipper while it is on the upper half of the dipper. In this way, on average the honey will remain confined. In the same way, the twisting of the magnetic field lines allows particles to remain close to a given magnetic surface. The twisting of magnetic field lines is quantified by the rotational transform, which counts the number of poloidal turns per toroidal turn of a field line, see Section \ref{sec:rotation_transform}. Rotational transform is required for confinement in tokamak and stellarator magnetic confinement devices and will be discussed further in Section \ref{sec:producing_rotational_transform}.

\section{Coordinate systems}
\label{sec:coordinates}

Toroidal geometry refers to a domain in $\mathbb R^3$ delimited by a genus one surface.
The position in toroidal geometry can be conveniently described by several coordinate systems. In particular, for a given static magnetic field under specific assumptions, coordinates adapted to the shape of the magnetic field are useful to simplify the geometric representation. A detailed introduction to flux coordinates for toroidal systems is provided in \cite{2012Dhaeseleer}.

A reminder on the canonical cylindrical coordinate system, often used to describe toroidal systems, is proposed in Section \ref{sec:toroidal_geom}. A discussion of non-orthogonal coordinates, which often arise in describing magnetic geometry, is presented in Section \ref{sec:non_orthogonal}. Section \ref{sec:magnetic_field_lines} presents a discussion of magnetic fields and toroidal magnetic surfaces, as a motivation. Section \ref{sec:flux_function} focuses on flux surfaces and how they are labeled. These surfaces form the basis for flux coordinate systems, described in Section \ref{sec:flux_coordinates}.

\subsection{Canonical cylindrical coordinates}
\label{sec:toroidal_geom}

\begin{figure}
\begin{center}
\includegraphics[clip,trim = 13cm 5cm 11cm 6cm, width=.6\textwidth]{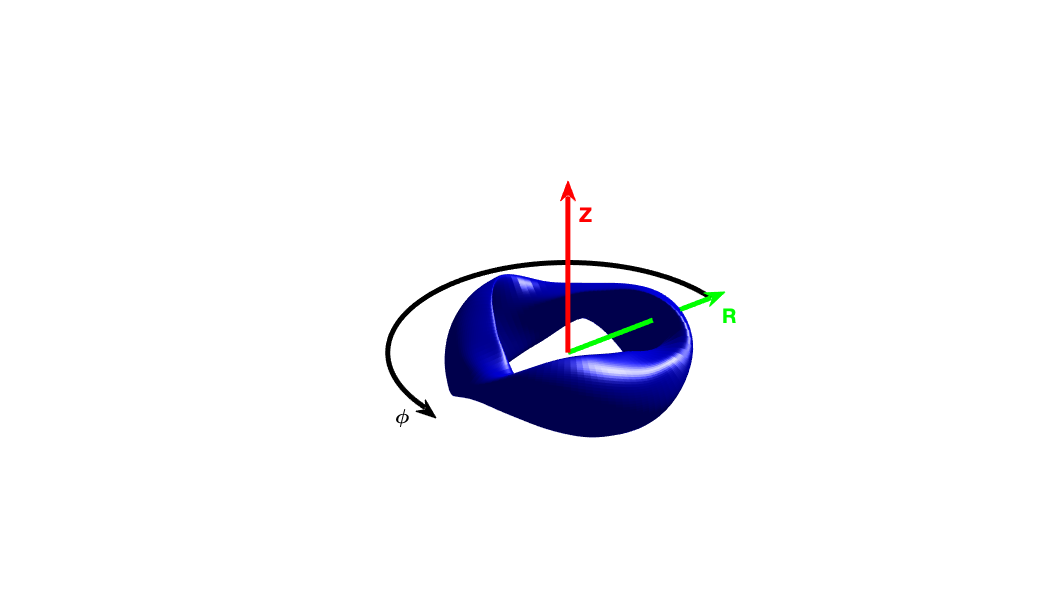}
\end{center}
\caption{The standard cylindrical coordinate system: $R$ measures distance from the $\hat{\bm{Z}}$ axis and $\phi$ is the standard angle of the cylindrical coordinate system such that $\hat{\bm{R}} \times \hat{\bm{\phi}} = \hat{\bm{Z}}$.}
\label{fig:toroidal_coord}
\end{figure}

Toroidal geometry can be described  by the classical cylindrical coordinates $(R,\phi,Z)\in \mathbb R^+\times[0,2\pi)\times \mathbb R$, see Figure  \ref{fig:toroidal_coord}. Given a reference axis $\hat{\bm{Z}}$ and a reference plane perpendicular to $\hat{\bm{Z}}$, the plane can be parameterized by Cartesian coordinates $X$ and $Y$, with $(\hat{\bm{X}},\hat{\bm{Y}}) = (\nabla X,\nabla Y)$, unit vectors such that $\hat{\bm{X}} \times \hat{\bm{Y}} = \hat{\bm{Z}}$. The toroidal angle $\phi$ is the standard cylindrical angle $\phi = \arctan\left(Y/X\right)$ while the major radius $R = \sqrt{X^2 + Y^2}$ measures the distance from the $\hat{\bm{Z}}$ axis. The unit vectors can be expressed in terms of gradients of the coordinates as $\hat{\bm{R}} = \nabla R$, $\hat{\bm{ \phi}} = R\nabla \phi$, and $\hat{\bm{Z}} = \nabla Z$. A poloidal plane is defined as a half plane at constant $\phi$, so that $(\hat{ \bm{R}},\hat{\bm{Z}})$ is an orthonormal basis of the poloidal plane while $\hat{\bm{ \phi}}$ is orthogonal to the poloidal plane. 
Cylindrical coordinates have a singularity  at $R = 0$, i.e. along the $\hat{\bm{Z}}$ axis, as $\phi$ is discontinuous across the axis. 
In general, it is convenient to apply this coordinate system to axisymmetric geometry, as $\phi$ is a symmetry direction. 

 This coordinate system is discussed in Section 4.6.1 of \cite{2012Dhaeseleer}. Cylindrical coordinates are useful in practice as the coordinate singularity at $R = 0$, lying along the $\hat{\bm{Z}}$ axis,  is typically outside the region of interest in toroidal systems. Note that it can be constructed for non-symmetric systems independently of the magnetic field geometry.  
In contrast, coordinate systems depending on the magnetic field geometry may have the advantage of simplifying the expression of quantities of interest, and we will now motivate and introduce such coordinate systems. However it is important to keep in mind that the existence of such coordinate systems will be restricted by additional assumptions on the magnetic field geometry.

\subsection{Non-orthogonal coordinates}
\label{sec:non_orthogonal}

While the classical cylindrical coordinates (see Section \ref{sec:toroidal_geom}) form an orthogonal system, a general coordinate system ($x_1$, $x_2$, $x_3$) may be non-orthogonal. As we will see in Section \ref{sec:flux_coordinates}, one such example is a flux coordinate system, which is particularly useful to stellarators. 

\begin{figure}
    \centering
    \includegraphics
    [trim=2cm 5cm 2cm 6cm,clip,width=.45\textwidth]{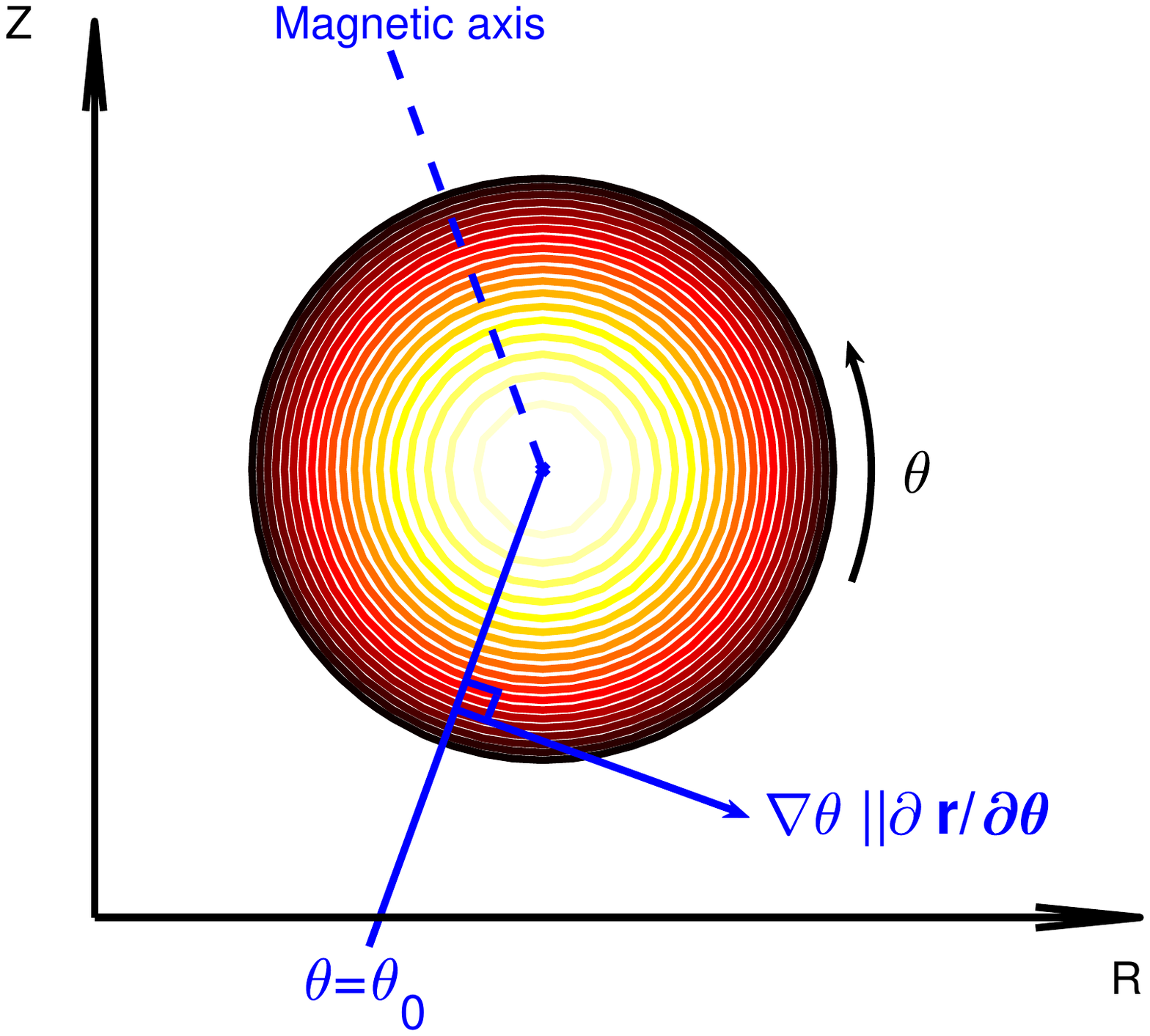}
    \includegraphics
    [trim=2cm 5cm 2cm 6cm,clip,width=.45\textwidth]{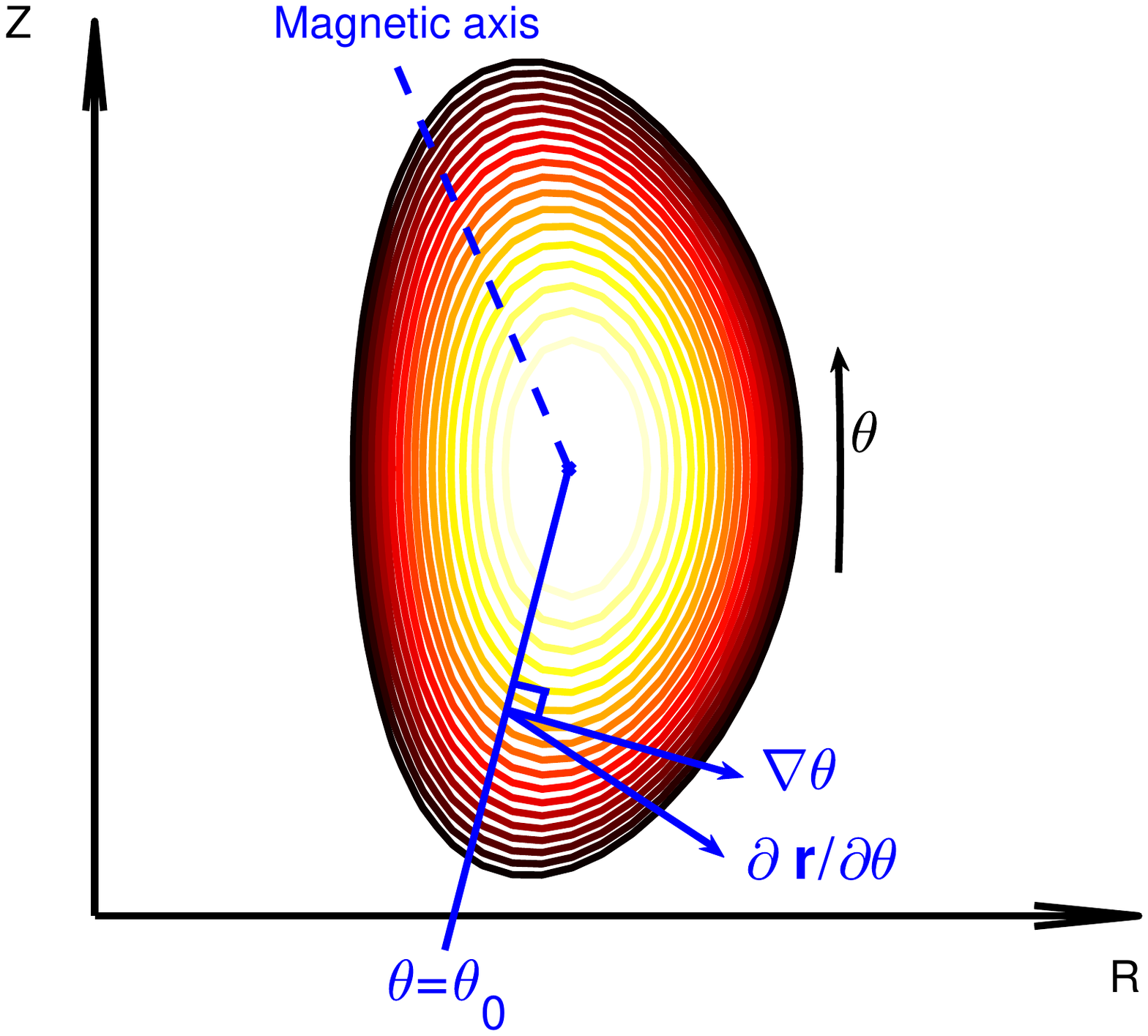}
    \caption{Comparison of orthogonal (left) and non-orthogonal (right) flux coordinate systems $(\psi,\theta,\phi)$ in a plane of constant $\phi$. Curves of constant coordinate $\psi$ for orthogonal and non-orthogonal coordinates are shown. In an orthogonal system (left), the covariant and contravariant basis vectors for a given coordinate are parallel, while, in a non-orthogonal system (right),  covariant and contravariant basis vectors are no longer parallel.}
    \label{fig:nonorthogonal}
\end{figure}

Considering a coordinate system $(x_1,x_2,x_3)$, two local bases can be defined at any point $\bm{ r} \in\mathbb R^3$:
\begin{itemize}
    \item the covariant basis $(\nabla x_1,\nabla x_2,\nabla x_3)$, which is  the basis of the gradients of the coordinates,
    \item the contravariant basis $\left(\partial \bm{r}/\partial x_1,\partial \bm{r}/\partial x_2,\partial \bm{r}/\partial x_3\right)$, which is the basis of the derivatives of the position vector.
\end{itemize}
In general, each basis vector depends on the position itself, however for the sake of clarity this is never expressed explicitly. 
The covariant basis vectors $\nabla x_i$ are perpendicular to isosurfaces of the coordinate $x_i$, while the contravariant basis vectors $\partial \bm{r}/\partial x_i$ point in the direction in which only $x_i$ changes (Figure \ref{fig:nonorthogonal}). They are related by the following expression:\,
\begin{gather}
   \partder{\bm{r}}{x_k} = \frac{\nabla x_i \times \nabla x_j}{\nabla x_i \times \nabla x_j \cdot \nabla x_k}. 
   \label{eq:basis_relation}
\end{gather}
The covariant and contravariant unit vectors satisfy the dual relation: $\nabla x_i\cdot \partial\bm{r}/\partial x_j = \delta^i_{j}$, for all indices $(i,j)\in\{ 1,2,3 \}^2$. 

In general these two bases are not orthogonal. However, for some coordinate systems, such as Cartesian or cylindrical coordinates, they are orthogonal at any point $\bm{ r} \in\mathbb R^3$. Note that the contravariant and covariant bases can only be orthogonal simultaneously. In particular, it is a direct consequence of \eqref{eq:basis_relation} that orthogonality of the covariant basis implies the orthogonality of the contravariant basis.
 In an orthogonal coordinate system, the covariant and contravariant basis vectors for each coordinate are parallel, that is to say that $\nabla x_i $ is parallel to $\partial\bm{r} /\partial x_i$ for all $i\in\{ 1,2,3 \}$. This is another consequence of \eqref{eq:basis_relation}. This is related to the fact that in an orthogonal coordinate system, at any point $(y_1,y_2,y_3)\in\mathbb R^3$ any two surfaces of constant coordinates, like $\{(x_1,x_2,x_3)\in\mathbb R^3,x_i=y_i\}$ and $\{(x_1,x_2,x_3)\in\mathbb R^3,x_j=y_j\}$ for $i\neq j$, have orthogonal tangent planes along their intersection. As opposed to this, in a non-orthogonal coordinate system, covariant and contravariant basis vectors are not necessarily parallel, and surfaces of constant coordinates do not necessarily have orthogonal tangent planes. 

A vector field $\bm{A}$ can be expanded in the covariant form,
\begin{gather}
    \bm{A}(x_1,x_2,x_3) = A_{1}(x_1,x_2,x_3) \nabla x_1 + A_{2}(x_1,x_2,x_3) \nabla x_2 + A_{3}(x_1,x_2,x_3) \nabla x_3,
\end{gather}
 or in the contravariant form,
\begin{multline}
    \bm{A}(x_1,x_2,x_3) = A^{1}(x_1,x_2,x_3) \partder{\bm{r}(x_1,x_2,x_3)}{x_1} + A^{2}(x_1,x_2,x_3) \partder{\bm{r}(x_1,x_2,x_3)}{x_2} \\ + A^{3}(x_1,x_2,x_3) \partder{\bm{r}(x_1,x_2,x_3)}{x_3}.
    \label{eq:Bfluxc}
\end{multline}
The jacobian in a general coordinate system is defined as
\begin{align}
    \sqrt{g} = \left(\partder{\bm{r}}{x_1} \times \partder{\bm{r}}{x_2}\right) \cdot \partder{\bm{r}}{x_3} = \left( \left(\nabla x_1 \times \nabla x_2\right) \cdot \nabla x_3 \right)^{-1}.
\end{align}

A further discussion of non-orthogonal coordinates can be found in Chapter 2 of \cite{2012Dhaeseleer}.
We provide here some of the basic formulas for integrating and differentiating in such coordinates. Consider a vector field $\bm{A}$ and a scalar function $q$. Here we assume that $(i,j,k)$ is either $(1,2,3)$ or one of its cyclic permutations.

\begin{table}
\begin{center}
{\renewcommand{\arraystretch}{3}%
    \begin{tabular}{|c|c|}
    \hline
    Covariant form & $\bm{A} = \sum_{i=1}^3 A_i \nabla x_i$ with $A_i = \bm{A}\cdot \partial \bm{r}/\partial x_i$ \\ \hline
    Contravariant form & $\bm{A} = \sum_{i=1}^3 A^i \partder{ \bm{r}}{ x_i}$ with $A^i = \bm{A}\cdot \nabla x_i$ \\  \hline
    Jacobian
    & $\sqrt{g} = \left(\partder{\bm{r}}{x_1} \times \partder{\bm{r}}{x_2}\right) \cdot \partder{\bm{r}}{x_3} = \left( \left(\nabla x_1 \times \nabla x_2\right) \cdot \nabla x_3 \right)^{-1} $ \\ \hline
    Relation between basis vectors 
    & $\partder{ \bm{r}}{ x_k} = \sqrt{g}\left(\nabla x_i \times \nabla x_j \right)$ \\ \hline 
    Relation between basis vectors 
    & $\nabla x_k  = \sqrt{g}^{-1}\left(\partder{\bm{r}}{x_i} \times \partder{\bm{r}}{x_j} \right)$ \\ \hline 
    Differential volume  &
    $d^3 x = |\sqrt{g}| d x_1 dx_2 dx_3$ \\ \hline
    Differential surface area (constant $x_k$) & $d^2x = |\sqrt{g}| |\nabla x_k| d x_i d x_j$ \\ \hline
    $\nabla \cdot \bm{A} = \sum_{i=1}^3 \frac{1}{\sqrt{g}} \partder{}{x_i} \left(\sqrt{g} 
    A^i
    \right) $ & $\nabla \times \bm{A} = \sum_{k=1}^3 \frac{1}{\sqrt{g}}  \left(\partder{A_j}{x_i} - \partder{A_i}{x_j} \right) \partder{\bm{r}}{x_k} $  \\ \hline
    $\nabla q = \sum_{i=1}^3 \partder{q}{x_i} \nabla x_i$ & $d\bm{r} = \sum_{i=1}^3 \partder{\bm{r}}{x_i} d x_i$ \\ \hline
    \end{tabular}}
\end{center}
\caption{Summary of formulas used to describe the geometry of a non-orthogonal coordinate system $(x_1,x_2,x_3)$. In the above, $\{i,j,k\}$ is a circular permutation of $\{1,2,3\}$.}
\label{table:non_orthogonal}
\end{table}

\subsection{Magnetic field lines and flux surfaces}
\label{sec:magnetic_field_lines}

The magnetic field is a vector field, and flow lines of this vector field (field lines) are often used for visualization and interpretation of physical phenomena. Particle confinement is related to the geometry of magnetic field lines. As described in Section \ref{sec:cylindrical}, in a straight magnetic field a particle will gyrate about field lines. When the field is curved or its magnitude varies in space, particles will exhibit a slow drift across field lines in addition to their motion along field lines, as described in Section \ref{gradBdrift}. As particles are free to move in the direction parallel to the field, the temperature tends to equilibrate along field lines. In magnetic confinement fusion devices, it is necessary to maintain a hot core that is not in thermal contact with the material walls. Therefore, field lines should not connect the plasma core to the material wall or the cooler edge of the plasma. If field lines do not intersect material surfaces, they must 
remain within a toroidal volume. Each field line can then either lie on a closed surface within the volume or fill a volume. A flux surface is a smooth surface such that at every point on the surface $\bm{B}\cdot\hat{\bm{n}}=0$, where $\hat{\bm{n}}$ is a normal vector to the surface. So, in particular, no magnetic field line crosses a magnetic surface: the field is tangent to the flux surface. 

As the Hopf-Poincar\'{e} theorem states that a non-vanishing, continuous tangential vector field cannot lie on a sphere in 3D, flux surfaces cannot be spherical surfaces. However, it is possible for a flux surface to be a toroidal surface. There may exist a set of toroidal surfaces within a given volume. All of these surfaces may be nested around a single closed field line, called a magnetic axis. We can distinguish between a `primary' magnetic axis, about which there are nested surfaces across the entire plasma volume. In between two surfaces nested about the primary magnetic axis, it is also possible for closed field lines to form, leading to `secondary' magnetic axes. In between two such surfaces, a secondary set of surfaces can also be nested about a secondary axis, forming what is called an island structure. 

\begin{figure}
    \centering
    \includegraphics[trim=6cm 5cm 3cm 3cm,clip,width= 0.7\textwidth]{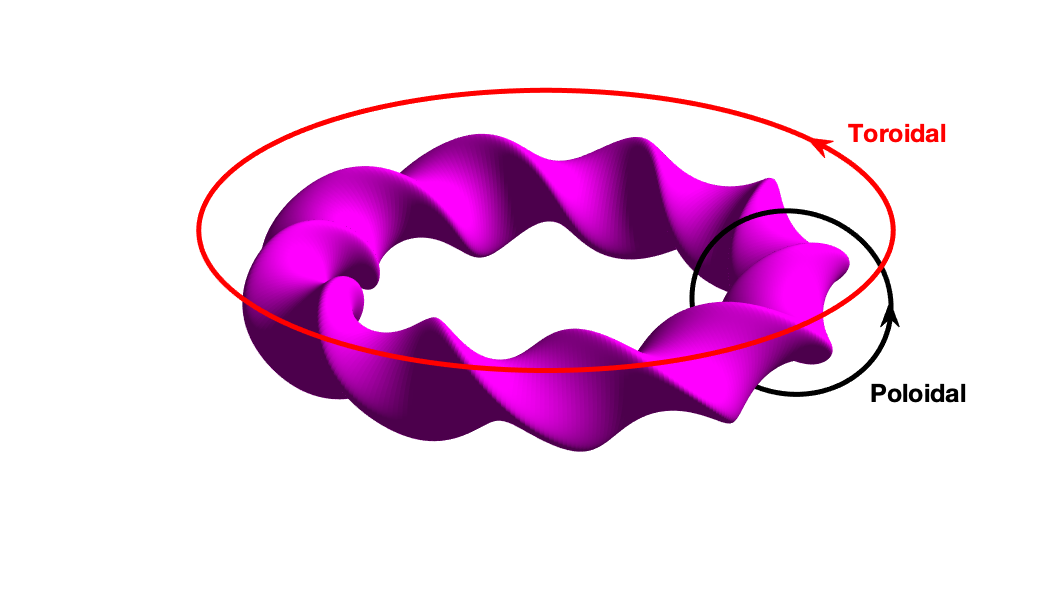}
    \caption{The position in a toroidal system is often described by two angles. A poloidal angle increases from 0 to 2$\pi$ on any closed poloidal loop (black) about the magnetic axis, the short way around the torus. A toroidal angle increases from 0 to 2$\pi$ on any closed toroidal loop (red) about the major axis of the coordinate system, the long way around the torus.}
    \label{fig:toroidal_poloidal_angles}
\end{figure}

As our desired geometry consists of toroidal surfaces, it is useful to define a toroidal coordinate system. We will use the term toroidal to refer to the direction the long way around the torus, while the term poloidal refers to the direction the short way around the torus. See Figure \ref{fig:toroidal_poloidal_angles} for a brief description of toroidal geometry.

\begin{figure}[ht]
\begin{subfigure}{0.49\textwidth}
\centering{\includegraphics[trim=17cm 2cm 30cm 2cm,clip,width=\textwidth]{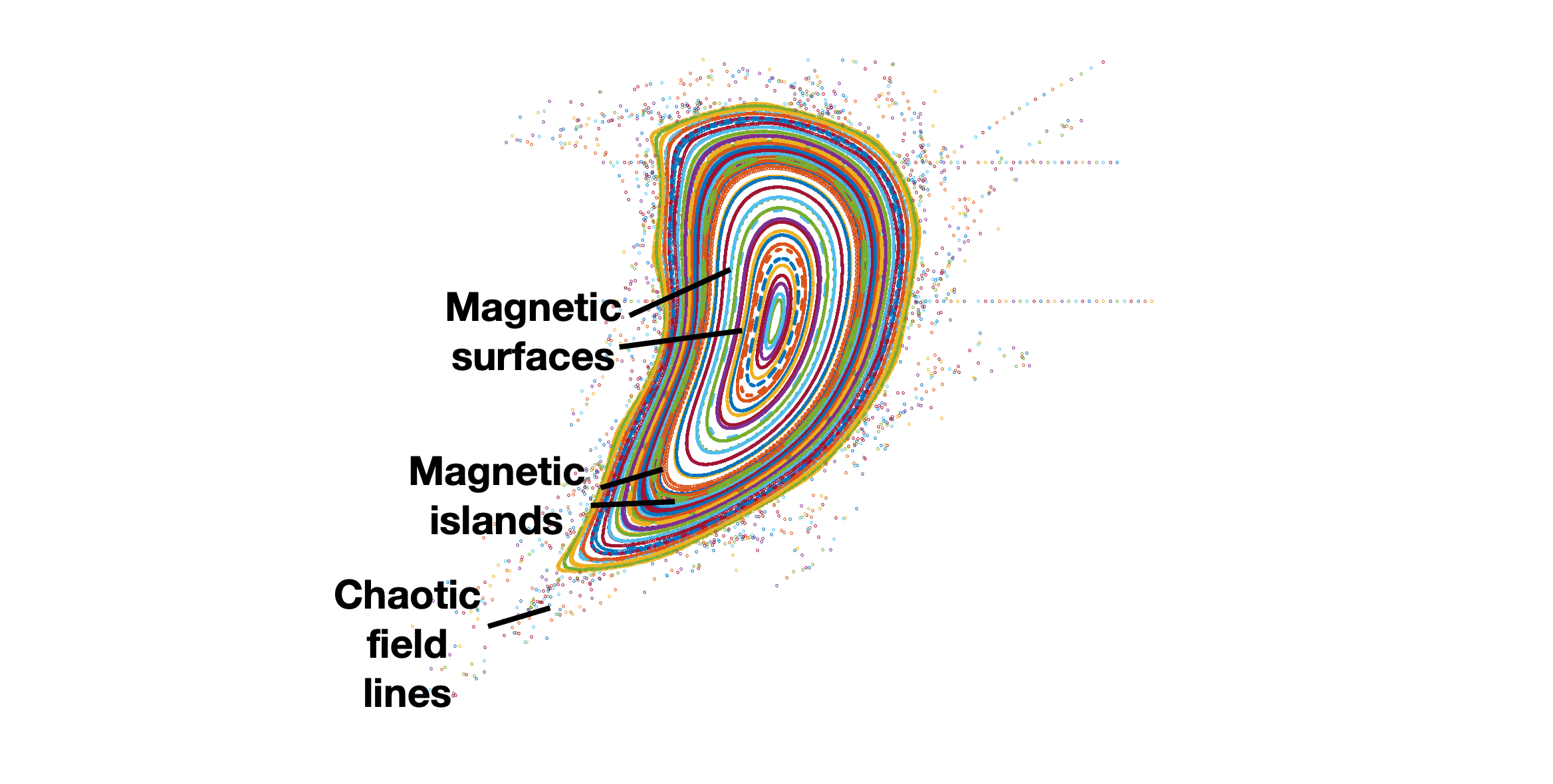}}
\end{subfigure}
\begin{subfigure}{0.49\textwidth}
\centering{\includegraphics[trim = 30cm 5cm 45cm 5cm,clip,width=\textwidth]{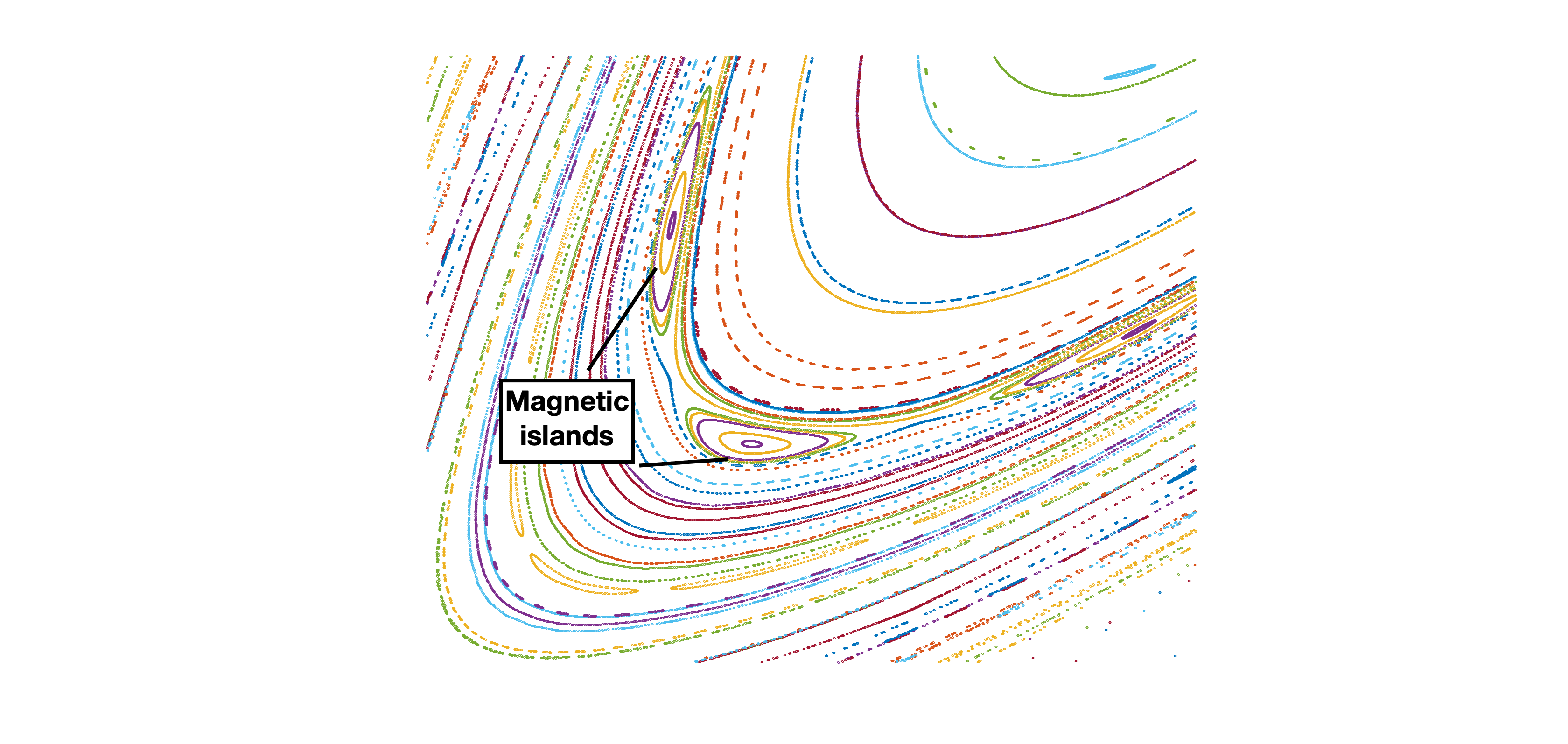}}
\end{subfigure}
\caption{To produce this Figure, field lines are followed, and each time they hit a plane at constant toroidal angle, a point is marked on the plot with colors indicating a given field line. This is often referred to as a Poincar\'e plot. The magnetic field is generated by the coils of the NCSX stellarator \cite{Zarnstorff2001}. Within the confinement region there are sets of magnetic surfaces as well as magnetic islands and chaotic field lines. 
\label{fig:PoincareGoodVsBad}
}
\end{figure}

For a given magnetic field $\bm B$, a common way to visualize its structure is through a Poincar\'{e} plot. The plot setting is a 2D plane, representing a given surface at constant toroidal angle $S$. The plot is produced by following a set of field lines through many full toroidal rotations around the device and placing a point wherever the line passes through $S$. Toroidally nested surfaces appear as points which fill out nested closed curves in the Poincar\'{e} plot, while island structures appear as a secondary set of closed curves in between two primary closed curves. For field lines which do not line on surfaces, called chaotic field lines, the Poincare\'{e} plot displays a set of points which do not fill out curves. Refer to Figure \ref{fig:PoincareGoodVsBad} for a Poincar\'{e} plot of the magnetic field produced by the NCSX coils \cite{Zarnstorff2001}.

In order to maintain a temperature gradient within the confinement region, we seek magnetic fields that minimize the volume occupied by chaotic regions and island structures, as the temperature is equilibrated rapidly within these structures. Ideally, we desire the magnetic field to lie on continuously nested surfaces (with a single magnetic axis). Additional motivation for magnetic surfaces is provided in Section \ref{confiningMagneitcField}, and justification for their existence is provided in Sections \ref{sec:field_line_flow}-\ref{sec:integrability}.

In perfect axisymmetry, closed, nested flux surfaces are guaranteed if there is a non-zero toroidal current in the plasma. This statement will be justified in Sections \ref{sec:field_line_flow}-\ref{sec:integrability} by demonstrating that magnetic field line flow can be described by a Hamiltonian system which possesses a conserved quantity under the assumption of axisymmetry. However, in 3-dimensional geometry (such as in a stellarator or a tokamak with 3D perturbations), field lines may become chaotic or may form islands in addition to forming nested flux surfaces in some regions of space (Figure \ref{fig:PoincareGoodVsBad}).


\subsection{Flux surface labels}
\label{sec:flux_function}

Assuming that a given magnetic field $\bm{B}$ is such that the magnetic field lines lie on continuously nested closed nested toroidal flux surfaces around a single magnetic axis in a given domain, certain physical quantities are constant on flux surfaces. We will denote such quantities as flux functions. For example, in practice particles are mostly confined to flux surfaces; therefore the temperature, density, and pressure are approximately flux functions. Certain flux functions can be used to define convenient coordinate systems, discussed in Section \ref{sec:flux_coordinates}. 

A flux surface label, generally denoted $\psi$, is a smooth one-to-one real-valued function defined on the set of flux surfaces and changes monotonically with distance from the axis, either increasing or decreasing. It is often assumed to vanish on the magnetic axis. Each flux surface can be uniquely labeled by a value of $\psi$: a flux surface label is a flux function. We will introduce the two most natural examples of common flux labels: the poloidal and toroidal fluxes.

\begin{figure}
\begin{center}
\begin{subfigure}{0.49\textwidth}
\includegraphics[trim=41cm 13cm 27cm 10cm,clip,width=1.0\textwidth]{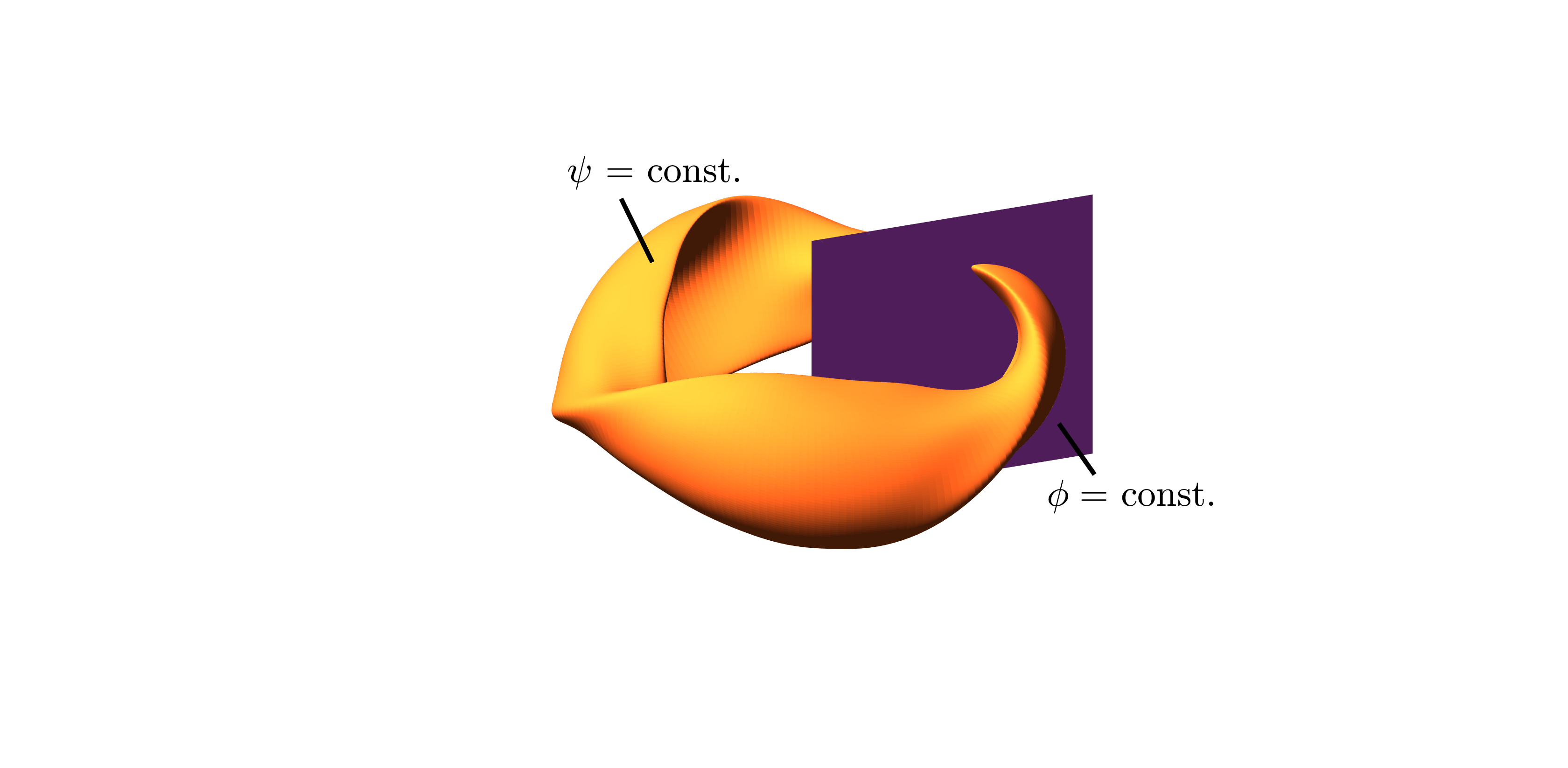}
\end{subfigure}
\begin{subfigure}{0.49\textwidth}
\includegraphics[trim=50cm 10cm 34cm 16cm,clip,width=1.0\textwidth]{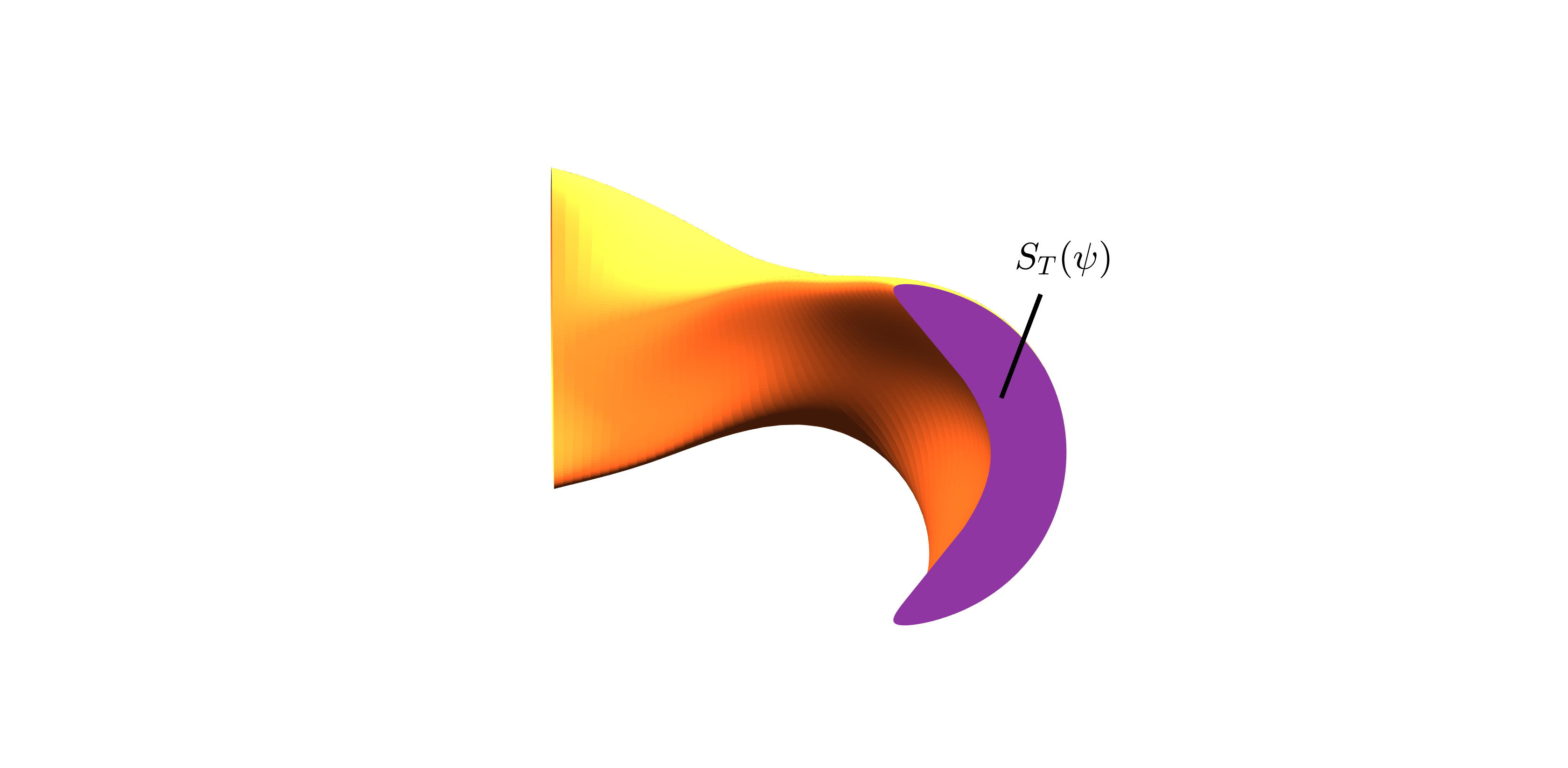}
\end{subfigure}
\caption{The toroidal flux, $\Psi_T(\psi)$, is the magnetic flux through a surface at constant $\phi$ bounded by the surface labeled by $\psi$.}
\label{fig:toroidal_flux}
\end{center}
\end{figure}

 Let $\phi$ be an angle which increases by $2\pi$ upon a toroidal loop. The toroidal flux, $\Psi_T$, of a given flux surface with flux label $\psi$, is the flux of magnetic field through a surface at constant $\phi$ bounded by the constant $\psi$ surface, which we call $S_T$ (see Figure \ref{fig:toroidal_flux}), 
\begin{gather}
\Psi_T(\psi) = \int_{S_T} \bm{B} \cdot \hat{\bm{n}} \, d^2 x, 
\end{gather}
where $\hat{\bm{n}}$ is an oriented unit normal and $d^2 x$ is the surface area element. 
 
\begin{figure}
\begin{center}
\begin{subfigure}{0.49\textwidth}
\includegraphics[trim=23cm 16cm 23cm 14cm,clip,width=1.0\textwidth]{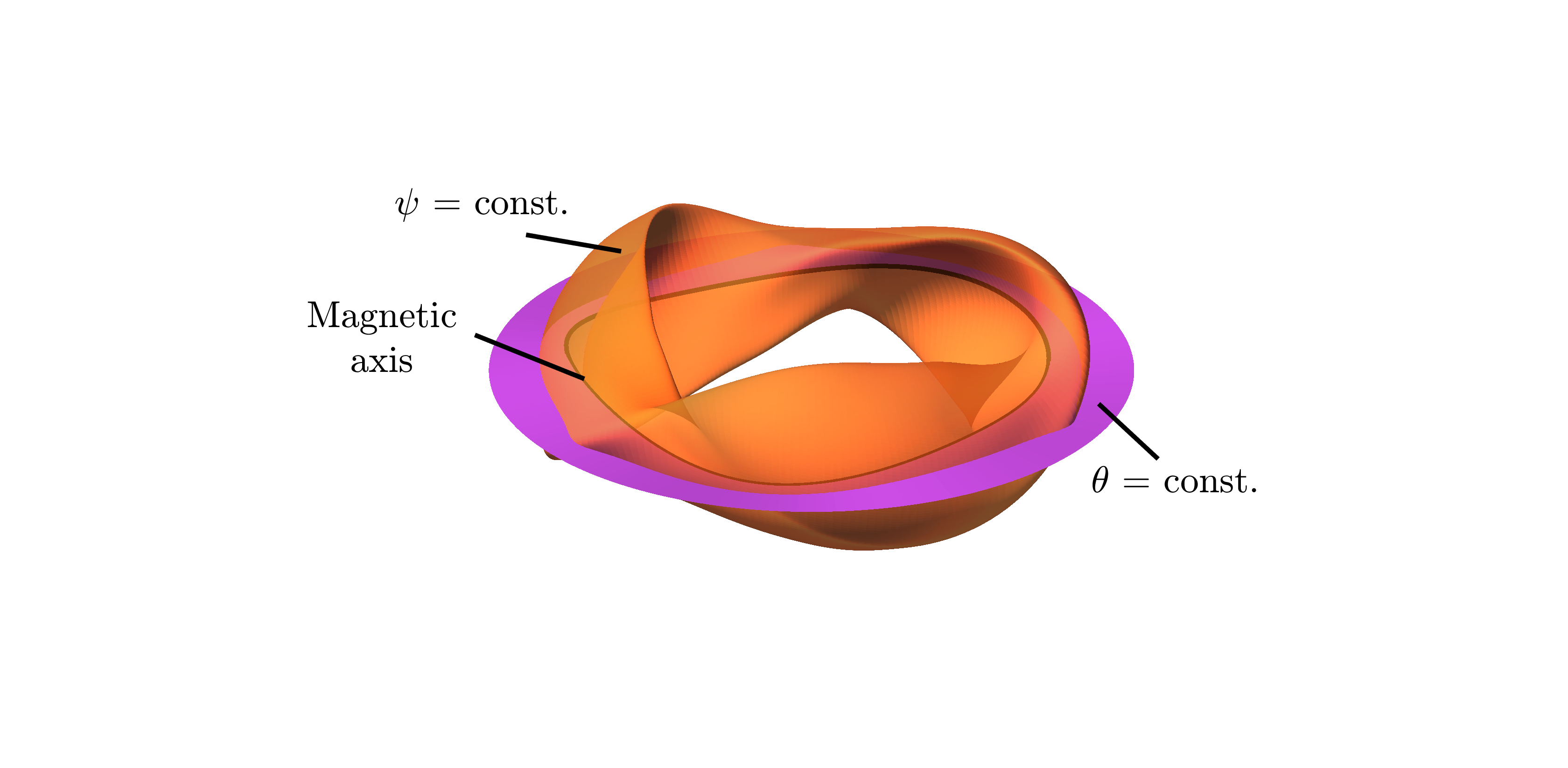}
\end{subfigure}
\begin{subfigure}{0.49\textwidth}
\includegraphics[trim=37cm 15cm 26cm 9cm,clip,width=1.0\textwidth]{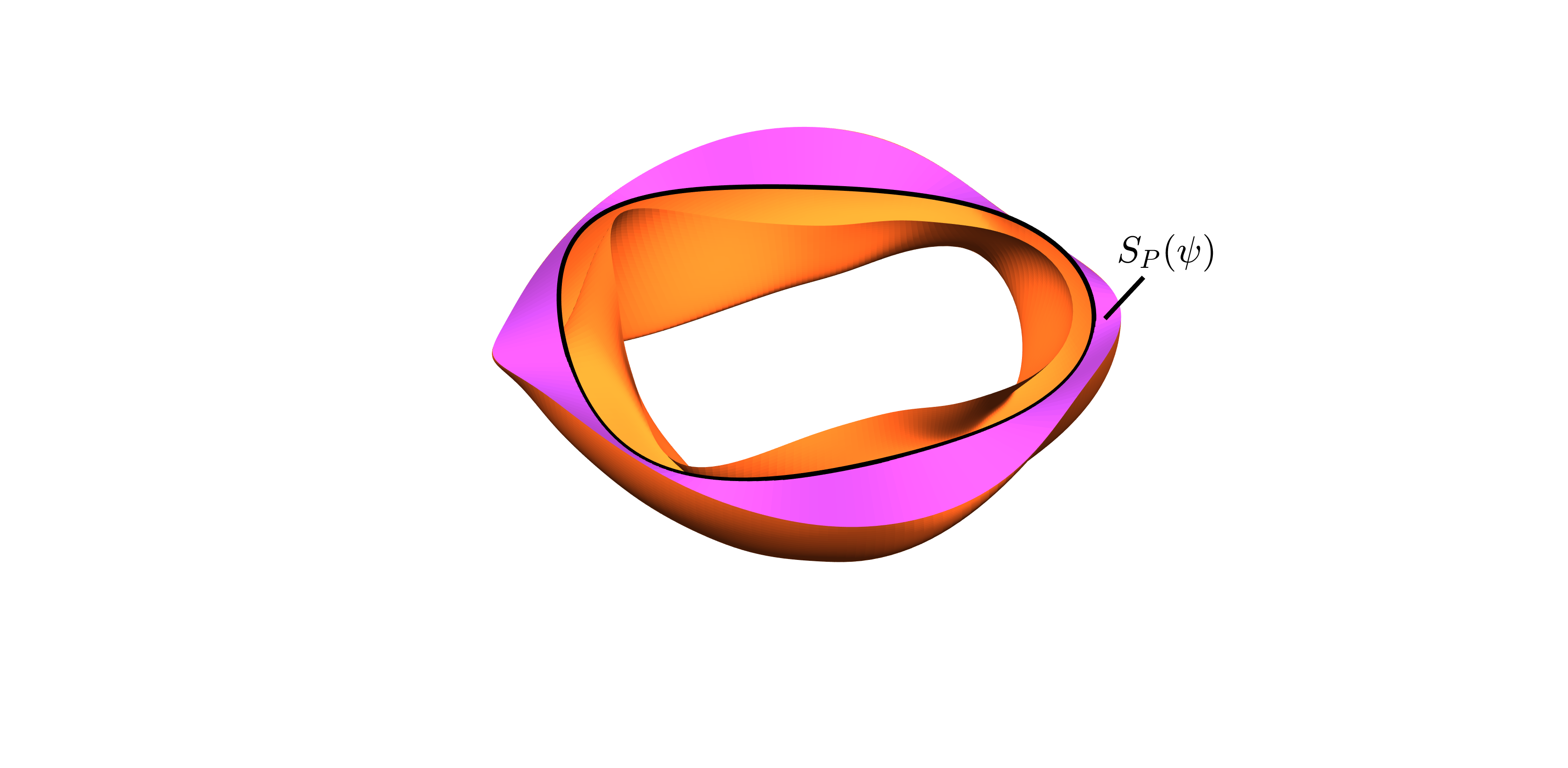}
\end{subfigure}
\caption{The poloidal flux, $\Psi_P(\psi)$, is the magnetic flux through a ribbon-like surface (pink) at constant $\theta$ bounded by the surface labeled by $\psi$ (orange) and the magnetic axis (black).}
\label{fig:poloidal_flux}
\end{center}
\end{figure}

 Suppose $\theta$ is an angle which increases by $2\pi$ upon a poloidal loop.
The poloidal flux of a given flux surface, $\psi$, is the flux of the magnetic field through a surface at constant $\theta$ bounded between the magnetic axis and the constant $\psi$ surface, which we call $S_P$ (see Figure \ref{fig:poloidal_flux}),
\begin{gather}
\Psi_P(\psi) = \int_{S_P} \bm{B} \cdot \hat{\bm{n}} \, d^2 x. 
\end{gather}

The rotational transform, which quantifies the number of poloidal turns of a field line per toroidal turn (see Section \ref{sec:rotation_transform}), can be defined in terms of these fluxes,
\begin{gather}
    \iota(\psi) = \frac{d\Psi_P(\psi)/d\psi}{d \Psi_T(\psi)/d\psi}.
    \label{eq:iota}
\end{gather}
This relation will become clear during the discussion of magnetic coordinates (Section \ref{sec:magnetic_coordinates}). The rotational transform is another example of flux label, it will play an important role in the design of the magnetic field. 

Further discussion of flux functions can be found in Chapter 4 of \cite{2012Dhaeseleer}. 

\subsection{Flux coordinates}
\label{sec:flux_coordinates}

If a set of continuously nested flux surfaces exist in a given domain, coordinate systems can be constructed based on such surfaces. These are known as flux coordinate systems, for which a flux label is used as one of the coordinates. The other two coordinates are angles that parameterize the position on each flux surface. Flux coordinate systems are practical for calculations, as physical processes along a surface are typically distinct in their spatial and time scales from those that occur across surfaces. Flux coordinates are convenient to use in practice, as physical quantities are periodic in the poloidal and toroidal angles. Therefore, Fourier series can be used for both analytic study of physical systems and for efficient numerical
discretization. However, if continuously nested flux surfaces do not exist in a whole domain, such as in the presence of magnetic islands, flux coordinates cannot be defined throughout the domain. On the other hand, in principle, within a magnetic island, it is possible to define a local set of flux coordinates.

\begin{figure}
\begin{center}
\includegraphics[trim=36cm 3cm 25cm 2cm,clip,width=.8\textwidth]{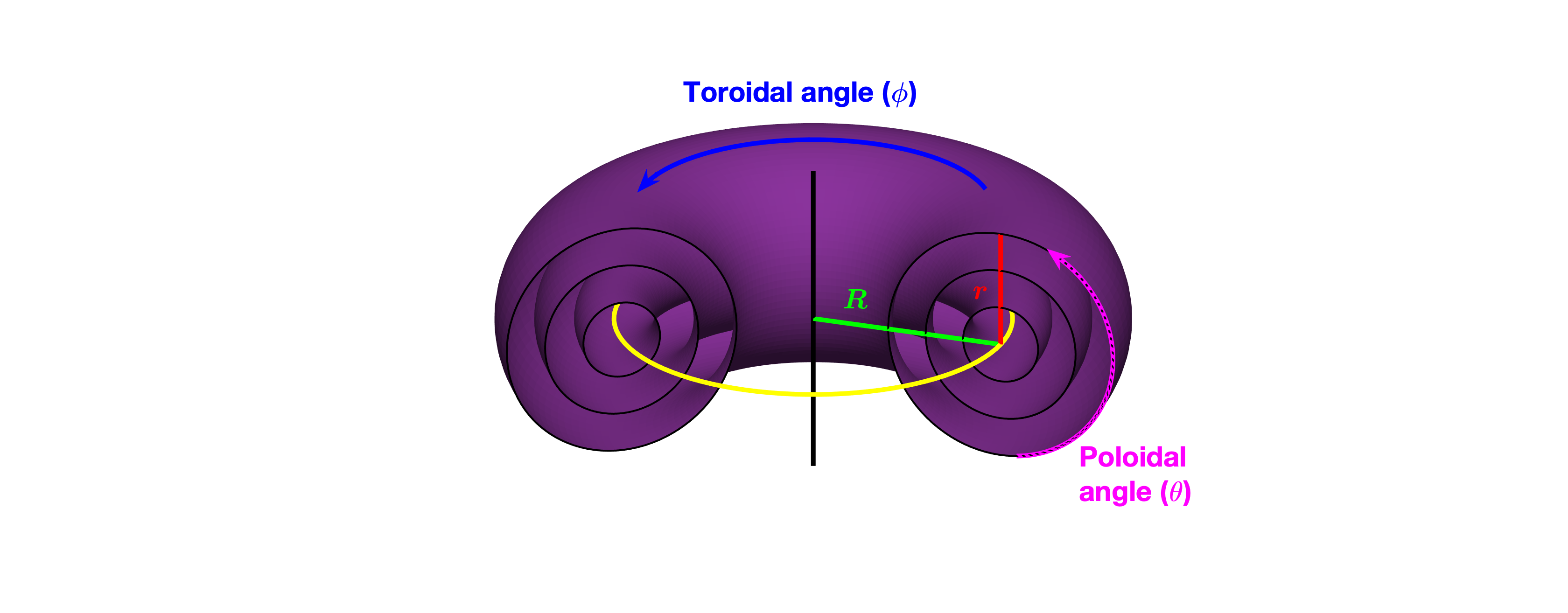}
\caption{We consider a flux coordinate system consisting of tori with circular cross-sections. Poloidal and toroidal angles describe the location on a magnetic surface. In the case of nested magnetic surfaces, the magnetic axis (yellow) is the line enclosed by all nested surfaces. The minor radius, $r$, is a measure of distance from the magnetic axis within a poloidal plane (half plane defined by $\phi$ = const.). The major radius, $R$, measures distance from the axis of rotation of the toroidal coordinate system. Similar coordinates can be constructed for non-axisymmetric system, where $r$ is a flux surface label.}\label{toroidal_poloidal}
\end{center}
\end{figure}

\begin{figure}
\begin{center}
\includegraphics[clip,trim=0.5cm 5cm 0.5cm 5cm,width=.47\textwidth]{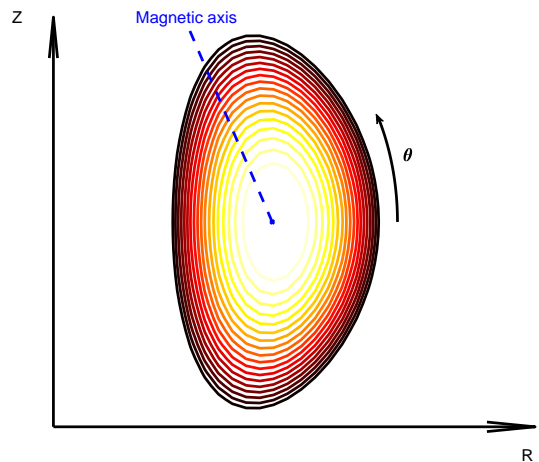} 
\end{center}
\caption{Poloidal angle $\theta$ and level curves of the flux label $\psi$ in a poloidal half plane. In the poloidal plane, the magnetic axis is the point enclosed by all the flux surfaces.}
\label{fig:cross_Section}
\end{figure}

The geometry of a set of toroidal surfaces nested around the magnetic axis (see Figure \ref{fig:cross_Section}) can be described by flux coordinates $(r,\theta,\phi)\in \mathbb R^+\times[0,2\pi)\times[0,2\pi)$, where $\phi$ is an angle which increases by $2\pi$ upon a toroidal loop, $\theta$ is an angle which increases by $2\pi$ upon a poloidal loop, and $r$ is a flux surface label. See Figure \ref{toroidal_poloidal} for an example consisting of nested toroidal surfaces with circular cross-sections, which we call a cylindrical torus.

Using flux coordinates, toroidal fluxes are expressed as, 
\begin{gather}
\Psi_T(r)  = \int_0^{r}  \int_0^{2\pi}  \frac{\bm{B} \cdot \nabla \phi}{\nabla r'\cdot \nabla \theta \times \nabla \phi} \,  d \theta d r',
\label{eq:psi_T}
\end{gather}
where $\nabla \phi/|\nabla \phi|$ is a unit normal and $|\nabla \phi| (\nabla r' \cdot \nabla \theta \times \nabla \phi)^{-1} d \theta d r'$ is the surface area element for the surface at constant $\phi$. The poloidal flux is,
\begin{gather}
\Psi_P(r)  = \int_0^{r}  \int_0^{2\pi} \, \frac{\bm{B} \cdot \nabla \theta}{\nabla r' \cdot \nabla \theta \times \nabla \phi} \,  d \phi d r',
\label{eq:psi_P}
\end{gather}
where $\nabla \theta/|\nabla \theta|$ is a unit normal and  $|\nabla \theta| (\nabla r' \cdot \nabla \theta \times \nabla \phi)^{-1} d \phi d r'$ is the surface area element for the surface at constant $\theta$.

We can estimate the scale of these fluxes under the assumption of a torus with circular toroidal cross-section with major radius $R$ and minor radius $r$. Given the toroidal magnetic field $B_T = \bm{B} \cdot \nabla \phi/|\nabla \phi|$, then the toroidal flux scales as $\Psi_T \approx \pi r^2 B_T$. Likewise if the poloidal magnetic field is $B_P = \bm{B} \cdot \nabla \theta/|\nabla \theta|$, then the poloidal flux scales as $\Psi_P \approx 2\pi R r B_P$. In practice $B_T$ and $B_P$ are not constant, but this provides a valid order of magnitude approximation. See Figure \ref{toroidal_poloidal}. 

While flux coordinates can be used with any flux label, in this text we will use the toroidal flux function $\psi = \Psi_T/2\pi$ as our flux surface label. 
Another example in a cylindrical torus is $r$, which measure the distance to the magnetic axis within a poloidal plane.
Other examples of flux functions are described in Section \ref{sec:flux_function}. 
Several examples of flux coordinate systems will be discussed in Section \ref{sec:magnetic_coordinates}. 

As flux coordinates are generally non-orthogonal (Section \ref{sec:non_orthogonal}), the magnetic field can be expressed in terms of the covariant,
\begin{align}
    \bm{B}(\psi,\theta,\phi) = B_{\psi}(\psi,\theta,\phi) \nabla \psi + B_{\theta}(\psi,\theta,\phi) \nabla \theta + B_{\phi}(\psi,\theta,\phi) \nabla \phi,
\end{align}
and contravariant basis vectors,
\begin{align}
     \bm{B}(\psi,\theta,\phi) =  B^{\theta}(\psi,\theta,\phi) \nabla \theta + B^{\phi}(\psi,\theta,\phi) \nabla \phi .
\end{align}
Note that the radial contravariant component of the magnetic field vanishes from the assumption that $\bm{B} \cdot \nabla \psi = 0$.


\FloatBarrier
\section{Toroidal magnetic confinement}
\label{sec:magnetic_confinement_devices}

We can now discuss confinement of particles in magnetic confinement configurations. In this section we focus on two leading approaches: the tokamak and the stellarator. In Section \ref{sec:axisymmetry} we propose a formal definition of axisymmetry, the fundamental property of the tokamak. In Section \ref{sec:confinement_axisymmetry} we will see that axisymmetry leads to the approximate conservation of the flux label, hence providing particle confinement. However, we will find that this confinement is associated with a major challenge: the necessity of a large current within the confinement region. We will revisit particle confinement without axisymmetry in Section \ref{sec:quasisymmetry}. An overview of the tokamak and stellarator concepts is provided in Sections \ref{sec:tokamak} and \ref{sec:stellarator}. 
Section \ref{sec:rotation_transform} focuses on a concept important to toroidal confinement, the rotational transform. In Section \ref{sec:producing_rotational_transform} we discuss how rotational transform can be produced with and without the assumption of axisymmetry.

\subsection{Axisymmetry}
\label{sec:axisymmetry}

Axisymmetry is the symmetry of vector or scalar fields with respect to the azimuthal toroidal angle, $\phi$, when expressed in canonical cylindrical coordinates (see Section \ref{sec:toroidal_geom}). 

Consider a vector field 
\begin{align}
    \bm{F}(R,\phi,Z) = F_R(R,\phi,Z) \hat{\bm{R}}(\phi) + F_{\phi}(R,\phi,Z) \hat{\bm{\phi}}(\phi) + F_Z(R,\phi,Z) \hat{\bm{Z}}.
\end{align}
Axisymmetry of $\bm{F}$ implies that 
\begin{align}
    \partder{F_R(R,\phi,Z)}{\phi} = \partder{F_{\phi}(R,\phi,Z)}{\phi} = \partder{F_Z(R,\phi,Z)}{\phi} = 0.
    \label{eq:axisymmetry}
\end{align}
This furthermore implies that the magnitude, $F = |\bm{F}|$, satisfies
\begin{align}
    \partder{F(R,\phi,Z)}{\phi} = 0.
    \label{eq:axisymmetric_scalar}
\end{align}
A vector field, $\bm{F}$, is said to be axisymmetric if it satisfies \eqref{eq:axisymmetry}, and a scalar field, $F$, is said to be axisymmetric if it satisfies \eqref{eq:axisymmetric_scalar}. 

Some magnetic confinement devices, such as tokamaks, are designed to have magnetic fields close to axisymmetry, according to the above definition.
With a suitable choice for gauge, the corresponding vector potential, $\bm{A}$, is also axisymmetric according to \eqref{eq:axisymmetry}. Similarly, the corresponding current density, $\bm{J}$, can be shown to be axisymmetric upon application of Ampere's law \eqref{eq:ampere_magnetostatic}. Therefore, many physical quantities of interest are axisymmetric if the magnetic field is axisymmetric.

Axisymmetry can also be expressed in a flux coordinate system $(r,\theta,\phi)$ (see Section \ref{sec:flux_coordinates}) if the toroidal angle is chosen to be the canonical azimuthal angle. Consider a vector field expressed in its covariant and contravariant forms
\begin{align}
    \bm{F}(r,\theta,\phi) &= F_r(r,\theta,\phi) \nabla r + F_{\theta}(r,\theta,\phi) \nabla \theta + F_{\phi}(r,\theta,\phi) \nabla \phi \nonumber \\
    &= F^r(r,\theta,\phi) \partder{\bm{r}(r,\theta,\phi)}{r} + F^{\theta}(r,\theta,\phi) \partder{\bm{r}(r,\theta,\phi)}{\theta} + F^{\phi}(r,\theta,\phi) \partder{\bm{r}(r,\theta,\phi)}{\phi}.
\end{align}
As flux coordinates are generally non-orthogonal, we need to consider the symmetry of both component forms. Axisymmetry implies that 
\begin{align}
    \partder{F_r(r,\theta,\phi)}{\phi} = \partder{F_{\theta}(r,\theta,\phi)}{\phi} = \partder{F_{\phi}(r,\theta,\phi)}{\phi} = \partder{F^r(r,\theta,\phi)}{\phi} = \partder{F^{\theta}(r,\theta,\phi)}{\phi} = \partder{F^{\phi}(r,\theta,\phi)}{\phi} = 0.
\end{align}
We will consider several implications of axisymmetry in Sections \ref{sec:producing_rotational_transform} and \ref{sec:confinement_axisymmetry}.

\subsection{Particle confinement in axisymmetry}
\label{sec:confinement_axisymmetry}
Axisymmetry is desirable because
it guarantees confinement of particles in the absence of collisions. This can be seen when considering the Lagrangian for the motion of a single particle with mass $m$ and charge $q$ in a magnetic field $\bm{B} = \nabla \times \bm{A}$ and an electric field $\bm{E} = - \nabla \Phi - \partial \bm{A}/\partial t$, as described in Section \ref{sec:lagrangian}.
We will express the Lagrangian \eqref{eq:lagrangian} in a cylindrical coordinate system, ($R$,$\phi$,$Z$) (see Section \ref{sec:toroidal_geom}),
\begin{multline}
L(R,\phi,Z,\dot{R},\dot{\phi},\dot{Z},t) = \frac{m}{2} \left(\dot{R}^2 + R^2 \dot{\phi}^2 + \dot{Z}^2\right) \\ + q \left( A_R(R,\phi,Z) \dot{R}+A_{\phi}(R,\phi,Z)R\dot{\phi} + A_Z(R,\phi,Z) \dot{Z} \right)
- q \Phi(R,\phi,Z,t).
\end{multline}
Here the scalar potential is decomposed in its components in cylindrical coordinates as $\bm{A} = A_R \hat{\bm{R}} +  A_{\phi} \hat{\bm{\phi}} + A_Z \hat{\bm{Z}}$. Under the assumption of axisymmetry, each of the components of $\bm{A}$ is independent of $\phi$ (see Section \ref{sec:axisymmetry}).
Furthermore, $\Phi$ is assumed to be axisymmetric.

The equations of motion implied by the Lagrangian come from the Euler-Lagrange equations \eqref{eq:euler_lagrange}. The Euler-Lagrange equation for the toroidal angle is,
\begin{gather}
\der{}{t} \left( \partder{L(R,\phi,Z,\dot{R},\dot{\phi},\dot{Z},t)}{\dot{\phi}} \right) = \partder{L(R,\phi,Z,\dot{R},\dot{\phi},\dot{Z},t)}{\phi}.
\end{gather}
We will call $p_{\phi}:= \partial L/\partial \dot{\phi} = mR^2\dot{\phi} + q RA_{\phi}$ the toroidal canonical momentum. Under the assumption of axisymmetry, the Lagrangian becomes independent of $\phi$, and a constant of motion is implied,
\begin{gather}
\der{p_{\phi}(t)}{t} = 0.
\end{gather}
This is an example of Noether's theorem: any continuous symmetry, in this case toroidal rotational symmetry, implies a conserved quantity, in this case $p_{\phi}$.

We will now estimate the size of each term in $p_{\phi}$ to gain additional insight by re-writing them in terms of physical quantities. As we will see in \eqref{eq:B_pol}, $(RA_{\phi})$ can be related to the magnitude of the poloidal magnetic field as $B_P = (1/R) |\nabla (RA_{\phi})|$. Assuming that the gradient length scale of $(RA_{\phi})$ scales as $|(RA_{\phi})^{-1}\nabla(RA_{\phi})|\sim R^{-1}$, we can approximate $RA_{\phi} \sim R^2 B_P$. We can also approximate $R\dot{\phi} \sim v_t$, where $v_t = \sqrt{2T/m}$ is the thermal velocity. So we find the ratio of the two terms in $p_{\phi}$ to be,
\begin{gather}
\frac{mR^2\dot{\phi}}{q RA_{\phi}} \sim \frac{m v_t }{q B_P R} \sim \frac{\omega_B}{\Omega_P}.
\end{gather}
Here $\omega_B = v_t/L_B$ is the frequency associated with the time variation of the magnetic field as introduced in Section \ref{ref:gyroaveraged_lagrangian}, the gradient length scale of the magnetic field is defined such that $L_B^{-1} = (1/B)|\nabla B|$, and $\Omega_P =  qB_P/m$ is the poloidal gyrofrequency, the analogous gyrofrequency ($\Omega = qB/m$) associated with the poloidal magnetic field. We have made the assumption that $L_B \sim R$. If $B_P$ is large enough, this ratio will be $\ll 1$. Therefore, we can make the approximation $p_{\phi} \approx qRA_{\phi}$. 

We note that $(RA_{\phi})$ is a flux function according to the discussion in Section \ref{sec:producing_rotational_transform}. As $p_{\phi}$ is then approximately constant on a flux surface, a particle will be confined to within a small distance of a flux surface whose magnitude scales with $\omega_B/\Omega_P \ll 1$. As this ratio scales inversely $B_P$, this is another way to understand the need for a poloidal magnetic field for confinement in a tokamak. See \cite{Littlejohn1983} and Chapter 7 in \cite{Helander2005} for more details.

In 3D geometry without a continuous symmetry, i.e. without axisymmetry, this result does not apply, so much more care must be taken in order to achieve good confinement. However, a ``hidden symmetry" can be exploited to obtain similar confinement properties in a stellarator, as will be explored in Section \ref{sec:quasisymmetry}.

\subsection{Tokamak}
\label{sec:tokamak}


A tokamak is a toroidal confinement device with genus one topology. Tokamaks are designed under the assumption of axisymmetry according to the definition provided in Section \ref{sec:axisymmetry}. In practice, the magnetic field of a tokamak is close enough to axisymmetry that this approximation is sufficient; thus many physical scalar quantities are independent of the toroidal angle, $\phi$. 

\begin{figure}
\begin{center}
\includegraphics[width=0.7\textwidth]{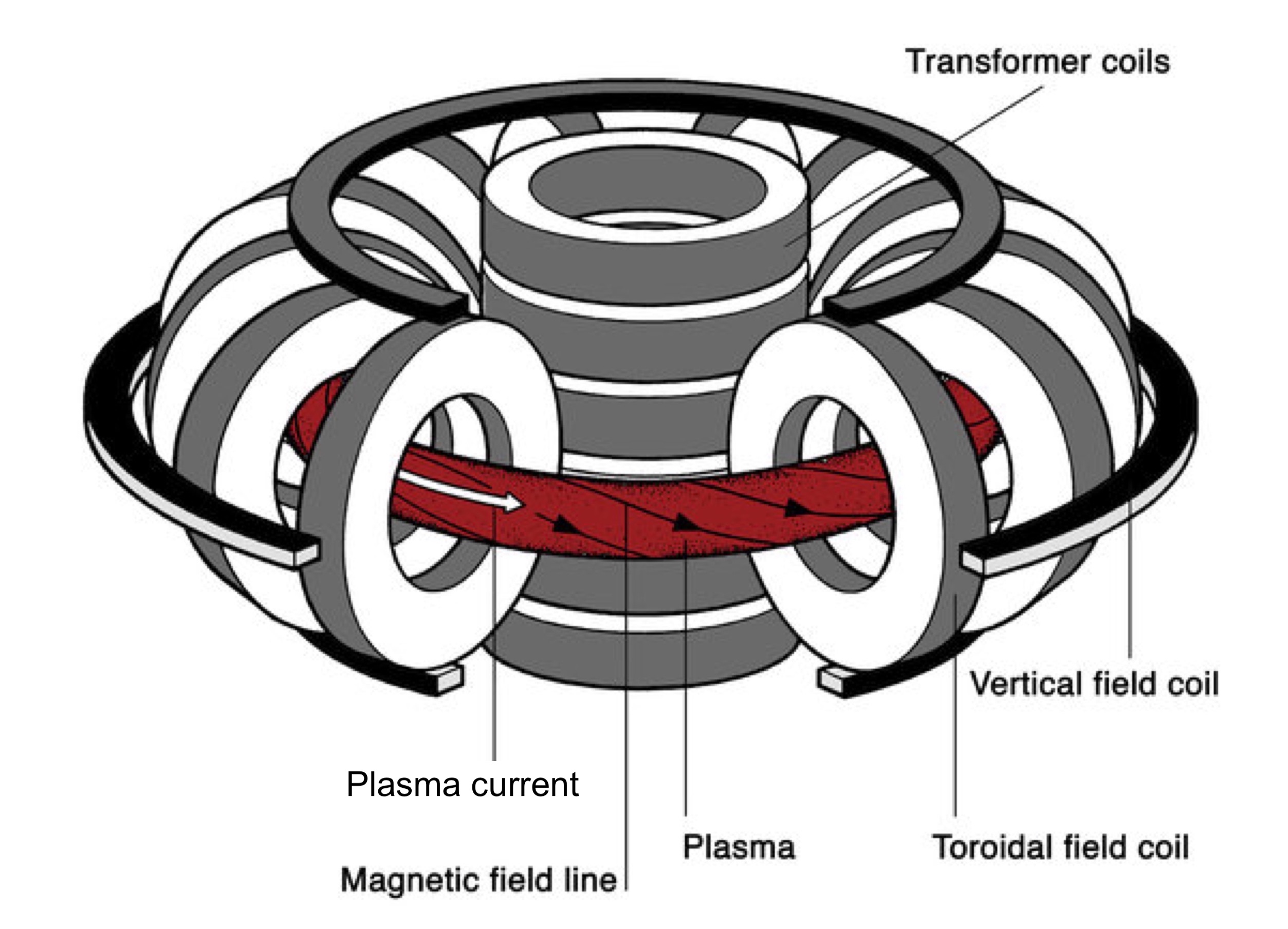}
\caption{The poloidal magnetic field of a tokamak is produced by toroidal plasma current, which is inductively driven by transformer coils. The toroidal field is produced by electromagnetic coils which are planar curves. Figure reproduced from \cite{ippweb}.}
\label{tokamak}
\end{center}
\end{figure}

The poloidal magnetic field of a tokamak is produced by the toroidal plasma current. We will see in Section \ref{sec:producing_rotational_transform} that this plasma current is \textit{required} to produce rotational transform in axisymmetry. While there is some self-generated plasma current, it is not sufficient for confinement, so current needs to be driven externally. Often this is done with a transformer through electromagnetic induction. By varying the current through a central transformer coil (see Figure \ref{tokamak}) an electric field is induced in the plasma. As the current through the transformer coil cannot be increased or decreased indefinitely, this cannot be used as a steady-state approach. 
Driving current also requires a significant amount of energy, so some of the energy produced by fusion must be recirculated for current drive in a tokamak reactor (Chapter 3 in \cite{wesson2011}). Many dangerous plasma instabilities are also driven by plasma current. The result of these instabilities is a sudden loss of confinement of the plasma, called a disruption.

Although the need for current has some disadvantages, the tokamak configuration is advantageous because of its simple geometry. The toroidal symmetry ensures that the collisionless particle orbits are confined when accounting for the drifts, as discussed in Section \ref{sec:confinement_axisymmetry}. The toroidal field coils of a tokamak are planar curves that are (relatively) easy to construct in contrast with those of a stellarator (see Figure \ref{tokamak}).

\subsection{Stellarator} 
\label{sec:stellarator}
\begin{figure}
\begin{center}
\includegraphics[width=0.7\textwidth]{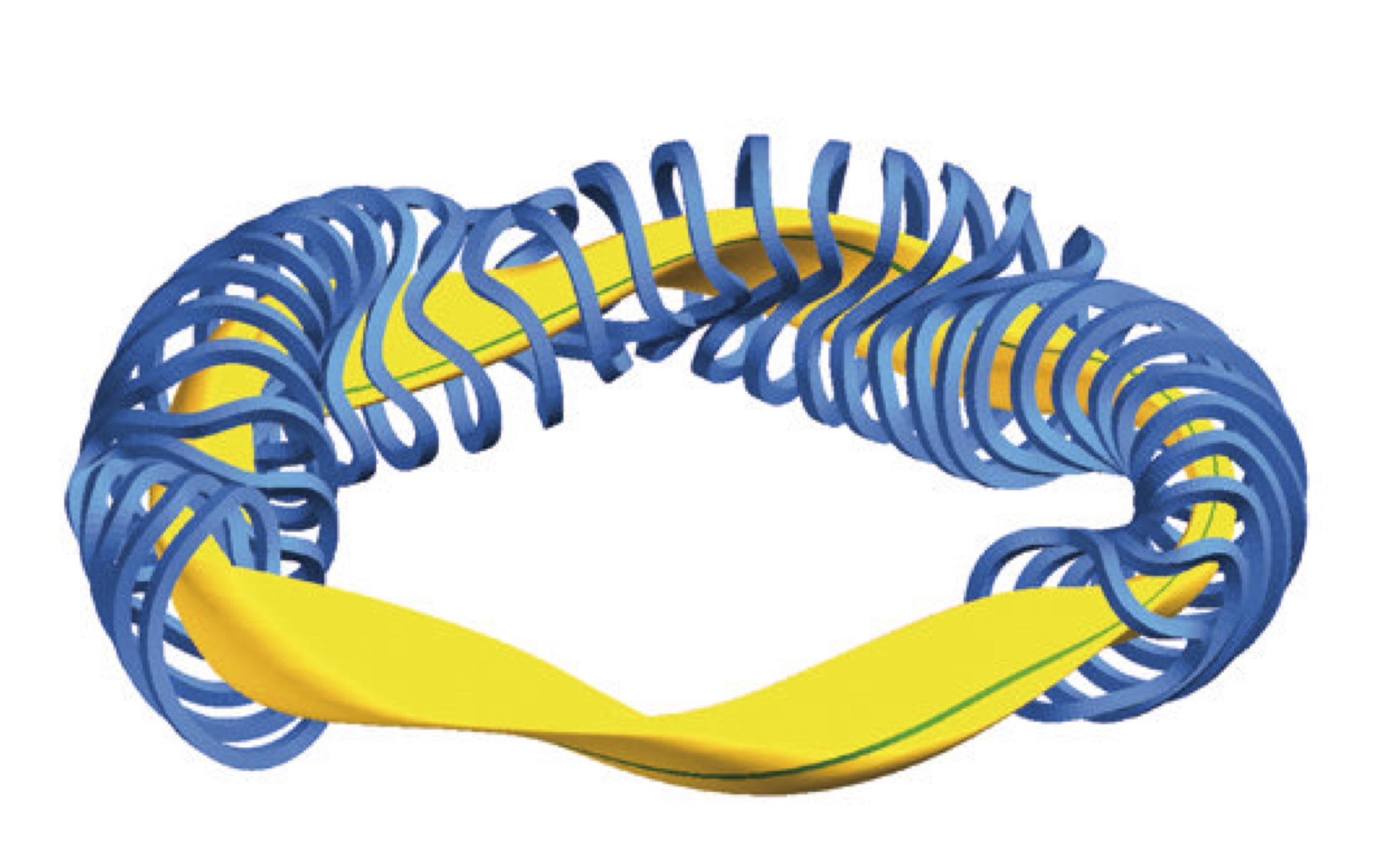}
\caption{A stellarator confines hot plasma with magnetic fields that do not exhibit a continuous toroidal symmetry. The coils of the Wendelstein 7-X device are shown along with the outer magnetic surface (yellow) and a field line (blue). The twisting of the field line around the surface is produced by the rather complicated electromagnetic coils. Figure reproduced from \cite{ippweb2}.}
\label{stellarator}
\end{center}
\end{figure}

While a stellarator is also a toroidal confinement device with genus one topology, unlike a tokamak, it does not exhibit axisymmetry according to the definitions in Section \ref{sec:axisymmetry}. Due to its lack of axisymmetry, plasma current is not required in order to produce rotational transform in a stellarator. The mechanisms by which rotational transform can be generated will be discussed in Section \ref{sec:producing_rotational_transform}. These mechanisms require shaping of the magnetic field structure, which ultimately needs to be produced by external coils. Although stellarators generally do not have externally driven current, there is a small amount of self-driven current that arises due to the average motion of ions and electrons. As stellarator coils have to produce a carefully shaped magnetic field in order to have sufficient rotational transform, stellarator coils tend to be much more complex than tokamak coils (see Figure \ref{stellarator}). This poses a major challenge for stellarator design. Because stellarators do not need driven current they can be run in a steady-state operation, as no inductive electric field is needed. Furthermore, stellarators tend to be more stable than tokamaks to current-driven instabilities. 

\subsection{Rotational transform}
\label{sec:rotation_transform}

As described in Section \ref{confiningMagneitcField}, a poloidal magnetic field is required for confinement, resulting in field lines which twist about flux surfaces. The twist in the field lines is quantified by the rotational transform, $\iota$, which indicates the number of poloidal turns of a field line around the magnetic axis for each toroidal turn around the $\hat{\bm{Z}}$ axis. The rotational transform can be defined in terms of the change in poloidal angle, $(\Delta \theta)_k$, after the $k^{\text{th}}$ toroidal turn along a field line, 
\begin{gather}
    \iota = \lim_{n \rightarrow \infty} \frac{\sum_{k=1}^n (\Delta \theta)_k}{2\pi n}.
\end{gather}
Even if flux surfaces do not exist, the rotational transform can be defined with respect to the poloidal and toroidal rotation of a field line with respect to the magnetic axis.

Often in the tokamak literature the safety factor $q = 1/\iota$ is used rather than the rotational transform. If flux surfaces exist, the rotational transform can also be defined in terms of the fluxes as described in Section \ref{sec:flux_function}.

\begin{figure}
\begin{center}
\begin{subfigure}{0.45\textwidth}
\includegraphics[trim=4cm 10cm 3cm 10cm,clip,width=1.0\textwidth]{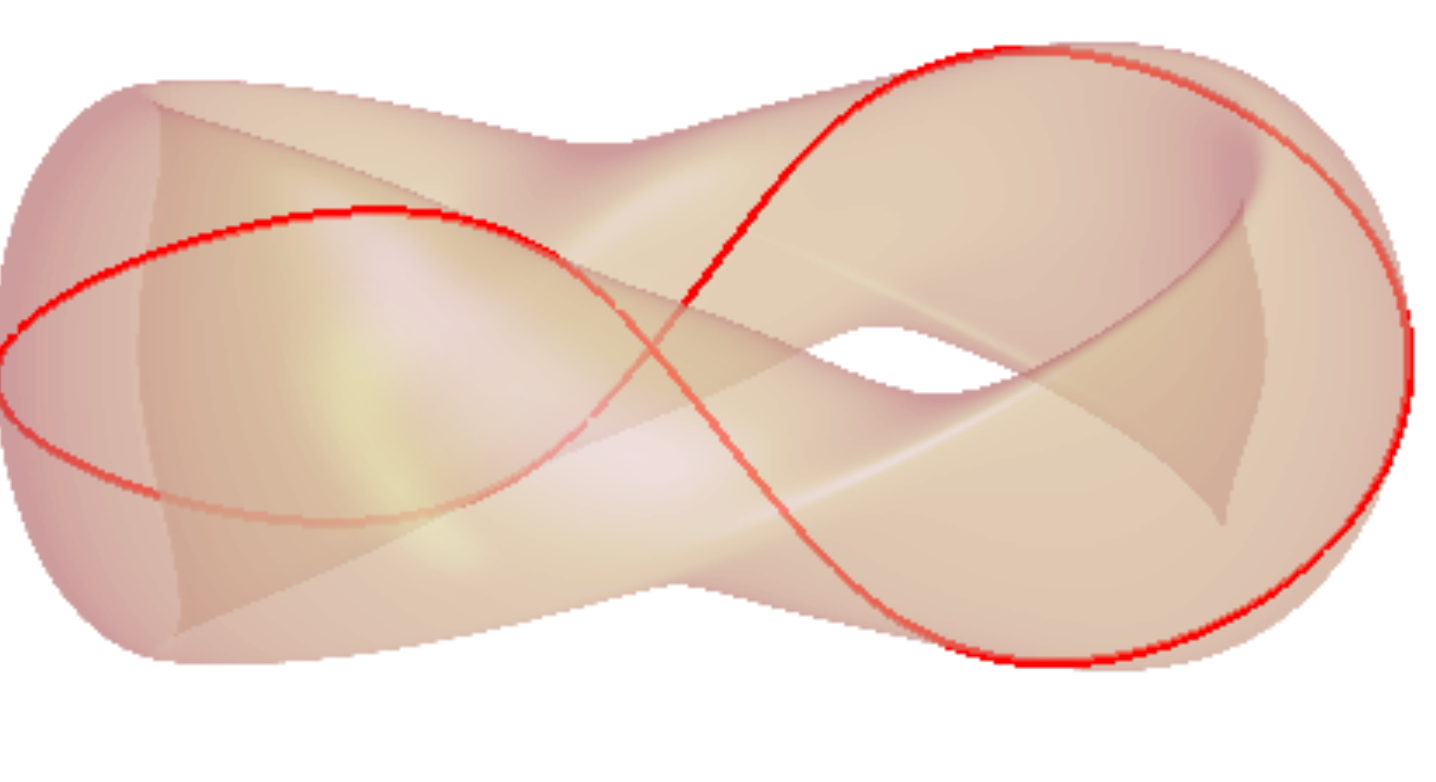}
\caption{$\iota =1$}\label{Figiota1}
\end{subfigure}
\begin{subfigure}{0.45\textwidth}
\includegraphics[trim=4cm 10cm 3cm 10cm,clip,width=1.0\textwidth]{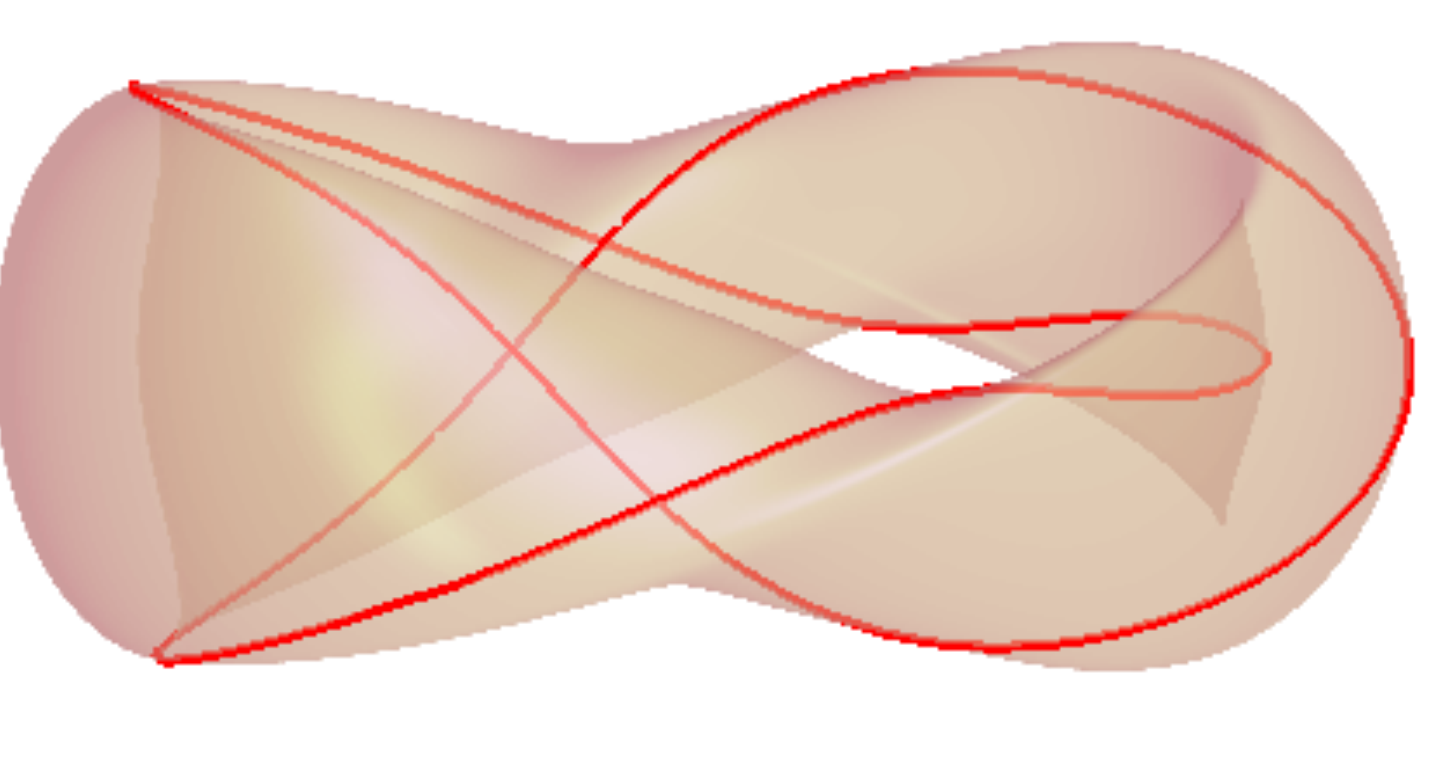}
\caption{$\iota = 3/2$}\label{Figiota.5}
\end{subfigure}
\caption{Two examples of field lines on a toroidal surface. An $\iota=1$ field line (left) makes one poloidal turn for each toroidal turn, while an $\iota = 3/2$ field line (right) makes three poloidal turns for two toroidal turns. As both are rational field lines, they close on themselves after a finite number of toroidal turns.}
\label{fig:rotational_transform}
\end{center}
\end{figure}

When the rotational transform is rational, $\iota = m/n$ for $m \in \mathbb{Z}$ and $n \in \mathbb{Z}$, a field line closes on itself after $n$ toroidal turns, having completed $m$ poloidal turns. Therefore, a single field line does not cover the entire surface (see Figure \ref{fig:rotational_transform}). Flux surfaces with rational values of $\iota$ are known as rational surfaces. When $\iota$ is irrational, a given field line comes arbitrarily close to every point on what is called an irrational surface. For details about the mechanisms by which rotational transform is produced, see Section  \ref{sec:producing_rotational_transform}. 

\subsection{Producing rotational transform}
\label{sec:producing_rotational_transform}

The rotational transform, $\iota$, is the number of poloidal turns of a field line for every toroidal turn of a field line. As described in Section \ref{confiningMagneitcField}, this twist in the field lines is necessary for toroidal confinement. We now explore how rotational transform can be produced with and without the assumption of axisymmetry. 

\subsubsection{With axisymmetry}
\label{sec:axisymmetry_iota}

The most straightforward way to produce rotational transform is with a toroidal plasma current, creating a poloidal magnetic field from Ampere's law \eqref{eq:ampere_magnetostatic}. In fact, this is the \textit{only} possibility under the assumption of axisymmetry. Throughout our discussion we will assume the existence of closed, nested flux surfaces. A further justification of this assumption is provided in Section \ref{sec:field_line_flow}.

To demonstrate this, we first express the magnetic field conveniently in terms of the poloidal flux. In the canonical cylindrical basis $(\hat{\bm{R}},\hat{\bm{\phi}},\hat{\bm{Z}})$, see Section \ref{sec:toroidal_geom}, the magnetic field reads
\begin{gather}
    \bm{B} = B_R \hat{\bm{R}} + B_Z \hat{\bm{Z}} + B_{\phi} \hat{\bm{\phi}}.
    \label{eq:r_z_phi}
\end{gather}
The magnetic field can be separated into a poloidal component, $\bm{B}_P = B_R \hat{\bm{R}} + B_{Z} \hat{\bm{Z}}$, and a toroidal component, $\bm{B}_T = B_{\phi} \hat{\bm{\phi}}$. As $\nabla \cdot \bm{B} = 0$, the magnetic field can be written in terms of a vector potential $\bm{A}$ as $\bm{B} = \nabla \times \bm{A}$. Under the assumption of axisymmetry the vector potential expressed in the canonical cylindrical coordinates satisfies $\partial A_{Z}/\partial \phi =0$, $ \partial A_{R}/\partial \phi = 0$, $ \partial A_{\phi}/\partial \phi = 0$, so in particular the first two terms in \eqref{eq:r_z_phi}, comprising the poloidal magnetic field, can be expressed in terms of the toroidal component of $\bm{A}$,
\begin{gather}
    \bm{B}_P(R,Z,\phi) = -\partder{A_{\phi}(R,Z)}{Z} \hat{\bm{R}}(\phi)  + \frac{1}{R} \partder{(R A_{\phi}(R,Z))}{R}\hat{\bm{Z}}.
    \label{eq:B_pol}
\end{gather}
As a result, since $\bm{B} \cdot \nabla (R A_{\phi}) = 0$, the quantity $(R A_{\phi})$ is a flux function.

More precisely, in order to relate $RA_\phi$ to the poloidal flux, consider a surface of constant $RA_\phi$, whose intersection with the horizontal plane $Z=0$ is made of two distinct curves. We denote by $C_{\psi_P}$ the one standing further away from the $\hat{\bm{Z}}$ axis. Consider the surface $S_P$ lying in the $Z = 0$ plane and bounded on one side by the magnetic axis, denoted $C_0$, and on the other side by $C_{\psi_P}$. Both of these curves are circular under the assumption of axisymmetry. The setting is illustrated in Figure \ref{fig:poloidal_flux}. The poloidal flux through $S_P$ can be computed thanks to Stokes' theorem  as
\begin{equation*}
    2\pi\psi_P = \int_{S_P} \bm{B} \cdot \hat{\bm{n}} \, d^2 x = \int_{C_{\psi_P}} \bm{A} \cdot d \bm{l}  -\int_{C_0} \bm{A} \cdot d \bm{l}. 
\end{equation*}
The two curves can be parameterized by $\phi$ so that in both integrals $\bm{A} \cdot d \bm{l} = R A_{\phi} \, d \phi$,
\begin{equation}
   2\pi\psi_P
    = 2\pi \left( (R A_{\phi})\rvert_{C_{\psi_P}}  - (R A_{\phi})\rvert_{C_{0}} \right).
\end{equation}
 As a result $(RA_{\phi})$ is the poloidal flux function $\psi_P$ up to an additive constant $(R A_{\phi})\rvert_{C_{0}}$, and therefore the poloidal magnetic field can be expressed in terms of $\psi_P$,
\begin{gather}
    \bm{B} = -\nabla \phi \times \nabla \psi_P + B_{\phi} \hat{\bm{\phi}}, 
    \label{eq:poloidal_axisymmetry}
\end{gather}
by expressing \eqref{eq:B_pol} in terms of $\psi_P$ and noting that $\nabla \phi = (1/R) \hat{\bm{\phi}}$.

We now consider the implications of the above expression. We compute the toroidal current through a surface $S_T(\psi_P)$ at constant $\phi$ bounded by a surface at constant $\psi_P$. This surface integral can be computed using Ampere's law \eqref{eq:ampere_magnetostatic},
\begin{align}
    I_T(\psi_P) &= \int_{S_T(\psi_P)} \bm{J} \cdot \hat{\bm{n}} \, d^2 x = \frac{1}{\mu_0} \oint_{\partial S_T(\psi_P)} \bm{B} \cdot d \bm{l}.
    \label{eq:IT_axisymmetry}
\end{align}
The line integral is explained in more detail in Section \ref{sec:magnetic_covariant}. We define a coordinate $l(R,Z)$ to parameterize the length along $\partial S_T(\psi_P)$. 

The integrand of the above is computed to be,
\begin{align}
   \bm{B} \cdot d\bm{l} = \bm{B} \cdot \partder{\bm{r}}{l} d l &= \left(- \frac{1}{R}\partder{\psi_P}{Z} \partder{l}{R} + \frac{1}{R}\partder{\psi_P}{R} \partder{l}{Z}\right) d l = \left(\nabla l \times \nabla \psi_P \cdot \nabla \phi\right) d l. 
\end{align}
By construction, the quantity $\nabla l \times \nabla \psi_P \cdot \nabla \phi$ does not vanish within the integrand, as $\nabla l$ points along the contour while $\nabla\psi_P$ is perpendicular to the contour within the surface of constant $\phi$. Thus the toroidal current $I_T(\psi_P)$ is non-zero from \eqref{eq:IT_axisymmetry}.

We conclude that, under the assumption of axisymmetry, closed nested flux surfaces can only exist if there is a non-zero toroidal current in the confinement region. Moreover, $\psi_P$ cannot be constant within a volume, as this implies that the poloidal magnetic field vanishes from \eqref{eq:poloidal_axisymmetry} but a purely toroidal field was ruled out for toroidal confinement in Section \ref{confiningMagneitcField}. 

Finally, if the rotational transform \eqref{eq:iota} $\iota = d \Psi_P(\Psi_T)/d \Psi_T$ is  non-zero in a volume then $\psi_P$ is non-constant in that volume and the toroidal current $I_T$ is non-zero from \eqref{eq:IT_axisymmetry}. Therefore, in order to produce rotational transform in an axisymmetric system with flux surfaces, a toroidal current is necessary. Thus tokamaks require current in the confinement region. The consequences are discussed further in Section \ref{sec:tokamak}.

\subsubsection{Without axisymmetry}

Surprisingly, rotational transform can also be produced without current in the confinement region. A classic result of Mercier \cite{Mericer1964,Helander2014}, using an asymptotic expansion of the magnetic field near the magnetic axis demonstrates the mechanisms by which rotational transform can be generated.

We will use a Frenet-Serret coordinate system to discuss this result. We define orthonormal unit vectors, at any point $P\in\mathbb{R}^3$: 
\begin{itemize}
    \item $\hat{\bm{e}}_1^P = \hat{\bm{b}}(P)$, the unit tangent vector in the direction of the magnetic field;
    \item $\hat{\bm{e}}_2^P =
\frac{\bm{\kappa}(P)}{|\bm{\kappa}(P)|}
$, the unit vector in the direction of the magnetic curvature $\bm{\kappa}(P) =( \hat{\bm{b}}(P) \cdot \nabla )\hat{\bm{b}}(P)$;
    \item $\hat{\bm{e}}_3^P = \hat{\bm{e}}_1^P \times \hat{\bm{e}}_2^P$. 
\end{itemize} 
 
\begin{figure}
    \centering
    \includegraphics[width=0.7\textwidth,trim=3cm 10cm 3cm 7cm,clip]{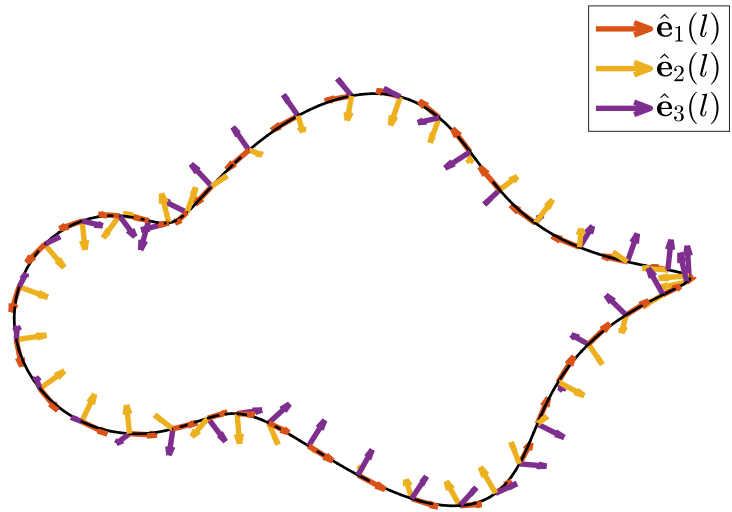}
    \caption{The magnetic axis of the TJ-II stellarator (black) is displayed with the orthonormal Frenet-Serret unit vectors.}
    \label{fig:Frenet_Serret}
\end{figure}

The magnetic axis can be defined by a line $\bm{r}_0(l)$, where $l$ is a parameter which measures distance along the curve.
Consider the Frenet-Serret basis defined along the magnetic axis $(\hat{\bm{e}}_1(l),\hat{\bm{e}}_2(l),\hat{\bm{e}}_3(l))$. The basis vectors $\hat{\bm{e}}_2$ and $\hat{\bm{e}}_3$ define a plane perpendicular to the magnetic axis. 
These vectors are related in the following way,
\begin{gather}\label{eq:torsion}
\der{\hat{\bm{e}}_3(l)}{l} = - \tau(l) \hat{\bm{e}}_2(l),
\end{gather}
where $\tau$ is the torsion of the magnetic axis. The torsion of a planar curve vanishes at all points, and $\tau$ can be thought of as a  measure of the non-planarity of the magnetic axis. The Frenet-Serret unit vectors on the magnetic axis of the TJ-II stellarator are shown in Figure \ref{fig:Frenet_Serret}.

As a result of the calculation which is discussed 
in detail in Section \ref{sec:axis_expansion}, we find that the cross-sections of the flux surfaces in the $\hat{\bm{e}}_2-\hat{\bm{e}}_3$ plane form ellipses. The angle of the major axis of the ellipse with respect to $\hat{\bm{e}}_2$ is denoted by $\delta(l)$. In order to have non-zero rotational transform in the absence of plasma current, either a non-zero value of $\delta'(l)$ or $\tau(l)$ are required (the functional dependence of the rotational transform on $\tau(l)$ and $\delta'(l)$ is given in \eqref{eq:iota_axis}). Both of these mechanisms require breaking of toroidal symmetry as defined in Section \ref{sec:axisymmetry}.

This idea was the basis for the first stellarator designed by Spitzer \cite{Spitzer1958}, which featured a magnetic axis shaped as a Figure-eight to produce torsion of the magnetic axis. Most stellarators use both torsion and ellipticity to produce rotational transform.

\FloatBarrier
\section{Coupling of particles and electromagnetic fields: MHD models}\label{coupling}
Magnetohydrodynamics (MHD) is a one-fluid model of plasma: it couples Maxwell's equations with a fluid model for the particles. It describes the plasma by a mass density $\rho$, center of mass fluid velocity $\bm{u}$, species-summed pressure $p$, and current density $\bm{J}$. Although many MHD models exist, in this document we will focus our attention on the ideal MHD model, as it provides a relatively simple, computationally tractable set of equations. This will be discussed in Section \ref{MHD}. Section \ref{sec:flux_freezing} describes how the magnetic topology is preserved by ideal MHD. Our ultimate goal is to understand the ideal equilibrium limit of these equations, see Section \ref{sec:mhd_eq}, such that the fluid flow and time dependence vanish. The axisymmetric MHD equilibrium is presented in Section \ref{sec:grad-shafranov}, before a more detailed discussion of 3D MHD models presented in Sections \ref{sec:3D_difficulties} and \ref{sec:equilibrium_fields}.

The MHD model can be used to describe the large-scale, global behavior of fusion plasmas. Although the ideal MHD model is derived under a set of approximations, described in Section \ref{sec:mhd_eq}, which are not necessarily valid in a fusion plasma, the ideal MHD stability predictions are more conservative than kinetic models (Chapter 2 in \cite{Freidberg2014}). A discussion on the limitations of MHD for equilibrium calculations is provided in Section \ref{sec:departures_MHD}. 

The set of PDEs which comprise the MHD model can be applied to the confinement region as well as the vacuum region surrounding the plasma.

\subsection{Ideal MHD}
\label{MHD}

We now outline the equations satisfied by the fluid mass density $\rho$, current density $\bm{J}$, and flow velocity $\bm{u}$ under the ideal MHD model. Mass conservation is ensured by the continuity equation,
\begin{gather}
\partder{\rho}{t} + \nabla \cdot \left( \rho \bm{u} \right) = 0. 
\label{eq:density_MHD}
\end{gather}
Momentum density obeys a similar conservation equation,
\begin{gather}
\rho \left( \partder{}{t} + \bm{u} \cdot \nabla \right) \bm{u} = \bm{J} \times \bm{B} - \nabla p,
\label{eq:force_balance_MHD}
\end{gather}
and the entropy conservation equation is,
\begin{gather}
\left(\partder{}{t} + \bm{u} \cdot \nabla \right)\left( \frac{p}{\rho^{\gamma}} \right) = 0.\label{eq:ideal_mhd_energy}
\end{gather}
Here $\gamma$ is the ratio of specific heats ($\gamma = 5/3$ for a monatomic gas system). 

The fields, $\bm{E}$ and $\bm{B}$, obey Maxwell's equations (see Section \ref{sec:maxwell}). Here the displacement current term, ($\partial \bm{E}/\partial t$) can be dropped for non-relativistic plasmas, 
\begin{subequations}
\begin{align}
\nabla \times \bm{B} &= \mu_0 \bm{J} \label{eq:curlB}\\
\nabla \times \bm{E} &= -\partder{\bm{B}}{t} \label{eq:maxwell} \\
\nabla \cdot \bm{B} &= 0. \label{divB}
\end{align}
\label{eq:maxwell_MHD}
\end{subequations}

In ideal MHD, we furthermore make the approximation that the electric field in a frame moving with the fluid is zero,
\begin{gather}
\bm{E} + \bm{u} \times \bm{B} = 0. 
\label{eq:ideal_ohms_law}
\end{gather}
This comes from the assumption that the plasma is perfectly conducting. Non-ideal MHD models include additional effects, such as resistivity which results from friction between electrons and ions. Resistive effects are needed to allow for changes in the magnetic topology, known as magnetic reconnection. In Section \ref{sec:flux_function} we will discuss the result that ideal MHD does not allow for changes in topology. 

The MHD equations can be considered to be a set of PDEs for $\rho$, $\bm{u}$, $p$, and $\bm{B}$, while $\bm{E}$ is determined from $\bm{u}$ and $\bm{B}$, and $\bm{J}$ is determined from $\bm{B}$. This model is derived under a series of assumptions, which are explained more completely in many references \cite{Nicholson1983,Freidberg2014,Goedbloed2004}. Specifically, the MHD equations can be obtained from kinetic or fluid models under certain limits. 
\begin{itemize}
\item Collisions are very strong, such that the electron and ion temperature have equilibrated. 
\item The gyroradius is small compared to wavelengths of interest.
\item The system is non-relativistic, such that the displacement current can be dropped from Ampere's law.
\item Frequencies faster than the electron plasma frequency, $\omega_{pe} = \sqrt{n_e e^2/(\mu_0m_e)}$, are not included.
\end{itemize}
The region of validity of ideal MHD is described in more detail in Chapter 2 of \cite{Freidberg2014}.

These assumptions have the following consequences.
\begin{itemize}
\item Since the displacement current is dropped from Ampere's law, MHD does not include light waves. 
\item As this is a fluid model, it does not account for velocity space effects such as particle trapping. 
\item As high frequency and small wavelengths are neglected, MHD describes the macroscopic, low-frequency behavior of plasmas.
\item An important result of ideal MHD is the frozen-in theorem, which states that the magnetic field is frozen into the fluid and must move with it. This will be discussed in Section \ref{sec:flux_freezing}.
\end{itemize}

\subsection{Flux freezing}
\label{sec:flux_freezing}
An important consequence of the ideal MHD equations is Alfv\'en's flux freezing theorem. This states that the magnetic flux through an open surface moving with an ideal MHD plasma does not change in time. Stated another way, magnetic field lines move with an ideal MHD plasma. As we will see, a consequence is that ideal MHD does not allow for changes in magnetic topology.

Consider the magnetic flux through a surface, $S$, 
\begin{align}
    \Phi_S = \int_S \bm{B} \cdot \hat{\bm{n}} \, d^2 x.
\end{align}
We will compute the change in $\Phi_S$ as the surface moves with the plasma and the field evolves according to Maxwell's equations,
\begin{align}
    \der{\Phi_S}{t} = \int_{S(t)} \partder{\bm{B}}{t} \cdot \hat{\bm{n}} \,  d^2 x  - \oint_{\partial S(t)} \bm{u} \times \bm{B} \cdot d \bm{l}.
\end{align}
The first term corresponds to the change in magnetic field at fixed position, while the second corresponds to the motion of the surface with velocity $\bm{u}$. For a proof of the above, we reference to Appendix C in \cite{Freidberg2014} and Chapter 4 in \cite{Goedbloed2004}. Applying \eqref{eq:ideal_ohms_law} and Stokes' theorem we obtain
\begin{align}
    \der{\Phi_S}{t} = \int_{S(t)} \hat{\bm{n}} \cdot \left( \partder{\bm{B}}{t} + \nabla \times \bm{E}  \right)\,  d^2 x. 
\end{align}
Thus from \eqref{eq:maxwell} we obtain $d \Phi_S/dt = 0$: the magnetic flux is conserved in the frame moving with the plasma. 

\begin{figure}
\centering
\includegraphics[trim=16cm 11cm 14cm 11cm, clip,width=0.8\textwidth]{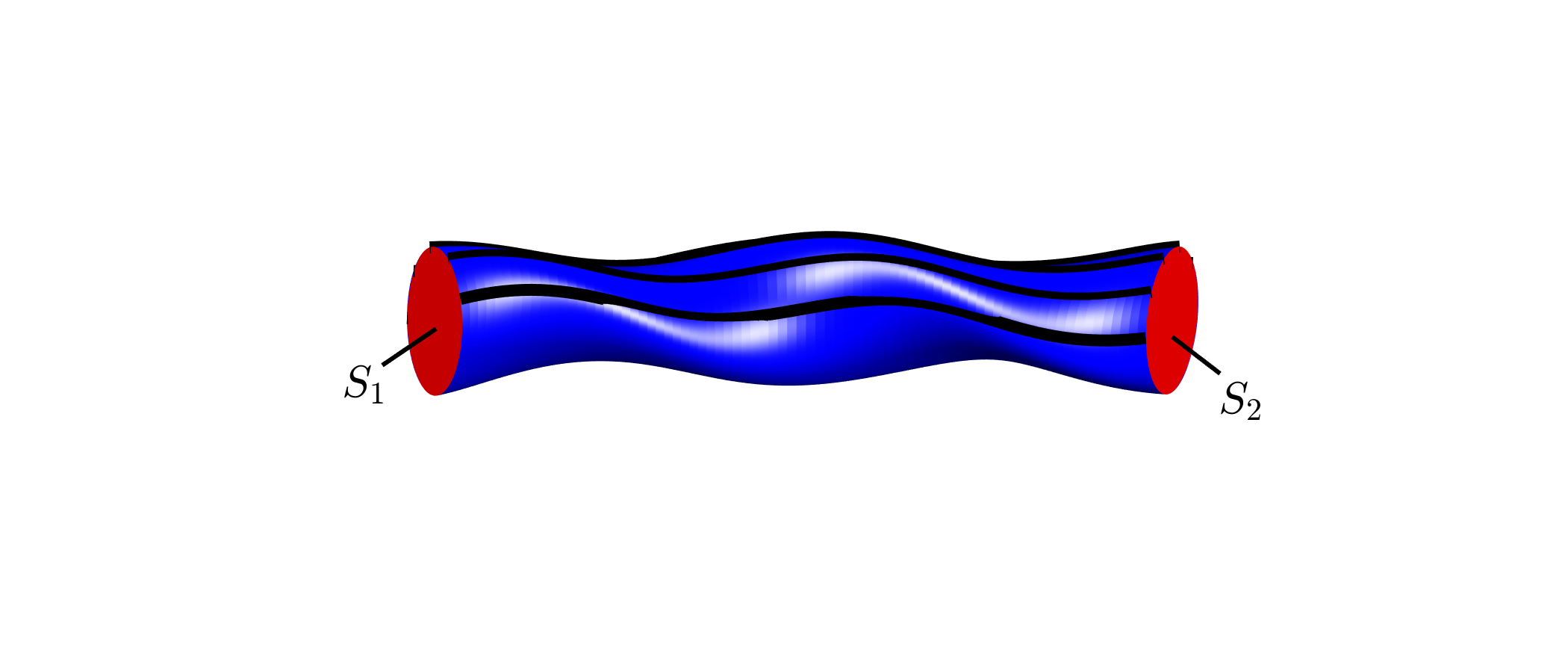}
\caption{A flux tube is shown (blue) on which field lines lie (black). The magnetic flux through the flux tube can be computed with any any surface as shown in \eqref{eq:flux_tube}.}
\label{fig:FluxTube}
\end{figure}

Consider a flux tube, a volume such that magnetic field lines lie tangent to its boundary surface. Consider two non-intersecting surfaces $S_1$ and $S_2$ cutting through the flux tube, as represented in Figure \ref{fig:FluxTube}, and the volume $V$ of the flux tube delimited by $S_1$ and $S_2$. We now demonstrate that the flux through a surface slicing through the flux tube, $\Phi_S$, is independent of the choice of surface. 
Since $\nabla\cdot \bm{B}=0$ from \eqref{divB}, then Gauss's law yields
\begin{align}
    0=\int_V \nabla \cdot \bm{B} \, d^3 x = \int_{S_1} \bm{B} \cdot \hat{\bm{n}} \, d^2 x - \int_{S_2} \bm{B} \cdot \hat{\bm{n}} \, d^2 x ,
    \label{eq:flux_tube}
\end{align}
where the unit normals, $\hat{\bm{n}}$ are chosen to be oriented in the same direction for both surface integrals. There is no contribution from the sides of the flux tube, as the field lines are tangent to this surface. As it holds for any choice of bounding surfaces, then for a given flux tube we conclude that $\Phi_S$ across a surface $S$ cutting through the flux tube is independent of the choice of surface.

\begin{figure}
\centering
\begin{subfigure}{0.49\textwidth}
\includegraphics[trim=36cm 11cm 36cm 9cm, clip,width=1.0\textwidth]{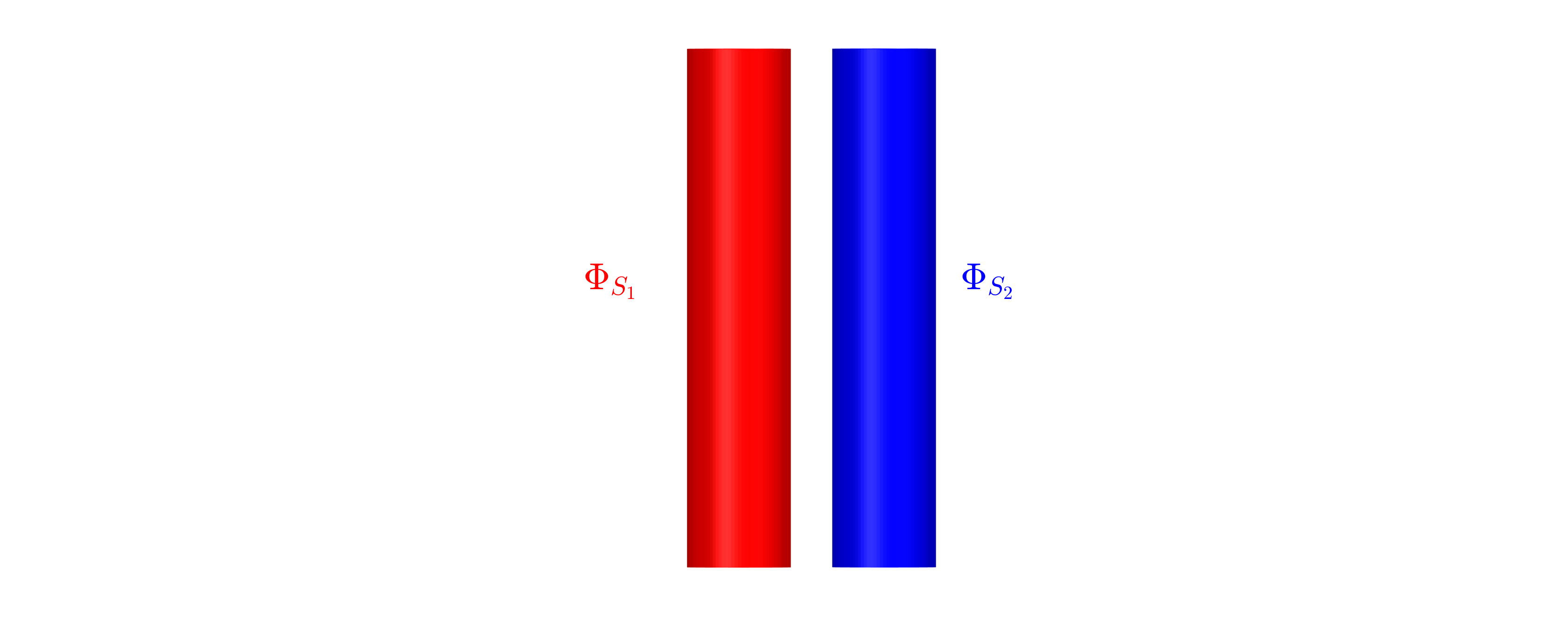}
\caption{Two flux tubes which are initially adjacent cannot merge under ideal MHD evolution, as their fluxes, $\Phi_{S_1}$ and $\Phi_{S_2}$, must be preserved.}
\label{fig:merge_flux}
\end{subfigure}
\begin{subfigure}{0.49\textwidth}
\includegraphics[trim=36cm 11cm 36cm 9cm, clip,width=1.0\textwidth]{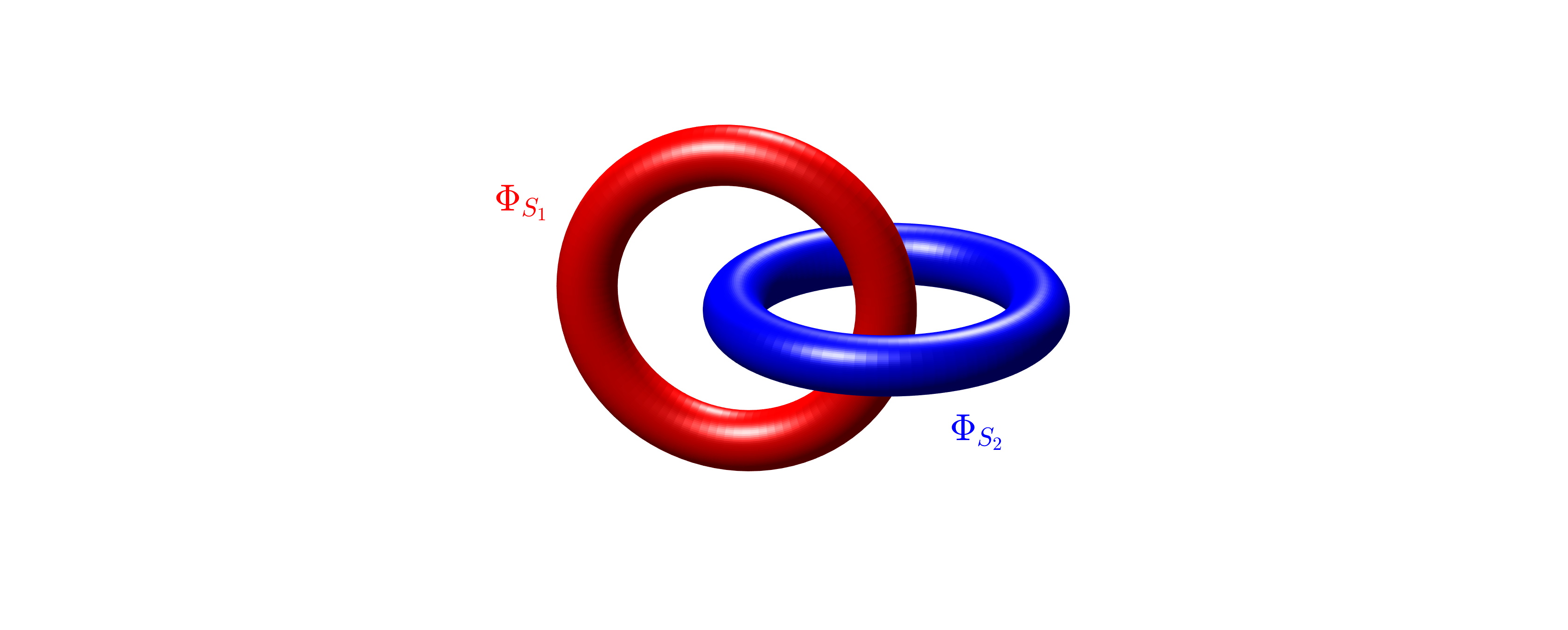}
\caption{Two flux tubes which are initially liked cannot unlike under ideal MHD evolution, as their fluxes, $\Phi_{S_1}$ and $\Phi_{S_2}$, must be preserved.}
\label{fig:linked_flux}
\end{subfigure}
\end{figure}

Furthermore, the flux-freezing theorem implies that the flux through a given flux tube, $\Phi_S$, does not change with time. This leads to the well-known result that ideal MHD plasmas do not allow changes in topology. A first example consists of two neighboring flux tubes with fluxes $\Phi_1$ and $\Phi_2$: under ideal MHD evolution, it is impossible for these two flux tubes to merge into one, as this would result in a single flux tube with flux $\Phi_1 + \Phi_2$ (see Figure \ref{fig:merge_flux}). A second example consists of two flux tubes that are initially linked to each other: as the flux through each tube must be preserved, the tubes cannot break and must remain linked (see Figure \ref{fig:linked_flux}). This can apply to the topology of a field line itself: we can imagine shrinking a flux tube down to a single field line, then the number of times this field line links another field line cannot change under ideal MHD evolution. It is in this sense that the topology of magnetic field lines is fixed under ideal MHD evolution. These considerations will be discussed in the application to 3D equilibrium models in Section \ref{sec:equilibrium_fields}.

\subsection{Ideal MHD equilibrium}
\label{sec:mhd_eq}
We now consider the ideal MHD equations under  the assumption of static ($\bm{u} = 0$) equilibrium ($\partial/\partial t = 0$). This limit is of interest when considering time scales longer than MHD time scales and when typical flow velocities, $|\bm{u}|$, are expected to be smaller than the sound speed, $c_s = \sqrt{\gamma p/\rho}$.

The conservation of momentum density \eqref{eq:force_balance_MHD} shows that the plasma pressure gradient balances electromagnetic forces,
\begin{gather}
\bm{J} \times \bm{B} = \nabla p. 
\label{equilibrium}
\end{gather}
The equilibrium fields must also satisfy,
\begin{subequations}
\begin{align}
\nabla \times \bm{B} &= \mu_0 \bm{J} \label{eq:equilibrium1} \\
\nabla \cdot \bm{B} &= 0.
\label{eq:equilibrium2}
\end{align}
\label{eq:equilibrium}
\end{subequations}
For magnetic confinement fusion, the resulting non-linear system of PDEs (\ref{equilibrium})-(\ref{eq:equilibrium}) is often solved in a toroidal domain $\Omega$. 

The force balance condition \eqref{equilibrium} is often considered under the assumption that $p(\psi)$, where $\psi$ is a flux surface label (see Section \ref{sec:flux_function}). Under these assumptions, \eqref{equilibrium} implies that $\bm{J} \cdot \nabla \psi = 0$, such that streamlines of both the current density and magnetic field line on surfaces of constant $\psi$.

The resulting equations in axisymmetry are described in the following Section \ref{sec:grad-shafranov}, and further details about computing 3D MHD equilibria will be discussed in Sections \ref{sec:3D_difficulties} and \ref{sec:equilibrium_fields}.

\subsection{MHD equilibrium in a tokamak: Grad-Shafranov}
\label{sec:grad-shafranov}

We now consider the MHD equilibrium equations under the assumption of axisymmetry (see Section \ref{sec:tokamak}). We will find that \eqref{equilibrium}-\eqref{eq:equilibrium} can be reduced to a two-dimensional, non-linear PDE. 

We begin with a convenient expression for the magnetic field in terms of the poloidal flux function $\psi_P$, discussed in Section \ref{sec:axisymmetry_iota},
\begin{gather}
    \bm{B} = -\nabla \phi \times \nabla \psi_P + B_{\phi} \hat{\bm{\phi}},
\end{gather}
and the current can be computed using Ampere's law \eqref{eq:equilibrium1},
\begin{align}
    \mu_0 \bm{J} &= -R\nabla \cdot \left(\frac{\nabla \psi_P}{R^2}\right) \hat{\bm{\phi}} +   \nabla \left(R B_{\phi}\right) \times \nabla \phi.
\end{align}
Here the vector identity $\nabla \cdot \left(\bm{a}\times\bm{b} \right) = \bm{b} \cdot \nabla \times \bm{a}-\bm{a} \cdot \nabla \times \bm{b}$ is used to compute the toroidal component of $\bm{J}$. We next note that the toroidal component of the magnetic field can be expressed in terms of the poloidal current, $I_P(\psi_P)$. Here $2\pi I_P(\psi_P)$ is the  current through a surface, $S_P(\psi_P)$, lying in the plane $Z=0$ and bounded by the $\psi_P$ surface,
\begin{align}
2\pi I_P(\psi_P) &= \int_{S_P(\psi_P)} \bm{J} \cdot \hat{\bm{n}} \, d^2 x \nonumber \\
&= \int_0^{2\pi} \int_0^{R(\psi_P)} R \bm{J} \cdot \hat{\bm{Z}} \, dR d \phi \nonumber \\
&= 2\pi R B_{\phi}.
\end{align}
The surface $S_P(\psi_P)$ used to computed the poloidal current is described in Figure \ref{fig:poloidal_current}. Therefore, in a tokamak, the vector unknowns, $\bm{B}$ and $\bm{J}$, can be expressed in terms of the scalar unknown, $\psi_P(R,Z)$,
\begin{subequations}
\begin{align}
\bm{B} &=   -\nabla \phi \times \nabla \psi_P +I_P(\psi_P) \nabla \phi \label{eq:tok_b} \\
\bm{J} &= \frac{1}{\mu_0} \left(- R\nabla \cdot \left(\frac{\nabla \psi_P}{R^2} \right) \hat{\bm{\phi}} -\nabla \phi \times \nabla I_P(\psi_P) \right).
\label{eq:tok_J}
\end{align}
\end{subequations}
Given this form for $\bm{B}$ and $\bm{J}$, dotting $\nabla \psi_P$ into the MHD equilibrium force balance equation \eqref{equilibrium} results in the Grad-Shafranov equation,
\begin{gather}
R^2 \nabla \cdot \left( \frac{\nabla \psi_P}{R^2} \right) = - \mu_0 R^2 \der{p(\psi_P)}{\psi_P} - I_P(\psi_P) \der{I_P(\psi_P)}{\psi_P}.
\label{eq:grad_shafranov}
\end{gather}
Often the notation $\Delta^* \psi_P = R^2 \nabla \cdot \left( R^{-2} \nabla \psi_P\right)$ is used. The Grad-Shafranov equation \eqref{eq:grad_shafranov} is solved in a toroidal domain $\Omega\subset\mathbb R^3$, bounded by a flux surface with poloidal flux function $\psi_{P0}\in\mathbb R$. By prescribing the constant Dirichlet boundary condition $\psi_{P}=\psi_{P0}$ on $\partial\Omega$, we fix the boundary to be the flux surface of label $\psi_{P0}$. This is a non-linear elliptic PDE for $\psi_P(R,Z)$, since $\psi_P$ is independent of $\phi$. Thus solving \eqref{eq:grad_shafranov} provides the shape of the flux surfaces $\psi_P(R,Z)$ bounded by the $\psi_{P0}$ surface. Generally, the flux functions $p(\psi_P)$ and $I_P(\psi_P)$ are given. In other words, given the shape of the outer boundary, we want to find the shape of the inner flux surfaces with specified pressure and current profiles. Once $\psi_P$ is determined, the magnetic field is known from (\ref{eq:tok_b}). 

For more details about the Grad-Shafranov equation, see Chapter 7 of \cite{Helander2005} and Chapter 6 of \cite{Freidberg2014}. For a discussion of computational methods for the Grad-Shafranov equation, see Chapter 4 in \cite{Jardin2010}.

\subsection{Summary}
Under various sets of hypotheses, the ideal MHD equations can be reduced to simpler models. Common reduced models are gathered in the following Table.  
{\renewcommand{\arraystretch}{2.0}
\begin{center}
\begin{tabular}{|c|c|c|c|}
\hline
&Ideal MHD & MHD Equilibrium & Grad-Shafranov \\\hline
& 
$\partial \bm{E}/\partial t = 0$
& $\bm{u} = 0$
& MHD equilibrium
\\
Hyp.  & Low frequency/long wavelength
& $\partial/\partial t = 0$
& $\partial/\partial \phi = 0$
\\
&
3D&
3D & 2D ($R$,$Z$)
\\\hline
&  $\partder{\rho}{t} + \nabla \cdot \left( \rho \bm{u} \right) = 0 $ 
& 
& Given $I_P(\psi_P)$ and $p(\psi_P)$,
\\ PDE&
  $\rho \left( \partder{}{t} + \bm{u} \cdot \nabla \right) \bm{u} = \bm{J} \times \bm{B} - \nabla p$
& $\bm{J} \times \bm{B} = \nabla p$
& $\nabla \cdot \left( \frac{\nabla \psi_P}{R^2} \right) 
= - \mu_0 p'(\psi_P)$
\\model&
  $\left(\partder{}{t} + \bm{u} \cdot \nabla \right)\left( \frac{p}{\rho^{\gamma}} \right) = 0$
& 
& \hspace{1.0cm}$- \frac{I_P(\psi_P)}{R^2}  I_P'(\psi_P)$
\\&
  $\nabla \times \bm{B} = \mu_0 \bm{J}$
& $\nabla \times \bm{B} = \mu_0 \bm{J}$
&
\\&
  $\nabla \times \bm{E} = -\partder{\bm{B}}{t}$
& 
&
\\&
  $\nabla \cdot \bm{B} = 0$
& $\nabla \cdot \bm{B} = 0$
& 
\\&
  $\bm{E} + \bm{u} \times \bm{B} = 0$
& 
&
\\\hline
Unkn.&
$\rho$, $\bm{u}$, $p$, $\bm{B}$&
$ \bm{B}$, $p$& $\psi_P(R,Z)$
\\\hline
 with& $\bm{E}$ function of $\bm{B}$, $\bm{u}$
&
& $\bm{B} = I_P(\psi_P) \nabla \phi -\nabla \phi \times \nabla \psi_P$ 
\\
&$\bm{J}$ function of $\bm{B}$
&$\bm{J}$ function of $\bm{B}$
& $\bm{J} = -\frac{1}{\mu_0} \bigg( \nabla \phi \times \nabla I_P(\psi_P) $ \\
& & & \hspace{1cm}$ + R^2 \nabla \cdot \left(\frac{\nabla \psi_P}{R^2}  \right) \nabla \phi \bigg)$
\\\hline
\end{tabular}
\end{center}
}

\FloatBarrier
\section{Magnetic coordinates}
\label{sec:magnetic_coordinates}

As magnetic coordinates are a specific form of flux coordinates, see Section  \ref{sec:flux_coordinates}, we assume again the existence of flux surfaces. To construct magnetic coordinates, it is also assumed that $\nabla \cdot \bm{B} = 0$, which holds for any physical magnetic field. A special property of magnetic coordinates with poloidal angle $\vartheta$ and toroidal angle $\varphi$ is that field lines appear straight in the $\vartheta$-$\varphi$ plane. The slope of the field lines is given by the rotational transform,
\begin{gather}
\iota = \frac{\bm{B} \cdot \nabla \vartheta}{\bm{B} \cdot \nabla \varphi} = \frac{d \vartheta}{d \varphi} \bigg \rvert_{\text{along field line}}.
\label{eq:magnetic_coordinates}
\end{gather}

Magnetic coordinates are constructed from general flux coordinates in Section \ref{sec:magnetic_contravariant}, leading to the desired contravariant form. The covariant representation is simplified under the assumption of MHD equilibrium in Section \ref{sec:magnetic_covariant}. A special form of magnetic coordinates, called Boozer coordinates, is introduced in Section \ref{sec:boozer_coordinates}.

\subsection{Contravariant form}
\label{sec:magnetic_contravariant}
Magnetic coordinates can be constructed from a general flux coordinate system $(\psi,\theta,\phi)$ where $\psi = \Psi_T/2\pi$ is the toroidal flux label (see Section \ref{sec:flux_function}), $\theta$ is any poloidal angle, and $\phi$ is any toroidal angle, as follows. As $\bm{B} \cdot \nabla \psi = 0$, the magnetic field can be written in the contravariant form as:
\begin{gather}
    \bm{B} = B^{\phi} \partder{\bm{r}}{\phi}   + B^{\theta} \partder{\bm{r}}{\theta},
\end{gather}
$(B^{\phi},B^{\theta})$ being the toroidal and poloidal contravariant components of the field.
Requiring that $\nabla \cdot \bm{B} = 0$, the following condition must hold, 
\begin{gather}
    0 = \frac{1}{\sqrt{g}} \left( \partder{(B^{\phi}\sqrt{g})}{\phi} + \partder{(B^{\theta}\sqrt{g})}{\theta} \right),
    \label{eq:div_B_magnetic}
\end{gather}
where $\sqrt{g} = \left( \nabla \psi \times \nabla \theta \cdot \nabla \phi\right)^{-1}$. Here $B^{\theta}\sqrt{g}$ and $B^{\phi}\sqrt{g}$ can be written in terms of some functions $A(\psi,\theta,\phi)$, $C(\psi,\theta,\phi)$, $j(\psi)$, and $h(\psi)$ such that,
\begin{subequations}
\label{eq:Bphitheta}
\begin{align}
    B^{\phi}\sqrt{g} &= A(\psi,\theta,\phi) + j(\psi), \\
    B^{\theta}\sqrt{g} &= C(\psi,\theta,\phi) + h(\psi).
\end{align}
\end{subequations}
As each of the contravariant components and the Jacobian must be periodic in $\theta$ and $\phi$, so must $A(\psi,\theta,\phi)$ and $C(\psi,\theta,\phi)$.
Defining the function $\lambda$ as, 
\begin{gather}
\label{eq:lambdaC}
    \lambda(\psi,\theta,\phi) = -\int_0^{\phi} C(\psi,\theta,\phi') \, d \phi',
\end{gather} then from \eqref{eq:div_B_magnetic}, we must have that
\begin{subequations}
\label{eq:AClambda}
\begin{align}
    A(\psi,\theta,\phi) &= \partder{\lambda(\psi,\theta,\phi)}{\theta} \\
    C(\psi,\theta,\phi) &= -\partder{\lambda(\psi,\theta,\phi)}{\phi}.
\end{align}
\end{subequations}
So $\bm{B}$ can be written as,
\begin{align}
    \bm{B} &= \nabla \psi \times \nabla \left(j(\psi)\theta - h(\psi) \phi + \lambda(\psi,\theta,\phi) \right),
    \label{eq:B_magnetic}
\end{align}
using the relation between basis vectors (Table \ref{table:non_orthogonal}) and noting that $\nabla q = \left(\partial q/\partial \theta\right) \nabla \theta + \left(\partial q/\partial \phi\right) \nabla \phi + \left(\partial q/\partial \psi\right) \nabla \psi$ for any $q$.
As we noted that $C(\psi,\theta,\phi)$ and $A(\psi,\theta,\phi)$ are periodic in the two angle, so must $\lambda(\psi,\theta,\phi)$. 

We can now compute the toroidal flux using \eqref{eq:psi_T} and \eqref{eq:B_magnetic},
\begin{gather}
    \Psi_T = 2 \pi \int_0^{\psi} j(\psi') \, d\psi',
\end{gather}
so $j(\psi)= 1$, since by definition of the toroidal flux label we have $\psi = \Psi_T/2\pi$. Similarly for the poloidal flux defined in \eqref{eq:psi_P} we obtain
\begin{gather}
    \Psi_P = 2 \pi \int_0^{\psi} h(\psi')\, d \psi',
\end{gather}
so from the definition of $\iota$ \eqref{eq:iota} we see that $h(\psi) = \iota(\psi)$. We can now define the new magnetic coordinates: let $\vartheta = \theta + \lambda$ and $\varphi = \phi$, then it is clear that 
\begin{gather}
    \bm{B} = \nabla \psi \times \nabla \vartheta - \iota(\psi) \nabla \psi \times \nabla \varphi
    \label{eq:magnetic_contravariant},
\end{gather}
and therefore we obtain the defining characteristic of magnetic coordinates \eqref{eq:magnetic_coordinates}: $\iota(\psi) = \bm{B} \cdot \nabla \vartheta /\bm{B} \cdot \nabla \varphi$, and so the field lines are straight in the coordinates $(\psi,\vartheta,\varphi)$.

\subsection{Covariant form}
\label{sec:magnetic_covariant}

The magnetic field can also be written in the basis of the gradients of the magnetic coordinates ($\psi$,$\vartheta$,$\varphi$), the covariant form,
\begin{gather}
    \bm{B} = B_{\psi} \nabla \psi + B_{\vartheta} \nabla \vartheta + B_{\varphi} \nabla \varphi.
    \label{eq:mag_covar}
\end{gather}
In this Section, we obtain expressions for the covariant components ($B_{\psi}$, $B_{\vartheta}$,$B_{\varphi}$) given the currents linking the plasma. More generally, the analysis in this Section applies not just to magnetic coordinates but to any flux coordinate system, only requiring the absence of radial current.

\begin{figure}
    \centering
    \includegraphics[width=1.0\textwidth]{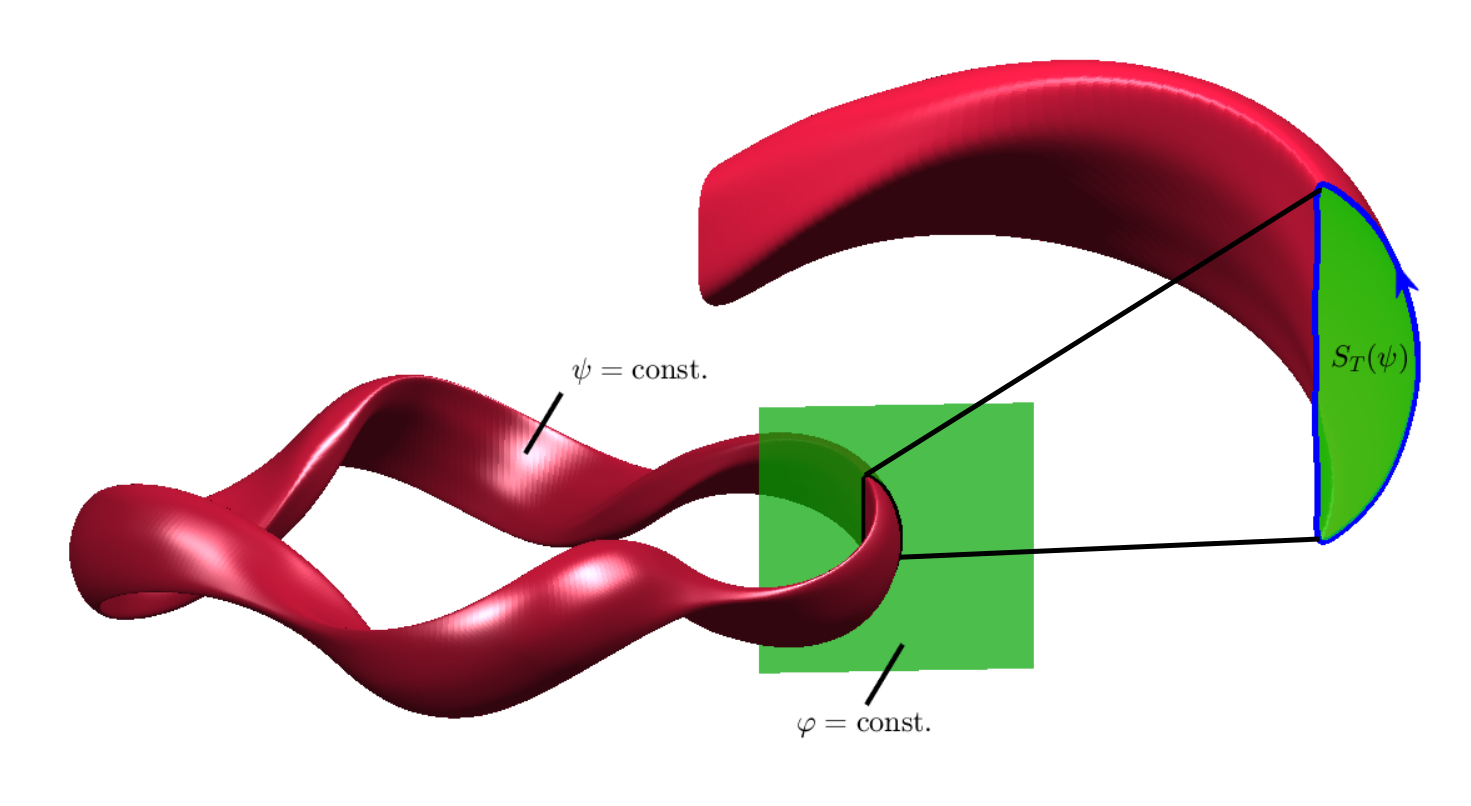}
    \caption{The integration curve at $\varphi =$ const. and $\psi =$ const. (black) encloses the surface $S_T(\psi)$ (green) through which the toroidal current is integrated.}
    \label{fig:toroidal_current}
\end{figure}

To find the expressions for $B_{\vartheta}$, $B_{\varphi}$, and $B_{\psi}$,
we substitute \eqref{eq:mag_covar} into Ampere's law \eqref{eq:ampere_magnetostatic},
\begin{gather}
    \mu_0 \bm{J} = \nabla B_{\vartheta} \times \nabla \vartheta + \nabla B_{\varphi} \times \nabla \varphi + \nabla B_{\psi} \times \nabla \psi.
\end{gather}
Assuming an MHD equilibrium with $p(\psi)$ \eqref{equilibrium} so that $\bm{J} \cdot \nabla \psi=0$ and taking the dot product of $\bm{J}$ with $\nabla\psi$ it follows that,
\begin{align}
    \mu_0 \sqrt{g} \bm{J} \cdot \nabla \psi &=  \partder{B_{\varphi}}{\vartheta} -\partder{B_{\vartheta}}{\varphi} \label{eq:rad_current} = 0. 
\end{align}
Consider a curve at constant $\varphi$ and $\psi$. The surface enclosed by this curve is denoted by $S_T(\psi)$, shown in Figure \ref{fig:toroidal_current}. From Ampere's law \eqref{eq:ampere_magnetostatic},
\begin{align}
   \mu_0 I_T(\psi) =\mu_0 \int_{S_T(\psi)}\bm{J} \cdot \hat{\bm{n}} \, d^2 x &= \oint_{\partial S_T(\psi)} \bm{B} \cdot d \bm{l} =
    \int_0^{2\pi}  \, B_{\vartheta} \, d \vartheta,
    \label{eq:B_theta}
\end{align}
where we have used $d \bm{l} = \left(\partial \bm{r}/\partial \vartheta\right) d \vartheta$. 
The requirement that $I_T$ be a flux function comes from integration of \eqref{eq:rad_current} with respect to $\vartheta$, noting that $B_{\varphi}$ must be periodic with respect to $\vartheta$.

\begin{figure}
    \centering
    \begin{subfigure}{0.7\textwidth}
    \includegraphics[trim=4cm 6cm 5cm 4cm, clip,width=1.0\textwidth]{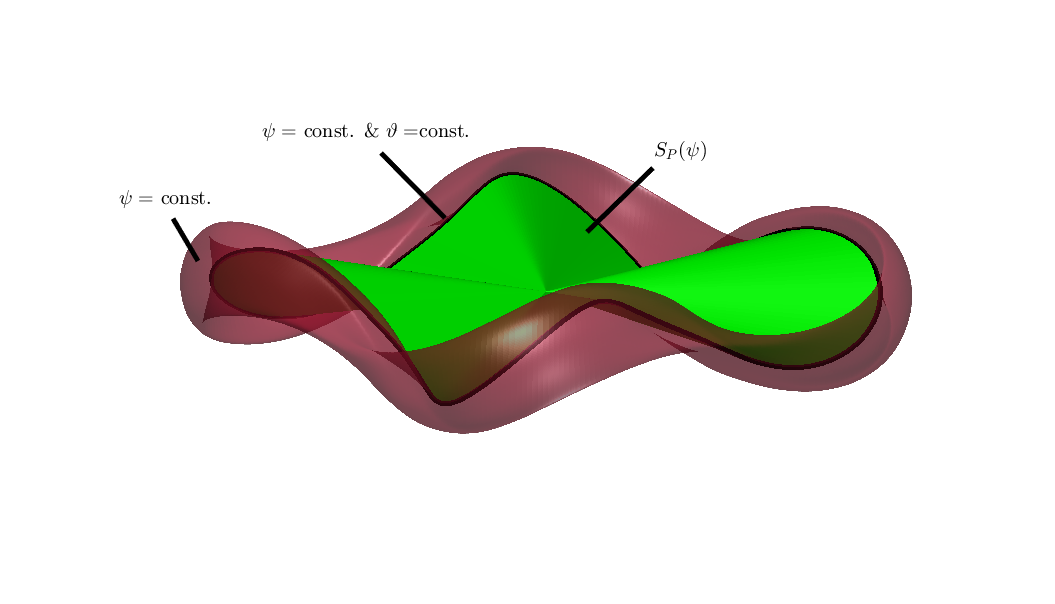}
    \caption{}
    \end{subfigure}
    \begin{subfigure}{0.29\textwidth}
    \includegraphics[trim=14cm 6cm 13cm 5cm, clip,width=1.0\textwidth]{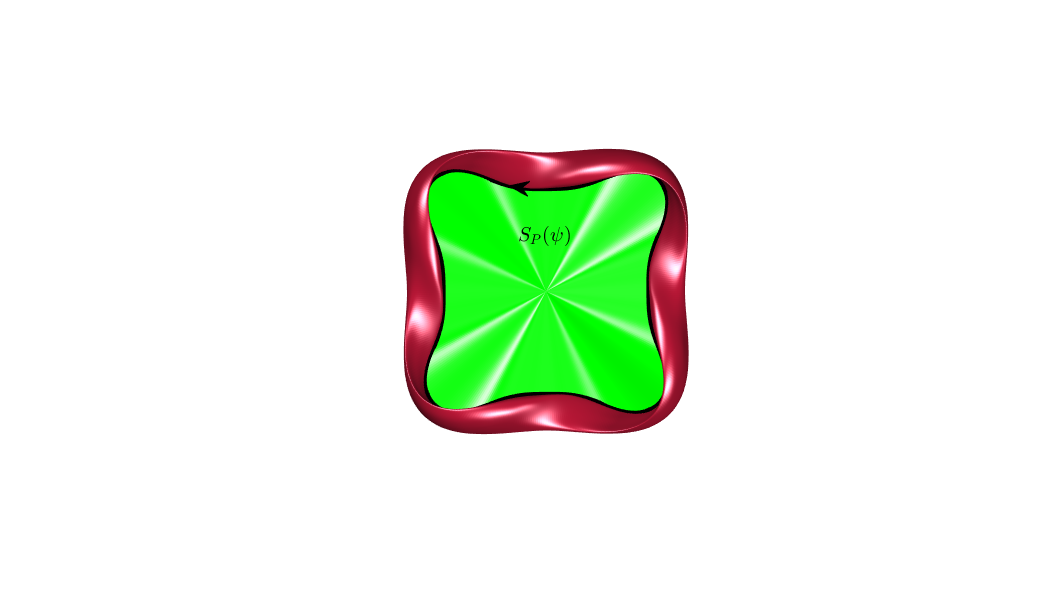}
    \caption{}
    \end{subfigure}
    \caption{The integration curve at $\vartheta =$ const. and $\psi= $ const. (black) encloses the surface $S_P(\psi)$ (green) through which the poloidal current is integrated.}
    \label{fig:poloidal_current}
\end{figure}

Now consider a curve at constant $\vartheta$ and $\psi$. The surface enclosed by this curve is denoted by $S_P(\psi)$, shown in in Figure \ref{fig:poloidal_current}. The current passing through this surface is computed from Ampere's law,
\begin{align}
\mu_0 I_P(\psi) = \mu_0 \int_{S_P(\psi)} \bm{J} \cdot \hat{\bm{n}} \, d^2 x = \oint_{\partial S_P(\psi)} \bm{B} \cdot d \bm{l} =
    \int_0^{2\pi}  B_{\varphi} \, d \varphi,
    \label{eq:B_phi}
\end{align}
where we have used $d \bm{l} = \left(\partial \bm{r}/\partial \varphi \right)d \varphi$. 
 
Here $I_P$ is the poloidal current outside the $\vartheta =$ const. surface, including contributions from the plasma current outside the $\psi$ surface and any coil current that passes through the center of the torus. The requirement that $I_P$ be a flux function comes from integration of \eqref{eq:rad_current} with respect to $\varphi$, noting that $B_{\vartheta}$ must be periodic with respect to $\varphi$. In general, we can write $B_{\varphi}$ and $B_{\vartheta}$ in terms of the flux functions, $I_T(\psi)$ and $I_P(\psi)$,
\begin{subequations}
\begin{align}
B_{\vartheta} &= \frac{\mu_0 I_T(\psi)}{2\pi} + H_1(\psi,\vartheta,\varphi) \\
B_{\varphi} &= \frac{\mu_0 I_P(\psi)}{2\pi} + H_2(\psi,\vartheta,\varphi).
\end{align}
\end{subequations}
Here we have separated out the components of $B_{\vartheta}$ and $B_{\varphi}$ that are averaged over the angles (from \eqref{eq:B_theta} and \eqref{eq:B_phi}) from the components that depend on the angles.  

According to \eqref{eq:rad_current}, we must have
\begin{subequations}
\begin{align}
H_1(\psi,\vartheta,\varphi) &= \partder{H(\psi,\vartheta,\varphi)}{\vartheta} \\
    H_2(\psi,\vartheta,\varphi) &= \partder{H(\psi,\vartheta,\varphi)}{\varphi}.
\end{align}
\end{subequations}
where
\begin{gather}
H(\psi,\vartheta,\varphi) = \int_0^{\vartheta} H_1(\psi,\vartheta',\varphi) \, d\vartheta'. 
\end{gather}
We can furthermore define $K(\psi,\vartheta,\varphi)$ in the following way,
\begin{gather}
    B_{\psi} = K(\psi,\vartheta,\varphi) + \partder{H(\psi,\vartheta,\varphi)}{\psi}.
\end{gather}
So, the covariant form of an equilibrium magnetic field is
\begin{gather}
    \bm{B}(\psi, \vartheta, \varphi) = I(\psi) \nabla \vartheta + G(\psi) \nabla \varphi + K(\psi, \vartheta, \varphi) \nabla \psi + \nabla H(\psi, \vartheta, \varphi),
    \label{eq:magnetic_covar}
\end{gather}
where $G(\psi) = \mu_0 I_P(\psi)/2\pi$ and $I(\psi) = \mu_0 I_T(\psi)/2\pi$.
Additional details regarding magnetic coordinates can be found in Chapter 6 of \cite{2012Dhaeseleer} and Section 2 of \cite{Helander2014}.  

\subsection{Boozer coordinates}
\label{sec:boozer_coordinates}
Boozer coordinates are  magnetic coordinates; thus flux surfaces are assumed to exist and the condition \eqref{divB} is assumed. In addition, the MHD equilibrium force balance condition \eqref{equilibrium} is assumed to hold. Boozer coordinates are constructed such that the magnetic field has a simple form when expanded in the basis of the gradients of the coordinates, $\{\nabla x_i,x_i \in \{\psi,\vartheta_B,\varphi_B\}\}$, the covariant form. This specific choice of magnetic coordinates are especially useful for some applications, as will be discussed further in Section \ref{sec:quasisymmetry}.
The angles $\vartheta_B$ and $\varphi_B$ are chosen such that the magnetic field can be written in the following way,
\begin{gather}
    \bm{B}(\psi,\vartheta_B,\varphi_B) = I(\psi) \nabla \vartheta_B + G(\psi) \nabla \varphi_B + K(\psi,\vartheta_B,\varphi_B) \nabla \psi.
    \label{eq:boozer_covariant}
\end{gather}
Comparing with the general covariant form for magnetic coordinates of equilibrium fields \eqref{eq:magnetic_covar}, we see that the Boozer angles must be chosen such that $H$ vanishes. 

We can also write the field in the contravariant form, expanded in the basis of the partial derivatives of the position vector with respect to the coordinates, $\partial \bm{r}/\partial x_i$
\begin{gather}
    \bm{B} = \sqrt{g}^{-1} \iota \partder{\bm{r}(\psi,\vartheta_B,\varphi_B)}{\vartheta_B} + \sqrt{g}^{-1} \partder{\bm{r}(\psi,\vartheta_B,\varphi_B)}{\varphi_B},
    \label{eq:boozer_contravariant}
\end{gather}
which follows from the expression for magnetic coordinates \eqref{eq:magnetic_contravariant}. The expression for the Jacobian, $\sqrt{g} = \left( \nabla \psi \times \nabla \vartheta_B \cdot \nabla \varphi_B \right)^{-1}$, can be obtained by dotting \eqref{eq:boozer_covariant} with \eqref{eq:boozer_contravariant},
\begin{gather}
    \sqrt{g}(\psi,\vartheta_B,\varphi_B) = \frac{G(\psi)+\iota(\psi) I(\psi)}{B^2(\psi,\vartheta_B,\varphi_B)}.
    \label{eq:jacobian_boozer}
\end{gather}
Thus, all components of the magnetic field are described by five scalar quantities: $I(\psi)$, $G(\psi)$, $\iota(\psi)$, $K(\psi,\vartheta_B,\varphi_B)$, and $B(\psi,\vartheta_B,\varphi_B)$. 

Note that an assumption implied by writing the magnetic field in Boozer coordinates is that the radial current, $\bm{J} \cdot \nabla \psi$, vanishes. This can also be seen by applying Ampere's law to the covariant form \eqref{eq:boozer_covariant}. This is one of the conditions implied by the MHD equilibrium force balance equation \eqref{equilibrium} assuming $p(\psi)$. The other equations of MHD equilibrium \eqref{eq:equilibrium} were also used in arriving at these results.

Boozer coordinates are especially useful for stellarators, as quasisymmetric magnetic fields exhibit a symmetry in this coordinate system. This results from the fact that Boozer coordinates have a particularly simple covariant form such that $B_{\varphi_B}$ and $B_{\vartheta_B}$ are flux functions, and the Jacobian only varies on a surface due to the field strength. Therefore, confinement properties can be inferred by simply considering the symmetry of the field strength. For this reason, Boozer coordinates will be used in Section \ref{sec:quasisymmetry_gc} to discuss quasisymmetry.

Boozer coordinates are discussed in Section 6.6 of \cite{2012Dhaeseleer} and Section 2.5 of \cite{Helander2014}. 

\subsubsection{Radial covariant component}
\label{sec:radial_covariant}

As we see from \eqref{eq:jacobian_boozer}, the Jacobian in Boozer coordinates is determined from the field strength, $B(\psi,\vartheta_B,\varphi_B)$,
and the net poloidal and toroidal  currents, $G(\psi)$ and $I(\psi)$. Given these quantities, we will obtain an expression for the radial covariant component, $K(\psi,\vartheta_B,\varphi_B)$, by applying the MHD equilibrium force balance condition. This analysis will enable us to infer symmetry properties in Section \ref{sec:quasisymmetry}.

Consider the MHD equilibrium condition with flux surfaces  for a given pressure profile $p(\psi)$,
\begin{gather}
    \left( \nabla \times \bm{B} \right) \times \bm{B} = \mu_0 \nabla p(\psi).
\end{gather}
Using the covariant form of $\bm{B}$ to evaluate the curl, and its contravariant form  when taking the cross product, we get,
\begin{multline}\left(\der{I}{\psi} \nabla \psi \times \nabla \vartheta_B +\der{G}{\psi} \nabla \psi \times \nabla \varphi_B +
    \nabla K \times  \nabla \psi  \right) \\ \times \bigg(\nabla \psi \times \nabla \vartheta_B - \iota(\psi) \nabla \psi \times \nabla \varphi_B\bigg) = \mu_0 \der{p}{\psi} \nabla \psi.
\end{multline}
From the vector identity $(\bm{A} \times \bm{B}) \times (\bm{C} \times \bm{D}) = \bm{C}(\bm{A} \times \bm{B} \cdot \bm{D}) - \bm{D}(\bm{A} \times \bm{B} \cdot \bm{C})$ and multiplying through by the Jacobian, $\sqrt g$, we find,
\begin{align}
    \iota(\psi) \partder{K}{\vartheta_B}(\psi,\vartheta_B,\varphi_B) + \partder{K}{\varphi_B}(\psi,\vartheta_B,\varphi_B) &= \sqrt{g}(\psi,\vartheta_B,\varphi_B) \mu_0 \der{p}{\psi}(\psi) + \der{G}{\psi}(\psi) + \iota(\psi)\der{I}{\psi}(\psi) . 
    \label{eq:magnetic_K}
\end{align}

Taking advantage of the periodic setting, we now seek the solution $K$ of this partial differential equation in terms of a Fourier series in the two periodic independent variables, namely $(\vartheta_B,\varphi_B)$, for fixed $\psi$. 

Let's comment first on the right hand side of this equation \eqref{eq:magnetic_K}.
We see that average over $\vartheta_B$ and $\varphi_B$ of the left hand side of \eqref{eq:magnetic_K} is always zero.
Therefore, for well-posedness, we require that the average of the right hand side over $\vartheta_B$ and $\varphi_B$ vanishes, that is to say:
\begin{gather}
    \frac{G(\psi) + \iota(\psi) I(\psi)}{B_0^2(\psi)} \mu_0 \der{p}{\psi}(\psi) + \der{G}{\psi}(\psi) +  \iota(\psi)\der{I}{\psi}(\psi) = 0, 
    \label{eq:average_force_balance}
\end{gather}
where we define $B_0^{-2}(\psi): = (2\pi)^{-2} \int_0^{2\pi}\int_0^{2\pi} B^{-2}(\psi,\vartheta_B,\varphi_B) d \vartheta_B  d \varphi_B$.
Moreover, since only the term  $\sqrt{g}(\psi,\vartheta_B,\varphi_B)= (G(\psi)+\iota(\psi) I(\psi))/B^2(\psi,\vartheta_B,\varphi_B)$ depends on the angles $(\vartheta_B,\varphi_B)$, while $B$ must be bounded and periodic in both angles, then so is $B^{-2}$ and it can be expressed as a Fourier series as follows:
\begin{gather}
    B^{-2}(\psi,\vartheta_B,\varphi_B) = B_0^{-2}(\psi) + \sum_{m,n,(m,n)\ne(0,0)} b_{m,n}e^{i(m\vartheta_B-n\varphi_B)},
    \label{eq:Fourier_1_over_B2}
\end{gather}
with complex Fourier coefficients $\{b_{m,n}\}_{(m,n)\in\mathbb Z^2, (m,n)\neq (0,0)}$.

We then seek the solution $K$ as a Fourier expansion  \begin{gather}
    K(\psi,\vartheta_B,\varphi_B) = \mathcal K(\psi) + \sum_{(m,n), (m,n)\neq (0,0)} a_{m,n}e^{i(m\vartheta_B-n\varphi_B)},
    \label{eq:K_Fourier}
\end{gather}
and from \eqref{eq:magnetic_K}, with the Fourier expansions \eqref{eq:Fourier_1_over_B2}-\eqref{eq:K_Fourier}, the Fourier coefficients satisfy
\begin{gather}
\text{ for all } (m,n)\in\mathbb Z^2\backslash \{ (0,0)\},\
a_{m,n} i(\iota(\psi)m-n ) 
= \mu_0 (G(\psi) + \iota(\psi) I(\psi))\der{p}{\psi}(\psi)b_{m,n}.
\label{eq:FourierCoefs}
\end{gather}
Depending on the value of $\iota(\psi)$, two situations can occur. On the one hand, when $\iota(\psi)$ is irrational, \eqref{eq:FourierCoefs} uniquely defines the coefficients $\{a_{m,n},(m,n)\in\mathbb Z^2, (m,n)\neq (0,0)\}$, and the solution $K$ is unique up to the constant value $\mathcal K(\psi)$. On the other hand, when $\iota(\psi)$ is rational, the set of equations corresponding to indices $S_0=\{(m,n)\in\mathbb Z^2, \iota = n/m\}$ in \eqref{eq:FourierCoefs} can only be satisfied under an extra constraint on the right hand side: since $(G + \iota I)$ can never vanish, as this would imply that $\sqrt{g} = 0$, then unless the conditions  $\{(d p(\psi)/d\psi )b_{m,n} = 0,(m,n)\in S_0\}$ are satisfied there exists no solution; whereas if these conditions are satisfied, then Equation \eqref{eq:magnetic_K} has solutions and the Fourier coefficients $\{a_{m,n},(m,n)\in S_0\}$ as well as $\mathcal K(\psi)$ are free. 

In both cases, the solutions $K$ to \eqref{eq:magnetic_K} are usually written as :
\begin{multline}
    K(\psi,\vartheta_B,\varphi_B) = 
    \mathcal{K}(\psi) + \sum_{m,n,(m,n)\ne(0,0)} \left(i \mu_0 \der{p}{\psi} \frac{G + \iota I}{n - \iota m} b_{m,n}+ \Delta_{mn} \delta(n-\iota m)\right)e^{i(m\vartheta_B-n\varphi_B)}  ,
    \label{eq:K_quasisymmetry}
\end{multline}
where $\delta$ is the Dirac delta function and the $\Delta_{mn}$ are arbitrary constants.  When $\iota(\psi)$ is irrational, $\delta(n-\iota m)=0$ for all $(m,n)\in\mathbb Z^2$ so $K$ is independent of the $\Delta_{mn}$'s. 
On the other hand, when $\iota\in \mathbb{Q}$,
$K$ depends on the additional free parameters $\{\Delta_{m,n},(m,n)\in\mathbb Z^2,\iota(\psi)=n/m\}$. This is referred to as a singularity with respect to $\iota(\psi)$, as it contributes to the solution only on surfaces with a rational $\iota(\psi)$. An interpretation of these parameters will be provided in the next subsection. Moreover, since  the denominator, $n-\iota m$, vanishes when $\iota(\psi) = m_0/n_0$ but also gets arbitrarily small even for any value of $\iota(\psi)$, it is said that there exists a $1/x$ type singularity. 
These singularities will be discussed further in Section \ref{sec:singularities}.

\subsubsection{Method of characteristics}
As an alternative approach to solving \eqref{eq:magnetic_K}, the solution $K$ can be sought thanks to the method of characteristics by searching for lines along which the partial differential equation \eqref{eq:magnetic_K} reduces to ordinary differential equations.  Using this approach, we will provide here an interpretation of the free parameters ($\Delta_{mn}$) in the solution \eqref{eq:K_quasisymmetry} when $\iota\in\mathbb Q$.

 For a given surface $\psi$, omitting $\psi$ to simplify the notation, the partial differential equation \eqref{eq:magnetic_K} is a transport equation for $K$ with respect to $(\vartheta_B,\varphi_B)$ of the form
\begin{equation}\label{eq:transport}
\iota \partder{K}{\vartheta_B}(\vartheta_B,\varphi_B) + \partder{K}{\varphi_B}(\vartheta_B,\varphi_B) =F(\vartheta_B,\varphi_B)
\text{ with }\iota\neq 0,
\end{equation}
with a $2\pi$-periodic right-hand side $F$ in both $\vartheta_B$ and $\varphi_B$, and with the particularity that we are interested in solutions that are $2\pi$-periodic in the two variables $(\vartheta_B,\varphi_B)$. A necessary condition for the existence of periodic solutions is then that 
\begin{equation}\label{eq:ave}
\int_0^{2\pi}\int_0^{2\pi} F(\vartheta_B,\varphi_B) \, d\vartheta_B d \varphi_B = 0.
\end{equation}
Note that independently of the value of $\iota$, the two-dimensional well-posedness conditions \eqref{eq:average_force_balance} and \eqref{eq:ave} are equivalent. However, we will see that the above condition is not  sufficient for existence of a periodic solution under the condition $\iota \in \mathbb{Q}$.

 The characteristics for this equation are defined as the curves $s\mapsto(\vartheta_B(s),\varphi_B(s))$, passing by the point $(\vartheta_B^0,\varphi_B^0)$, by
\begin{align}
\left\{
\begin{array}{l}
\vartheta_B'(s)=\iota\\
\varphi_B'(s)=1
\end{array}
\right.
\Leftrightarrow
\left\{
\begin{array}{l}
\vartheta_B(s)=\iota s+\vartheta_B^0\\
\varphi_B(s)=s +\varphi_B^0
\end{array}
\right. .
\end{align}
Along each characteristic the equation reduces to an ordinary differential equation for the unknown $y(s):=K(\vartheta_B(s),\varphi_B(s))$, which is just $K$ along the characteristic. This ordinary differential equation reads:
\begin{equation}\label{eq:ODEChar}
\frac{d}{ds} y(s) = F(\vartheta_B(s),\varphi_B(s))
\quad\text{ or equivalently }\quad
\frac{d}{ds} y(s) = F(\iota s+\vartheta_B^0,s +\varphi_B^0).
\end{equation}
Along each characteristic, 
there is a unique solution for any given initial value $y(0)$, which can be written as:
\begin{equation}\label{eq:FLsol}
\text{ for all } s\geq 0, y(s)=y(0)+
\int_0^{s} F(\vartheta_B(\tilde s),\varphi_B(\tilde s)) \, d\tilde s.
\end{equation}
The solution has $\mathcal C^1$ regularity with respect to $s$ if $F$ is continuous along the characteristic. However the periodic setting of the two-dimensional problem will now lead to further considerations.

In order to relate the characteristic solutions to the two-dimensional equation \eqref{eq:transport}, we
consider the condition \eqref{eq:ave} in the context of characteristics. The change of variable $(\vartheta_B,\varphi_B) = (\iota s,s+\varphi_B^0)$, between $\{(s,\varphi_B^0)\in [0,2\pi/\iota)\times[0,2\pi)\}$ and $\{(\vartheta_B,\varphi_B)\in[0,2\pi)^2\}$, yields
\begin{equation}
\label{eq:FLave}
\int_0^{2\pi}\int_0^{2\pi} F(\vartheta_B,\varphi_B) \, d\vartheta_B d \varphi_B
=
\iota \int_0^{2\pi}\int_0^{2\pi/\iota}
F(\iota s,s+\varphi_B^0) \, dsd\alpha.
\end{equation}

We now consider the case when $\iota\in\mathbb Q$, represented by the irreducible fraction $\iota = N/D$. Then each characteristic, starting at a point $\left(\vartheta_B^0,\varphi_B^0\right)$ is periodic (it closes onto itself) and $s_D=2\pi D$ is the smallest non-zero value of $s$ along the characteristic such that
\begin{align}
\left\{
\begin{array}{l}
\vartheta_B(s_D)=\vartheta_B^0 \text{ mod } 2\pi \\
\varphi_B(s_D)=\varphi_B^0  \text{ mod } 2\pi.
\end{array}
\right.
\end{align}
Let's turn to the periodicity of the solution along characteristics. Along each characteristic, for any initial condition $y(0)$ the solution \eqref{eq:FLsol} is $s_D$-periodic under the assumption 
\begin{equation}\label{eq:aveChar} \int_0^{s_D} F(\vartheta_B(s),\varphi_B(s)) \, ds = 0.
\end{equation}
In order to define a solution at each point $(\vartheta_B,\varphi_B)\in[0,2\pi)^2$, it is then sufficient to provide one initial condition per closed characteristic: for instance providing an initial value at the points $\{(0,\varphi_B^0),\text{ for all }  \varphi_B^0\in[0,2\pi/\iota)\}$.

Note that \eqref{eq:aveChar} is different from \eqref{eq:ave}, since it concerns the average along the closed characteristic rather than a two-dimensional average. However, if \eqref{eq:aveChar} is satisfied for each characteristic then \eqref{eq:ave} is true since, combining \eqref{eq:FLave} with the periodicity of $F$, we get:
\begin{align}
\int_0^{2\pi}\int_0^{s_D}
F\left(\iota s,s+\varphi_B^0\right)\, ds
d\varphi_B^0
&= N
 \int_0^{2\pi}\int_0^{2\pi/\iota}
F\left(\iota s,s+\varphi_B^0\right)\, ds d\varphi_B^0\nonumber \\
&= D
\int_0^{2\pi}\int_0^{2\pi} F(\vartheta_B,\varphi_B) \, d\vartheta_B  d \varphi_B.
\end{align}

In terms of the partial differential equation \eqref{eq:transport}, no assumption on the initial condition $\varphi_B^0\in[0,2\pi/\iota)\mapsto K(0,\varphi_B^0)$ is required for the solution $K$ to be defined for any $s$ by
\begin{equation}
    K(\vartheta_B(s),\varphi_B(s))
    =
    K\left(\vartheta_B^0,\varphi_B^0\right)
    +
    \int_0^{s} F(\vartheta_B(\tilde s),\varphi_B(\tilde s)) \, d\tilde s.
\end{equation}
If $F$ is continuous, this solution is smooth along each characteristic but it could have a singularity across characteristics, depending on the initial condition $\varphi_B^0\in[0,2\pi/\iota)\mapsto K(0,\varphi_B^0)$. Such a solution is periodic only if the assumption \eqref{eq:aveChar} is satisfied. For a given surface $\psi$, any physical quantity, in particular $K$, is necessarily periodic with respect to $(\vartheta_B,\varphi_B)$.

    
On each characteristic, labeled by $(0,\varphi_B^0)$, the initial condition is arbitrary in the transport approach, so initial condition $\varphi_B^0
\mapsto K(0,\varphi_B^0)$ can be any arbitrary $2\pi/\iota$-periodic function, whereas in the Fourier discussion the initial conditions across characteristics are assumed to be at least integrable. To relate the two approaches, let's focus on the free parameters in \eqref{eq:K_quasisymmetry}, that is $\mathcal K$ and the $\{\Delta_{mn}\}$. We can express along a characteristic 
\begin{align}
\begin{array}{l}\displaystyle
\mathcal K(\psi) +\sum_{m,n,(m,n)\ne(0,0)}  \Delta_{m,n} \delta(n-\iota m)e^{i(m\vartheta_B(s)-n\varphi_B(s))}
\\\displaystyle
=
\sum_{p\in\mathbb Z}
\alpha_{p}e^{i p (D\vartheta_B(s)-N\varphi_B(s))}
\text{ with }  \alpha_0 = \mathcal K(\psi), \text{ and for all }  p\in\mathbb Z^* \ \alpha_p:=\Delta_{pD,pN}.
\end{array}
\end{align}
This function appears to be constant along characteristics, where the value of $D\vartheta_B(s)-N\varphi_B(s) = D\vartheta_B^0-N\varphi_B^0$ is independent of $s$. Moreover it represents a $2\pi/N$-periodic function 
as a Fourier series,
since choosing to label characteristics with $\vartheta_B^0=0$ we can write
\begin{equation}
    \sum_{p\in\mathbb Z}
\alpha_{p}e^{ip(D\vartheta_B(s)-N\varphi_B(s))}
=
\sum_{p\in\mathbb Z}
\alpha_{p}e^{-i Np\varphi_B^0}
.
\end{equation}
This is in particular $2\pi/\iota$-periodic.  
As a result, fixing the free parameters $\mathcal K$ and $\{ \Delta_{mn}, n=\iota m \}$ is equivalent to fixing an initial condition for the method of characteristics.



\subsection{Summary}

Below we include a Table which summarizes the important properties of magnetic and Boozer coordinate systems. 
\begin{center}
{\renewcommand{\arraystretch}{1.7}%
    \begin{tabular}{|c|c|c|}
    \hline
    & Magnetic & Boozer \\ \hline 
   Assumptions & $\nabla \cdot \bm{B} = 0$ & $\nabla \cdot \bm{B} = 0$ \\ 
   & &  $\bm{J} \cdot \nabla \psi = 0$ \\ \hline
   Contravariant & $\bm{B} = \nabla \psi \times \nabla \vartheta - \iota(\psi) \nabla \psi \times \nabla \varphi$ & $\bm{B} = \nabla \psi \times \nabla \vartheta_B - \iota(\psi) \nabla \psi \times \nabla \varphi_B$ \\ \hline
   Covariant & $\bm{B} = B_{\vartheta}(\psi,\vartheta,\varphi) \nabla \vartheta + B_{\varphi}(\psi,\vartheta,\varphi) \nabla \varphi  $ & $\bm{B} = I(\psi) \nabla \vartheta_B + G(\psi) \nabla \varphi_B$  \\ 
   & $\hspace{0.5cm} + B_{\psi}(\psi,\vartheta,\varphi) \nabla \psi$  & \hspace{0.5cm} + $K(\psi,\vartheta_B,\varphi_B) \nabla \psi$\\
   & 
   & $G(\psi) = \mu_0 I_P(\psi)/2\pi$ \\ 
   & 
   & $I(\psi) = \mu_0 I_T(\psi)/2\pi$
   \\ \hline 
   Jacobian & $\sqrt{g} = \frac{B_{\varphi}(\psi,\vartheta,\varphi) + \iota(\psi) B_{\vartheta}(\psi,\vartheta,\varphi)}{B^2}$ & $\sqrt{g} = \frac{G(\psi) + \iota(\psi) I(\psi)}{B^2}$ \\ \hline
    \end{tabular}}
\end{center}

\FloatBarrier
\section{Challenges associated with 3D equilibrium fields}
\label{sec:3D_difficulties}

Under the assumption of axisymmetry,
there are only two non-trivial spatial variables. Without this assumption, a physical system is therefore said to be 3D.
As we will see, the resulting loss of a conserved quantity has important implications, namely that the existence of continuously nested equilibrium flux surfaces is no longer guaranteed.
This is particularly important for devices such as stellarators which rely on equilibrium magnetic fields that lie on closed nested surfaces.
 
Sections \ref{sec:field_line_flow}-\ref{sec:integrability} consider questions related to the existence of surfaces in 2D and 3D. The equations of motion which follow magnetic field line trajectories in a toroidal system can be cast as a Hamiltonian system. Using the concept of integrability, the existence of flux surfaces arises as a fundamental consequence of the Hamiltonian nature of the system under specific assumptions.
Finally, Section \ref{sec:singularities} describes the implications, and potential complications, which can arise from presuming the existence of continuously nested magnetic surfaces in the calculation of 3D MHD equilibria. In spite of these challenges, Section \ref{sec:axis_expansion} demonstrates the existence of 3D equilibrium solutions with magnetic surfaces by considering an asymptotic expansion near the magnetic axis.

\subsection{From existence to non-existence of magnetic surfaces}
\label{sec:field_line_flow}
In axisymmetric geometry, we will find that the equations describing the position of field lines possess a conserved quantity which implies that the field lines lie on surfaces. When axisymmetry is broken, however, the conserved quantity may be lost leading to the possibility of non-existence of surfaces.

\subsubsection{Analysis of the vector potential}

To begin the discussion on the existence of surfaces, we will analyze the vector potential with and without the assumption of axisymmetry.
Consider a coordinate system $(r,\theta,\phi)$, where $\theta$ is a poloidal angle, $\phi$ is a toroidal angle, and $r$ labels toroidal nested surfaces, not necessarily flux surfaces. The vector potential is not uniquely defined since the addition of a curl-free component, the gradient of any scalar function, does not affect the value of $\bm{B} = \nabla \times \bm{A}$. The standard by which $\bm{A}$ is chosen is called the gauge. 
The vector potential $\bm{A}$ can generally be expressed in the $(r,\theta,\phi)$ coordinate system as 
\begin{gather}
    \bm{A}(r,\theta,\phi) = \uppsi_1(r,\theta,\phi) \nabla \theta - \uppsi_2(r,\theta,\phi) \nabla \phi + \nabla g(r,\theta,\phi),
\end{gather}
and we choose the scalar function $g$ so that $\nabla g(r,\theta,\phi) = 0$, which yields the magnetic field,
\begin{gather}
    \bm{B} = \nabla \uppsi_1 \times \nabla \theta - \nabla \uppsi_2  \times \nabla \phi. 
    \label{eq:B_covariant}
\end{gather}

The form of \eqref{eq:B_covariant} is similar to that of magnetic coordinates except that surfaces of constant $\uppsi_1$ do not overlap with those of constant $\uppsi_2$. This form for the magnetic field can be applied with or without the assumptions of surfaces.
If we assume that surfaces of $\uppsi_1$ form closed, nested toroidal surfaces, then 
$2\pi \uppsi_1$ will denote the toroidal flux enclosed by a surface of constant $\uppsi_1$. This can be seen by computing the toroidal flux using \eqref{eq:psi_T} with the flux coordinate system $(\uppsi_1,\theta,\phi)$, 
\begin{align}
    2\pi \uppsi_1 &= \int_0^{\uppsi_1} \int_0^{2\pi} \frac{\bm{B} \cdot \nabla \phi}{\nabla \uppsi_1 ' \times \nabla \theta \cdot \nabla \phi}\,  d \theta d \uppsi_1'. 
    \label{eq:uppsi1}
\end{align}
Thus we will denote $\uppsi_1$ by $\psi_T$ in analogy with \eqref{eq:magnetic_contravariant}. Similarly, we can denote $\uppsi_2$ by $\psi_P$, noting that $2\pi \uppsi_2$ is the poloidal flux enclosed by a surface of constant $\uppsi_2$ under the assumption of flux surfaces. This can be seen by computing the poloidal flux using \eqref{eq:psi_P} with the flux coordinate system $(\uppsi_2,\theta,\phi)$,
\begin{gather}
    2\pi \uppsi_2 = \int_0^{\uppsi_2} \int_0^{2\pi} \frac{\bm{B} \cdot \nabla \theta}{\nabla \uppsi_2' \times \nabla \theta \cdot \nabla \phi} \, d \phi d \uppsi_2'.
\end{gather}
With $\uppsi_1=\psi_T$ and $\uppsi_2=\psi_P$ and the gauge defined above, the vector potential can now be expressed in the $(\psi_T$,$\theta$,$\phi)$ system as, 
\begin{gather}
    \bm{A}(\psi_T,\theta,\phi) = \psi_T \nabla \theta - \psi_P(\psi_T,\theta,\phi) \nabla \phi. 
    \label{eq:vector_potential_Hamiltonian}
\end{gather}
In using this coordinate system, we have made the assumption that $\nabla \psi_T \times \nabla \theta \cdot \nabla \phi \ne 0$ in the domain of interest. This is equivalent to the assumption that $\bm{B} \cdot \nabla \phi \ne 0$ from \eqref{eq:B_covariant}. 

In what follows, we will not be assuming that flux surfaces exist, but we will simply use the form of the vector potential given in \eqref{eq:vector_potential_Hamiltonian} assuming that $\nabla \psi_T \times \nabla \theta \cdot \nabla \phi \ne 0$; in that case surfaces of constant $\psi_T$ and $\psi_P$ may not overlap to form a set of closed, nested, toroidal surfaces. 
However, for
axisymmetry, each of the vector components of the vector potential must be independent of the toroidal angle, implying that $\partial \psi_P/\partial \phi = 0$. If this condition holds, then it follows from \eqref{eq:B_covariant} that $\bm{B} \cdot \nabla \psi_P = \nabla \psi_T \times \nabla \theta \cdot \nabla \phi \left(\partial \psi_P/\partial\phi\right) = 0$. Under the assumption that $\psi_P$ is differentiable, the level sets of $\psi_P$ form surfaces except for at critical points where $\nabla \psi_P = 0$. We are not interested in solutions for which $\psi_P$ is constant throughout a volume, as this implies that the poloidal component of the field, the second term in \eqref{eq:B_covariant}, vanishes. Thus the level sets of $\psi_P$ form surfaces except at isolated points or lines. One example of such a line is the magnetic axis when it exists. We conclude that the magnetic field is tangent to surfaces of constant $\psi_P$. Under the assumption that the isosurfaces of $\psi_P$ are closed and compact, the Poincar\'e-Hopf theorem implies that their topology is toroidal (see ch. 4 in \cite{Dodson1997}). Thus we conclude that a set of toroidal magnetic surfaces exists; these are the flux surfaces.

\subsubsection{Variational principle for field line flow}

While we have seen that axisymmetry implies the existence of surfaces by simply analyzing the vector potential in toroidal coordinates, as we will now see, this property turns out to be
a consequence of the Hamiltonian nature of field line flow. Thus we will arrive at the same conclusion by considering the equations of field line flow as a Hamiltonian system. 

We begin by showing that field lines obey equations of motion which can be derived from a variational principle analogous to Hamilton's principle in classical mechanics. The equation of motion describing the position of a field line, $\bm{r}$, parameterized by distance along a field line, $l$, is given by,
\begin{gather}
   \hat{\bm{b}}(l) = \der{\bm{r}(l)}{l}. 
   \label{eq:field_line_flow}
\end{gather}
We will show that this equation of motion can be obtained from a variational principle involving the following functional,
\begin{gather}
W[\bm{r}] := \int \bm{A}(\bm{r}(l)) \cdot \der{\bm{r}(l)}{l} \, dl, 
\end{gather}
where $\bm{A}$ is the vector potential \eqref{eq:vector_potential}.
The first variation of  $W[\bm{r}]$ reads,
\begin{align}
    \delta W[\bm{r};\delta \bm{r}] &= \int  \bm{A}(\delta \bm{r}(l)) \cdot \der{\bm{r}(l)}{l} + \bm{A}(\bm{r}(l))\cdot \der{\delta \bm{r}(l)}{l} \, dl \nonumber \\
    &= \int  \delta \bm{r}(l) \cdot \left(\nabla \bm{A} \cdot \der{\bm{r}(l)}{l}  - \der{\bm{r}(l)}{l} \cdot \nabla \bm{A} \right) \, dl,
\end{align}
where the second line is obtained by integration by parts, using the boundary condition $\bm{A} \cdot \delta \bm{r} = 0$ at the endpoints of the trajectory and applying the notation $\nabla = \partial/\partial \bm{r}$. We use the vector identity $(\nabla \times \bm{A}) \times \bm{B} = \left(\bm{B} \cdot \nabla\right) \bm{A} - \left(\nabla \bm{A}\right) \cdot \bm{B}$ to obtain the following condition,
\begin{gather}
    (\nabla \times \bm{A}) \times \der{\bm{r}(l)}{l} = 0,
\end{gather}
which implies that $d \bm{r}/d l$ is parallel to $\hat{\bm{b}}$ since $\bm{B}=\nabla\times\bm{A}$. Furthermore, $d \bm{r}(l)/dl$ is a unit vector by definition, thus we recover the equation of motion describing a field line \eqref{eq:field_line_flow}. 

\subsubsection{Relation to Hamiltonian dynamics}\label{sec:relation_to_hamiltonian}

We now demonstrate explicitly the connection between Hamiltonian mechanics and the variational principle for field line flow by casting the latter as a Hamiltonian system.

We express the functional $W$ defined by \eqref{eq:field_line_flow} along a trajectory $\bm{r}$, expressed in toroidal coordinates $(\psi_T,\theta,\phi)$ and parametrized by $l$, as follows:
\begin{align}
    W[\bm{r}] &= \int  \left( \bm{A}(\bm{r}(l)) \cdot \partder{\bm{r}(\theta(l))}{\theta} \der{\theta(l)}{l} + \bm{A}(\bm{r}(l)) \cdot \partder{\bm{r}(\phi(l))}{\phi} \der{\phi(l)}{l} \right)\, dl. 
\end{align}
From the assumption that $(\psi_T,\theta,\phi )$ is a coordinate system, then $\bm{B}\cdot\nabla \phi\neq 0$ so field lines can be parametrized by $\phi$. Hence $(d \theta/dl)dl = (d \theta/d\phi)d \phi$ and $(d \phi/dl) dl = d \phi$. 
With the expression of the vector potential \eqref{eq:vector_potential_Hamiltonian} we can then write
\begin{align}
    W[\psi_T,\theta]   &= \int \, \left( \psi_T(\phi) \der{\theta(\phi)}{\phi} - \psi_P\Big(\psi_T(\phi),\theta(\phi),\phi\Big) \right) \, d\phi.
\end{align}
This is analogous to 
\eqref{eq:hamiltonian_functional} where $(\theta,\psi_T,\phi)$ correspond to $(\bm{q},\bm{p},t)$, and $\psi_P$ corresponds to $H$.

The resulting Euler-Lagrange equations are given by, 
\begin{subequations}
\begin{align}
    \der{\theta(\phi)}{\phi} &= \partder{\psi_P(\theta(\phi),\psi_T(\phi),\phi)}{\psi_T} \label{eq:hamiltonian1} \\
    \der{\psi_T(\phi)}{\phi} &= - \partder{\psi_P(\theta(\phi),\psi_T(\phi),\phi)}{\theta}.\label{eq:hamiltonian}
\end{align}
\label{eq:hamiltonian_subequations}
\end{subequations}
Thus, by analogy with \eqref{eq:Hamiltons_subequations}, if $\psi_P$ does not depend explicitly on $\phi$, meaning the system is axisymmetric, then $\psi_P$ is a constant of the motion from \eqref{eq:Hamiltonian_time},
\begin{gather}
    \der{\psi_P(\theta(\phi),\psi_T(\phi),\phi)}{\phi} = 0.
    \label{eq:field_line_constant}
\end{gather}
In an axisymmetric system,
field lines are therefore confined to surfaces of constant $\psi_P$. Thus magnetic surfaces exist.
Further discussion of the Hamiltonian nature of field line flow can be found in many references \cite{Cary1983,Boozer1983,Elsasser1986,Helander2014}. 

\subsection{Integrability of a Hamiltonian system}\label{sec:integrability}
We have seen that the behavior of magnetic field lines can be described by a Hamiltonian system, and we will now employ the framework of Hamiltonian dynamics to understand the fundamental challenges associated with the behavior of magnetic field lines in 3D, that is, without the assumption of axisymmetry.
The properties, which follow from the Hamiltonian nature of the system, have important implications for stellarator physics and the existence of continuously nested flux surfaces. This is particularly relevant for finding MHD equilibria in 3D, as the assumption of the existence of surfaces leads to several consequences which will be discussed in Section \ref{sec:singularities}.

\subsubsection{Defining integrability}
\label{sec:integrability_conditions}

If a system is not integrable, some trajectories of the Hamiltonian are ergotic, meaning they may eventually fill out a finite volume of phase space rather than being confined to surfaces. In the context of field line flow, this implies that in a non-integrable system, some field lines do not lie on toroidal surfaces. We will define the concept of integrability in a generic way before considering the application to magnetic fields in 3D.

To begin, we define precisely a Hamiltonian system as follows. We consider coordinates and momenta $\bm{q}\in \mathbb{R}^N$, $\bm{p}\in \mathbb{R}^N$; the space $\mathbb R^N\times\mathbb R^N$ is referred to as the phase space. The space spanned by $(\bm{q}(t),\bm{p}(t))$ satisfying Hamilton's equations \eqref{eq:Hamiltons_subequations} is denoted by $M\in \mathbb R^N\times\mathbb R^N$. Let the Hamiltonian $H(\bm{q},\bm{p},t)$ be a real scalar function. A trajectory is defined as a solution $t\mapsto(\bm{q}(t),\bm{p}(t))$ to Hamilton's equations.

Consider any scalar function $F$ of ($\bm{q}$,$\bm{p}$,$t$). The time derivative of this function along the trajectory ($\bm{q}(t),\bm{p}(t)$) is given by,
\begin{equation}
\der{F(\bm{q}(t),\bm{p}(t),t)}{t}=\{F(\bm{q}(t),\bm{p}(t),t),H\} + \partder{F(\bm{q}(t),\bm{p}(t),t)}{t} , 
\label{equation:hamiltoniansys}    
\end{equation}
where $\{f,g\}$ is the Poisson bracket which will be defined subsequently in a local coordinate system. This is obtained by simply applying the chain rule and Hamilton's equations.

In a small neighbourhood of any point of $M$ there are local coordinates $q_1\dots q_N$
and $p_1 \dots p_N$ such that,
\begin{equation}
\{f,g\}=\sum_{i=1}^{N}\left(\frac{\partial f}{\partial q_i}\frac{\partial g}{\partial p_i}-\frac{\partial f}{\partial p_i}\frac{\partial g}{\partial q_i}\right),
\label{equation:poissonb}    
\end{equation}
where $q_i$ and $p_i$, which we call respectively the canonical coordinate and canonical momentum. 
The Poisson bracket satisfies several algebraic properties which are discussed in detail in \cite{Kozlov1983}.

For $i=1,\dots,N$, each conjugate pair $(q_i,p_i)$ corresponds to one degree of freedom (dof).
A $2N$ dimensional system has $N$ conjugate pairs and can thus be described by an $N$-dof Hamiltonian.

A quantity, $I(\bm{q}(t),\bm{p}(t),t)$, is said to be a constant of the motion if $dI/dt=0$: $I$ is a conserved quantity along any trajectory.
Two functions $f$ and $g$ are said to be in involution if,
\begin{equation}
\{f,g\}=0,    
\end{equation}
meaning that they commute.
An autonomous Hamiltonian system (where $H$ does not depend explicitly on time) with $N$-degrees of freedom is then said to be integrable if there exist $N$ independent, smooth constants of the motion, $\{I_i\}_{1\leq i\leq N}$, which are in involution. This implies that the constants of motion are independent in the sense that none of the $I_i$ can be expressed as a function of the other $(N-1)$. An integrable Hamiltonian system means that solutions can be found by evaluating integrals of known functions from a given set of initial conditions. Integrable systems exhibit regular motion meaning that trajectories are confined to $N$ different surfaces of dimension $2N-1$ in a $2N$ dimensional phase space, defined by the constants of motion.
It can be shown that these surfaces are $N$-tori under the assumption that the phase space $M$ is compact and connected \cite{Arnold2009}; thus they are often referred to as invariant tori.
In the context of axisymmetric field line flow, which is an integrable $1$-dof Hamiltonian \eqref{eq:hamiltonian_subequations}, only one constant of the motion is required to guarantee integrability. As the Hamiltonian, $\psi_P$, is one such constant, it is sufficient, and magnetic field lines are thus confined to lie on invariant tori.

Integrability can be similarly defined for non-autonomous Hamiltonian systems ($H$ depends explicitly on time). The procedure involves extending the non-autonomous Hamiltonian to a higher dimensional autonomous one for which integrability can be defined as above \cite{Angelo2012,Lichtenberg2013}.
A non-autonomous Hamiltonian system with $N$-dof can
also be shown to be integrable if there exist $N$, possibly time-dependent, constants of the motion \cite{Bouquet1998}.

In the following Sections, we limit our attention to the Hamiltonian associated with field line flow.
With the correspondence $(\theta,\psi_T,\phi)$  to $(\bm{q},\bm{p},t)$, and $\psi_P$  to $H$, we have seen that, with axisymmetry, the magnetic field line flow corresponds to a $1$-dof Hamiltonian system.
In 3D, since $d\psi_p/d\phi\neq0$, the field line flow Hamiltonian becomes non-autonomous.
In the literature, this is sometimes referred to as a $1.5$-dof Hamiltonian.
Therefore, as we will see, in 3D systems, integrability of the field line flow is no longer guaranteed, as the Hamiltonian is not a constant of the motion.

\subsubsection{Hamilton-Jacobi method}
As we are interested in the 1-dof field line flow Hamiltonian, for the remaining discussion, we limit our consideration to $1$-dof autonomous and non-autonomous Hamiltonian systems.
However the subsequent discussion regarding non-integrability can be generalized to higher dimensional systems.

One approach to solving the equations of motion \eqref{eq:Hamiltons_subequations}, known as the Hamilton-Jacobi method \cite{Kozlov1983,Morrison1998,Goldstein2002}, involves the use of canonical transformations to find ignorable coordinates, meaning that the Hamiltonian is independent of this coordinate.
Stated precisely, a canonical transformation is a diffeomorphism, $\varphi: (q,p)\in M \to (Q(q,p),P(q,p))\in M$ 
which preserves the Poisson bracket such that $\{f(q,p),g(q,p)\} = \{\widetilde{f}(Q,P),\widetilde{g}(Q,P)\}$ where $\widetilde{f}(Q(q,p),P(q,p)) = f(q,p)$ and similarly for $\widetilde{g}$.
In particular, both the mapping and its inverse are smooth and differentiable, ensuring it is possible to transform from one set of phase space coordinates to another. 

Our goal is to choose a transformed coordinate system in which the motion becomes very simple.

We start by assuming that a generating function $S(q,P,t):\mathbb{R}^{3}\to\mathbb{R}$ and a coordinate transformation $(q,p)\to(Q,P)$ satisfy,
\begin{subequations}
\begin{eqnarray}
p(t)&=&\frac{\partial S(q(t),P(t),t)}{\partial q} \label{eq:generating1} \\
Q(t)&=&\frac{\partial S(q(t),P(t),t)}{\partial P}.
\label{eq:generating}
\end{eqnarray}
\label{eq:generating_subequation}
\end{subequations}
Hamilton's equations \eqref{eq:Hamiltons_subequations} can then be expressed in the transformed coordinates
\begin{subequations}
\begin{eqnarray}
\der{P(t)}{t} &=&-\frac{\partial K(Q(t),P(t),t)}{\partial Q} \label{eq:hamiltonian_canonicala} \\
\der{Q(t)}{t} &=&\frac{\partial K(Q(t),P(t),t)}{\partial P},
\label{eq:hamiltonian_canonical}
\end{eqnarray}
\label{eq:hamiltonian_canonical_subequation}
\end{subequations}
where the new Hamiltonian is
\begin{gather}\label{eq:newHam}
    K(Q,P,t):=H(q(Q,P,t),p(Q,P,t),t) + \partder{S(q(Q,P,t),P,t)}{t}.
\end{gather}
It can be shown that \eqref{eq:hamiltonian_canonical_subequation} is equivalent to \eqref{eq:Hamiltons_subequations}. 

In order to find such a generating function $S(q,P,t)$, for a desired new Hamiltonian $K$, we consider \eqref{eq:newHam} as the non-linear Hamilton-Jacobi equation in $S$,
\begin{equation}
H\left(q,\frac{\partial S(q,P,t)}{\partial q},t\right) + \partder{S(q,P,t)}{t}=K\left(\partder{S(q,P,t)}{P},P,t\right).
\label{eq:hamilton_jacobi}
\end{equation}
This is a first order PDE for $S(q,P,t)$ in $q$, $P$, and $t$. 

 We make a particular choice of canonical transformation 
 under the assumption of an autonomous system which exhibits periodic motion, either $p$ is periodic in $q$ or both $p$ and $q$ are periodic in time with the same frequency. This transformation, known as action-angle variables, has a particularly simple form: the momentum becomes constant in time while the position coordinate grows linearly. Moreover, the new Hamiltonian becomes independent of $Q$.
 We will first discuss the action-angle variables, in order to apply them to the field line flow system under the assumption of axisymmetry, as the momentum $\psi_T$ is periodic in the coordinate $\theta$. 

From the assumption that the new momentum is constant, $\dot{P} = 0$, we conclude that $K$ is independent of $Q$ from \eqref{eq:hamiltonian_canonical}. Under the assumption that $\dot{Q}$ is a constant of the motion, we can conclude from \eqref{eq:hamiltonian_canonicala} that $\partial K/\partial P$ is only a function of $P$, which is itself a constant of the motion. As any function of only $t$ can be added to $K$ without consequence to Hamilton's equations \eqref{eq:hamiltonian_canonical_subequation}, we therefore seek a new Hamiltonian which depends on only the new momentum, $K(P)$. Furthermore, we note that for an autonomous system the Hamiltonian is a constant of the motion, so it can be expressed as a function of $P$ alone which we denote $E(P)$. From \eqref{eq:hamilton_jacobi} we find the generating function satisfies
\begin{align}
     \partder{ S(q,P,t)}{t} = K(P) - E(P).
\end{align}
We are free to choose the generating function such that $E(P) = K(P)$;
thus we can express it in the functional form $S(q,P,t) = W(q,P)$. 

We will now define a choice for $P$ such that the Hamiltonian becomes a function of $P$ alone. The new momentum, often called the action variable, is defined as,
\begin{align}
    P = \frac{1}{2\pi} \oint p \, dq =  \frac{1}{2\pi} \oint p(t) \der{q(t)}{t} \, dt.
    \label{eq:action}
\end{align}
The integral is computed along a trajectory ($p(t)$, $q(t)$) for a single period of the motion. Hence $P$ is constant along a trajectory as it only depends on $H(q(t),p(t)) = E$, a constant of the motion for an autonomous system. The action-angle coordinate transformation is obtained from the following generating function,
\begin{align}
    S(q,P,t) = \int_0^q p(q',P) \, dq',
\end{align}
where the integral is computed at constant $P$ along a trajectory. Hamilton's equations in the canonical coordinates \eqref{eq:hamiltonian_canonical_subequation} therefore become,
\begin{subequations}
\begin{align}
    \der{P}{t} &= 0 \label{eq:canonicala} \\
    \der{Q}{t} &= \partder{K(P)}{P} = \omega(P),\label{eq:canonical}
\end{align}
\end{subequations}
where $\omega(P)$ can be interpreted as a constant angular frequency.
Solving the ODE to obtain an explicit expression for $Q$, it is apparent that the new coordinate changes linearly with time,
\begin{align}
    Q(t) = Q_0 + \omega(P) t,
\end{align}
where $Q_0$ is a constant of integration associated with the initial condition of the trajectory.
We can note that $Q$ can be interpreted as an angle by computing the change in this coordinate, $\Delta Q$, during one period of the motion,
\begin{align}
    \Delta Q &= \oint \partder{Q}{q} \, d q = \partder{}{P} \oint \partder{S(q,P,t)}{q} \, dq = \partder{}{P} \oint p \, dq = 2\pi.
\end{align}
Therefore we expect many physical properties to be $2\pi$ periodic in $Q$.

The field line flow Hamiltonian system is integrable under the assumption of axisymmetry, as it is an autonomous system with 1-dof and there exists one constant of the motion, \eqref{eq:field_line_constant}. Furthermore, the momentum, $\psi_T$, is periodic in the coordinate, $\theta$.
Thus there exists a transformation into action-angle coordinates for this system. 
The action integral for the field line flow Hamiltonian system is given by,
\begin{align}
    P &= \frac{1}{2\pi}\int_0^{2\pi} \psi_T(\psi_P,\theta) \, d \theta
    = \frac{1}{2\pi} \oint  \bm{A}(\bm{r}) \cdot d \bm{r}.
\end{align}
The integration is performed at constant $\psi_P$ along a closed poloidal loop. Here we have used \eqref{eq:vector_potential_Hamiltonian} such that $\psi_T = \bm{A} \cdot \partial \bm{r}/\partial \theta$.
Upon application of Stokes' theorem, we see that $P$ is related to the magnetic flux through $S_T$, a surface enclosed by the isosurface of $\psi_P$ at constant toroidal angle,
\begin{align}
    P = \frac{1}{2\pi} \int_{S_T} \bm{B} \cdot \hat{\bm{n}} \, d^2 x.
\end{align}
Therefore $P = (1/2\pi) \Psi_T$, the toroidal flux function. By a similar argument we note that $\psi_P = (1/2\pi) \Psi_P$, the poloidal flux function. 

The action-angle transformation allows us to write the Hamiltonian as a function of $P$ alone, which we denote $\overline{\psi_P}(P)$. Since the angle satisfies $d Q/t = \partial\overline{\psi_P}(P)/\partial P$, the equation of motion for the canonical coordinate is given by \eqref{eq:hamiltonian_canonical},
\begin{equation} \der{Q}{t} = \partder{\overline{\psi_P}(P)}{P} = \der{\psi_P(\psi_T)}{\psi_T} = \iota(\psi_T),
\end{equation}
where $\iota(\psi_T)$ is the rotational transform  \eqref{eq:iota}, a constant of the motion. Thus the canonical coordinate, $Q$, increases linearly with the toroidal angle,
\begin{align}
    Q(\phi) = Q_0 + \iota(\psi_T) \phi.
    \label{eq:coordinate_actiona_angle}
\end{align}
We note that $Q$ can be interpreted as a poloidal angle, as the quantity $(\theta - \iota \phi)$ is constant along a field line. Thus $Q_0$ is a constant which labels a field line.

The contours of $\psi_P$, the conserved quantity, define invariant tori in the system's phase space on which all motion is confined.
The magnetic field line trajectories thus lie on concentric tori, yielding the continuously nested flux surfaces. 

\subsubsection{Perturbations about integrability}
\label{sec:perturbation_integrability}
Many Hamiltonian systems are, in fact, non-integrable. In the absence of axisymmetry the Hamiltonian of the field line flow system, $\psi_P$,
is no longer independent of $\phi$ and thus non-autonomous. In particular, there may not exist a solution by the Hamilton-Jacobi method such that $\dot{\bm{P}}=0$ as required for action-angle coordinates, since this implies that there exist $N$ independent constants of the motion and therefore the Hamiltonian is integrable.

Although many physical systems are not integrable, the need for analytic tractability led historically to the development of a perturbation theory approach \cite{Kozlov1983,Ferrell1971}.
A Hamiltonian is described as nearly integrable if it can be constructed as a perturbation series about an autonomous integrable Hamiltonian, $H_0(q,p)$,
for which there exists a canonical transformation to the action-angle coordinates $(Q_0,P_0)$.
The corresponding generating function is then denoted $S^*(q,P_0)$
and the transformed Hamiltonian  $H_0(P_0)$.  We seek a canonical transformation from $(Q_0,P_0)$ to $(Q,P)$, as in \eqref{eq:hamilton_jacobi}, for the perturbed Hamiltonian such that 
\begin{align}\label{eq:HJcan}
    H\left(Q_0(Q,P,t),\partder{S(Q_0(Q,P,t),P,t)}{Q_0},t\right) + \partder{S(Q_0(Q,P,t),P,t)}{t} = K(Q,P,t).
\end{align}
Formally, the perturbed Hamiltonian and generating function expressed in the unperturbed action-angle coordinates are expanded with respect to   $\epsilon \ll 1$ as follows,
\begin{subequations}
\begin{align}
        H(Q_0,P_0,t)&= H_0(P_0)+\sum_{i=1}^\infty \epsilon^i H_i(Q_0,P_0,t)
        \label{eq.hamiltonian3.0}
        \\
     S(Q_0,P,t)&=\sum_{i=0}^\infty \epsilon^i S_i(Q_0,P,t).
    \label{eq.hamiltonian3}
\end{align}
\label{eq:epsilon_subequations}
\end{subequations}
Assuming that $H$ can be written as \eqref{eq.hamiltonian3.0}, if there exists $S_i$ for all $i\geq1$, satisfying \eqref{eq:HJcan}, such that the perturbation series \eqref{eq.hamiltonian3} for  $S$ converges, then the perturbed Hamiltonian $H$ is nearly integrable and the corresponding nearly integrable system has a solution. Our goal is to obtain conditions under which the motion remains periodic although there is a perturbation away from integrability.

Recall that, for the canonical transformation associated with action-angle coordinates, the generating function is chosen such that $\dot{P}=0$ and $\dot{Q}=\omega(P)$: the momentum is constant while the coordinates change linearly with time. As we saw, this transformation provided insight into periodic motion. While the action-angle transformation assumes an autonomous Hamiltonian, this assumption is not satisfied for the field line flow Hamiltonian without axisymmetry.

We now aim to obtain a different coordinate transformation under which the motion becomes very simple: 
we seek a generating function, $S(Q_0,P,t)$ such that both $\dot{Q}$ and $\dot{P}$ vanish. The constant coordinate and momentum can be determined from the initial conditions of the problem. Thus the generating function \eqref{eq:generating_subequation} provides a mapping from the initial conditions and time to the old coordinates. Together $\dot{Q}=0$ and $\dot{P} = 0$ imply that both $\partial K/\partial Q$ and $\partial K/\partial P$ vanish. Such a transformation implies $K(Q,P,t)$ depends only on $t$. Observing that the equations of motion remain invariant under the addition of any function of only $t$ to $K(Q,P,t)$, we seek a transformation such that $K=0$.
The lowest orders of the perturbation series obtained from the Hamiltonian-Jacobi equation \eqref{eq:HJcan} combined with \eqref{eq:epsilon_subequations} read,
\begin{multline}
H_0\left(\partder{S_0(Q_0,P,t)}{Q_0}\right) + \partder{S_0(Q_0,P,t)}{t} + \epsilon \partder{H_0(P_0)}{P_0} \partder{S_1(Q_0,P,t)}{Q_0} + \epsilon H_1 \left(Q_0,\partder{S_0(Q_0,P,t)}{Q_0},t\right) \\ + \epsilon \partder{S_1(Q_0,P,t)}{t}
+ \mathcal{O}(\epsilon^2)= 0. \label{eq.hamiltonian4}
\end{multline}
The $\mathcal{O}(\epsilon^0)$ terms provide the following equation for $S_0$,
\begin{align}
    H_0\left( \partder{S_0(Q_0,P,t)}{Q_0} \right) + \partder{S_0(Q_0,P,t)}{t} = 0.
\end{align}
In other words, $S_0(Q_0,P,t)$ provides a coordinate transformation from $(Q_0,P_0)$ to $(Q,P)$ such that the new Hamiltonian is zero; therefore $Q$ and 
$P$ are invariants with respect to $H_0$ as in \eqref{equation:hamiltoniansys}.

The $\mathcal{O}(\epsilon^1)$ terms provide an equation for $S_1$ given $S_0$ and $H_1$,
\begin{align}\label{eq:O(e)}
   \omega_0 \partder{S_1(Q_0,P,t)}{Q_0} + H_1 \left(Q_0,\partder{S_0(Q_0,P,t)}{Q_0},t \right) + \partder{S_1(Q_0,P,t)}{t} = 0,
\end{align}
since $\partial H_0(P_0)/\partial P_0 = \dot{Q}_0 = \omega_0$, which is the canonical unperturbed frequency.

We consider a perturbed Hamiltonian that is periodic in $Q_0$, the unperturbed angle, and in time, 
\begin{align}
    H_1(Q_0,P,t) &= \sum_{m,n} H_{1}^{m,n}(P) e^{i (m Q_0-n t)}. 
\end{align}
This assumption is justified for the field line flow Hamiltonian, as we note that $Q_0$ and time ($\phi$) are poloidal and toroidal angles, respectively from \eqref{eq:coordinate_actiona_angle}. The Hamiltonian for this system, $\psi_P$, is one of the covariant components of the vector potential \eqref{eq:vector_potential_Hamiltonian}. Thus any physically relevant Hamiltonian is doubly periodic in $t$ and $Q_0$.

We now attempt to construct a solution by expressing $S_1(Q_0,P,t)$ as a Fourier series in $(Q_0,t)$: 
\begin{align}
    S_1(Q_0,P,t) &= \sum_{m,n} S_1^{m,n}(P) e^{i (m Q_0-n t)}.
\end{align}
Inserting these expressions into \eqref{eq:O(e)} we obtain
\begin{align}
    S_1(Q_0,P,t) &= i \sum_{m,n} \left(\frac{H_{1}^{m,n}(P)}{m\omega_0-n} + \Delta_{m,n} \delta(m - \omega_0 n) \right)e^{i( m Q_0-nt)}.
    \label{eq:series_solution}
\end{align}
As long as $\omega_0\notin\mathbb Q$, then this solution is unique since the $\delta$ term is always zero. If $\omega_0\in\mathbb Q$ then there exists an infinite set of solutions parameterized by the $\Delta_{m,n}$, and a series solution $S_1$ can only be expressed in the above form if $H_1^{m,n}(P) = 0$ for all $(m,n)\in\mathbb Z^2$ such that $\omega_0 = n/m$. 
In fact, the series \eqref{eq:series_solution} may diverge, even for $\omega_0 \in \mathbb{R}\setminus\mathbb{Q}$, if $\omega_0$ is sufficiently well approximated by a rational number. We discuss some implications in the following Section.
This is known as the problem of small divisors in classical mechanics perturbation theory and has been recognized since the time of Poincar\'e \cite{Kozlov1983,Yoccoz1992}. In the context of the field line flow Hamiltonian, recall that $\omega_0$ is $\iota(\psi_P)$.
The preceding results demonstrate that the magnetic field may not be integrable in the neighborhood of rational values of the rotational transform and thus these flux surfaces may not exist.

In order to construct a solution as a perturbation series, the $\mathcal{O}(\epsilon^n)$ terms provide equations for the corresponding $S_n$.
In general, we do not expect a non-integrable system to be close enough to an integrable system to be represented as such a convergent perturbation series in $\epsilon$. The following Section will discuss conditions under which a perturbative solution exists. 

\subsubsection{Persistence of some flux surfaces: KAM theory}
\label{sec:KAM}

In Section \ref{sec:integrability_conditions} we have seen that the non-existence of continuously nested flux surfaces in 3D MHD is a consequence of fundamental properties of the associated Hamiltonian system.
That is not to say, however, that an infinitesimal deviation away from axisymmetry results in a complete loss of all flux surfaces.
In this Section, we briefly describe the persistence of some surfaces, even in the presence of perturbations away from axisymmetry, which follows the results of Kolmogorov \cite{Kolmogorov1954}, Arnold \cite{Arnold2009}, and Moser \cite{Moser1962}, known as KAM theory. We refer to several other sources \cite{Ott2002,Goldstein2002,Hudson2012}.

We continue our discussion from Section \ref{sec:perturbation_integrability}, considering bounded, periodic motion of an integrable Hamiltonian with canonical frequency $\omega_0$. KAM theory addresses the question of which invariant surfaces persist as a perturbation is applied, quantified by non-zero values of $\epsilon$ in the expansion of $H$ \eqref{eq.hamiltonian3.0}. 
In the context of magnetic field line flow in a torus, the result identifies some conditions under which flux surfaces persist in the presence of perturbations away from integrability, or axisymmetry. 

An invariant surface associated with frequency $\omega_0$ is said to persist if it can be transformed continuously with $\epsilon$ into an invariant surface of the perturbed system for a frequency $\omega(\epsilon)$. 
While the perturbation may deform the invariant surface in phase space, the surfaces must be deformed continuously from the unperturbed invariant surfaces. 

Assuming perturbations are sufficiently small ($\epsilon\ll1$), for an invariant surface to persist, it is sufficient for the unperturbed frequency $\omega_0$ to satisfy the Diophantine condition,
\begin{align}\label{eq:Dioph}
\exists (r,k)\in\mathbb R^+\times \mathbb N, k\geq 2, \text{ s.t. for all }(m,n)\in\mathbb Z^*\times\mathbb Z,
    \left|\omega_0-\frac{n}{m}\right|>\frac{r}{m^k}.
\end{align}
This condition excludes regions surrounding each rational value of the frequency  $\omega_0=n/m$.
For magnetic field lines, where $\omega_0$ is $\iota$, the condition \eqref{eq:Dioph} means that the rotational transform must be sufficiently irrational.
Another sufficient condition is that the frequencies of the unperturbed Hamilitonian are non-degenerate,
\begin{align}\label{eq:condND}
    \bigg \rvert \partder{\omega_0(P)}{P}\bigg \rvert > 0,
\end{align}
implying that there does not exist multiple unperturbed periodic orbits with the same frequency. In the context of magnetic field line flow, the magnetic shear, $d \iota(\psi_T)/d\psi_T$, cannot vanish at any point. As there are applications in which this assumption cannot be made, for example stellarators sometimes are designed to have small magnetic shear, a similar theory has been developed for degenerate systems \cite{Del1996,Morrison1998}. 

An important result of measure theory is that the set of frequencies satisfying the Diophantine condition \eqref{eq:Dioph} is of finite measure in $\mathbb{R}$, in the sense of Lebesgue. KAM theory implies that for a Hamiltonian system  sufficiently close to integrability satisfying \eqref{eq:condND}, a non-zero Lebesgue measure of the tori of the unperturbed Hamiltonian system survive. For non-axisymmetric magnetic fields satisfying the above conditions, this means that these invariant tori (i.e.\ flux surfaces) occupy a non-zero volume in phase space.

As the perturbation from integrability, $\epsilon$, increases, invariant surfaces may break up. The trajectories of field lines may no longer lie on nested toroidal surfaces but instead fill an area on the Poincar\'{e} surface of section or form magnetic island structures. 

\begin{figure}
    \centering
\begin{subfigure}{0.49\textwidth}
     \includegraphics[trim=1cm 0cm 3cm 2cm,clip,width=1.0\textwidth]{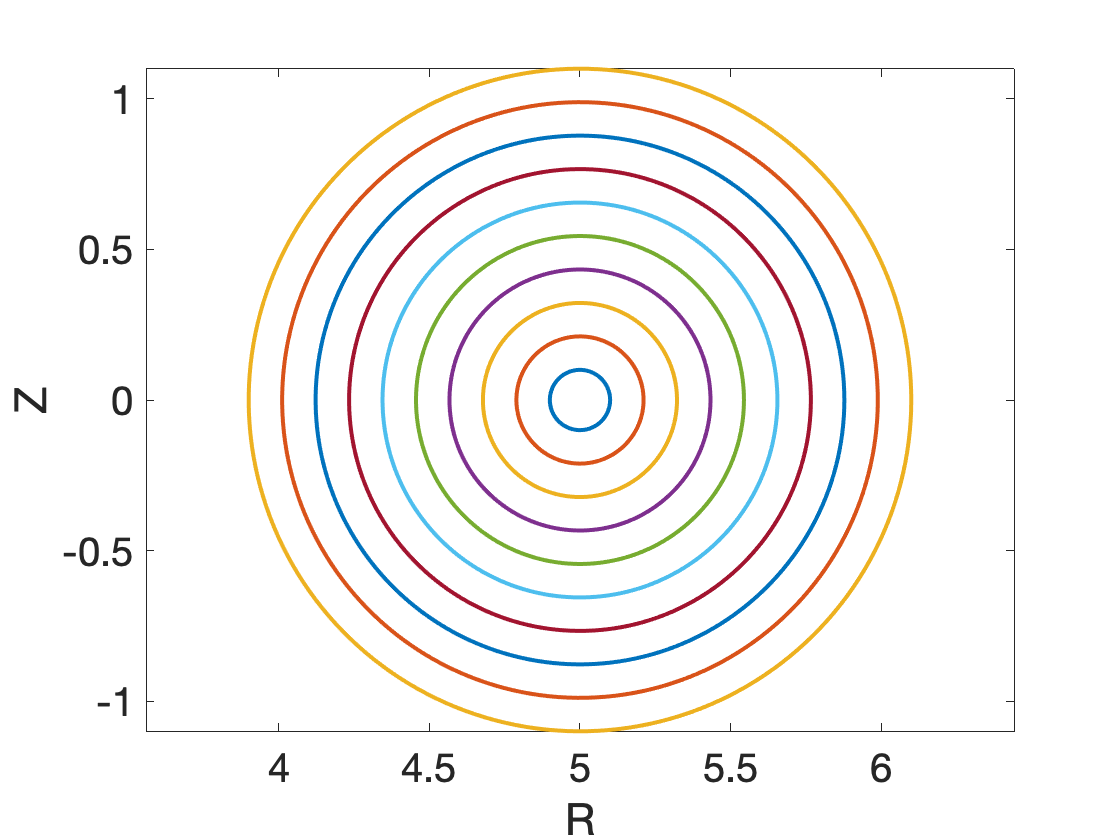}
     \caption{$\epsilon = 0$}
\end{subfigure}
\begin{subfigure}{0.49\textwidth}
     \includegraphics[trim=1cm 0cm 3cm 2cm,clip,width=1.0\textwidth]{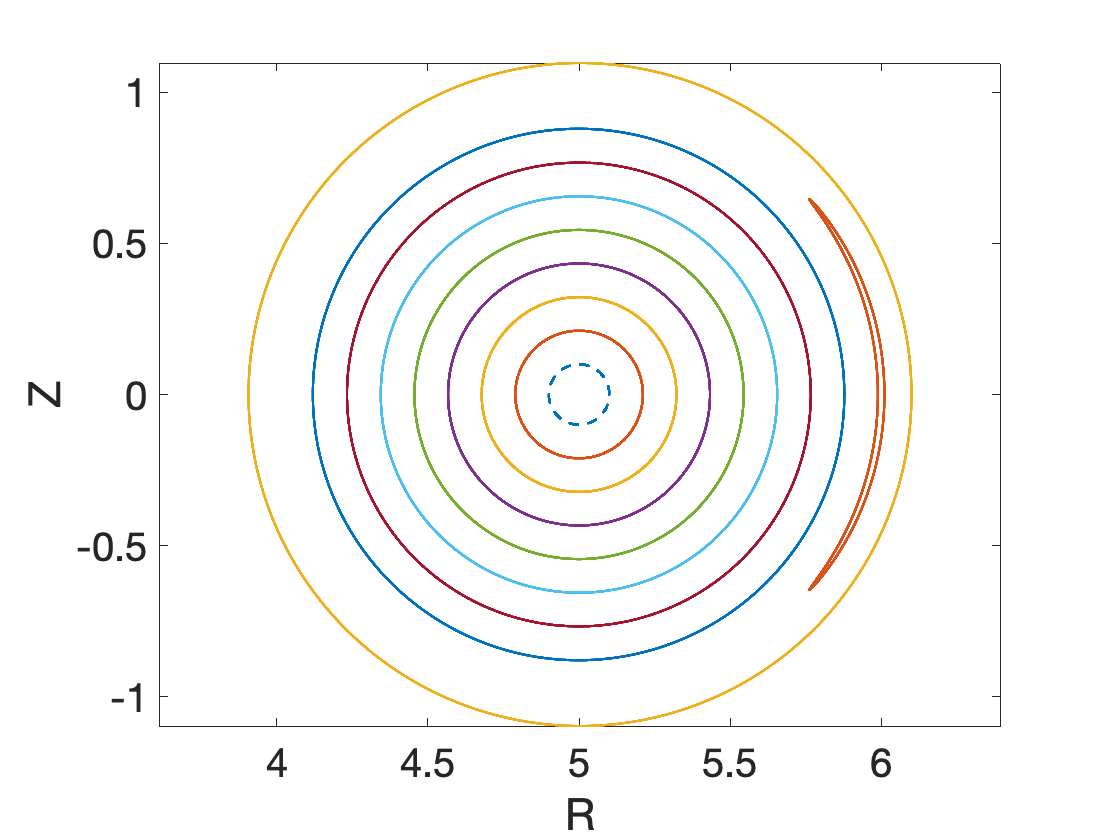}
     \caption{$\epsilon = 0.0005$}
\end{subfigure}
\begin{subfigure}{0.49\textwidth}
     \includegraphics[trim=1cm 0cm 3cm 2cm,clip,width=1.0\textwidth]{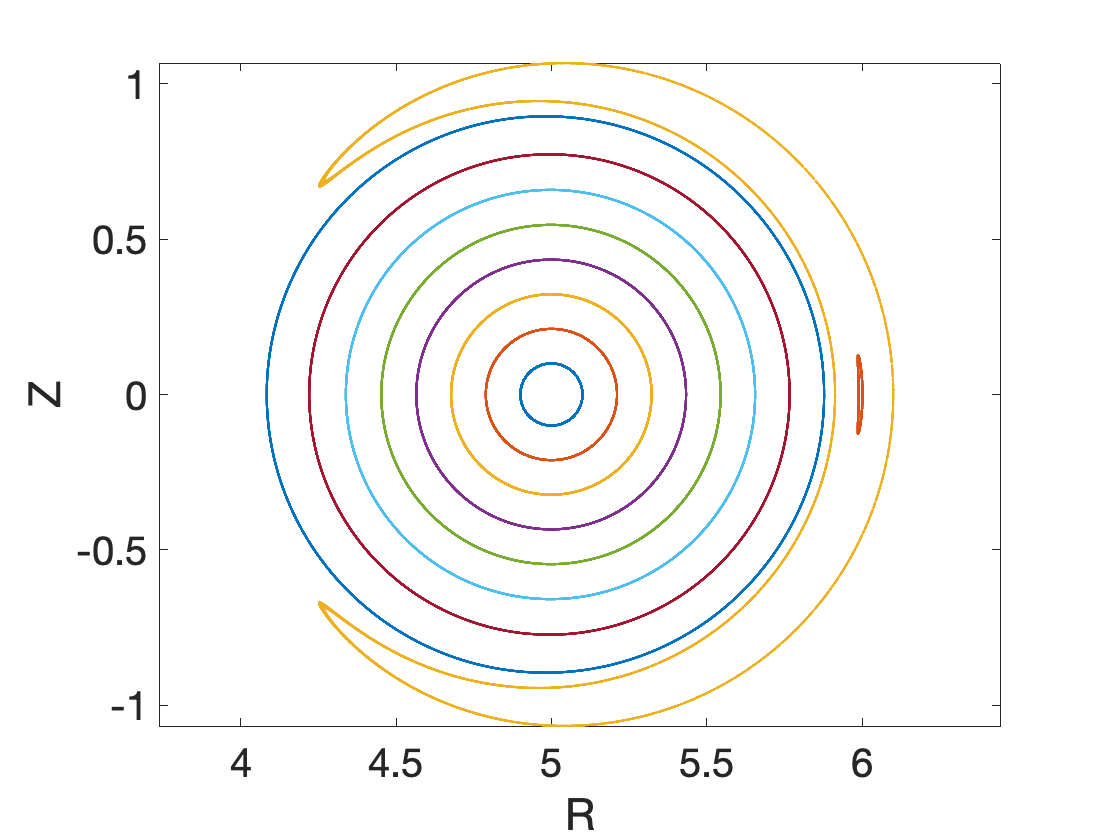}
     \caption{$\epsilon = 0.005$}
\end{subfigure}
\begin{subfigure}{0.49\textwidth}
     \includegraphics[trim=1cm 0cm 3cm 2cm,clip,width=1.0\textwidth]{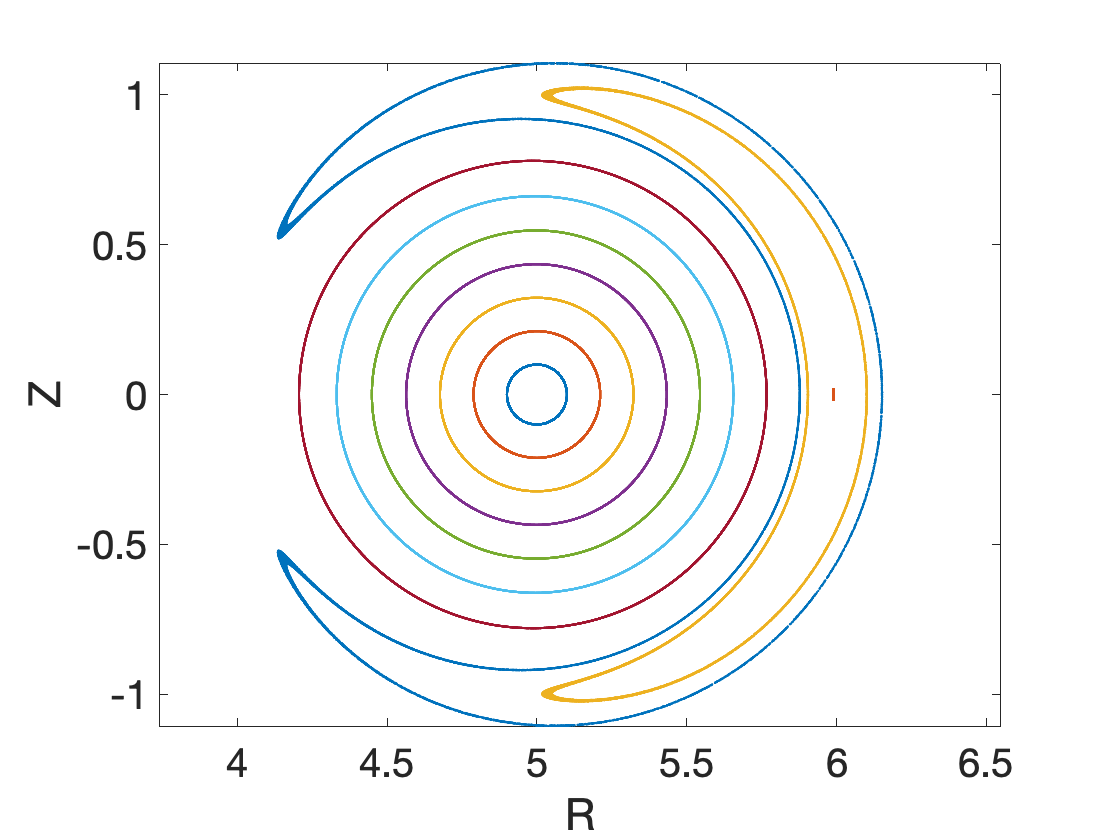}
     \caption{$\epsilon = 0.01$}
\end{subfigure}
\caption{Poincar\'e surfaces of section are shown for the model magnetic field \eqref{eq:model_KAM} with several values of $\epsilon$. Here magnetic field lines are followed in the model field, and when a field line hits a plane at constant toroidal angle $\phi$, a mark is made with color indicating the field line.}
    \label{fig:KAM}
\end{figure}

We will demonstrate by considering a model magnetic field given by, 
\begin{align}
    \bm{B}(r,\theta,\phi) = \nabla r \times \nabla \theta - \nabla \chi(r,\theta,\phi)\times \nabla \phi,
    \label{eq:model_KAM}
\end{align}
where $\chi(r,\theta,\phi) = r^2/2 (1 + \epsilon \cos(\theta-\phi))$. Here, considering the fixed curve defined by ($R_0 = 5$, $Z_0 =0$), $r$ is a radial coordinate measuring distance from the fixed curve within the poloidal plane, $\theta = \arctan\left((R-R_0)/(Z-Z_0)\right)$ is a poloidal angle, and $\phi$ is the standard azimuthal angle. When $\epsilon = 0$, this magnetic field is integrable, lying on surfaces of constant $r$ with rotational transform $\iota(r) = r$. When $\epsilon$ is non-zero, a non-axisymmetric perturbation is added that causes an island to form. When the perturbation is small, a large volume of magnetic surfaces remain. For larger values of $\epsilon$, a large magnetic island near the $\iota = 1$ surface begins to develop. Poincar\'{e} surfaces of section are shown in Figure \ref{fig:KAM} for several values of $\epsilon$. 

An important practical consequence of these results is that confinement is not completely destroyed upon an infinitesimal deviation away from axisymmetry, as necessarily occurs in all real tokamak and stellarator fusion devices. 
In closing we note that the existence of some flux surfaces (as shown above) is distinct from the existence of continuously nested flux surfaces (as occurs in axisymmetric magnetic fields).
This behavior is a consequence of the Hamiltonian nature of the flow of static magnetic field lines, and is thus is independent of choice of physical model used to describe the plasma itself. 
In seeking to use and develop models for plasma physics, it is therefore important to note the compatibility of imposed properties of the magnetic field with assumptions of the model.
Next we discuss some implications in the context of 3D MHD equilibria.

\subsection{Singularities and surface currents in 3D MHD}\label{sec:singularities}

As we saw in Section \ref{sec:integrability}, the existence or otherwise of magnetic surfaces in 3D is a direct consequence of the Hamiltonian nature of the system. In general 3D magnetic fields may not be integrable.
Nonetheless, as described in Section \ref{sec:magnetic_field_lines}, flux surfaces are required for good confinement properties.
Consequently, 
we are typically interested in configurations that possess at least a region with good flux surfaces.
A common tactic in 3D equilibrium calculations, including some approaches which apply a variational principle (see Sections \ref{sec:variational_principle} and \ref{sec:3d_mhd_surf}), is to assume the existence of continuously nested flux surfaces.

As we will see, the ideal MHD model requires certain constraints to be satisfied on rational surfaces. This is a fundamental challenge associated with constructing 3D MHD equilibria. We also find that assuming the existence of continuously nested flux surfaces in 3D ideal MHD equilibria gives rise to sheet currents on rational surfaces. We will show now that the constraints can be satisfied given assumptions on either the pressure profile or the magnetic geometry. Here we discuss two strategies to overcoming these challenges. The first motivates non-smooth equilibrium models which are discussed in detail in Section \ref{sec:MRXMHD}. The second involves choosing the boundary of the domain so as to enforce constraints on the pressure.


\subsubsection{Computation of the parallel current density}
Given an equilibrium solution for $\bm{B}$ with closed nested flux surfaces and a pressure profile, we now explicitly compute the parallel current density, $J_{\parallel}$, to illustrate the appearance of the current singularity and surface current associated with rational surfaces in 3D.
We will begin by considering the PDE for the parallel current
by expressing $\bm{J} = J_{||} \hat{\bm{b}} + \bm{J}_{\perp}$ and noting that  \eqref{eq:ampere_magnetostatic} implies that $\nabla \cdot \bm{J} = 0$, which gives 
\begin{align}
    \bm{B} \cdot \nabla \left( \frac{J_{||}}{B} \right) = - \nabla \cdot \bm{J}_{\perp}.
    \label{eq:parallel_current}
\end{align}
This type of PDE is known as a magnetic differential equation, as it involves derivatives along the magnetic field. The perpendicular current can be evaluated from equilibrium force balance \eqref{equilibrium},
\begin{align}
    \bm{J}_{\perp} = \frac{\bm{B} \times \nabla p}{B^2}.
\end{align}
In order for \eqref{eq:parallel_current} to have solutions, the following constraint must be satisfied,
\begin{align}
    \oint \frac{dl}{B} \, \left(\nabla \cdot \bm{J}_{\perp}\right) = p'(\psi) \oint \frac{dl}{B} \, \bm{B} \times \nabla \psi \cdot \nabla \left( \frac{1}{B^2} \right) = 0. 
\end{align}
This condition is obtained by integrating both sides of \eqref{eq:parallel_current} along a closed field line parameterized by length, $l$. This operation annihilates the left hand side due to periodicity. 
We can express this condition in the coordinate system $(\psi,\alpha,l)$, where $\psi$ is the toroidal flux label, $\alpha = \vartheta - \iota(\psi) \varphi$ is a coordinate constant on field line, and $\vartheta$ and $\varphi$ are the poloidal and toroidal angles comprising a magnetic coordinate system, as discussed in Section \ref{sec:magnetic_coordinates}. The integral constraint can now be expressed as 
\begin{align}
    p'(\psi) \partder{}{\alpha} \left( \oint \frac{dl}{B} \right) = 0.
    \label{eq:dl_over_B}
\end{align}
We note that \eqref{eq:dl_over_B} is satisfied if the pressure gradient vanishes on every closed field line, or on all rational surfaces. Alternatively, a constraint is placed on the magnetic field such that $\partial/\partial \alpha \left( \oint \frac{dl}{B} \right) = 0.$ It can be shown that the corresponding constraint for irrational field lines is automatically satisfied.

We can now seek a solution of \eqref{eq:parallel_current} in magnetic coordinates $(\psi,\vartheta,\varphi)$,
\begin{align}
\label{eq:mag_eqJperp}
    \partder{}{\vartheta} \left( \frac{J_{||}}{B} \right) + \iota(\psi)  \partder{}{\varphi} \left(\frac{J_{||}}{B} \right) &= - \sqrt{g} \nabla \cdot \bm{J}_{\perp},
\end{align}
noting that $\bm{B} \cdot \nabla \varphi = \sqrt{g}^{-1}$ and $\bm{B} \cdot \nabla \vartheta = \sqrt{g}^{-1} \iota(\psi)$. This equation has a similar structure as \eqref{eq:magnetic_K}, and will be solved in a similar manner. Under the assumption of periodicity, we can apply Fourier analysis,
\begin{align}
    \left(\frac{J_{||}}{B}\right) (\psi,\vartheta,\varphi) &= \sum_{m,n}\left(\frac{J_{||}}{B}\right)_{m,n} (\psi) e^{i(m\vartheta-n\varphi)} \\
    \left(\sqrt{g} \nabla \cdot \bm{J}_{\perp}\right) (\psi,\vartheta,\varphi) &= \sum_{m,n} \left(
\sqrt{g}\nabla \cdot \bm{J}_{\perp} \right)_{m,n} (\psi) e^{i(m\vartheta-n\varphi)},
\end{align}
 to obtain the following condition on the Fourier components
 \begin{align}
i \left(\iota(\psi)m - n \right) \left(\frac{J_{||}}{B} \right)_{m,n} = - \left(\sqrt{g} \nabla \cdot \bm{J}_{\perp} \right)_{m,n} . 
 \end{align}
 This implies that $(\sqrt{g} \nabla \cdot \bm{J}_\perp)_{m,n}$ must vanish for all $n/m = \iota(\psi)$, otherwise \eqref{eq:mag_eqJperp} would not have a solution. This is equivalent to the integral condition on rational surfaces given above, \eqref{eq:dl_over_B}. If this constraint is satisfied, then we can write a solution in the following form,
 \begin{align}
     \left(\frac{J_{||}}{B} \right) &= \sum_{m,n} \left( i \frac{(\sqrt{g}\nabla \cdot \bm{J}_{\perp} )_{m,n} }{\iota(\psi)m - n} + \Delta_{m,n} \delta(\iota(\psi)m-n)\right) e^{i(m\vartheta-n\varphi)},
     \label{eq:parallel_current_sum}
 \end{align}
 where $\Delta_{m,n}$ are undetermined constants. 
 
 The first term in \eqref{eq:parallel_current_sum}, known as the Pfirsch-Schl\"{u}ter current, is said to contain a singularity of type $1/x$, referring to the fact that there is a division by the quantity $x = \iota(\psi)-n/m$. This quantity  vanishes for rational values of $\iota(\psi)$, but also gets arbitrarily small even for any value of $\iota(\psi)$. 
 
 The second term in \eqref{eq:parallel_current_sum} is said to contain a singularity with respect to $\iota(\psi)$, or equivalently with respect to $\psi$, referring to the fact that this term can be non-zero only at rational values of $\iota(\psi)$ because of the $\delta$-function. This term represents a sheet current appearing only on surfaces with a rational value of the $\iota$.
 
The physical quantity associated with the current density is the current through any surface $S$. It is the surface integral of the current density, $\bm{J}$, given by 
 \begin{align}
     I = \int_{S} \bm{J} \cdot \hat{\bm{n}} \, d^2 x .
 \end{align}
We note that the $1/x$ singularity associated with the Pfirsch-Schl\"{u}ter current still yields a singularity after integration independently of the choice of surface $S$, as opposed to the $\delta$ singularity. Therefore, the $1/x$ Pfirsch-Schl\"{u}ter term is not a physical singularity, as it would imply an infinite amount of current, or charge moving through a surface per unit time. 

We will next discuss sufficient conditions to guarantee the existence of a physical solution, that is a solution with a non-singular current $I$.

\subsubsection{Overcoming singular currents at rational surfaces}
One approach to obtaining 3D equilibria which satisfy the required constraint is thus to avoid pressure gradients at rational surfaces, as seen in \eqref{eq:dl_over_B}.
In this case, in analogy with the convergence of the Fourier series in Section \ref{sec:perturbation_integrability}, pressure gradients may be avoided not only exactly at rational surfaces, but also in nearby neighborhoods, at irrationals which recurrently approximate rationals, to avoid singularities in the sum \eqref{eq:parallel_current_sum}, see \cite{grad1967}. Such a pressure profile, if it could be constructed, would be continuous but lead to highly non-smooth pressure gradients that may seem unnatural. 

A pressure profile which is constant across the entire plasma+vacuum domain does satisfy the above constraint but these are typically not of interest for realistic magnetic confinement fusion scenarios.
Generally, we are interested in equilibrium solutions with pressure gradients.   
The challenges associated with constructing 3D MHD equilibria with continuously nested flux surfaces thus motivates alternative models which assume the existence of only some flux surfaces leading to discontinuous but globally non-uniform pressure profiles. Some models are discussed in Sections \ref{sec:force_free} and \ref{sec:MRXMHD}.

Alternatively, instead of such seemingly unnatural profile, it has been shown that the constraints can be satisfied by careful construction of the magnetic geometry such that \eqref{eq:dl_over_B} is satisfied on rational surfaces \cite{Weitzner2014,Zakharov2015,Weitzner2016}. While this approach is advantageous as it does not require flattening of the pressure profile, it restricts the class of allowed boundaries of the MHD equilibria. It also requires a rotational transform close to a rational value and low shear (the rotational transform stays close to this rational value). 

In addition to the Pfirsch-Schl\"{u}ter current, recall the $\delta(x)$-function singularity which, when integrated, produces a current sheet, a non-zero current supported on a surface.
 Since, on integration, the $\delta(x)$-function current density yields a physically well-behaved, though discontinuous, current, the existence of current sheets in the equilibrium solutions is not generally considered to be problematic. 
Their existence near rational surfaces has been verified numerically in 3D equilibrium codes \cite{Loizu2015,Lazerson2016,Mikhailov2019}.

Although ideal MHD is associated with the difficulties discussed in this Section, it remains the primary tool for determining stellarator equilibrium magnetic fields. In the following we will briefly describe models which incorporate departures from ideal MHD. In Section \ref{sec:axis_expansion} we will discuss a 3D equilibrium solution with magnetic surfaces obtained by analytic construction. In Section \ref{sec:equilibrium_fields} we provide an overview of several 3D equilibrium models applied in numerical calculation.

\subsubsection{Islands and surface currents in a simplified model}
Recall that, in addition to the unphysical $1/x$ current, finite surface currents can also arise and are a consequence of the ideal MHD constraint which requires the topology of the magnetic field lines to be preserved exactly. In this section, we use a simple model to illustrate different equilibrium solutions which can exist when the ideal MHD constraints are relaxed.
When the current density is localised to a plane in $\mathbb R^3$, we call this a current sheet.
We consider here the model studied by Hahm and Kulsrud \cite{hahm1985}. They  illustrate different possible long time behaviors of solutions converging to a steady state either with or without a current sheet. Here we will focus on the steady state description, 
constructing solutions of the linear ideal MHD equilibrium model evidencing these two different magnetic field structures. Starting by building a smooth equilibrium $\bm B_0$ for a given boundary value problem, we will proceed by building solutions to a perturbation of this boundary value problem.

Here we solve the ideal MHD model using a different approach, which does not involve prescribing $p$. Taking the curl of the momentum conservation \eqref{eq:force_balance_MHD} to eliminate $\nabla p$, we consider a problem in which $p$ does not appear, so no prescription of the pressure gradient is necessary, while on the contrary $\nabla p$ can be recovered in a post-processing step.

In order to build equilibria, we will consider the ideal MHD model in the static, time-independent limit, where $\bm{u}=0$ and $\partial_t\to0$. Starting with the static MHD force balance $\nabla p=\bm{J}\times\bm{B}$ \eqref{equilibrium}
and using Ampere's law \eqref{eq:equilibrium1} to eliminate $\bm{J} = (\nabla \times \bm{B})/\mu_0$, the total magnetic field $\bm B$ then satisfies
\begin{equation}
\nabla \times \left(\left(\nabla \times \bm{B}\right)\times \bm{B} \right) = 0,
\end{equation}
which does not involve the pressure profile.
    
\begin{figure}
    \begin{center}
    \includegraphics[width=0.8\textwidth]{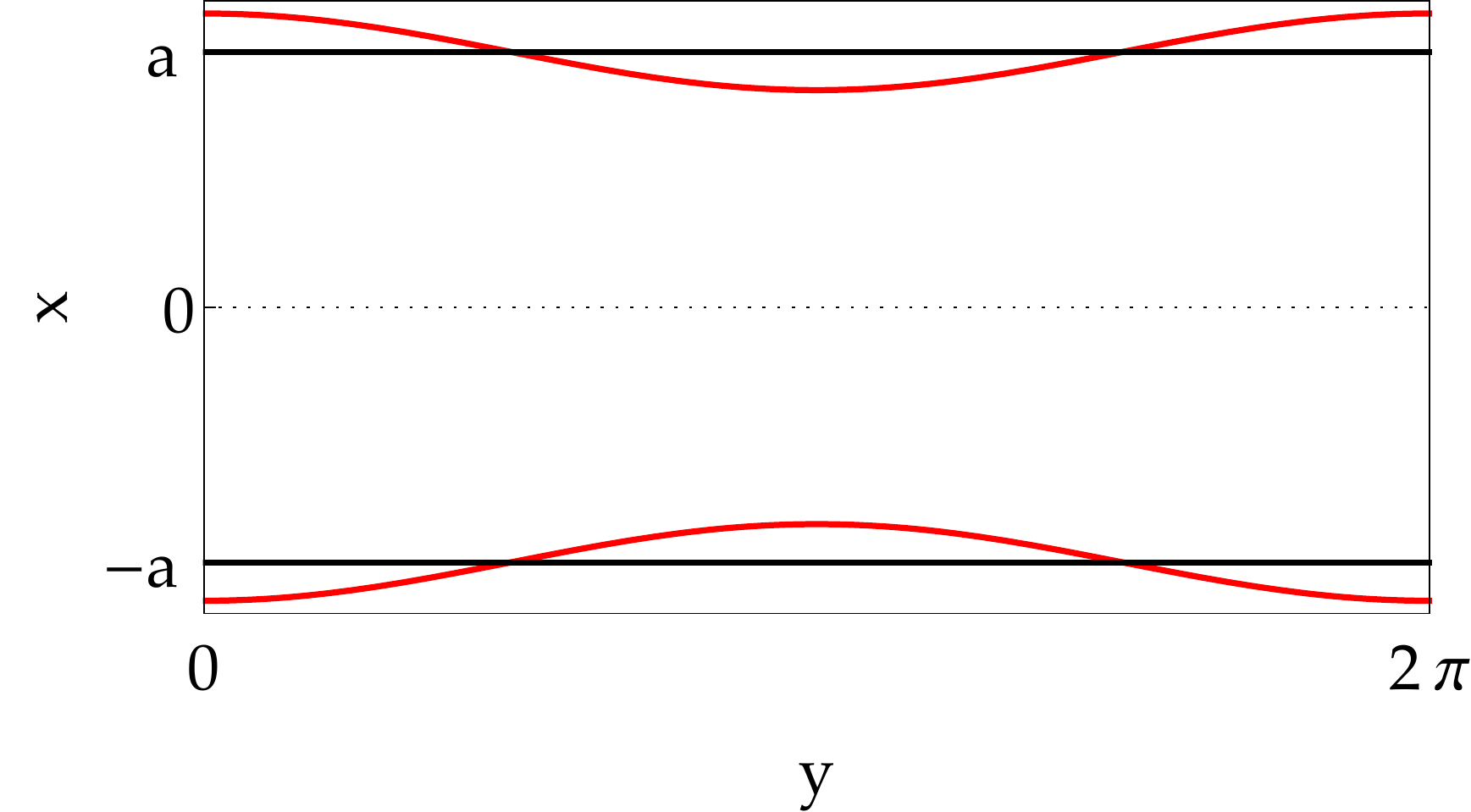}
    \caption{The reference ($\delta=0$) and perturbed ($\delta>0$) domains in the $(y,x)$-plane are shown. The black and red lines correspond to $x=\pm a$ and $x_{\pm a}(y)$, respectively.
    } \label{fig:OmegaDelta}
       \end{center}
    \end{figure}

Consider a domain in $\mathbb R^3$ defined by a rectangle in Cartesian coordinates $(x,y)$, with $x\in[-a,a]$ for some constant $a$, and $y \in [0,2\pi]$, 
together with $z\in\mathbb R$. All physical quantities are assumed to be independent of the $z$ coordinate and $2\pi$-periodic with respect to the $y$ coordinate. Suppose we perturb the initial domain into a deformed domain that is symmetric with respect to the $y$-axis. This is referred to as a mirror symmetry. More precisely, the top and bottom boundaries $x=\pm a$ are replaced by  $x_{\pm a}(y)=
\pm (a-\delta\cos ky)$ with $\delta\ll a$. In the following discussion, we follow the original procedure of \cite{hahm1985} in which the perturbation analysis is performed using $\delta$ as a small parameter. We note, however, that this definition is scale dependent since its validity depends on $\delta\ll a$. In subsequent work, e.g.\ \cite{dewar2013}, the expansion is performed in terms of $\delta/a$ which is dimensionless.
\TBD
Both the perturbed ($\delta>0$) and reference ($\delta=0$) resulting domain $\Omega_\delta$ are represented in the $(y,x)$-plane on Figure \ref{fig:OmegaDelta}. The top and bottom boundary will be referred to as $\Gamma_\delta$, in other words
\begin{equation}
    \Gamma_\delta=
    \{ (x_{+a}(y),y),y\in[0,2\pi] \}
    \cup 
    \{ (x_{-a}(y),y),y\in[0,2\pi] \} .
\end{equation}
As a result we will assume that all the solutions of the perturbed problem are all even functions of the $x$ variable.

The boundary value problem of interest to build equilibrium solutions is then
\begin{subequations}
\begin{align}
\nabla \times \left(\left(\nabla \times \bm{B}\right)\times \bm{B} \right) = 0\ \text{ in } \Omega_\delta
\label{eq:force_balance_condition} 
\\
\bm{B}\cdot\hat{\bm{n}}=0\ \text{ on } \Gamma_\delta.
\end{align}
\label{eq:BVPCurrentSheetEquil}
\end{subequations} 
This problem may not have a single solution, but we are only interested in possible behavior of solutions. 
As we will show, to linear order in $\delta$, \eqref{eq:BVPCurrentSheetEquil} is equivalent to solving \eqref{eq:force_balance_condition} on $\Omega_0$ with a modified boundary condition.
Hence, in what follows, we will not obtain solutions to \eqref{eq:force_balance_condition} in $\Omega_\delta$, but rather solutions in the unperturbed domain with a modified boundary condition that is consistent with a perturbation of the boundary to linear order in $\delta$.

A strong out-of-plane magnetic field is imposed, with a constant magnitude $B_T$. We assume that there is no perturbation to the $\hat{\bm{z}}$-component of the field, i.e. $(\bm{B}-B_T \hat{\bm{z}})\cdot\hat{\bm{z}}=0$ at every order in $\delta/a$. 
Under the assumption that $\nabla \cdot \bm{B} = 0$, we can then define a scalar function $\psi$ by $\psi(x,y) := \int_0^x B_y(x',y) \, dx'$ and hence write the components of the magnetic field $\bm{B} = B_x \hat{\bm{x}} + B_y \hat{\bm{y}} + B_z \hat{\bm{z}}$ as
\begin{subequations}
\begin{align}
B_x &= - \partder{}{y}\int_0^x B_y(x',y) \, dx' =- \partder{\psi(x,y)}{y} \\
B_y &= \partder{}{x} \int_0^x B_y(x',y) \, dx' = \partder{\psi(x,y)}{x}.
\end{align}
\label{eq:BxBypsi}
\end{subequations}
As a result, the magnetic field can be expressed in terms of the flux function $\psi$ as
\begin{align}
\bm{B}(x,y,z)=B_T\hat{\bm{z}}+\hat{\bm{z}}\times\nabla\psi(x,y)   .
\label{eq:hahm_kulsrud_potential}
\end{align}
So defining the flux function $\psi$ is sufficient to define the field $\bm B$

Consider the equilibrium magnetic field, $\bm{B}_0$, given by,
\begin{align}
\bm{B}_0=B_{T}\hat{\bm{z}}+\hat{\bm{z}}\times\nabla\psi_0(x), 
\label{eq:equilibrium_field}
\end{align}
where $\psi_0(x)=B_0x^2/(2a)$ for some constant $B_0$. 
The field $\bm{B}_0$ clearly solves \eqref{eq:force_balance_condition}  on the unperturbed domain $\Omega_0$, as well as the associated boundary condition  on $x\in\{-a,a\}$ since $\hat{\bm{n}}=\pm\hat{\bm{x}}$ and $\psi_0$ is independent of $y$, hence $\bm{B}_0\cdot\hat{\bm{n}}=(B_0)_x=0$. 

We now build a solution to \eqref{eq:BVPCurrentSheetEquil} with a perturbation of the boundary condition, for a small perturbation amplitude $\delta$. As the field is perturbed, we assume that the out-of-plane magnetic field does not change under the perturbation. We express a flux function $\psi$ as a perturbation of $\psi_0$ according to    
\begin{align}
    \psi(x,y)=\psi_0(x)+\delta\psi_1(x,y)+\mathcal{O}(\delta^2),
    \label{def:psidelta}
\end{align}
which implies that, 
\begin{align}
\bm{B} = \bm{B}_0+ \delta \bm{B}_1 + \mathcal{O}(\delta^2).    
\label{eq:perturbed_field}
\end{align}
As the field is perturbed, we assume that the out-of-plane magnetic field does not change under the perturbation. This implies $\bm{B}\cdot\hat{\bm{z}}\approx B_T$ to at least $\mathcal{O}(\delta^2)$ and is valid when $B_T$ is sufficiently large. Under these assumptions, the $\mathcal{O}(\delta)$ component of the perturbed magnetic field can be written as,
\begin{equation}
\bm{B}_1 = \hat{\bm{z}} \times \nabla \psi_1(x,y).\label{eq:hk_perturbed_flux_representation}
\end{equation}

From the definition of $\bm B_0$ \eqref{eq:equilibrium_field} and the expression of $\bm B_1$ \eqref{eq:hk_perturbed_flux_representation}, it is clear that both $\nabla\times \bm B_0$ and $\nabla\times \bm B_1$ are in the $\hat{\bm z}$ direction, and we easily get that $(\nabla\times\bm{B}_1)\cdot\nabla\bm{B}_0=0$ and $(\nabla\times\bm{B}_0)\cdot \nabla \bm{B}_1 =0$, while $(\bm{B}_1\cdot\nabla)\left(\nabla\times\bm{B}_0\right)=0$. The MHD force balance  \eqref{eq:force_balance_condition} for a field expressed as the perturbation expansion  \eqref{eq:perturbed_field} therefore reads,
\begin{equation}
\delta(\bm{B}_0\cdot\nabla)\left(\nabla\times\bm{B}_1\right)+\mathcal{O}(\delta^2)=0,
\label{eq:curlequilibrium}
\end{equation}
which, in terms of flux functions, is equivalent to, 
\begin{equation}
\frac{d\psi_0}{d x}(x)\frac{\partial}{\partial y}\Delta\psi_1(x,y)=0 .\label{eq:hk_perturbed_flux_pde}
\end{equation}

Focusing on solutions such that $\bm B$ satisfies the mirror symmetry, which can be expressed as $\psi_1(-x,y) = \psi_1(x,y)$, we will simply consider the equation on $\psi_1$ for $x>0$, and the corresponding solution for $x\le0$ will obtained from continuity and mirror symmetry. We therefore define $\Omega_\delta^+:=\Omega_\delta\cap\{(x,y,z)\in\mathbb R^3,x> 0\}$ and $\Gamma_\delta^+:=\Gamma_\delta\cap\{(x,y,z)\in\mathbb R^3,x> 0\}$.

Let's now turn to the associated boundary conditions.
We enforce continuity and mirror symmetry of $\psi$ along the line $x=0$ at all orders in $\delta/a$. 
We also enforce the constraint that $\psi$ on the boundary does not change as we perturb the boundary. This is expressed as the requirement that 
\begin{align}
    \psi_0 (x_a(y)) - \psi_0(a) + \delta \psi_1 (a,y) =  - \delta \cos(ky) \partder{\psi_0(x)}{x} \bigg\rvert_a + \delta \psi_1(a,y) + \mathcal{O}(\delta^2) = 0.
    \label{eq:const_psi_boundary}
\end{align}
Under a linear approximation
and using $\psi_0(a)=B_0a/2$ yields,
\begin{equation}
\psi_1( a,y)=B_0\cos(ky).\label{eq:hk_boundaryder}    
\end{equation}
This boundary condition can also be derived in the following way. On the top part of the boundary, $x=x_{a}(y)$, the boundary condition on the magnetic field is $\bm{B}\cdot\hat{\bm{n}}=0$ while the normal, defined by $\hat{\bm{n}} = (\delta k\sin(ky)\hat{\bm{y}} - \hat{\bm{x}}) / (1+\delta^2k^2\sin^2(ky))$, satisfies $\hat{\bm{n}} = - \hat{\bm{x}} + \delta k\sin(ky)\hat{\bm{y}} + \mathcal O(\delta^2)$. The resulting $\mathcal O(\delta^0)$ term of $\bm B\cdot \hat{\bm n}$ is zero from the definition of $\bm B_0$, while the resulting $\mathcal O(\delta^1)$ term reads
$[\partial_y\psi_1]( x,y)=-B_0k\sin(ky)$ for all $(x,y)\in\Gamma_{\delta}^+$. Since moreover $[\partial_y\psi_1]( a-\delta\cos(ky),y) = [\partial_y\psi_1]( a,y)+\mathcal O(\delta)$, we recover the first order term boundary condition \eqref{eq:hk_boundaryder}.

We will then turn to the construction of solutions to the problem,
\begin{subequations}
\begin{align}
    \frac{d\psi_0}{d x}(x)\frac{\partial}{\partial y}\Delta\psi_1(x,y)=0 \text{ on } \Omega^+_0
    \\
    \psi_1(a,y) = B_0 \cos(ky)\text{ on } \Gamma^+_0
    \label{eq:hk_boundary}
\end{align}
\label{eq:psi1}
\end{subequations}
 
On $\Omega_{0}^+$, we choose to expand $\psi_1(x,y)$ as a Fourier series in $y$,
\begin{equation}
\psi_1(x,y)=\widetilde{\psi}_{0}(x)+\sum_{l=1}^{\infty}\widetilde{\psi}_{l}(x)\cos(l y).\label{eq:hkfourier}    
\end{equation}
A sine series is not needed as the boundary condition is even with respect to $y$.
Plugging the Fourier expansion \eqref{eq:hkfourier} into the equation \eqref{eq:hk_perturbed_flux_pde} yields,
\begin{equation}
\sum_{l=1}^\infty l\left[\frac{d^2\widetilde{\psi}_{l}(x)}{dx^2}-l^2\widetilde{\psi}_{l}(x)\right]\sin(ly)=0.\label{eq:hk_pertpsi_ode}
\end{equation}
The $l = 0$ mode is unconstrained by \eqref{eq:hk_pertpsi_ode}, so for simplicity we choose to set it to zero and focus on the $l\neq 0 $ terms. The general solution of \eqref{eq:hk_pertpsi_ode} is given by $\widetilde{\psi}_l(x) = A_l\cosh(lx)+B_l\sinh(lx)$ for any $l \ne 0$, where $A_l$ and $B_l$ are some constants that we will determine thanks to the boundary condition. 

More precisely, the boundary condition \eqref{eq:hk_boundary} can then be satisfied by retaining only the $l = k$ mode of \eqref{eq:hkfourier}, so the corresponding solution to \eqref{eq:psi1} reads,
\begin{equation}
    \psi_1(x,y) = \left(A_k\cosh(kx)+B_k\sinh(kx)\right)\cos(ky).
\end{equation}

Therefore, getting back to the expansion \eqref{def:psidelta} of $\psi$ with respect to $\delta$, we will now focus on the first order approximation,
\begin{align}
\psi(x,y)=\frac{B_0}{2a}x^2+\delta\left(A_k\cosh(kx)+B_k\sinh(k|x|)\right)\cos(ky).\label{eq:hahmkulsrud_psi}  
\end{align}
The absolute value follows from imposing the mirror symmetry condition, $\psi(x,y)=\psi(-x,y)$. 
We can also eliminate one of the constants to parametrize the solution  by $A_k = \widetilde{\psi}_k(0)$, which is related to the amount of flux crossing the surface at $x=0$. The flux (per unit length in $z$) crossing $x=0$ between $y=0$ and $y=Y$ is $\int_{0}^{Y}B_y dy= \psi(0, Y) - \psi(0, 0)$, and it can be related to $\widetilde \psi_k$ from \eqref{eq:hahmkulsrud_psi} evaluated at $x=0$:
\begin{equation}
\psi(0,y)=\delta\widetilde{\psi}_k(0)\cos(ky).    
\end{equation}
To $\mathcal{O}(\delta)$ we have then built a solution given by,
\begin{align}
\psi(x,y)=\frac{B_0}{2a}x^2+\delta\left[\widetilde{\psi}_k(0)\left(\cosh(kx)-\frac{\sinh(k|x|)}{\tanh (ka)}\right)+B_0 \frac{\sinh(k|x|)}{\sinh(ka)}\right]\cos(ky) .\label{eq:hahmkulsrud_psi2}
\end{align}
It actually describes a family of solutions parameterized by the constant $\widetilde{\psi}_k(0)$. We are now interested in discussing some properties of these solutions.

\begin{figure}
    \begin{center}
    \includegraphics[width=0.8\textwidth]{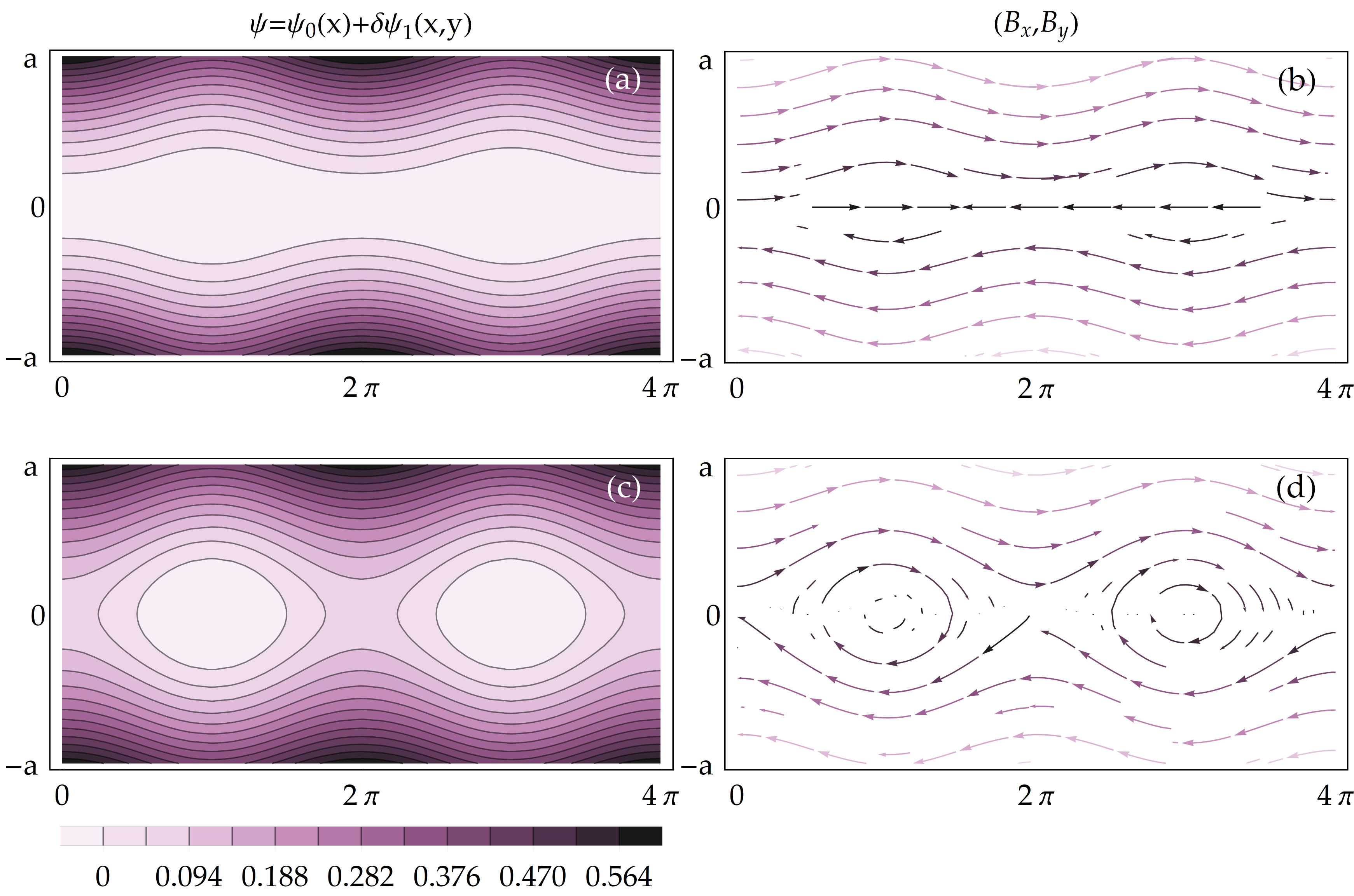}
    \caption{Illustrating the solutions of \eqref{eq:hahmkulsrud_psi} in the $(y,x)$ plane for $\widetilde{\psi}(0)=0$ (a) and $\widetilde{\psi}(0)=B_0/\cosh(ka)$ (c), with $B_0=1,~a=1,~k=1$ and $~\delta=0.1$.
    The stream plots of the corresponding magnetic field in the $(y,x)$ plane are shown in (b) and (d), respectively.}
     \label{fig:hahm_kulsrud}
       \end{center}
    \end{figure}

 Choosing the value of $\widetilde{\psi}_k(0)$ will lead to different topological structures of the constant $\psi$ surfaces, which are the flux surfaces of $\bm B$. This can be seen by considering local extrema of $\psi(x,y)$. In the unperturbed problem, the lines of constant $\psi$ wrap around the domain in the $y$ direction. If island structures are present, then several local extrema of $\psi$ occur, each of which is reached at a point corresponding to the island center, so the topology of the surfaces is different from the unperturbed case. To demonstrate that both of these solution types can occur, we consider two special cases, where $\widetilde{\psi}_{k}(0)=0$ and $\widetilde{\psi}_{k}(0)=B_0/\cosh(ka)$. For these two cases, the solutions for \eqref{eq:hahmkulsrud_psi2} are illustrated in Figure \ref{fig:hahm_kulsrud}. When $\widetilde{\psi}_k(0)=0$, to $\mathcal{O}(\delta)$, $\psi(x,y)$ has the same topology as the unperturbed function, $\psi_0(x)$. We next consider the case in which $\widetilde{\psi}_{k}(0) = B_0/\cosh(ka)$, for which the global minimum of $\psi$ occurs on the line $x = 0$ at $y = (2N +1) \pi/k$ for integers $N$. Thus several point minima arise, and the topology of the surfaces is not preserved. 

 Choosing the value of $\widetilde{\psi}_k(0)$ will may also lead to the presence of a current sheet, as some solutions possess a discontinuous magnetic field on the line $x = 0$. Indeed, to evaluate $\partial\psi/\partial x$, which corresponds to $B_y$ and may not be differentiable at $x=0$ due to mirror symmetry, we compute the one-sided derivatives by taking the limits $x\to0^{+}$ and $x\to0^{-}$ as follows,
\begin{subequations}
\begin{align}
\frac{\partial\psi(x,y)}{\partial x}\bigg \rvert_{x\rightarrow 0^+}&= \delta \left( -  \frac{\widetilde{\psi}_k(0) k}{\tanh(ka)} + \frac{B_0 k}{\sinh(ka)}  \right)\cos(ky) \\
\frac{\partial\psi(x,y)}{\partial x}\bigg \rvert_{x\rightarrow 0^-}&= \delta \left(  \frac{\widetilde{\psi}_k(0) k}{\tanh(ka)} - \frac{B_0 k}{\sinh(ka)}  \right)\cos(ky),
\end{align}
\end{subequations}
so that the jump in $B_y$ across $x=0$ is given by,
\begin{equation}
\left[\left[B_y\right]\right]_{x = 0}\equiv\frac{\partial\psi(x,y)}{\partial x}\bigg \rvert_{x\rightarrow 0^+}-\frac{\partial\psi(x,y)}{\partial x}\bigg \rvert_{x\rightarrow 0^-}=\frac{2k\delta\cos(ky)}{\sinh(ka)}\left[B_0-\widetilde{\psi}(0)\cosh(ka)\right].\label{eq:hk_perturbed_jump}   
\end{equation}
On the one hand, if $\widetilde{\psi}_k(0)=B_0/\cosh(ka)$ then the jump in the magnetic field \eqref{eq:hk_perturbed_jump} vanishes.
On the other hand, when $\widetilde{\psi}_k(0)=0$ then the  jump in the margnetic field \eqref{eq:hk_perturbed_jump} is non-zero.
Using the integral form of Ampere's law \eqref{eq:ampere_surface_int}, it follows that tangential discontinuity in $\bm{B}$ across $x=0$ when $\widetilde{\psi}_k(0)=0$ leads to a current density along $x=0$, $\bm{J}_S$, given by $\hat{\bm{n}}\times[\![\bm{B}]\!]=\mu_0\bm{J}_S$ where $[\![\bm{B}]\!]=\bm{B}_{x\to0^{+}}-\bm{B}_{x\to0^{-}}$. 
Using \eqref{eq:hk_perturbed_jump} we see that $\bm{J}_S=\mu_0^{-1}[\![B_y]\!]\hat{\bm{z}}$.
Since $\bm{J}_S$ is localized to a plane, it is known as a current sheet.
As we are solving \eqref{eq:hk_perturbed_flux_pde} for $x>0$, there is no guarantee that the solution constructed by symmetry across $x=0$ satisfies force balance. It turns out, however, that the current sheet does satisfy force balance at $x=0$, as can be seen by evaluating $[\![p+B^2/(2\mu_0)]\!]$ and observing it vanishes across $x=0$.

We have indeed built a family of solutions to the perturbed problem, both with and without a current sheet, hence emphasizing different possible behavior of linear MHD equilibria. Determining the conditions under which these equilibria correspond to steady state solutions of a physical system would require solving a time-dependent MHD model, initialized with $\bm{B}_0$ given by \eqref{eq:equilibrium_field}. This analysis is presented in \cite{hahm1985}.

\subsection{Beyond ideal MHD}
\label{sec:departures_MHD}




While the ideal MHD equilibrium model is advantageous in its (relative) simplicity, it makes considerable simplifying assumptions which can be insufficient in some contexts. In this section we discuss some modifications to and extensions of the ideal MHD model describing additional physical effects, in particular through the addition of dissipative terms to the ideal MHD equations.

If a time-dependent function converges as $t\rightarrow \infty$ to a steady-state, then trivially an approximation to the steady-state can be computed by evaluating the time-dependent solution at a large but finite time. 
Instead of solving the ideal MHD equilibrium equations, steady-state solutions (i.e.\ $\partial_t\to0$) of modified time-dependent MHD equations, with additional physics such as resistivity and viscosity effects, have been proposed to approximate equilibria.
Such relaxation approaches form the basis for some 3D MHD codes, such as the HINT and HINT2 \cite{Suzuki2006} discussed in Section \ref{sec:3d_mhd_no_surf}. A similar iterative method, which allows for the transport of pressure along field lines, is discussed in Section \ref{sec:iterative_pressure_gradient}. In Section \ref{sec:resisitivity}, we discuss the resistive MHD equations and their application to approximating equilibrium states. In Section \ref{sec:viscosity}, we discuss the inclusion of diffusion terms in the MHD equations.


\subsubsection{An iterative method for approximate MHD equilibria}
\label{sec:iterative_pressure_gradient}
From the ideal MHD force balance, it follows that $\bm{B} \cdot \nabla p = 0 $, i.e.\ that the pressure is constant along magnetic field lines.
Where the magnetic field is stochastic and a field line fills up a finite volume, an implication of the ideal MHD model is that the pressure must also be constant in these regions.
In practice, however, finite diffusion of heat along field lines can lead to pressure gradients in stochastic field regions. 
This has been experimentally confirmed in the  W7-AS \cite{Zarnstorff2004,Reiman2007} and LHD \cite{Sakakibara2008,Suzuki2009,Suzuki2013} stellarators. Thus non-ideal effects are required to more accurately model these regions.

In ideal MHD, it is assumed that pressure equilibrates along magnetic field lines instantaneously. Multiple approaches have been proposed to modify the governing system of equations to accommodate large but finite parallel pressure transport together with non-zero perpendicular diffusion, such that pressure gradients may be supported in stochastic field regions.
In the following, we discuss a selection of these models and some considerations associated with their numerical implementation.

It has been proposed \cite{1986Reiman} in that a small pressure gradient along field lines can be balanced by viscous forces, which can be modeled by considering a modified force balance equation, 
\begin{align}
    \bm{J} \times \bm{B} -\nabla p + \bm{F} = 0,
    \label{eq:non_ideal_force_balance}
\end{align}
where $\bm{F}$ is assumed to be a small correction to the ideal MHD equations such that $|\bm{F}| \ll |\nabla p|$. For example, $\bm{F}$ could account for the effects of flows, $\rho \bm{u} \cdot \nabla \bm{u}$, or a pressure tensor, $-\nabla \cdot \bm{\pi}$.
Taking the dot product with $\bm{B}$ we find 
\begin{align}
    \bm{B} \cdot \nabla p = \bm{B} \cdot \bm{F}. 
    \label{eq:parallel_gradient}
\end{align}
This implies that a small pressure gradient along magnetic field lines can be supported, thus accommodating 
pressure gradients in stochastic field regions. Taking the cross product of \eqref{eq:non_ideal_force_balance} with $\bm{B}$ we find,
\begin{align}
    \bm{J}_{\perp} &= \frac{\bm{B} \times \nabla p}{B^2} - \frac{\bm{B}\times \bm{F}}{B^2}. 
\end{align}
In practice $|\bm{B} \times \bm{F}| \ll |\bm{B} \times \nabla p|$, so we can approximate
\begin{align}
    \bm{J}_{\perp} \approx \frac{\bm{B} \times \nabla p}{B^2}.
    \label{eq:J_perp}
\end{align}
Since quasineutrality implies that $\nabla \cdot \bm{J} = 0$ in equilibrium, and the current density can be expressed as $\bm{J} = J_{||} (\bm{B}/B) + \bm{J}_{\perp}$,
we then obtain the condition,
\begin{align}
    \bm{B} \cdot \nabla \left( \frac{J_{||}}{B} \right) = - \nabla \cdot \bm{J}_{\perp},
    \label{eq:J_par}
\end{align}
using \eqref{eq:equilibrium2}. 

A set of equations for the current density as a function of the magnetic field and pressure is given by \eqref{eq:J_perp} and \eqref{eq:J_par}. In addition, Ampere's law \eqref{eq:equilibrium1} can be solved for $\bm{B}$ given $\bm{J}$. Hence, for a given pressure, which can be obtained from experimental or transport modeling, we can consider the system of two vector equations, \eqref{eq:equilibrium1} and \eqref{eq:J_perp}, and one scalar equation \eqref{eq:J_par},  
for the unknowns $\bm{B}$ and $\bm{J}$. An approximate solution can be constructed by iterating between \eqref{eq:J_perp}-\eqref{eq:J_par} and \eqref{eq:equilibrium1}. Given $\bm{B}$ and $p$, \eqref{eq:J_perp} provides an equation for $\bm{J}_{\perp}$, and this is used to compute $J_{||}$ from \eqref{eq:J_par}. Given $\bm{J}$ the magnetic field is then updated from \eqref{eq:equilibrium1}.
Additional challenges are introduced in obtaining solutions to the magnetic differential equation \eqref{eq:J_par}. While $\nabla \cdot \bm{J}$ should vanish in any physical system, as discussed in Section \ref{sec:singularities}, inverting the $\bm{B} \cdot \nabla$ operator is singular on rational field lines, and periodic solutions to \eqref{eq:J_par} only exist under the restriction that \eqref{eq:dl_over_B} be satisfied for all closed field lines. 
However, possible numerical solutions have been proposed. In regions of good flux surfaces, \eqref{eq:J_par} can be solved using magnetic coordinates similar to the analysis in Section \ref{sec:singularities}. In stochastic regions a statistical averaging approach can be employed, as described in \cite{Krommes2009}. This is the method used in the PIES \cite{1986Reiman} code, which will be discussed further in Section \ref{sec:3d_mhd_no_surf}.

\subsubsection{Resistive MHD}
\label{sec:resisitivity}

As discussed in Section \ref{sec:flux_freezing}, another consequence of ideal MHD is that the magnetic field lines are ``frozen-in'' meaning no topological changes of the magnetic field are permitted. An implication, for example, is that equilibrium codes which utilize ideal MHD constraints can find, in principle, only solutions with the same magnetic field topology as the configuration prescribed during the initialization.

In Section \ref{sec:singularities} we saw that seeking solutions of the 3D ideal MHD equations with sufficiently smooth pressure results in non-smooth solutions. This motivates the search for solutions with greater regularity, i.e. seeking more differentiable solutions. Various models have been proposed and studied to regularize the time-dependent ideal MHD equations, as in \cite{larios2014} and references therein. The properties of the solutions to these models have been the subject of both theoretical and numerical analysis \cite{catania2011,kuberry2012}. The singular currents of 3D ideal MHD can also be regularized by incorporating additional physics into the governing equations, or in other words, by adding dissipative terms to the ideal MHD model.


The most important term associated with the smoothing of singular currents is the addition of resistivity, acting as a dissipation term for the current density. Resistivity is incorporated into the MHD equations thanks to the replacement of \eqref{eq:ideal_ohms_law} by
\begin{align}
    \bm{E} + \bm{u} \times \bm{B} = \eta \bm{J}.
    \label{eq:resistive_ohms}
\end{align}
In plasma physics, there are two commonly employed models for resistivity, corresponding to choosing the parameter $\eta$.
One model, known as Spitzer resistivity \cite{Spitzer1953}, is based on electron-ion collisions and such that $\eta\propto T_e^{-3/2}$, where $T_e$ is the electron temperature.
In analogy with the study of non-ideal (i.e\ viscous) neutral fluids, another common approach is to introduce a resistive boundary layer in the region of interest so that the plasma is considered ideal (i.e.\ $\eta =0$) everywhere except in a small volume about the region of interest, for example where there are large local gradients in the magnetic field.
In many foundational analyses which employ a boundary layer approach \cite{furth1963,hahm1985}, $\eta$ is taken to be constant and non-zero in the boundary layer.
Since plasmas are generally highly conducting in fusion relevant conditions, this is a reasonable approximation for macroscopic MHD behaviour.

The entropy equation \eqref{eq:ideal_mhd_energy} is also modified to include the effects of resistivity,
\begin{align} 
\partder{p}{t} + \bm{u} \cdot\nabla p + \gamma p \nabla \cdot \bm{u} = (\gamma-1)\eta J^2.
\end{align}
This implies that a stationary equilibria ($\partial/\partial t = \bm{u}=0$) exists only if $J = 0$. As such states are typically not of interest, the entropy equation can be modified to include a pressure source term, $S_P$,
\begin{align} 
\partder{p}{t} + \bm{u} \cdot \nabla p + \gamma p \nabla \cdot \bm{u} = (\gamma-1)\eta J^2 + S_P.
\label{eq:pressure_resistive}
\end{align}
In order to reach a steady state, the pressure source term is often negative, $S_P<0$, acting as a sink. 
Clearly, any source terms must be chosen carefully to ensure the model remains consistent with other assumptions that are made.
One approach is to define the sources implicitly to satisfy an imposed assumption, for example, that the value of a plasma parameter (such as temperature) is fixed at the plasma boundary.
The time-dependent resistive MHD equations are comprised of \eqref{eq:pressure_resistive} along with \eqref{eq:density_MHD}, \eqref{eq:force_balance_MHD}, \eqref{eq:maxwell_MHD}, and \eqref{eq:resistive_ohms}. To summarize, the resistive MHD model is
\begin{align}
\left\{
\begin{array}{rl}
\bm{E} + \bm{u} \times \bm{B} &= \eta \bm{J} \\
\partder{p}{t} + \bm{u} \cdot \nabla p + \gamma p \nabla \cdot \bm{u} &= (\gamma-1)\eta J^2 + S_P \\
    \partder{\rho}{t} + \nabla \cdot \left( \rho \bm{u} \right) &= 0 \\
    \rho \left( \partder{}{t} + \bm{u} \cdot \nabla \right) \bm{u} &= \bm{J} \times \bm{B} - \nabla p \\
    \nabla \times \bm{B} &= \mu_0 \bm{J} \\
    \nabla \times \bm{E} &= -\partder{\bm{B}}{t} \\
    \nabla \cdot \bm{B} &= 0.
    \end{array}
\right.
\label{eq:resistive_mhd}
\end{align}
%
%
They form a coupled, non-linear set of equations \cite{Shadid2016}, the numerical solution of which is highly non-trivial and the subject of active research. 
We remark that the global existence and regularity of solutions to the dissipative MHD equations in 2D and 3D have also been the subject numerous studies in mathematics. Some of these considerations are discussed further in references such as \cite{wu2003, sermange1983}. 

As ideal MHD does not allow for changes in magnetic topology, the resistive MHD equations are sometimes used to compute equilibrium states by evolving to a steady state \cite{Reiman2007}. This allows for the evolution from an initial condition with integrable fields towards an equilibrium with structures such as magnetic islands and field line chaos.  
Loosely speaking, the solutions may be of three types. (i) The system reaches a steady state and all quantities become independent of time. (ii) The system is not strictly in a steady state and changes in time due to effects such as the resistive diffusion of the magnetic field. However, the evolution occurs on sufficiently long timescales that the system can be thought of as being in a quasi-steady state. (iii) The solution may not come to an effective steady state, but may become unbounded or exhibit cyclic behavior.
Since solutions of the system can exhibit such qualitatively different behaviours, not all of them are consistent with approximating an equilibrium. 
The case of (i) is, of course, consistent with approximating an equilibrium. Even so, however, in practice, if the convergence of the solution to a steady-state is very slow, then approximating this steady-state by numerically evolving the system in time might become too costly. 
Similarly, whether (ii) is a suitable approximation depends on the properties of the particular problem under consideration. In practice, when computing numerical solutions to \eqref{eq:resistive_mhd} to approximate an equilibrium, cases (i) and (ii) may be indistinguishable, as one must choose a finite time scale over which the solution is effectively in a steady state.

The non-linearity of the resistive MHD model leads to challenges in terms of establishing the existence of steady-state solutions, but also in terms of sensitivity to initial conditions. For example, although a steady-state solution may exist, this alone is not sufficient to guarantee convergence to such a solution for arbitrary initializations. Solving \eqref{eq:resistive_mhd} to determine time evolution and approximate non-ideal steady states is routine for tokamak applications where the domain is axisymmetric, even if the solutions may not be. 
In stellarator geometry, the non-axisymmetric domain poses additional challenges to obtaining solutions of \eqref{eq:resistive_mhd}. 
Over the years, different numerical schemes have been proposed to either solve \eqref{eq:resistive_mhd} directly, or to solve an augmented system of equations which are argued to be equivalent, in a stellarator domain. 
This includes the approaches used by the M3D code \cite{Sugiyama2001} and the HINT2 code \cite{Suzuki2009} which is discussed in greater detail in Section \ref{sec:3d_mhd_no_surf}. 





\subsubsection{Other diffusion effects}
\label{sec:viscosity}

Resistivity primarily arises due to collisions in a plasma. Collisions also gives rise to further modifications of the MHD equations, such as diffusion of momentum and pressure.

Viscous effects can be incorporated into the momentum balance equation \eqref{eq:resistive_mhd} by considering the hydrostatic pressure tensor which, in a general form, can be written as \cite{hosking2016}, 
\begin{align}
\bm{p}=\left[p-\left(\mu_v-\frac{2}{3}\mu\right)\nabla\cdot\bm{u}\right]\overleftrightarrow{\bm{I}}-\mu\left[\nabla\bm{u}+(\nabla\bm{u})^T\right],\label{eq:pressure_tensor}    
\end{align}
where $\mu$ is the shear viscosity coefficient and $\mu_v$ is the volume viscosity coefficient. The pressure tensor, $\bm{p}$, has the same interpretation as in hydrodynamics, and so we consider briefly the translation into magnetohydrodynamics. The volume viscosity coefficient, $\mu_v$, is related to thermodynamic changes associated with expansion and contraction of the fluid. The total hydrostatic pressure is $(p-\mu_v\nabla\cdot\bm{u})$. 

When $(\mu_v-2\mu/3)\nabla\cdot\bm{u}$ is negligible, which is valid, for example, when $\nabla\cdot\bm{u}\approx0$, \eqref{eq:pressure_tensor} reduces to $\bm{p}=p\overleftrightarrow{\bm{I}}-2\mu\bm{s}$ where $\bm{s}=\frac12(\nabla \bm{u} +\left(\nabla \bm{u}\right)^T)$ is the rate of deformation tensor \cite{hosking2016}. Taking the divergence of $\bm{p}=p\overleftrightarrow{\bm{I}}-2\mu\bm{s}$ with $\mu=0$ gives the pressure gradient term, $\nabla p$, of the usual momentum balance, \eqref{eq:resistive_mhd}. If $\mu\neq0$ and $\nabla\cdot\bm{u}\approx0$ then $\nabla\cdot\bm{p}\approx\nabla p-\mu\Delta\bm{u}$ hence \eqref{eq:resistive_mhd} can be modified to yield, 
\begin{align}
    \rho \left(\partder{\bm{u}}{t} + \bm{u} \cdot \nabla \bm{u} \right) = \bm{J} \times \bm{B} - \nabla p + \mu \Delta \bm{u},
\end{align}
which includes a simple model for diffusion of momentum. The inclusion of such a term will tend to damp out gradients in $\bm{u}$ to converge to a steady-state solution with a smoother flow profile.


The entropy equation can also be modified to include diffusion of pressure,
\begin{align}
\partder{p}{t} + \bm{u} \cdot \nabla p + \gamma p \nabla \cdot \bm{u} = (\gamma-1)\eta J^2 + S_P +  \nabla \cdot \left(\rho\overleftrightarrow {\bm{\kappa}} \cdot \nabla \left(\frac{p}{\rho} \right) \right),
\end{align}
where $\overleftrightarrow{\bm{\kappa}}$ is a thermal conductivity tensor. 
\TBD 
By choosing $\overleftrightarrow{\bm{\kappa}}$ to be diagonal and setting $\rho=\text{const}$., we arrive at a simple model for anisotropic pressure diffusion where,
\begin{equation}
\nabla \cdot \left(\rho\overleftrightarrow {\bm{\kappa}} \cdot \nabla \left(\frac{p}{\rho} \right) \right)=\nabla\cdot\left(\kappa_\parallel\nabla_\parallel p+\kappa_\perp\nabla_\perp p\right).    
\end{equation}
This model for anisotropic pressure diffusion has been used to study the stochastic field regions \cite{Hudson2010a}, as it allows for temperature equilibration both parallel and perpendicular to the magnetic field.
In practice, diffusion coefficients can be motivated by both physical and numerical considerations. For example, the viscosity coefficients, $\mu$, may be tuned to aid numerical stability. Thermal conductivity coefficients may be determined empirically or calculated from different transport models.

Resistive MHD and other time-dependent non-ideal MHD models are also, of course, used to investigate transient and other dynamical phenomena.
Resistive MHD models can form a subset of more generalised MHD models, which can be referred to collectively as extended MHD \cite{Schnack2009}. 
There exist several numerical implementations of the 3D extended MHD equations such as the initial value codes M3D-C1 \cite{Jardin2007}, NIMROD \cite{Sovinec2004} and JOREK \cite{czarny2008}.

\subsection{Constructing 3D MHD equilibria near the magnetic axis}
\label{sec:axis_expansion}

While computing \textit{global} 3D MHD equilibria with surfaces is associated with difficulties, these can be circumvented by restricting our attention to a \textit{localized} region.
We will construct a class of \textit{local} solutions to the vacuum field equations with surfaces near the magnetic axis. The calculation will be performed by asymptotic expansion with respect to the distance from the axis. As a result, we obtain the expression for the rotational transform on the magnetic axis discussed in Section \ref{sec:producing_rotational_transform}. 

The magnetic axis $\bm{r}_0$ is parameterized by the length, $l$. The corresponding Frenet-Serret orthonormal basis introduced in Section \ref{sec:producing_rotational_transform} is denoted by  $(\hat{\bm{e}}_1(l),\hat{\bm{e}}_2(l),\hat{\bm{e}}_3(l))$, and $(\rho,\vartheta)\in \mathbb R^+\times [0,2\pi)$ denote polar coordinates in the $\hat{\bm{e}}_2$-$\hat{\bm{e}}_3$ plane. The position $\bm{r}$ can then be expressed as:
\begin{gather}
\bm{r}(\rho, \vartheta, l ) = \bm{r}_0(l) +  \rho \cos (\vartheta)\hat{\bm{e}}_2(l) + \rho \sin (\vartheta)\hat{\bm{e}}_3(l) .
\label{eq:r_rho_vartheta_l}
\end{gather}
In order to construct an orthogonal system, consider the angle $\omega = \vartheta + \int_0^l \tau(l') dl'$, where $\tau$ is the torsion of the magnetic axis \eqref{eq:torsion}.
In the coordinate system $(\rho,\omega,l)$ the Jacobian is given by $\sqrt{g}(\rho,\omega,l) = \rho\left(1-\kappa(l) \rho \cos \left( \omega - \int_0^l \tau(l') dl' \right)\right)$ where $\kappa(l) = |\hat{\bm{e}}_1 (l)\cdot \nabla \hat{\bm{e}}_1(l)|$ is the magnitude of the curvature. 

We seek a vacuum field solution (see Section \ref{sec:vacuum}) so $\bm{B}$ is both divergence and curl free. From the curl-free condition we can represent the field by a scalar potential, $\bm{B} = \nabla \Phi(\bm{r})$. From the divergence-free condition the potential must satisfy Laplace's equation, following Section \ref{sec:vacuum}, $\Delta \Phi(\bm{r}) = 0$.
Under the assumption that a field with continuously nested flux surfaces exists, 
the magnetic field can be described in terms of flux coordinates by a toroidal flux label $\psi$ and a function which labels field lines, $\alpha$, satisfying
\begin{gather}
     \nabla \Phi(\bm{r}) = \nabla \psi(\bm{r}) \times \nabla \alpha(\bm{r}),
    \label{eq:field_axis}
\end{gather}
as described in Section \ref{sec:magnetic_coordinates}, see \eqref{eq:B_magnetic}. 
In order to build the vacuum magnetic field with surfaces, we now seek the unknown scalar functions $\Phi$, $\psi$, and $\alpha$.
There is obviously no unique solution to this problem for two reasons: (1) each unknown can be defined up to an additive constant without altering \eqref{eq:field_axis}, and (2) the right-hand side being bilinear in $(\psi,\alpha)$, each of these two unknowns can be multiplied respectively by a constant and its inverse without altering the equation. This second point will be addressed through the constant $\mu$ defined below, by imposing that $\psi(\bm{r})$ be the toroidal flux label. The functions $\Phi(\bm{r})$ and $\alpha(\bm{r})$ need not be periodic functions of $l$ and $\omega$, while $\psi(\bm{r})$ must be periodic as it is related to the magnetic flux. 

Expressed in our orthogonal coordinate system $(\rho,\omega,l)$, \eqref{eq:field_axis} is equivalent to the following set of equations
\begin{subequations}
\begin{align}
       \frac{1}{\rho} \left( \partder{\psi}{\rho} \partder{\alpha}{\omega} - \partder{\psi}{\omega} \partder{\alpha}{\rho} \right) &= \frac{1}{h} \partder{\Phi}{l}  \label{eq:l_axis} \\
   \frac{1}{\rho h}\left(\partder{\psi}{\omega} \partder{\alpha}{l} - \partder{\alpha}{\omega}\partder{\psi}{l} \right)&= \partder{\Phi}{\rho} \label{eq:rho_axis} \\
 \frac{1}{h} \left(\partder{\psi}{l} \partder{\alpha}{\rho} - \partder{\alpha}{l} \partder{\psi}{\rho} \right) &= \frac{1}{\rho} \partder{\Phi}{\omega} \label{eq:omega_axis},
\end{align}
\label{eq:axis}
\end{subequations}
and Laplace's equation is expressed as
\begin{align}
        \frac{1}{h \rho } \partder{}{\rho} \left( h \rho \partder{\Phi}{\rho} \right) + \frac{1}{h\rho^2}\partder{}{\omega} \left( h \partder{\Phi}{\omega} \right) + \frac{1}{h} \partder{}{l} \left(\frac{1}{h} \partder{\Phi}{l} \right) = 0,
       \label{eq:axis_laplace}
\end{align}
where $h(\rho,\omega,l) = 1 - \kappa(l) \rho \cos \left( \omega - \int_0^l \tau(l') dl' \right)$. 

Rather than search for a general solution to \eqref{eq:axis}-\eqref{eq:axis_laplace}, we focus on a local solution near the magnetic axis. 
To this end, we consider the following asymptotic series expansions in $\rho$,
\begin{subequations}
\begin{align}
    \Phi(\rho,\omega,l) &= \Phi_{[0]}(\omega,l)  + \Phi_{[1]}(\omega,l) \rho + \Phi_{[2]}(\omega,l) \rho^2 + \mathcal{O}(\rho^3) \\
    \psi(\rho,\omega,l) &= \psi_{[2]}(\omega,l) \rho^2 + \mathcal{O}(\rho^3)  \\
    \alpha(\rho,\omega,l) &= \alpha_{[0]}(\omega,l) + \alpha_{[1]}(\omega,l) \rho + \mathcal{O}(\rho^2), 
\end{align}\label{eq:expansions}
\end{subequations}
where $\psi_{[0]}=0$ as the magnetic flux vanishes on the magnetic axis, and $\psi_{[1]} = 0$ since we require that $\nabla \psi$ vanish on axis as it is a coordinate singularity.

The $\mathcal{O}(\rho^{-1})$ expression of \eqref{eq:omega_axis} gives
\begin{align}
    \partder{\Phi_{[0]}}{\omega} = 0.
\end{align}
Thus $\Phi_{[0]}$ is an unknown function of $l$ only: $\Phi_{[0]}(l)$. 
From \eqref{eq:field_axis}, this implies that the $\mathcal{O}(\rho^0)$ magnetic field strength on axis is $B_{[0]}(l) = \Phi_{[0]}'(l)$. 
The expression of \eqref{eq:rho_axis} to $\mathcal{O}(1)$ is 
\begin{align}
    \Phi_{[1]} = 0. 
\end{align}
The expression of \eqref{eq:axis_laplace} to $\mathcal{O}(1)$ is 
\begin{gather}
     4\Phi_{[2]} + \partder{^2\Phi_{[2]}}{\omega^2}  + \partder{^2 \Phi_{[0]}}{l^2} = 0,
\end{gather}
then $\Phi_{[2]}$ can be expressed in terms of $\Phi_{[0]}$ as
\begin{align}
    \Phi_{[2]}(\omega,l) = - \frac{\Phi_{[0]}''(l)}{4} + \Phi_c(l) \cos(2u(\omega,l)) + \Phi_s(l) \sin(2u(\omega,l)),
    \label{eq:phi_2}
\end{align}
where $u(\omega,l) =  \omega - \int_0^l \tau(l') \, dl' + \delta(l)$ and $\delta(l)$, $\Phi_c(l)$, and $\Phi_s(l)$ are integration constants with respect to $\omega$.

The expressions of \eqref{eq:omega_axis} to $\mathcal{O}(\rho)$ and  \eqref{eq:l_axis} to $\mathcal{O}(1)$ give
\begin{subequations}
\begin{align}
    \partder{\alpha_{[0]}}{l}  &= - \partder{\Phi_{[2]}}{\omega} \frac{1}{2 \psi_{[2]}}
    \label{eq:alpha_l}
\\
    \partder{\alpha_{[0]}}{\omega}& = \partder{\Phi_{[0]}}{l} \frac{1}{\psi_{[2]}}. 
    \label{eq:alpha_omega}
\end{align}
\end{subequations}
By requiring that the mixed partial derivatives of $\alpha_{[0]}$ commute, we arrive at a PDE for $1/\psi_{[2]}$,
\begin{align}
    \partder{}{\omega} \left( \partder{\Phi_{[2]}}{\omega} \frac{1}{\psi_{[2]}} \right) + \partder{}{l} \left(\partder{\Phi_{[0]}}{l} \frac{1}{\psi_{[2]}} \right) = 0.  
    \label{eq:consistency}
\end{align}
Expressed as a linear PDE for $\psi_{[2]}$ in terms of $\Phi_{[2]}$ and $\Phi_{[0]}$, subject to boundary conditions of periodicity in $\omega$ and $l$, a solution can be written
\begin{align}
   \psi_{[2]}(\omega,l) = \Phi_{[0]}'(l)\mu  \left(e^{\eta(l)} \cos^2 u(\omega,l) + e^{-\eta(l)} \sin^2 u(\omega,l)\right),
   \label{eq:psi_2}
\end{align}
for constants of integration $\eta(l)$ and $\mu$.

Equation \eqref{eq:consistency} also implies the following forms for $\Phi_s(l)$ and $\Phi_c(l)$,
\begin{subequations}
\begin{align}
    \Phi_s(l) &= \frac{\Phi_{[0]}'(l) (\delta'(l)-\tau(l))}{2} \tanh(\eta(l)) 
    \label{eq:Phi_s}
    \\
    \Phi_c(l) &= -\frac{\Phi_{[0]}'(l) \eta'(l)}{4}.
    \label{eq:Phi_c}
\end{align}
\end{subequations}
Thanks to the expression of $\Phi_{[2]}$ \eqref{eq:phi_2} with the integration constants $\Phi_s$ \eqref{eq:Phi_s} and $\Phi_c$ \eqref{eq:Phi_c}, together with the expression of $\psi_{[2]}$  \eqref{eq:psi_2}, $\alpha_{[0]}$ can be computed from \eqref{eq:alpha_l} as
\begin{align}
    \alpha_{[0]}(\omega,l) = \overline{\alpha} + \frac{1}{2 \mu}\arctan\left( e^{-\eta(l)} \tan u(\omega,l) \right)-\frac{1}{2 \mu} \int_0^l \frac{\delta'(l')-\tau(l')}{\cosh(\eta(l'))} \, dl',
    \label{eq:alpha_solution}
\end{align}
where $\overline{\alpha}$ in a constant of integration.

We can now fix the constant of integration, $\mu$. We define, $\Psi_{[2]} = \rho^2 \psi_{[2]}$, the lowest order approximation of the toroidal flux label $\psi$. We use the flux coordinate system $(\Psi_{[2]},\vartheta,l)$ to compute the toroidal flux from  \eqref{eq:psi_T} using the lowest order magnetic field, $\nabla \Psi_{[2]} \times \nabla \alpha_{[0]}$,
\begin{align}
    2\pi \Psi_{[2]} &= \int_0^{\Psi_{[2]}} \int_0^{2\pi} \frac{\nabla \Psi_{[2]}' \times \nabla \alpha_{[0]} \cdot \nabla l}{\nabla \Psi_{[2]}' \times \nabla \vartheta \cdot \nabla l} d \vartheta d\Psi_{[2]}'= \int_0^{\Psi_{[2]}} \int_0^{2\pi} \partder{\alpha_{[0]}}{\vartheta} \bigg \rvert_{l} \, d \vartheta d \Psi_{[2]}' \nonumber \\
    &= \frac{\Psi_{[2]}}{2\mu} \int_0^{2\pi} \frac{d \vartheta}{\cosh(\eta(l))+ \sinh(\eta(l))\cos(2(\vartheta + \delta(l)))}
    = \frac{\pi \Psi_{[2]}}{\mu}. 
\end{align}
Thus we find $\mu = 1/2$. In the above expression, we have used \eqref{eq:alpha_solution} to compute $\partial \alpha_{[0]}/\partial \vartheta$ by expressing $\omega(\vartheta,l) = \vartheta + \int_0^l \tau(l')\, dl'$.

\begin{figure}
    \begin{center}
    \begin{subfigure}{0.49\textwidth}
     \includegraphics[trim=28cm 17cm 23cm 7cm,clip,width=1.0\textwidth]{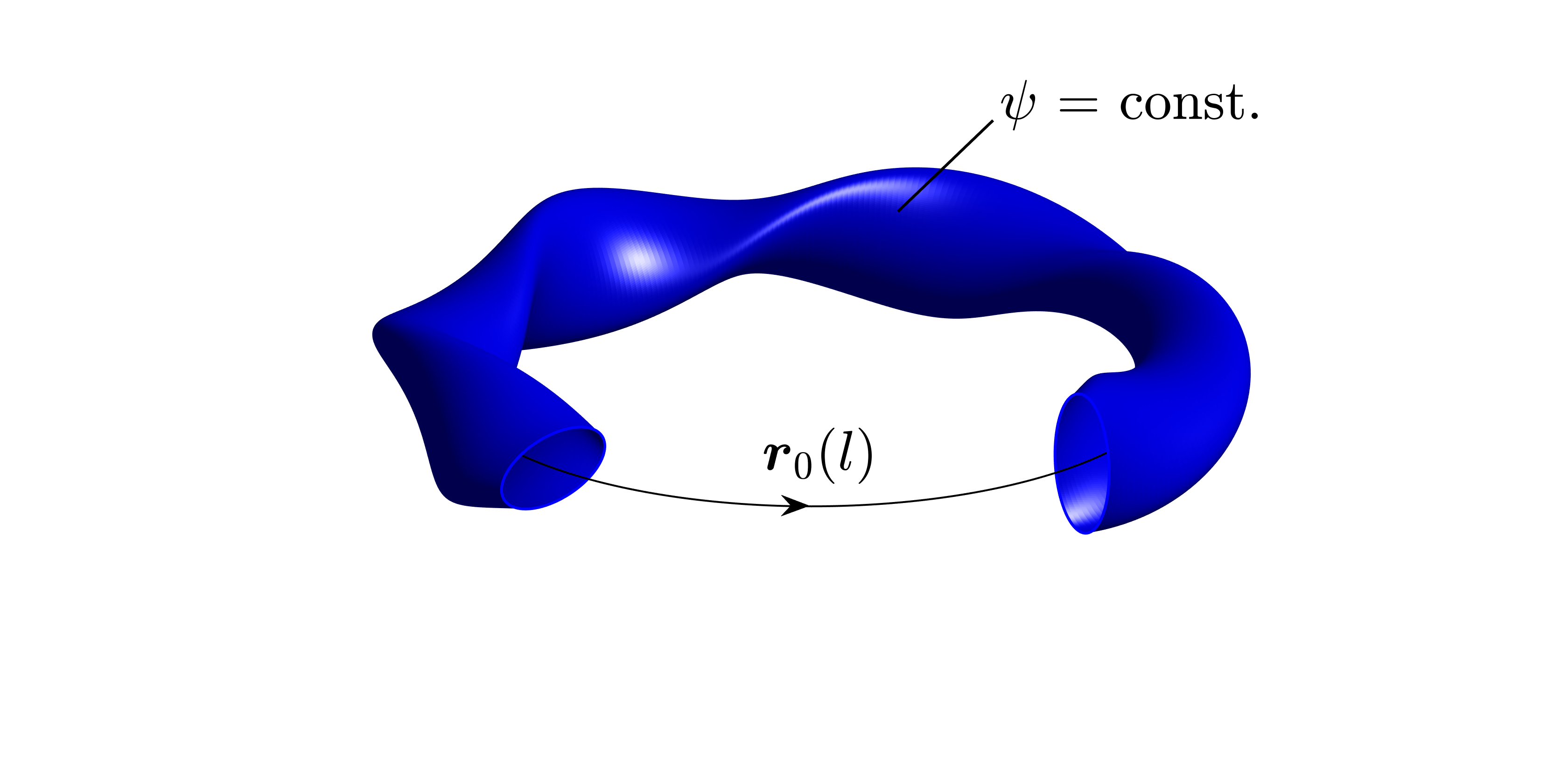}
     \caption{}
     \end{subfigure}
     \begin{subfigure}{0.49\textwidth}
     \includegraphics[trim=40cm 5cm 40cm 4cm,clip,width=1.0\textwidth]{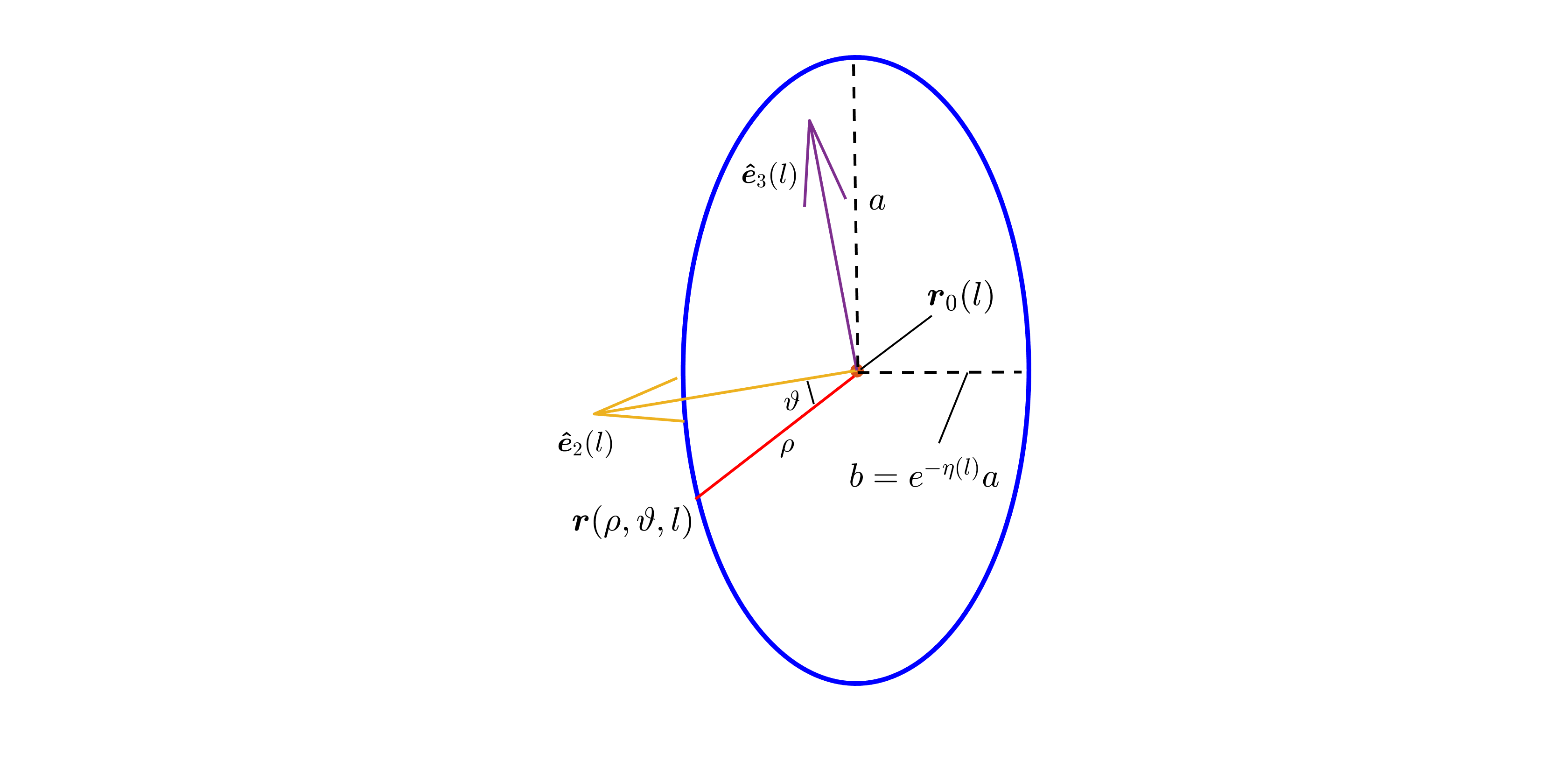}
     \caption{}
     \end{subfigure}
       \caption{We consider a surface of constant $\psi$ near the magnetic axis, $\bm{r}_0(l)$ (a). The cross-section of such a surface in the plane spanned by $\hat{\bm{e}}_1(l)$ and $\hat{\bm{e}}_2(l)$ is shown in (b). A point in this plane is given in the $(\rho,\vartheta,l)$ coordinate system by \eqref{eq:r_rho_vartheta_l}.
        Near the axis, the magnetic surfaces take the form of an ellipse with major axis $a$ and minor axis $b$.} \label{fig:axis_expansion}
       \end{center}
    \end{figure}

In consideration of \eqref{eq:psi_2}, we see that at fixed $l$, the boundary of $\psi_{[2]}$ describes an ellipse, with major axis $a$ and minor axis $b$ satisfying,
\begin{subequations}
\begin{align}
   a(\omega,l) &= \sqrt{\frac{\Phi_{[0]}'(l)}{2}} e^{\eta(l)/2} \cos u(\omega,l)  \\
   b(\omega,l) &= \sqrt{\frac{\Phi_{[0]}'(l)}{2}} e^{-\eta(l)/2} \sin u(\omega,l). 
\end{align}
\end{subequations}
The quantity $u$ measures the angle from the $a$ axis, the area of the ellipse is $\pi \Phi_{[0]}'(l)/2$, and  $e^{\eta(l)}$ is a measure of the ellipticity of the cross-sections of surfaces of constant flux. The quantity $\Phi_{[0]}'(l)$ is non-negative, as it is the $\mathcal{O}(\rho^0)$ magnetic field strength $B_{[0]}(l)$. As the angle with respect to the major axis is given by $u = \vartheta + \delta(l)$ and $\vartheta$ denotes an angle in the poloidal plane with respect to the curvature vector, $\delta(l)$ can be interpreted as the angle between the major axis of the ellipse and the curvature vector. The coordinate system is shown in Figure \ref{fig:axis_expansion}.

As discussed in Section \ref{confiningMagneitcField}, rotational transform is important for toroidal confinement, and we will now compute the rotational transform generated by the lowest order terms in the expansions \eqref{eq:expansions}, corresponding to the rotational transform on the axis.
Through this calculation we will demonstrate the geometric dependence of the rotational transform on the axis shape and flux surface shapes near the axis. Importantly, we will find that this class of equilibrium solutions allows for non-zero rotational transform.

To do so, we express the $\mathcal{O}(\rho^0)$ magnetic field, $\bm{B}_{[0]} = \nabla \Psi_{[2]} \times \nabla \alpha_{[0]}$, in flux coordinates $(\Psi_{[2]},\vartheta,\zeta)$, where $\zeta := 2\pi l/L$ defines a toroidal angle increasing by $2\pi$ upon a toroidal loop.
Following the discussion in Section \ref{sec:magnetic_coordinates}, the field line label $\alpha_{[0]}$ can be identified in \eqref{eq:B_magnetic} as:
\begin{align}
    \alpha_{[0]}(\psi_{[0]},\vartheta,\zeta) = \vartheta - \iota_{[0]}\zeta + \widetilde{\lambda}(\psi_{[0]},\vartheta,\zeta),
    \label{eq:alpha}
\end{align}
where $\iota_{[0]}$ is the $\mathcal{O}(\rho^0)$ rotational transform of interest, while the function $\widetilde{\lambda}$ is $2\pi$ periodic in $\zeta$. The rotational transform to $\mathcal{O}(\rho^0)$ is then computed by integration with respect to $\zeta$:
\begin{align}
 2\pi \iota_{[0]}\left(\psi_{[2]}\right) &= -\alpha_{[0]}(\psi_{[2]},\vartheta=0,l=L) + \alpha_{[0]}(\psi_{(2)},\vartheta = 0,l=0). 
\end{align}
We now apply expression \eqref{eq:alpha_solution} for $\alpha_{[0]}$, noting  that $\delta(L)-\delta(0) = m\pi$ for $m\in \mathbb{Z}$, as the elliptical cross-sections must make a half integer number of rotations in one toroidal turn, to get:
\begin{align}
    \iota_{[0]}\left(\psi_{[2]}\right) &= \frac{1}{2\pi} \left(- \arctan\left(e^{-\eta(l)} \tan (\delta(l)) \right) \bigg \rvert_0^L + \int_0^L \frac{\delta'(l)-\tau(l)}{\cosh(\eta(l))} \, dl \right) \nonumber \\
    &= \frac{1}{2\pi} \left(- \delta(L)+\delta(0) + \int_0^L \frac{\delta'(l)-\tau(l)}{\cosh(\eta(l))}\, dl \right). 
    \label{eq:iota_axis}
\end{align}
As a consequence, under the assumption of vacuum field, i.e. in the absence of current in the plasma, rotational transform can be produced due to torsion $\tau(l)$ of the magnetic axis and ellipticity $\delta'(l)$ of the flux surfaces. That is, if the magnetic axis is non-planar or the flux surfaces are ellipses which twist as one moves toroidally, there may be a non-zero rotational transform. While tokamaks use only current to produce rotational transform $\iota$, the plasma current in stellarators is very small. However, stellarators can still generate a non-zero rotational transform due to torsion and rotating ellipticity, which may be present in 3D geometry. 

In conclusion, we have constructed  3D equilibrium solutions with magnetic surfaces valid locally in a neighborhood of the magnetic axis with a possibly non-zero rotational transform.
This classical result of Mercier \cite{Mericer1964} is also discussed in \cite{Solov1970,Helander2014,Sengupta2019,Jorge2020a}. While this section has focused on solutions of the vacuum equations, similar techniques can be applied in the presence of current and pressure gradients. Understanding 3D equilibria with near-axis expansions continues to be an active area of research \cite{Garren1991,Landreman2018a,Landreman2018b}, and near-axis solutions have been shown to be consistent with 3D numerical equilibrium calculations of quasisymmetric configurations \cite{Landreman2019}. These solutions can be used in the context of stellarator optimization, as will be discussed further in Section \ref{sec:new_frontiers}. 

\FloatBarrier
\section{Models of 3D equilibrium magnetic fields}
\label{sec:equilibrium_fields}

To model the steady-state behavior of a stellarator it makes sense to look for equilibrium solutions. In principle the plasma modifies the background magnetic field, thus in some situations it is necessary to include coupling between the particles and the fields. Due to the separation of scales in magnetic confinement devices, it is often a good approximation to model the large-scale behavior with ideal MHD models. While other models exist, in practice the MHD equilibrium model has been shown to be relatively reliable for fusion plasmas with strong magnetic fields (Chapter 2 in \cite{Freidberg2014}).

Once the time-independent magnetic field is obtained, other properties of the plasma in the presence of this field can be computed. As described in Section \ref{sec:magnetic_confinement}, charged particle motion depends on the geometry of the magnetic fields. For computing particle trajectories, as well as other  such as stability and transport, the equilibrium magnetic field must be determined. Here we will focus our attention on ideal MHD equilibrium models as introduced in Section \ref{sec:mhd_eq}. Ideal MHD equilibrium provides a relatively simple set of equations and is computationally tractable. For this reason, ideal MHD is often applied to study the global, large-scale behavior of toroidal plasmas, including both the plasma and the vacuum region surrounding the plasma.

We will consider several levels of approximation. Pressure gradients and currents in the plasma can be included, as in Sections \ref{sec:3d_mhd_surf} and \ref{sec:3d_mhd_no_surf}. An important subset of MHD models is the force-free model, presented in Section  \ref{sec:force_free}, where all currents are parallel to the magnetic field while the pressure is constant throughout the domain of interest.  An extension of the force-free model is provided by MRxMHD (section \ref{sec:MRXMHD}), which allows for annular regions each with a force free magnetic field.
 In the vacuum model, presented in Section \ref{sec:vacuum_toroidal}, currents and pressure gradients are not included in the equilibrium model, i.e. the pressure is constant while the current is zero. Thus some models include a feedback of the plasma on the magnetic field however the vacuum model does not.

There is also a distinction between models based on the assumption of surfaces. As discussed in Section \ref{sec:3D_difficulties}, the existence of flux surfaces cannot generally be assumed in 3D. However, models based on the assumption of closed nested surfaces are often applied because of their computational efficiency. These are discussed in Sections \ref{sec:variational_principle} and \ref{sec:3d_mhd_surf}. An alternative model assumes the existence of only some surfaces, see Section \ref{sec:MRXMHD}, and others rely on the existence of only one surface which serves as the boundary of the computational domain, see Sections \ref{sec:force_free} and \ref{sec:vacuum_toroidal}. Models not assuming the existence of any surfaces are presented in Sections \ref{sec:3d_mhd_no_surf}
and \ref{sec:vacuum_toroidal}.

 We will discuss the benefits and related challenges of the various 3D equilibrium models since they can each be useful in different applications.

\subsection{Variational principle for MHD equilibria}
\label{sec:variational_principle}

In this Section we will discuss the variational principle for MHD equilibria with surfaces.
Physically, this follows from the intuition that the plasma will tend towards a state which minimizes the energy.
As we will see, equilibria are found by using techniques of variational calculus to extremize the plasma energy, subject to a set of constraints. These constraints include the existence of a set of closed, nested flux surfaces. This is not to say, however, that the variational principle provides any information on the \textit{evolution to} an equilibrium state. 
The idea of finding ideal MHD equilibria via energy minimization was first studied over 60 years ago \cite{Kruskal1958}, but remains widely used today.

We will now show that MHD force balance,
\begin{gather}
    \bm{J} \times \bm{B} = \nabla p(\psi),
    \label{eq:force_balance_variational}
\end{gather}
on a computational domain $\Omega$ is equivalent to finding stationary points of $W$,
\begin{gather}
    W[\bm{B},p] = \int_{\Omega}\left(\frac{B^2(\bm{r})}{2\mu_0} - p(\bm{r})\right) \, d^3 x,
    \label{eq:mhd_energy}
\end{gather}
with respect to perturbations of $\bm{B}$ and $p$, subject to several constraints:
 \begin{enumerate} 
     \item There exists a set of flux surfaces such that $\bm{B} \cdot \nabla \psi=0$, labeled by a toroidal flux label $\psi$;
     \item $\nabla \cdot \bm{B} = 0$;
     \item the pressure as a function of flux is fixed $\delta p(\psi) = 0$;
     \item the rotational transform as a function of flux is fixed, $\delta \iota(\psi) = 0$;
     \item the total toroidal flux enclosed by the toroidal domain is fixed, $\delta \psi = 0$ on $\partial \Omega$.
\end{enumerate}
Constraints 1 and 4 correspond to constraints of ideal MHD and preclude changes to the magnetic field topology (see Section \ref{sec:flux_freezing}).
Constraint 3 follows as a consequence of assuming the plasma evolves adiabatically. Working directly with the assumption of adiabatic evolution, however, leads to minor differences in the energy functional.
For simplicity, here we begin directly
from the assumption that $\delta p(\psi) = 0$. Constraint 5 arises under the assumption that the plasma is surrounded by a perfectly conducting boundary such that the magnetic field lines must lie tangent to this surface. Although we assume that flux surfaces exist, their shapes are not fixed and will be determined from the variational principle.

We will start by deriving a convenient expression for the field $\bm{B}$. We choose a flux coordinate system $(\psi,\theta,\phi)$, where the position of surfaces of constant $\psi = \Psi_T/2\pi$ are not yet determined, while $\theta$ and $\phi$ are the poloidal and toroidal angles that are given functions of space. As shown in Section \ref{sec:magnetic_coordinates} (see \eqref{eq:B_magnetic}), the magnetic field in a toroidal system under assumptions 1 and 2 can generally be expressed as
\begin{gather}
    \bm{B} = \nabla \psi \times \nabla \left(\theta - \iota(\psi) \phi + \lambda(\psi,\theta,\phi) \right).
\end{gather}
Therefore, we will consider variations of $\{\lambda,\psi\}$ rather than $\bm{B}$, as the angles $\theta$ and $\phi$ and the rotational transform $\iota(\psi)$ are held fixed. 

The linear perturbation to the magnetic field can be expressed as 
\begin{multline}
    \delta \bm{B}[\lambda,\psi;\delta \lambda,\delta \psi] = \nabla \delta \psi \times \nabla \left( \theta - \iota(\psi) \phi + \lambda(\psi,\theta,\phi) \right)\\ + \nabla \psi \times \nabla \left(- \iota'(\psi) \delta \psi \phi + \delta \lambda(\psi,\theta,\phi) + \partder{\lambda(\psi,\theta,\phi)}{\psi} \delta \psi \right). 
\end{multline}
We see that $\lambda$ varies due to its dependence on $\psi$ and also through explicit variations. Moreover,  constraint 3 is enforced by expressing the perturbation to the pressure at a given position in terms of the perturbation to the flux label,
\begin{gather}
    \delta p(\bm{r}) = p'(\psi) \delta \psi,
\end{gather}
in particular it is independent of $\lambda$.

The first variation of $W$ with respect to $\lambda$ is now computed,
\begin{align}
    \delta W[\lambda,\psi;\delta \lambda] &= \int_{\Omega} \frac{\bm{B} \times \nabla \psi \cdot \nabla \delta \lambda(\psi,\theta,\phi)}{\mu_0} d^3 x \nonumber \\
    &= -\int_{\Omega} \delta \lambda(\psi,\theta,\phi) \bm{J} \cdot \nabla \psi \, d^3 x,
    \label{eq:dlambda}
\end{align}
where we have integrated by parts, noting that if $\partial \Omega$ is a flux surface then the boundary term vanishes. Thus we obtain a condition that stationary points of $W$ satisfy 
\begin{align}
    \bm{J} \cdot \nabla \psi = 0.
    \label{eq:J_psi}
\end{align}
We note that this condition is equivalent to applying $\bm{B} \times \nabla \psi \cdot \left(...\right)$ to \eqref{eq:force_balance_variational}; thus we recover one component of force balance.

We now consider the first variation of $W$ with respect to $\psi$,
\begin{align}
    \delta W[\lambda,\psi;\delta \psi] &= \int_{\Omega} \left( \frac{\bm{B} \cdot \left(\nabla \delta \psi \times \nabla (\theta -\iota \phi + \lambda) + \nabla \psi \times \nabla \left( \left(- \iota'(\psi) \phi + \partder{\lambda}{\psi}\right)\delta \psi \right)\right)}{\mu_0} - p'(\psi) \delta \psi \right) \,  d^3 x \nonumber \\
    &= \int_{\Omega} \delta \psi \left(\frac{\nabla \times \bm{B} \cdot \nabla (\theta - \iota \phi + \lambda) + \left(\iota'(\psi) \phi -\partder{\lambda}{\psi} \right) \nabla \times \bm{B} \cdot \nabla \psi}{\mu_0} - p'(\psi) \right) d^3 x. 
\end{align}
Here we have integrated by parts, noting that the boundary terms will vanish  as $\delta \psi = 0$ on $\partial \Omega$. The second term in the above expression will vanish from \eqref{eq:J_psi}, so we have the condition
\begin{align}
\frac{1}{\mu_0}\nabla \times \bm{B} \cdot \nabla(\theta - \iota \phi + \lambda(\psi,\theta,\phi)) - p'(\psi)  &= 0 \nonumber \\
\bm{J} \times \bm{B} \cdot \partder{\bm{r}}{\psi} - p'(\psi) &= 0,
\label{eq:force_balance_2}
\end{align}
since $\bm{J} \times \bm{B} = \nabla \psi \left(\bm{J} \cdot \nabla \alpha\right)$ due to \eqref{eq:J_psi}. Thus we recover from the variational principle the $\bm{B} \times \nabla \psi$ component \eqref{eq:J_psi} and $\partial \bm{r}/\partial \psi$ component \eqref{eq:force_balance_2} of the force balance equation \eqref{eq:force_balance_variational}. Force balance parallel to $\bm{B}$ is implied from the assumption of $\bm{B}\cdot \nabla \psi = 0$. We have therefore recovered each of the vector components of \eqref{eq:force_balance_variational}. Therefore, finding stationary points of $W$ with respect to $\lambda$ and $\psi$ is equivalent to \eqref{eq:force_balance_variational} under the above assumptions. This implies that an equilibrium magnetic field can be obtained efficiently from a variational method if the assumptions can be applied. Applications for 3D MHD calculations will be discussed in the following Section.

The discussion in this Section is similar to that in \cite{Kruskal1958} and \cite{Helander2014}. A discussion of numerical applications of energy principles for MHD equilibria is given in Section 4.5 of \cite{Jardin2010}.

\subsection{3D MHD equilibrium with assumption of surfaces}
\label{sec:3d_mhd_surf}

As discussed in Section \ref{sec:3D_difficulties}, the assumption of continuously nested flux surfaces cannot generally be made in 3D. In addition, the assumption of continuously nested flux surfaces together with smooth pressure profiles gives rise to the possibility of singular currents at rational surfaces.
Under these assumptions, however, equilibrium fields on a given domain $\Omega$ can be found by applying the variational approach explained in the previous section by obtaining stationary points of $W$ \eqref{eq:mhd_energy} via a gradient-descent method.
This is the basis for the NSTAB \cite{Garabedian2002} and VMEC \cite{Hirshman1983} codes. Because of its computational efficiency, the VMEC code is routinely applied to design stellarator equilibria within numerical optimization procedures \cite{Spong2001}.
A possible justification is to argue that, since the 3D solutions of physical interest must have good confinement properties, the desired solutions should have a set of magnetic surfaces of non-zero measure.
The assumption should ultimately be checked with other methods described in the following sections which do not assume the existence of surfaces.

For variational methods, several quantities need to be supplied:
\begin{itemize}
    \item The pressure $p(\psi)$ is a given function depending only on the flux surface label (see Section \ref{sec:flux_function}). Thus from \eqref{equilibrium} it must be that $\bm{J} \cdot \nabla \psi = 0$ and $\bm{B} \cdot \nabla \psi = 0$.
    \item A second flux function, typically the toroidal current enclosed by a flux surface, $I_T(\psi)$, or the rotational transform, $\iota(\psi)$, is also given.
    \item 
    A Neumann boundary condition is prescribed as $\bm{B}\cdot \hat{\bm n}=0$, guaranteeing that the boundary $\partial \Omega$ of the computational domain is a flux surface.
The value of the toroidal flux \eqref{eq:psi_T} on $\partial \Omega$, $\Psi_T$, is also specified.
\end{itemize}

The shape of the flux surfaces within the computational domain must be determined from the equilibrium equations. Thus $\psi(\bm{r})$ is a result of the calculation.

Depending on whether the shape of the plasma is prescribed as the computational domain or not is referred to as either a fixed- or free-boundary calculation, respectively. 

For the fixed-boundary approach, the region outside $\Omega$ is assumed to be a vacuum such that $p = 0$, along with the three assumptions previously itemized.
Therefore the boundary of the computational domain is the boundary of the plasma. 

 For a free-boundary approach, the location of electromagnetic coils and their currents are prescribed.  We then seek a plasma boundary $\partial \Omega$ for which the total field in the vacuum region, $\bm{B}_V$, satisfies several conditions: the total pressure should be continuous across the boundary between the vacuum and plasma region, $\left(p +  B^2/2\mu_0 \right)_{\partial \Omega} = \left( B_V^2/2 \mu_0 \right)_{\partial \Omega}$, while the boundary should define a flux surface, $\bm{B}_V \cdot \hat{\bm{n}} \rvert_{\partial \Omega} = 0$.  The assumption of continuity of total pressure arises from imposing force balance at the boundary. 
 This is achieved through the following iterative process, searching for a boundary that is consistent with the coils. 
 \begin{enumerate}
    \item Given the coil currents and positions, the vacuum magnetic field $\bm{B}_0$, that exists in the absence of plasma currents, is determined. 
     \item Given an initial guess for the plasma boundary $\partial\Omega$:
 \begin{enumerate}
 \item The magnetic field (and thus current) can be determined in $\Omega$ using \eqref{equilibrium}-\eqref{eq:equilibrium} and the same boundary conditions for the fixed boundary case, $\bm{B} \cdot \hat{\bm{n}}\rvert_{\partial \Omega} = 0$.
     \item Given the plasma current from \eqref{eq:equilibrium1} in $\Omega$, the magnetic field in the region outside of $\Omega$ due to the plasma current, $\bm{B}_P = \nabla\Phi_B$, can be computed. Here $\Phi_B$ satisfies Laplace's equation for vacuum fields (see Section \ref{sec:vacuum_toroidal}).
     \item The total field in the vacuum region is then obtained as $\bm{B}_V = \bm{B}_0 + \nabla \Phi_B$, and at the boundary we can evaluate  $\left(p +  B^2/2\mu_0 \right)_{\partial \Omega} $, and $\bm{B}_V \cdot \hat{\bm{n}} \rvert_{\partial \Omega}$. 
     \end{enumerate}
     \item While $\left(p +  B^2/2\mu_0 \right)_{\partial \Omega} \neq \left( B_V^2/2 \mu_0 \right)_{\partial \Omega}$ and $\bm{B}_V \cdot \hat{\bm{n}} \rvert_{\partial \Omega} \neq 0$, repeat:
     \begin{enumerate}
         \item Update the boundary $\partial \Omega$ of the computational domain.
        \item Perform steps (a)-(b)-(c).
     \end{enumerate}
 \end{enumerate}
 This is the method used in the free-boundary VMEC code \cite{Hirshman1986}.

 \subsection{3D MHD equilibria without assumption of surfaces}
 \label{sec:3d_mhd_no_surf}
 
As discussed in Section \ref{sec:integrability}, in general 3D geometry continuously nested flux surfaces are not guaranteed to exist. In this Section we discuss models for computing equilibrium solutions with $\nabla p \ne 0$ without the assumption that magnetic surfaces exist.

 The force balance condition \eqref{equilibrium} implies that $\bm{B} \cdot \nabla p = 0$ and $\bm{J} \cdot \nabla p = 0$, so that pressure is constant along field lines and streamlines of the current density. While models which assume surfaces can satisfy this condition by enforcing that pressure is a given flux function, in the absence of surfaces the pressure cannot be imposed, as it will not generally satisfy these conditions. Several numerical approaches for finding solutions in the absence of continuously nested flux surfaces have been developed, some of which are outlined in the proceeding discussion.
 
The approach employed by the PIES code \cite{1986Reiman} is to provide an initial guess for $\bm{B}$ and $p$ in $\Omega$ (for example, the field computed with the variational approach could be used) such that $\bm{B} \cdot \nabla p = 0$. Then, \eqref{equilibrium}-\eqref{eq:equilibrium} are solved iteratively, as follows:
\begin{itemize}
 \item The current density perpendicular to the magnetic field is known from \eqref{equilibrium},
 \begin{gather}
     \bm{J}_{\perp} = \frac{1}{B}\hat{\bm{b}} \times \nabla p.
     \label{eq:perp_current}
 \end{gather}
 \item The parallel current can be computed by enforcing that $\nabla \cdot \bm{J} = 0$, which follows from \eqref{eq:equilibrium2},
 \begin{gather}
     \bm{B} \cdot \nabla \left( \frac{J_{||}}{B} \right) = - \nabla \cdot \left(\frac{1}{B} \hat{\bm{b}} \cdot \nabla p \right). 
 \end{gather}
 We note that this is an example of a magnetic differential equation which may have singular behavior at rational surfaces (see Section \ref{sec:singularities}).
 \item Given $\bm{J} = \bm{J}_{\perp} +J_{||} \hat{\bm{b}} $, the magnetic field can be determined from \eqref{eq:equilibrium1}. 
 \item The perpendicular and parallel currents are again computed as described above, and the iteration continues until convergence.
\end{itemize}

The SIESTA \cite{Hirshman2011} code, on the other hand, uses a variational method similar to that described in Section \ref{sec:variational_principle}. Here a small term is added to \eqref{equilibrium} such that there is a small perturbation away from ideal MHD. In this way, the magnetic topology is no longer constrained, allowing the break-up of surfaces and island formation.

Another example is the HINT code \cite{Harafuji1989}, which evolves a set of time-dependent equations in order to arrive at a steady state solution which is then interpreted as an equilibrium state. As a first step, the magnetic field is held fixed while the pressure is allowed to evolve to satisfy $\bm{B} \cdot \nabla p = 0$. Second, the pressure is held fixed while the field evolves to satisfy force balance. Some small non-ideal terms are added to allow for changes in magnetic topology. These two steps are iterated until convergence is reached and force balance is achieved within a tolerable error.

\subsection{Force-free fields}
\label{sec:force_free}

The magnetic fields in a stellarator are sometimes described by a force-free model. Force-free refers to the fact that $\bm{J} \times \bm{B}$ and $\nabla p$ vanish independently. As a consequence, the plasma current is everywhere parallel to the magnetic field.
Such situations can arise in both laboratory and space plasmas. In this Section we provide some motivation for employing this model in magnetic confinement configurations as well as a variational principle for computing such states.

If $\bm{J}\times \bm{B} = 0$ in a given domain $\Omega$, then from \eqref{eq:equilibrium1} we must have
\begin{align}
        \nabla \times \bm{B} = \lambda \bm{B},
        \label{eq:force_free}
\end{align}
for some scalar function $\lambda$. Note that $\lambda$ is proportional to the parallel current density. Force-free magnetic fields are thus Beltrami fields, as will be discussed in Section \ref{sec:equilsum}. The value of the pressure is not constrained by this model, as long as it is a constant in $\Omega$. Vacuum fields are a subset of force-free fields in the limit $\lambda = 0$. 

In addition, from \eqref{eq:equilibrium2} and \eqref{eq:force_free} we require that 
\begin{align}
    \bm{B} \cdot \nabla \lambda = 0.
    \label{eq:B_lambda}
\end{align}
Often $\lambda$ is taken to be a constant in $\Omega$ to satisfy the above, sometimes called the linear force-free model. We will make this assumption throughout this discussion.

The force-free equation \eqref{eq:force_free} with constant $\lambda$ can be solved in a toroidal domain with the following conditions 
\begin{subequations}
\begin{align}
    \bm{B} \cdot \hat{\bm{n}} &= 0 \hspace{1cm} \text{on } \partial \Omega \label{eq:normal_force_free} \\
    \int_{\mathcal{S}_T} \bm{B} \cdot \hat{\bm{n}} \, d^2 x &= \Psi_T \hspace{0.7cm} \text{on } \partial\Omega. \label{eq:flux_force_free}
\end{align}
\end{subequations}
 Thus the boundary of the domain must be a magnetic surface. We note that the overall scale factor of $\bm{B}$ is not determined from \eqref{eq:force_free}. Therefore, the toroidal flux enclosed by $\Omega$ can be specified to determine the overall scale. In \eqref{eq:flux_force_free}, $\mathcal{S}_T$ is a surface at constant poloidal angle bounded by $\partial \Omega$. Often instead of $\lambda$, the magnetic helicity is prescribed,
 \begin{gather} 
    K = \int_{\Omega} \bm{A} \cdot \bm{B} \, d^3 x ,
    \label{eq:helicity}
\end{gather}
where $\bm{A}$ is the vector potential such that $\bm{B} = \nabla \times \bm{A}$. 
Then $\lambda$ is chosen to achieve the desired value of the helicity. Rather analogously to helicity in fluid dynamics, $K$ is a measure of the knottedness of magnetic field lines (see \cite{Berger1984} for further discussion). Although $K$ depends explicitly on $\bm{A}$, if $\partial \Omega$ is a flux surface such that $\bm{B} \cdot \hat{\textbf{n}}\rvert_{\partial \Omega} = 0$, then $K$ is independent of the choice of gauge. $K$ can be interpreted as the Gauss linking number \cite{Berger1999} which is studied in fields such as knot theory, algebraic topology, and differential geometry.

The force-free equations can also be solved within toroidal annuli rather than a full toroidal domain. We now consider the solution of \eqref{eq:force_free} in a toroidal annulus $\Omega = \Omega_{\text{outer}} \backslash \Omega_{\text{inner}}$, where $\Omega_{\text{outer}}$ and $\Omega_{\text{inner}}$ are toroidal volumes such that $\Omega_{\text{inner}}\subset \Omega_{\text{outer}}$. These volumes are limited by toroidal surfaces, $\Gamma_{\text{outer}} = \partial \Omega_{\text{outer}}$ and $\Gamma_{\text{inner}} = \partial \Omega_{\text{inner}}$. Conditions \eqref{eq:normal_force_free}-\eqref{eq:flux_force_free} must be imposed as in the case of a toroidal domain, with the addition of a flux constraint,
\begin{subequations}
\begin{align}
    \bm{B} \cdot \hat{\bm{n}} &= 0  \hspace{1cm} \text{on } \Gamma_{\text{outer}} \text{ and } \Gamma_{\text{inner}}  \\
    \int_{\mathcal{S}_{T}} \, \bm{B} \cdot \hat{\bm{n}} \, d^2 x  &= \Psi_T(\Gamma_{\text{outer}})-\Psi_T(\Gamma_{\text{inner}}) \\
    \int_{\mathcal{S}_{P}} \bm{B} \cdot \hat{\bm{n}} \, d^2 x &= \Psi_P(\Gamma_{\text{outer}})-\Psi_P(\Gamma_{\text{inner}}).
\end{align}
\end{subequations}
Here $\mathcal{S}_{T}$ is a surface at constant toroidal angle bounded by surfaces $\Gamma_{\text{outer}}$ and $\Gamma_{\text{inner}}$, while $\mathcal{S}_{P}$ is a surface at constant poloidal angle bounded by surfaces $\Gamma_{\text{outer}}$ and $\Gamma_{\text{inner}}$.

Force-free equilibria in a toroidal domain, $\Omega$, can be found by minimizing an energy functional as in Section \ref{sec:variational_principle}, but with a different set of constraints. In particular, we seek stationary points of,
\begin{gather}
    W = \int_{\Omega} \frac{B^2}{2 \mu_0} \, d^3 x,
\end{gather}
subject to the following constraints:
\begin{enumerate}
    \item the total magnetic helicity \eqref{eq:helicity}
    is given and constant;
    \item the boundary of the domain, $\partial \Omega$, is a magnetic surface such that $\bm{B} \cdot \hat{\bm{n}}|_{\partial\Omega}=0$;
    \item the boundary of the domain is fixed such that the perturbation to the magnetic field satisfies $\delta \bm{B} \cdot \hat{\bm{n}} |_{\partial \Omega} = 0$;
    \item the toroidal magnetic flux through $\Omega$ is fixed,
    \begin{align}
        \int_{\mathcal{S}_T} \bm{B} \cdot \hat{\bm{n}} \, d^2 x &= \Psi_T \hspace{1cm} \text{on }\partial\Omega.
    \end{align}
    This leads to the result that the perturbation to the vector potential satisfies $\delta \bm{A} \times \hat{\bm{n}} = 0$ on $\partial \Omega$.\footnote{This can be seen by noting that the perturbed magnetic field can be written in magnetic coordinates as,
    \begin{align}
        \delta \bm{B} = \nabla\delta  \psi \times \nabla \alpha + \nabla \psi \times \nabla \delta \alpha = \nabla \times \left(\delta \psi \nabla \alpha - \delta \alpha \nabla \psi \right),
    \end{align}
    where $\alpha = \vartheta - \iota \varphi$ is the field line label and $\delta \psi = 0$ on $\delta  \Omega$ from assumption 4. Thus the perturbed vector potential can be taken to be $\delta \bm{A} = -\delta \alpha \nabla \psi$ on $\partial \Omega$.}
\end{enumerate}

Imposing constraint 1 with a Lagrange multiplier $\alpha$ and computing the first variation in $W$ with respect to $\bm{A}$ yields,
\begin{align}
    \delta W[\bm{A};\delta \bm{A}] &= \int_{\Omega} \left( \frac{\bm{B} \cdot \delta \bm{B}}{\mu_0} - \alpha \left(\bm{A} \cdot \delta \bm{B} - \delta \bm{A} \cdot \bm{B}\right) \right) \, d^3 x,
\end{align}
where $\delta \bm{B} = \nabla \times \delta \bm{A}$.
Integrating by parts, we obtain,
\begin{align}
    \delta W[\bm{A};\delta \bm{A}] &= \int_{\Omega}\delta \bm{A} \cdot \left(\frac{(\nabla \times \bm{B})}{\mu_0} - 2\alpha \bm{B} \cdot  \right) \, d^3 x \, + \int_{\partial \Omega} \delta \bm{A} \times \hat{\bm{n}} \cdot \left(\alpha \bm{A} - \frac{\bm{B}}{\mu_0} \right)  \, d^2 x.
    \label{eq:W_force_free}
\end{align}
The surface term vanishes under assumption 4. Thus the field satisfies \eqref{eq:force_free}
with $\lambda = 2\alpha/\mu_0$. 

The key assumption of force-free fields as energy minimizing equilibria is the conservation of total helicity (constraint 1). The veracity of this assumption as well as the dynamical process by which such states may form is a topic of some debate. The famous conjecture by J.B. Taylor \cite{Taylor1974,Taylor1986} argues that plasmas relax to such force-free states via turbulent processes under which $K$ is approximately conserved. The states are therefore sometimes referred to as Taylor states within the literature. While this type of process is not relevant for stellarator confinement, this model has several features which may become important for computing 3D fields. These considerations will be described below.

Under ideal MHD evolution (Section \ref{sec:force_free}), changes in magnetic topology are not allowed. Thus if a system begins in a state with closed, nested surfaces, the final state must also have the same topology, and no islands or stochastic regions can form. While the variational principle presented in Section \ref{sec:variational_principle} requires that the local topology of the magnetic field is fixed during variations, only the \textit{global} helicity is fixed when performing variations of \eqref{eq:W_force_free}. Thus this model allows the calculation of energy minimized equilibria which differ in magnetic field topology from the initial state, and can allow for the formation of islands or stochastic field regions.

We also note that the force-free model is consistent with the assumption of a non-integrable magnetic field. If a given field line comes arbitrarily close to every point within a domain $\Omega$, then from \eqref{equilibrium} we conclude that pressure must be constant within $\Omega$. Therefore the magnetic field must satisfy \eqref{eq:force_free}. From the condition that the magnetic field must be divergence-free, \eqref{eq:B_lambda} implies that $\lambda$ is constant along field lines. Under the assumption that a given field line comes arbitrarily close to any point in $\Omega$, we conclude that $\lambda$ is a constant within $\Omega$. We emphasize that this assumption on the field line structure is not required to invoke a force-free model, simply that it is consistent.

The existence and uniqueness of solutions to \eqref{eq:force_free} has been shown for toroidal domains and toroidal annuli \cite{Kress1986}. For more details on the constraints necessary for solving this set of PDEs in toroidal annuli and its implementation for axisymmetric geometry, see \cite{Oneil2018}. This formalism has been applied to stellarator geometry in the Boundary
Integral Equation Solver for Taylor states (BIEST) code \cite{Malhotra2019}. The extension of this model to describe multiple regions simultaneously is presented in the following Section.

\subsection{Stepped pressure equilibrium (MRxMHD)}
\label{sec:MRXMHD}

The multi-region relaxed MHD (MRxMHD) equilibrium model \cite{Hole2006,Dewar2015} generalizes the single-volume force-free state by allowing discontinuous (stepped) pressure profiles and thus permitting the pressure to vary throughout the plasma volume.
The plasma is partitioned into nested volumes in which the plasma is force-free (i.e.\ $p$ is constant) and separated by interfaces which accommodate jumps in $p$.
The geometry of the interfaces is not known \textit{a priori} and must satisfy a specified set of jump and flux conditions which are described below. Note that these interface conditions are 
similar to those used in immiscible fluid models or capillary interface models where the interfaces between phases or fluids are unknown\footnote{See for instance \cite{heywood} for the introduction of a flux condition on an artificial boundary within the domain to ensure well-posedness for a Navier-Stokes problem, \cite{turek} for numerical aspects of a flow through an aperture in an infinite wall, or \cite{turekBenchmark,turek2D} for a jump condition modeling force balance at an unknown interface.}.

Consider a toroidal domain, $\Omega$, in which $p$ and $\lambda$ \eqref{eq:force_free} are piecewise constant.
The domain is partitioned into $m$ nested toroidal sub-regions $\Omega_i$ for all $i$ from $1$ to $m$.
The innermost volume, $\Omega_1$, is a genus one torus, otherwise $\Omega_i$ for $i\geq2$ are genus two tori. 
Excepting $\Omega_1$, each $\Omega_i$ is bounded by two non-intersecting toroidal surfaces, $\Gamma_{i-1}$ and $\Gamma_{i}$ for $i$ from $2$ to $m$.
Note that $\Omega_1$ is bounded by a single toroidal surface, $\Gamma_1$, and denote the outer most boundary $\Gamma_m=\partial\Omega$.
The discontinuities which can arise in this model, including in $p$ and $\lambda$, occur at the surfaces, $\Gamma_i$.

The components of the model can be described as follows:
\begin{itemize}
    \item the parameters $\{ \lambda_i\}_{1\leq i \leq m}$ define the $\lambda$ profile as $\lambda=\lambda_i $ in $\Omega_i$ for all $i$ from $1$ to $m$, while the magnetic field satisfies in each $\Omega_i$
    \begin{equation}\label{eq:volB}
        \nabla\times\bm{B}_i = \lambda_i\bm{B}_i \text{ in } \Omega_i,
    \end{equation}
where $\bm{B}_i$ denotes the solution in $\Omega_i$;
    \item the parameters $\{ p_i \}_{1\leq i \leq m}$ define the stepped pressure profile as $p=p_i $ in $\Omega_i$ for all $i$ from $1$ to $m$, while the total pressure balance (including the plasma pressure $p$ and the magnetic pressure $B^2/2\mu_0$) 
    at the interfaces is expressed as:
\begin{equation}\label{eq:jumpcond}
[[p + B^2/2\mu_0]]_{\Gamma_i} = 0 , \text{ for all }i \text{ between } 1 \text{ and } m-1;
\end{equation}
\item the flux surface conditions hold at the interfaces:
\begin{equation}
    \bm{B} \cdot \hat{\bm{n}} = 0   \text{ on } \Gamma_i, \text{ for all } i \text{ between } 1 \text{ and } m;
\end{equation}
\item the parameters $\{ \Psi_T^i,\Psi_P^i\}_{2\leq i \leq m}$ define the toroidal and poloidal fluxes across each annular subdomain:
\begin{align}
     \int_{\mathcal{S}_{T_i}} \bm{B} \cdot \hat{\bm{n}} \, d^2 x &= \Psi_T^i , \text{ for all } i \text{ between } 2 \text{ and } m \\
     \int_{\mathcal{S}_{P_i}} \bm{B} \cdot \hat{\bm{n}} \, d^2 x &= \Psi_P^i , \text{ for all } i \text{ between } 2 \text{ and } m.
\label{eq:fluxinterfaces}
\end{align}
Here ${\mathcal S_{T_i}}$ is a surface at constant toroidal angle bounded by surfaces $\Gamma_i$ and $\Gamma_{i-1}$, and ${\mathcal S_{P_i}}$ is a surface at constant poloidal angle bounded by surfaces $\Gamma_i$ and $\Gamma_{i-1}$;
\item the parameter $\Psi_T^1$ defines the toroidal flux across the innermost toroidal domain $\Omega_1$:
\begin{align}\label{eq:T1}
    \int_{\mathcal{S}_{\mathcal{T}_1}} \bm{B} \cdot \hat{\bm{n}} \, d^2 x = \Psi_T^1,
\end{align}
where $\mathcal{S}_{T_1}$ is a surface at constant toroidal angle bounded by $\Gamma_1$.

\end{itemize}

In summary, given $\{ \lambda_i,p_i,\Psi_T^i\}_{1\leq i \leq m}$ and $\{ \Psi_P^i\}_{2\leq i \leq m}$, \eqref{eq:volB} can be solved, subject to the constraints \eqref{eq:jumpcond}-\eqref{eq:T1}, to determine the position of free surfaces $\{ \Gamma_i \}_{1\leq i \leq m}$ and value of the magnetic field $\bm{B}_i$ in $\cup_{1\leq i\leq m}\Omega_i$ which defines an MHD equilibrium.
In place of $\{ \lambda_i\}_{1\leq i \leq m}$, the total helicity in each $\Omega_i$ may be prescribed and, in place of ${\psi_T^1}\cup\{\Psi_T^i, \Psi_P^i\}_{2\leq i \leq m}$, the value of the rotational transform on each $\Gamma_{i}^{-,+}$ for $1\leq i\leq m$ may be fixed.

An important characteristic of the model is that $\Gamma_i$, referred to as ideal interfaces and across which $p$ is discontinuous, are magnetic surfaces.
Recall from Section \ref{sec:singularities} that, in order to avoid the appearance of unphysical $1/x$ singular currents, pressure gradients must be avoided except at surfaces corresponding to sufficiently irrational values of the rotational transform. In the MRxMHD model, the pressure jump can be chosen to occur at interfaces which satisfy the Diophantine condition \eqref{eq:Dioph} and thus do not produce unphysical currents in 3D. Moreover, these magnetic surfaces may persist according to KAM theory (Section \ref{sec:KAM}) if the perturbation from integrability is sufficiently small.
Combined, this ensures the model produces well-defined equilibrium solutions, even in the absence of axisymmetry.

This model has been implemented numerically as Stepped Pressure Equilibrium Code (SPEC) \cite{Hudson2011}, which solves for the magnetic field, iterating on the position of the interfaces until the constraints are satisfied.

\begin{figure}
    \centering
    \begin{subfigure}{0.49\textwidth}
    \includegraphics[trim=11cm 2cm 7cm 5cm,clip,width=1.0 \textwidth]{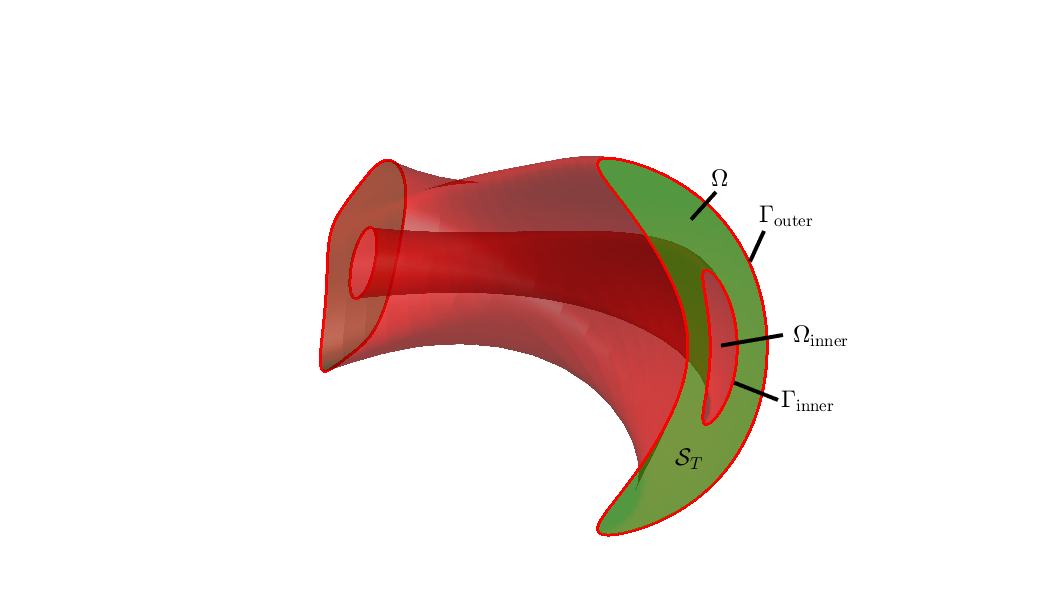}
    \caption{}
    \end{subfigure}
    \begin{subfigure}{0.49\textwidth}
    \includegraphics[trim=11cm 5cm 7cm 3cm,clip,width=1.0 \textwidth]{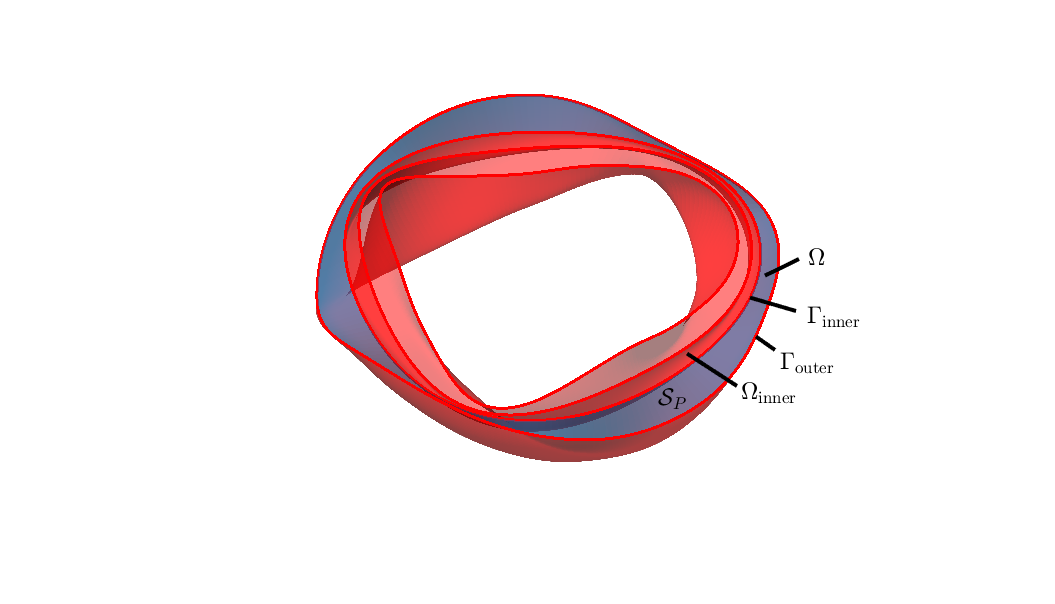}
    \caption{}
    \end{subfigure}
    \caption{A force-free equilibrium can be computed in an annular region $\Omega$ bounded by toroidal surfaces $\Gamma_{\text{outer}}$ and $\Gamma_{\text{inner}}$ given the flux of magnetic field through a surface at constant toroidal angle, $\mathcal{S}_T$, (green in (a)) and the flux through a surface at constant poloidal angle, $\mathcal{S}_P$, (blue in (b)) bounded by $\Gamma_{\text{inner}}$ and $\Gamma_{\text{outer}}$}
    \label{fig:annular}
\end{figure}

\subsection{Vacuum fields}
\label{sec:vacuum_toroidal}

As described in Section \ref{sec:vacuum}, the vacuum model for magnetic fields can be used if there is no current in a given domain. In order for the vacuum approximation to be valid, an additional assumption of small pressure gradients must be made, as pressure gradients cause a parallel current due to collisional effects (the bootstrap current) and a perpendicular current due to MHD force balance (the diamagnetic current) \cite{Helander2014}. Even if these assumptions are not valid within the confinement region, the vacuum model can be used to describe the region outside the confinement region. Under the vacuum assumption, $\bm{B}$ is curl-free from \eqref{eq:equilibrium1}, so it can be written in terms of a scalar potential
\eqref{eq:phi_B}, which must satisfy Laplace's equation \eqref{eq:laplace}.

In this Section we assume 
Laplace's equation is solved in a toroidal domain, $\Omega$. The toroidal angle $\phi$, will be taken to be the standard cylindrical angle (see Section \ref{sec:toroidal_geom}). The poloidal angle, $\theta$, will be some angle which increases by $2\pi$ on a poloidal loop about the torus. In general, the scalar potential can be separated into secular and periodic pieces, 
\begin{gather}
    \Phi_B = \widetilde{\Phi}_B + A \phi + C \theta, 
    \label{eq:phi_B2}
\end{gather}
here $\widetilde{\Phi}_B$ is periodic in $\theta$ and $\phi$ and $A$ and $C$ are constants. The scalar potential can generally be written as the form given in \eqref{eq:phi_B2}, as $\bm{B} = \nabla \Phi_B$ must be periodic in $\phi$ and $\theta$. However, $\Phi_B$ can contain secular terms as it is not a physical quantity. We now enforce that the current enclosed by a poloidal loop about the torus vanishes, as there is no current inside $\Omega$. We apply Ampere's law \eqref{eq:equilibrium1}, using a loop at constant $\phi$ on the boundary of the toroidal domain,  $\partial \Omega$,
\begin{gather}
    \oint_{\phi = \text{const.}} \,  \bm{B} \cdot d \bm{l} = \mu_0 I_{T} = 0.
    \label{eq:I_pol_vacuum}
\end{gather}
We can see that \eqref{eq:I_pol_vacuum} implies that $C=0$, using $d \bm{l} = \left(\partial \bm{r}/\partial \theta\right) d \theta$. The constant $A$ can be determined by considering a loop at constant $\theta$ on $\partial\Omega$,
\begin{align}
    \oint_{\theta = \text{const.}} \bm{B} \cdot d\bm{l} &= \mu_0 I_{P}. 
    \label{eq:I_tor_vacuum}
\end{align}
The integrals in \eqref{eq:I_pol_vacuum} and \eqref{eq:I_tor_vacuum} are described in Section \ref{sec:magnetic_covariant}. 
From \eqref{eq:I_tor_vacuum}, we can see that $A = \mu_0 I_{P}/2\pi$, using $d \bm{l} = \left(\partial \bm{r}/\partial \phi\right)d\phi$. As there is no current in $\Omega$, $I_P$ is the total coil current linking the plasma poloidally. The boundary condition for $\Phi_B$ on $\partial \Omega$ is determined by specifying $\bm{B}\cdot \hat{\bm{n}}$ on $\partial \Omega$, 
\begin{gather}
\hat{\bm{n}} \cdot \nabla \widetilde{\Phi}_B + \frac{\mu_0 I_{P}}{2\pi} \hat{\bm{n}} \cdot \nabla \phi = \bm{B} \cdot \hat{\bm{n}} \hspace{1cm} \text{on } \partial \Omega.
\end{gather}
If $\partial \Omega$ is a magnetic surface, then $\bm{B} \cdot \hat{\bm{n}} = 0$. Otherwise, it can be computed from the Biot-Savart law \eqref{eq:biot_savart} if all external currents are known. 
Laplace's equation 
subject to this Neumann boundary condition can be solved using a Green's function method, as is done in the NESTOR code \cite{Merkel1986}.


Instead of solving Laplace's equation, the fields in a vacuum region can also be computed using the Biot-Savart law \eqref{eq:biot_savart} applied to sources of current outside the vacuum region. Thus if the currents in electromagnetic coils are provided, the magnetic field can be determined without the assumption of any magnetic surfaces under the vacuum model.

\subsection{Summary and analogy with steady Euler flow}\label{sec:equilsum}

Below is a Table which summarizes the models used to describe equilibrium stellarator magnetic fields in a toroidal domain $\Omega$, with flux surface boundary conditions on the boundary of $\Omega$. 

{\renewcommand{\arraystretch}{2.0}
\begin{center}
    \begin{tabular}{|c|c|c|c|}
    \hline
&  MHD equilibrium &  Force-free fields & Vacuum fields \\ 
& (surfaces assumed)  &  &  \\ \hline
   Hyp.  & $\bm{J} \times \bm{B} \neq 0$& $\bm{J}\times\bm{B}=0$ & $\bm{J}=0$\\ 
     &  $\nabla p \neq 0$ &  $\nabla p = 0$ and $\lambda=\text{const.}$ & $\nabla p=0$ \\ \hline
         & $\bm{J} \times \bm{B} = \nabla p$
         & $\nabla \times \bm{B} = \lambda \bm{B}$ &$\Delta \widetilde{\Phi}_B$ = 0  \\ 
   PDE   & $\nabla \cdot \bm{B} = 0$  &
   & 
   \\ 
   model & $\mu_0 \bm{J} = \nabla \times \bm{B}$  & $\mu_0 \bm{J} = \nabla \times \bm{B}$ &
   \\ \hline 
   Given & $p(\psi)$, $\iota(\psi)$, $\Psi_T$ & $\lambda$, $\Psi_T$ & $I_{P}$  \\ \hline
   Unkn. & $\bm{B}$ & $\bm{B}$ & $\widetilde{\Phi}_B$ \\ \hline
   With & $\bm{J}$ function of $\bm{B}$  & $\bm{J}$ function of $\bm{B}$ &  $\bm{B} = \nabla \Big( \widetilde{\Phi}_B + \left(\mu_0 I_{P}/2\pi \right) \phi\Big)$
   \\ \hline
   BC & $\bm{B} \cdot \hat{\bm{n}} = 0$  & $\bm{B} \cdot \hat{\bm{n}} = 0$ & $\hat{\bm{n}} \cdot \nabla \widetilde{\Phi}_B + \left(\mu_0 I_{P}/2\pi\right) \hat{\bm{n}} \cdot \nabla \phi = \bm{B} \cdot \hat{\bm{n}}$ \\
   \hline
    \end{tabular}
\end{center}
}
  
Connections exist between MHD and fluid dynamics.
In particular, the MHD equilibrium equations share many similarities with the steady Euler flow equations.
Over the years, this has facilitated the exchange of ideas between the fluid dynamics and plasma physics communities (e.g. \cite{Moffatt1985,Hegna1991,Bruno1996}).

The steady ($\partial/\partial t=0$)  incompressible Euler equations with constant and uniform density, $\rho_0$, are equations for the flow velocity $\bm{u}$ and pressure $P$. 
They can be written,
\begin{subequations}
\begin{align}
\bm{u} \cdot \nabla \bm{u} = -\nabla \left(\frac{P}{\rho_0}\right)
\Leftrightarrow
(\nabla\times\bm{u})\times\bm{u} = -\nabla \left( \frac{P}{\rho_0} + \frac{u^2}{2} \right)
    \label{eq:steady_euler}\\
    \nabla \cdot \bm{u} = 0. \label{eq:incompressible_euler}
\end{align}
\end{subequations}
  Conservation of momentum density is expressed by \eqref{eq:steady_euler}, while \eqref{eq:incompressible_euler} expresses incompressibility of the flow. The vorticity is defined as $\bm{\omega} = \nabla \times\bm{u}$.
 Beltrami flows describe states in which the vorticity is parallel to $\bm{u}$; they are analogous to force-free fields. Flows with vanishing vorticity can be expressed in terms of a potential; they are analogous to vacuum fields.
 {\renewcommand{\arraystretch}{2.0}
\begin{center}
    \begin{tabular}{|c|c|c|}
    \hline
Hyp.&Steady Euler models &  MHD equilibrium models\\ \hline\hline
&$(\nabla\times\bm{u})\times\bm{u} = -\nabla \left( \frac{P}{\rho_0} + \frac{u^2}{2} \right)$
&
$(\nabla \times \bm{B}) \times \bm{B} = \mu_0 \nabla p$
\\
&$\nabla \cdot \bm{u} = 0$
&
$\nabla \cdot \bm{B} = 0$
\\\hline\hline
Beltrami flows/fields
&
$\nabla \times \bm{u} = \alpha \bm{u}$
&
$\nabla \times \bm{B} = \lambda \bm{B}$
\\
(vorticity parallel to field)
   &
   $\nabla \cdot \bm{u} = 0$
   &
   $\nabla \cdot \bm{B} = 0$
\\\hline\hline
Potential flows
&
$\bm{u} = \nabla \varphi$
&
 $\bm{B} = \nabla \Phi_B$ 
\\
(zero vorticity)
&
$\Delta \varphi = 0$
&
$\Delta \Phi_B = 0$
\\\hline
    \end{tabular}
\end{center}
}

\FloatBarrier
\section{Symmetries in stellarators}
\label{sec:symmetry}

Symmetries have a long history in physics.
Due to the strong connection between symmetries and conserved quantities as described in Section \ref{sec:classical_mechanics}, leveraging symmetries can provide new physical insight. For example, axisymmetry implies conservation of the canonical angular momentum, as in Section \ref{sec:confinement_axisymmetry}, and existence of magnetic surfaces, as in Section \ref{sec:field_line_flow}.
Although stellarators are non-axisymmetric, it is often possible to identify other symmetries which can become useful for confinement or efficient numerical discretization. In this Section we discuss several of these. These symmetries are properties of the equilibrium magnetic field and have been implemented in many stellarator configurations. In Section \ref{sec:quasisymmetry}, an important symmetry which provides confinement of guiding center trajectories, quasisymmetry, is discussed. In Section \ref{sec:Np_symmetry}, a periodicity in the number of field periods, often referred to as $N_P$ symmetry, is presented. In Section \ref{sec:stellarator_symmetry}, a discrete reflection symmetry, known as stellarator symmetry, is described. While $N_P$ symmetry and stellarator symmetry may not result in improved confinement as quasisymmetry does, they are present in almost all stellarator configurations to date. 

\subsection{Quasisymmetry}
\label{sec:quasisymmetry}

Quasisymmetry \cite{Boozer1995,Nuhrenberg1988} is expressed in terms of Boozer coordinates, described in Section \ref{sec:boozer_coordinates}.
Quasisymmetry is defined as a symmetry of the field strength, $B$, with respect to a linear combination of Boozer angles, $\vartheta_B$ and $\varphi_B$. More precisely, quasisymmetry means that there exists a change of coordinates $(\psi,\vartheta_B,\varphi_B) \rightarrow (\psi,\chi,\eta)$, where $\chi = M \vartheta_B - N \varphi_B$ and $\eta = M' \vartheta_B - N'\varphi_B$ with $M'/N' \ne M/N$,
\begin{gather}
    B(\psi,\vartheta_B,\varphi_B) = \widetilde B(\psi,\chi,\eta),
\end{gather}
such that the magnetic field amplitude is independent of the coordinate $\eta$,
\begin{gather}    \partder{\widetilde B(\psi,\chi,\eta)}{\eta} = 0. 
    \label{eq:quasisymmetry}
\end{gather}
Thus there exists a symmetry direction when the field strength is expressed in Boozer coordinates.
For example, quasi-axisymmetry corresponds to $M = 1$ and $N = 0$, and
the field strength is independent of $\varphi_B$, 
\begin{gather}
    \partder{B(\psi,\vartheta_B,\varphi_B)}{\varphi_B} = 0.
\end{gather}

We can furthermore note that under the assumption of \eqref{eq:quasisymmetry}, the radial Boozer covariant component expressed in the $(\psi,\chi,\eta)$ system, $\widetilde{K}(\psi,\chi,\eta) = K(\psi,\vartheta_B,\varphi_B)$, satisfies,
\begin{align}
    \partder{\widetilde{K}(\psi,\chi,\eta)}{\eta} = 0.
\end{align}
This can be seen considering the expression obtained assuming an MHD equilibrium, \eqref{eq:K_quasisymmetry}. For the remainder of this Section, we consider $\iota\in\mathbb{R}\setminus \mathbb{Q}$ so that the $\delta$-function term is zero, in which case,
\begin{gather}
    K(\psi,\vartheta_B,\varphi_B) =  
     \mathcal{K}(\psi) +i \mu_0 \der{p(\psi)}{\psi} \sum_{m,n,mn\ne0} \left(\frac{G(\psi) + \iota(\psi) I(\psi)}{n - \iota m} b_{m,n}\right)e^{i(m\vartheta_B-n\varphi_B)},
\end{gather}
where $b_{m,n}$ are the Fourier harmonics of $1/B^2$ defined in \eqref{eq:Fourier_1_over_B2}.

We now see that if $B(\psi,\vartheta_B,\varphi_B)$ varies on a surface only through $\chi = M \vartheta_B - N\varphi_B$, then so must $K$, as $b_{m,n}$ vanishes for $m/n \ne M/N$. Therefore, $K$ has the same symmetry properties as $B$. This condition will become important for studying the guiding center motion in the following Section.

\subsubsection{Guiding center motion in quasisymmetry}
\label{sec:quasisymmetry_gc}

We will study the consequences of this symmetry on guiding center motion. The analysis will be similar to that in Section \ref{sec:confinement_axisymmetry}, where we saw that axisymmetry provided particle confinement under the assumption of a strong magnetic field. Again, the existence of a coordinate which does not appear in the Lagrangian will result in a constant of motion. Throughout this Section we assume the electrostatic potential, $\Phi$, is a flux function, where the electric field $\bm{E} = -\nabla \Phi$. In practice, this is a good approximation for stellarator configurations \cite{Helander2014}.

In this Section we will assume that the fields are time-independent.
We express the gyroaveraged Lagrangian, repeated here for convenience,
\begin{gather}
    \mathcal{L}(\bm{R},\rho,\dot{\bm{R}},\dot{\varphi}) = \frac{m}{2}\left(\left(\dot{\bm{R}} \cdot \hat{\bm{b}}(\bm{R}) \right)^2 + \rho^2 \dot{\varphi}^2 \right) + q \bm{A}(\bm{R}) \cdot \dot{\bm{R}} - \frac{\dot{\varphi} \rho^2}{2} B(\bm{R}) - q \Phi(\bm{R}),
\end{gather}
in Boozer coordinates ($\psi$,$\vartheta_B$,$\varphi_B$). 
The time derivative of the guiding center position can be expressed in the Boozer contravariant form as
\begin{gather}
    \dot{\bm{R}} = \dot{\psi} \partder{\bm{r}}{\psi} + \dot{\vartheta}_B \partder{\bm{r}}{\vartheta_B} + \dot{\varphi}_B \partder{\bm{r}}{\varphi_B}.
    \label{eq:R_contravariant}
\end{gather}
To compute the quantity $\dot{\bm{R}} \cdot \hat{\bm{b}}(\bm{R})$, we express $\dot{\bm{R}}$ as \eqref{eq:R_contravariant}
and $\hat{\bm{b}} = \bm{B}/B$ in the covariant form using \eqref{eq:boozer_covariant}. To compute the quantity $\bm{A}(\bm{R}) \cdot \dot{\bm{R}}$, we first see that $\bm{B}(\bm{R}) = \nabla \times \bm{A}(\bm{R})$ implies that the vector potential $\bm{A}$ can be written
\begin{gather}
 \bm{A}(\psi,\vartheta_B,\varphi_B) = \psi\nabla \vartheta_B - \psi_P(\psi) \nabla \varphi_B,
\end{gather}
where $\psi_P = \Psi_P/2\pi$ is the poloidal flux function such that $\psi_P'(\psi) = \iota(\psi)$. Upon application of a curl, we recover the Boozer contravariant form of $\bm{B}$ \eqref{eq:boozer_contravariant}. We will again express $\dot{\bm{R}}$ as in \eqref{eq:R_contravariant} to compute $\bm{A}(\bm{R})\cdot \dot{\bm{R}}$. Doing so we obtain the following form for the gyroaveraged Lagrangian in Boozer coordinates,
\begin{multline}
    \mathcal{L}(\psi,\vartheta_B,\varphi_B,\rho,\dot{\psi},\dot{\vartheta}_B,\dot{\varphi}_B,\dot{\varphi}) = \frac{m}{2B^2(\psi,\vartheta_B,\varphi_B)}\left(\dot{\psi}K(\psi,\vartheta_B,\varphi_B) + \dot{\vartheta}_B I(\psi) + \dot{\varphi}_B G(\psi) \right)^2  \\+ m \frac{\rho^2 \dot{\varphi}^2}{2} + q  \left(\psi \dot{\vartheta}_B - \psi_P(\psi) \dot{\varphi}_B \right) - \frac{\dot{\varphi} \rho^2}{2} B(\psi,\vartheta_B,\varphi_B) - q \Phi(\psi).
    \label{eq:lagrangian_boozer}
\end{multline}
We now will assume that $B$ and $\Phi$ are independent of $\chi = M \vartheta_B - N \varphi_B$. As discussed in Sections \ref{sec:confinement_axisymmetry} and \ref{sec:quasisymmetry_gc}, if the Lagrangian is independent of one coordinate, this implies the existence of a conserved quantity. Our aim will be to compute this conserved quantity and discuss the implications. 

We perform a change of coordinates such that
$(\psi,\vartheta_B,\varphi_B) \rightarrow (\psi,\chi,\eta)$ where $\chi = M \vartheta_B - N \varphi_B$ and $\eta = M' \vartheta_B - N'\varphi_B$ with $M'/N' \ne M/N$,
\begin{gather}
    \mathcal{L}(\psi,\vartheta_B,\varphi_B,\rho,\dot{\psi},\dot{\vartheta}_B,\dot{\varphi}_B,\dot{\varphi}) = \widetilde{\mathcal{L}}(\psi,\chi,\eta,\rho,\dot{\psi},\dot{\chi},\dot{\eta},\dot{\varphi}). 
\end{gather}
Here we note that $\vartheta_B = (N \eta -N'\chi)/(M'N-MN')$ and $\varphi_B = (M\eta-M'\chi)/(M'N-MN')$, so
\begin{multline}
    \widetilde{\mathcal{L}}(\psi,\chi,\eta,\rho,\dot{\psi},\dot{\chi},\dot{\eta},\dot{\varphi})\\ = \frac{m}{2\widetilde{ B}^2(\psi,\chi)} \left( \dot{\psi} \widetilde{ K}(\psi,\chi) - \frac{\dot{\chi}\left(N'I(\psi)+M'G(\psi) \right)}{M'N-MN'}  + \frac{\dot{\eta}\left(NI(\psi)+MG(\psi)\right)}{M'N-MN'}  \right)^2 \\
    + m \frac{\rho^2 \dot{\varphi}^2}{2} + q \left(\frac{\dot{\chi}(M'\psi_P(\psi)-N'\psi)}{M'N-MN'}+ \frac{\dot{\eta}(N\psi-M\psi_P(\psi))}{M'N-MN'}\right) - \frac{\dot{\varphi} \rho^2}{2} \widetilde B(\psi,\chi) - q \widetilde \Phi(\psi).
\end{multline}
The Euler-Lagrange equation corresponding to $\eta$ is,
\begin{gather}
    \partder{\mathcal{\widetilde{L}}}{\eta} = \der{}{t} \left(\partder{\mathcal{\widetilde{L}}}{\dot{\eta}}\right)  \Rightarrow
 \der{p_{\eta}(t)}{t} = 0,
 \label{eq:euler_Lagrange}
\end{gather}
where $p_{\eta}$ is the canonical momentum corresponding to $\eta$,
\begin{gather}
    p_{\eta}(\psi,\chi) := \frac{1}{M'N-MN'} \left( \frac{m \widetilde{V_{||}}(\psi,\chi)}{\widetilde B(\psi,\chi)} (NI(\psi)+MG(\psi)) + q \left(N \psi - M \psi_P(\psi)\right) \right),
\end{gather}
and the parallel guiding center velocity, $\widetilde{V}_{||}$, is defined in the following way,
\begin{equation}
    \widetilde{V_{||}}(\psi,\chi) \widetilde{B}(\psi,\chi)= \dot{\widetilde{\bm{R}}} \cdot \widetilde{\bm{B}}(\widetilde{\bm{R}}) = \dot{\psi} \widetilde{ K}(\psi,\chi) - \frac{\dot{\chi}\left(N'I(\psi)+M'G(\psi) \right)}{M'N-MN'}  + \frac{\dot{\eta}\left(NI(\psi)+MG(\psi)\right)}{M'N-MN'}.
\end{equation}

Similar to the analysis in Section \ref{sec:confinement_axisymmetry}, we will consider the relative size of each term in $ p_{\eta}$. Here we will approximate $ V_{||} \sim v_t$ where $v_t$ is the thermal velocity, as was assumed in Section \ref{sec:cylindrical}, and $\psi \sim r^2 B_T/2$, where $r$ is an approximate scale of the minor radius and $B_T \sim \bm{B} \cdot \nabla \varphi_B/|\nabla \varphi_B|$ is an approximate scale of the toroidal field. In stellarators, the toroidal plasma current is much smaller than the poloidal current in the coils so we can assume that $I(\psi) \ll G(\psi)$, as the covariant components are defined such that $G(\psi) = \mu_0 I_P(\psi)/2\pi$ and $I(\psi) = \mu_0 I_T(\psi)/2\pi$. To compute an approximate scaling for the toroidal field, we assume $B_T \sim \mu_0 I_P/(2\pi R)$ where $R$ is the approximate scale of the major radius (see Section \ref{confiningMagneitcField}), assuming that all of the poloidal current flows through the coils so that $I_P$ is independent of $\psi$.
Since the plasma current in a stellarator is often small, this is typically a valid approximation.
Assuming the integers ($N$, $M$, $N'$, $M'$) are $\mathcal{O}(1)$ and that $\psi_P \sim \psi$, it follows that the 
ratio of the two terms in $p_{\eta}$ scales as,
\begin{gather}
    \frac{mV_{||} (N I(\psi)+M G(\psi))/B}{q(N\psi -M\psi_P(\psi))} \sim \frac{v_t (\mu_0 I_P/(2\pi))}{\Omega (r^2 \mu_0 I_P/(4\pi R))} \sim \frac{\rho}{r} \sim \epsilon \ll 1,
\end{gather}
where $\Omega = qB/m$ is the gyrofrequency and $\rho = v_t/\Omega$ is the gyroradius. In the above we have made the assumption that $r \sim R$, and $\epsilon \ll 1$ is the small parameter considered in Section \ref{ref:gyroaveraged_lagrangian}, expressing the fact that the gyroradius is much smaller than typical length scales of our system.  

Thus, we can approximate $p_{\eta} \sim q (N\psi- M\psi_P(\psi))$ to lowest order in $\epsilon$. Since $p_\eta$ is conserved along trajectories from the Euler-Lagrange equation \eqref{eq:euler_Lagrange}, so is $\psi$ to lowest order in $\epsilon$. We can therefore conclude that guiding center orbits stay close to a flux surface, as they do in axisymmetry. In this way, guiding center motion exhibits good confinement properties when the magnetic field strength and electrostatic potential only varies on a surface through $\chi$.

\subsubsection{Remark on ``quasi''}

As seen from the analysis in Section \ref{sec:quasisymmetry_gc}, quasisymmetry refers to configurations for which the magnetic field strength, $B$, only varies on a surface through a given linear combination of Boozer angles. This symmetry implies that there exists a symmetry coordinate, $\eta$, (i.e. the field strength does not depend on $\eta$), which results in a conserved quantity upon application of the Euler-Lagrange equations. This is referred to as \textit{quasi}symmetry, as it is a property of the field strength $B$ rather than the vector field $\bm B$ (that is, it is not a property of each of the vector components). In this way, quasisymmetry is distinct from axisymmetry, which implies a symmetry of each of the vector components of the field (see Section \ref{sec:axisymmetry}). Furthermore, some results suggest that quasisymmetry cannot be achieved globally, but only locally on a flux surface \cite{Garren1991}. Finally, perfect quasisymmetry can never be achieved in practice, as other physics considerations must be accounted for when designing the magnetic field, and coils cannot be engineered to perfectly reproduce the desired field. However, several stellarators have been designed to be close to quasisymmetry due to its beneficial confinement properties. Several classes of quasisymmetric stellarators will be discussed in the following Section.

\subsubsection{Types of quasisymmetry}\label{sec:symmetry_types}

\begin{figure}
    \centering
    \begin{subfigure}[b]{0.49\textwidth}
    \includegraphics[trim=1cm 7cm 0cm 7cm, clip,width=\textwidth]{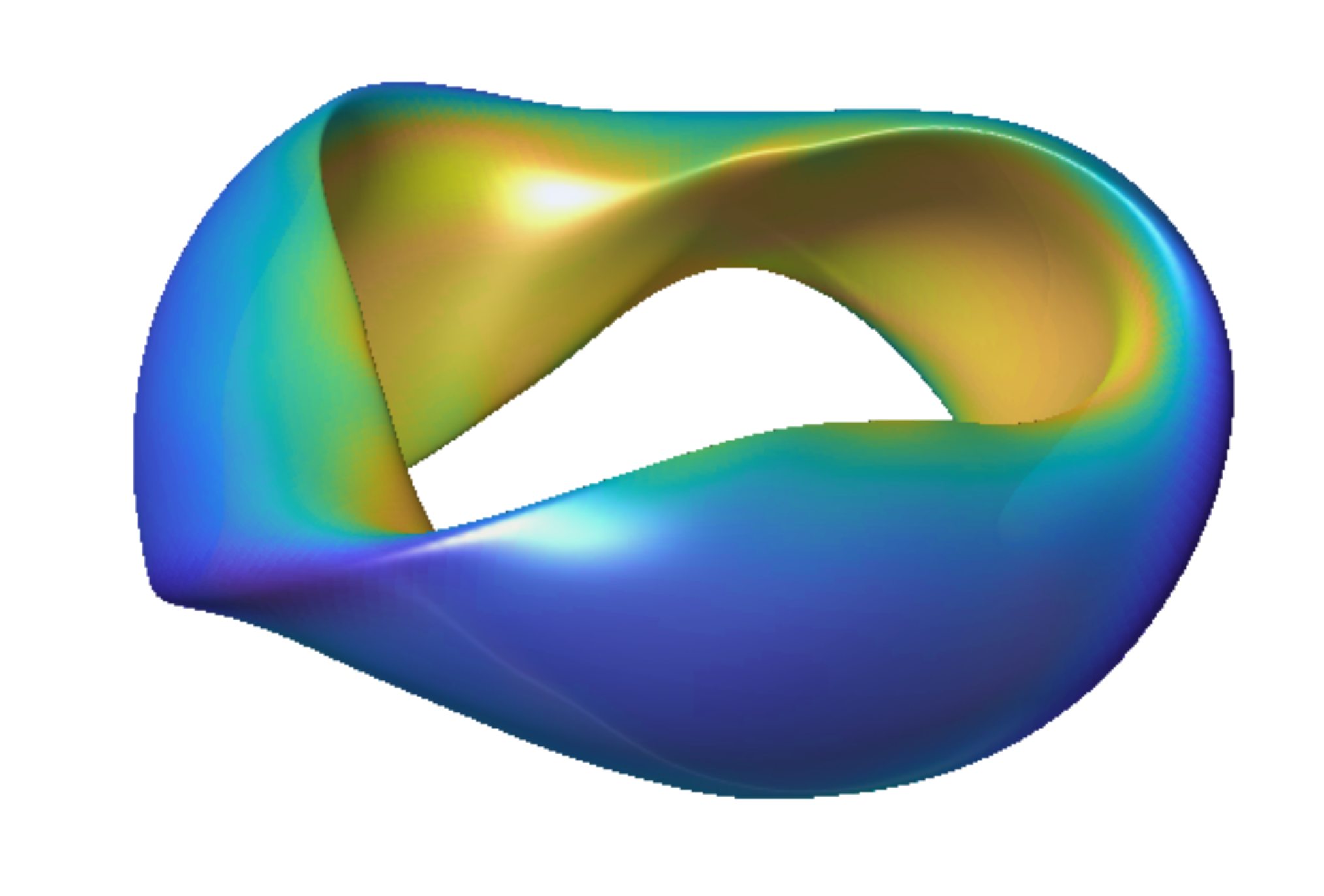}
    \caption{}
    \end{subfigure}
    \begin{subfigure}[b]{0.49\textwidth}
    \includegraphics[trim=1cm 6cm 1cm 6cm, clip,width=\textwidth]{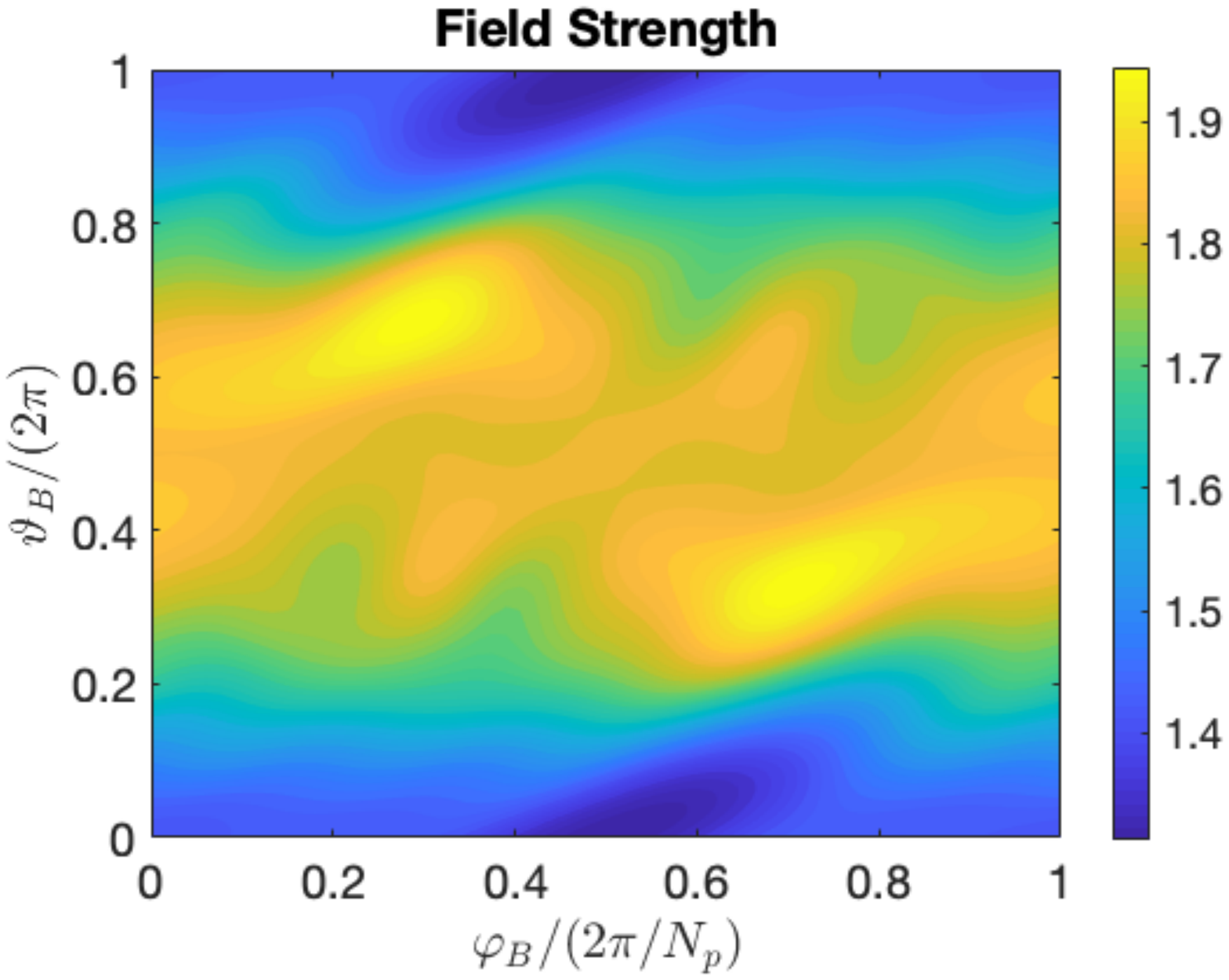}
    \caption{}
    \end{subfigure}
    \caption{(a) The last magnetic surface of NCSX is shown with the colorscale indicating field strength. (b) The field strength on the last magnetic surface is plotted as a function of the two Boozer angles. As NCSX is quasi-axisymmetric, $B$ is nearly constant along lines of constant $\varphi_B$.}
    \label{fig:NCSX}
\end{figure}

Quasi-axisymmetry (QA) refers to the case when $N = 0$. Thus contours of $B$ close toroidally without wrapping poloidally around the flux surface (see Figure \ref{fig:NCSX}). While there are currently no QA experiments in operation, the QA stellarator, NCSX \cite{Zarnstorff2001}, was designed and partially constructed at the Princeton Plasma Physics Laboratory. 
The CFQS \cite{Shimizu2018} configuration has been designed and will be constructed at Southwest Jiaotong University in China. 
In addition, there have been several other configurations designed to be close to QA, including Aries-CS \cite{Najmabadi2008}, ESTELL \cite{Drevlak2013}, and QuASDEX \cite{Henneberg2019}.

Quasi-poloidal (QP) symmetry refers to the case when $M = 0$ so that contours of $B$ close poloidally without wrapping toroidally around the flux surface. It should be noted that this type of symmetry cannot be achieved in practice near the axis due to the requirement that the pressure gradient vanishes on axis (see Section 2.7 in \cite{Landreman2011}). However, away from the axis, it may be possible to get close to QP symmetry. An example of a configuration designed for QP symmetry is the Quasi Poloidal Stellarator \cite{Strickler2004}.

\begin{figure}
    \centering
    \begin{subfigure}[b]{0.49\textwidth}
    \includegraphics[trim=1cm 7cm 0cm 7cm, clip,width=\textwidth]{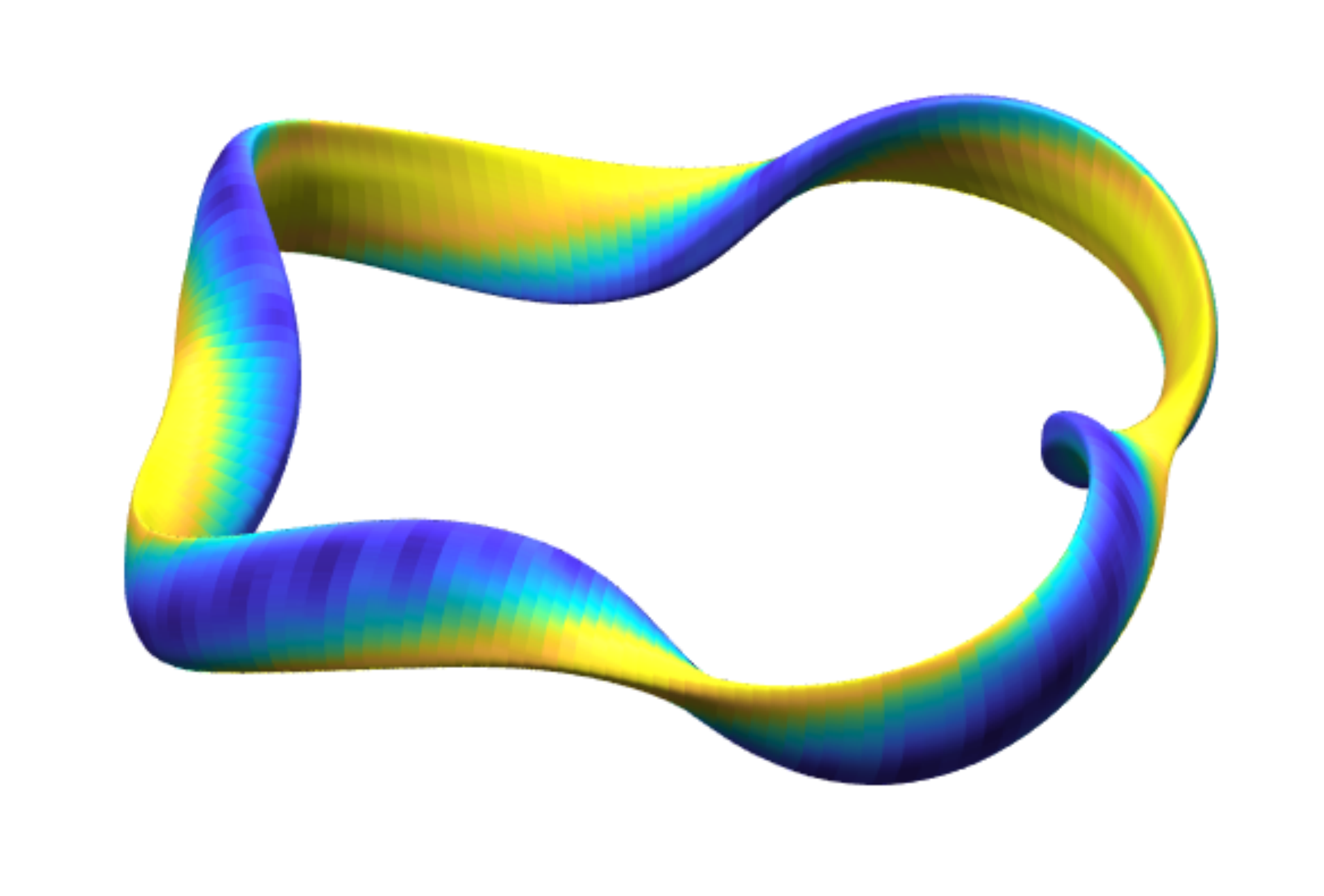}
    \caption{}
    \end{subfigure}
    \begin{subfigure}[b]{0.49\textwidth}
    \includegraphics[trim=1cm 6cm 1cm 6cm, clip,width=\textwidth]{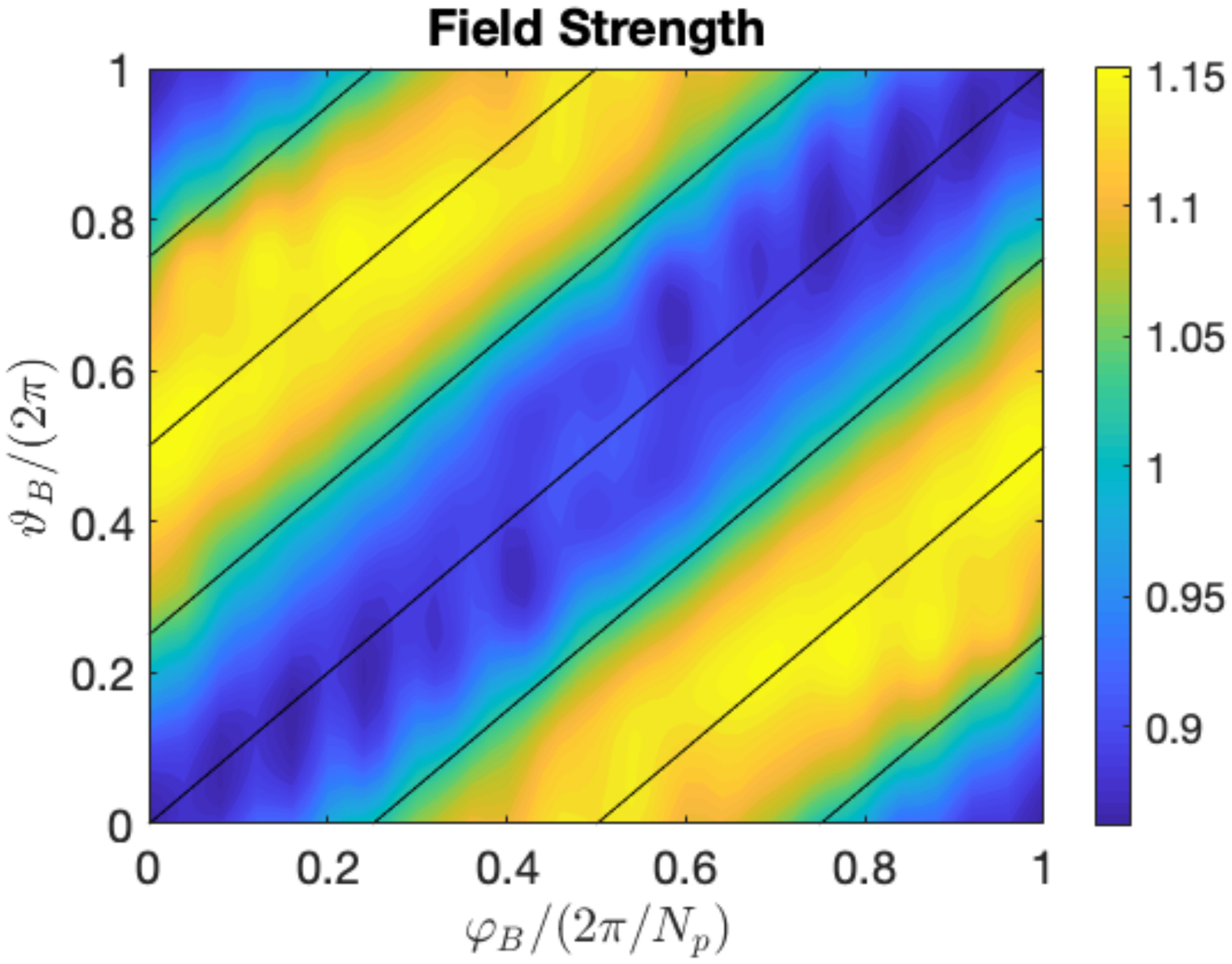}
    \caption{}
    \end{subfigure}
    \caption{(a) The last magnetic surface of HSX is shown with the colorscale indicating field strength. (b) The field strength on the last magnetic surface is plotted as a function of the two Boozer angles. As HSX is quasi-helically symmetric, $B$ is nearly constant along lines of constant $\vartheta_B + N_P\varphi_B$ (black), where $N_P$ is the number of field periods (see Section \ref{sec:Np_symmetry}).}
    \label{fig:HSX_QS}
\end{figure}

Quasi-helical (QH) symmetry implies that the field strength can be expressed as $B(\psi,M \vartheta_B-N\varphi_B)$ with $M \ne 0$ and $N \ne 0$. Thus contours of $B$ close both toroidally and poloidally, or helically (see Figure \ref{fig:HSX_QS}). The Helically Symmetric Experiment (HSX) at the University of Wisconsin \cite{Anderson1995} is the only existing quasisymmetric stellarator. Several other QH configurations have been designed \cite{Nuhrenberg1988,Ku2010} but not constructed.

\subsubsection{Other definitions}

There are several equivalent definitions of quasisymmetry that do not rely upon a transformation to Boozer coordinates. We will discuss one such representation here \cite{Helander2008},
\begin{align}
    \bm{B} \times \nabla \psi \cdot \nabla B = F(\psi) \bm{B} \cdot \nabla B,
    \label{eq:quasisymmetry2}
\end{align}
where $F(\psi)$ is any flux function. This expression of quasisymmetry will become useful to study the motion of trapped particles in Section \ref{sec:omnigeneity}.

We can see that the definition of quasisymmetry presented in Section \ref{sec:quasisymmetry} implies \eqref{eq:quasisymmetry2} with \begin{align}
    F(\psi) = \frac{N I(\psi) + M G(\psi)}{\iota(\psi) M-N},
\end{align}
for quasisymmetry with poloidal mode $M$ and toroidal mode $N$. 

The converse is also true: \eqref{eq:quasisymmetry2} implies \eqref{eq:quasisymmetry}. Expressing \eqref{eq:quasisymmetry2} in Boozer coordinates, we find that 
\begin{align}
    -I(\psi) \partder{B}{\varphi_B} + G(\psi) \partder{B}{\vartheta_B}  = F(\psi) \left( \partder{B}{\varphi_B} + \iota(\psi) \partder{B}{\vartheta_B} \right).
    \label{eq:quasisymmetry3}
\end{align}
We write $B$ in a Fourier series in the two angles,
\begin{align}
    B(\psi,\vartheta_B,\varphi_B) = \sum_{m,n}  B_{m,n}(\psi) e^{i(m \vartheta_B - n \varphi_B)}, 
\end{align}
to express \eqref{eq:quasisymmetry3} as 
\begin{align}
    \left(I(\psi)n  + G(\psi) m + n F(\psi) - \iota(\psi) m F(\psi) \right) B_{m,n}(\psi) = 0.
\end{align}
If $B_{m,n}(\psi) \ne 0$, we find that 
\begin{align}
    F(\psi) = \frac{n I(\psi) + m G(\psi)}{m\iota(\psi) - n}. 
\end{align}
This implies that either $m = 0$, $n = 0$, or $m/n$ is a fixed ratio on the surface. Thus we recover the definition in \eqref{eq:quasisymmetry}. 

Several other equivalent definitions of quasisymmetry are described in other references \cite{Helander2014,Landreman2019b}. 

\subsection{Omnigeneity}
\label{sec:omnigeneity}

While quasisymmetry implies particle confinement in the absence of collisions or fluctuations, it is not a necessary condition to obtain confinement in a stellarator. Another more general property is omnigeneity, meaning that the time-averaged magnetic drift off of a magnetic surface vanishes for all particles.

Consider $\rho$ is the gyroradius and $L$ is a gradient scale length,
possibly the equilibrium magnetic field, the major or minor radius.
To lowest order in $\epsilon = \rho/L$, guiding centers simply move along field lines (see Section \ref{ref:gyroaveraged_lagrangian}). As discussed in Section \ref{sec:adiabatic_invariant}, the magnetic moment $\mu = v_{\perp}^2/(2B)$ is conserved along a trajectory, where $v_{\perp}$ is the magnitude of the velocity perpendicular to the magnetic field. In addition, the energy of a particle, $E = (V_{||}^2 + v_{\perp}^2)/2 + q \Phi$, is conserved in time-independent fields (see Section \ref{sec:energy_conservation}). We will make the assumption that the electrostatic potential is a flux function, $\Phi(\psi)$ (this approximation is often valid in stellarators, see for example \cite{Helander2014}). As discussed in Section \ref{sec:particle_trapping}, this results in trapping of some particles in regions of low magnetic field strength, as the parallel velocity,
\begin{align}
    V_{||} = \pm \sqrt{\frac{2(E - q\Phi(\psi))}{m} - 2 \mu B},
\end{align}
can vanish at points where $B = B_{\text{crit}} := (E-q\Phi)/\mu m$. For other particles, their $B_{\text{crit}}$ value exceeds the maximum magnetic field strength on a surface; thus they are said to be passing particles. 

To next order in $\epsilon = \rho/L$, the trapped and passing particles experience cross-field drifts \eqref{eq:drifts}, repeated here for convenience,
\begin{align}
    \dot{\bm{R}}_{\perp} = \frac{V_{||}^2}{\Omega} \hat{\bm{b}} \times \bm{\kappa} + \frac{\mu}{\Omega} \hat{\bm{b}} \times \nabla B + \frac{\bm{E} \times \bm{B}}{B^2}.
\end{align}
As $\bm{E} = - \nabla \Phi(\psi)$, the $\bm{E} \times \bm{B}$ drift will not result in any radial drift out of the device, as $\bm{E} \times \bm{B} \cdot \nabla \psi = 0$. However, the curvature and grad-$B$ drifts will result in motion off of a surface of constant $\psi$. The radial drift can be evaluated by noting that $\hat{\bm{b}} \times \bm{\kappa} \cdot \nabla \psi = \left(\nabla \psi \times \hat{\bm{b}} \cdot \nabla B\right)/B$ under the assumption that $\nabla \times \bm{B} \cdot \nabla \psi$ vanishes,
\begin{align}
    \dot{\bm{R}}_{\perp} \cdot \nabla \psi = \frac{V_{||}^2 + \mu B}{\Omega B^2} \bm{B} \times \nabla B \cdot \nabla \psi  = \frac{V_{||}}{\Omega} \nabla \cdot \left(\frac{V_{||}}{B}\bm{B} \times \nabla \psi  \right),
\end{align}
where in the second equality the gradient is computed at fixed $E$ and $\mu$ and we have used the identity $\nabla \cdot (\bm{B} \times \nabla \psi) = 0$.
We can furthermore evaluate this expression in a coordinate system $(\psi,\alpha,l)$, where $\alpha$ is a field line label such that $\bm{B} = \nabla \psi \times \nabla \alpha$ and $l$ measures length along field lines. In magnetic coordinates (see Section \ref{sec:magnetic_coordinates}), we can take $\alpha = \vartheta - \iota(\psi) \varphi$. We can therefore define a coordinate system $(\psi,\alpha,l)$ whose Jacobian is given by $\sqrt{g} = \left(\nabla \psi \times \nabla \alpha \cdot \nabla l\right)^{-1} = \left(\bm{B} \cdot \nabla l\right)^{-1} =B^{-1}$. In such a coordinate system, the radial drift evaluates to, 
\begin{align}
   \dot{\bm{R}}_{\perp} \cdot \nabla \psi = \frac{mV_{||}}{q} \left[\partder{V_{||}}{\alpha}  + \partder{}{l} \left(\frac{V_{||} \hat{\bm{b}} \times \nabla \psi \cdot \nabla l}{B} \right) \right].
    \label{eq:radial_drift}
\end{align}

 If $\dot{\bm{R}}_{\perp} \cdot \nabla \psi$ vanished for all particles, then in the collisionless limit, particles would be perfectly confined to magnetic surfaces. In practice, obtaining a magnetic field such that this constraint is satisfied is very restrictive. Instead, we would like the \textit{average} motion of guiding centers off the surface to vanish such that particles remain within a small distance of a given surface. In order to do so, we now wish to compute the averaged radial drift along a field line for both trapped and passing particles, 
\begin{align}
    \left\langle \dot{\bm{R}}_{\perp} \cdot \nabla \psi  \right\rangle = \frac{\oint \frac{dl}{V_{||}} \, \dot{\bm{R}}_{\perp} \cdot \nabla \psi}{\oint \frac{dl}{V_{||}}}.
    \label{eq:passing_drift}
\end{align}
Here integration is taken along a closed orbit along a field line. For passing particles on rational surfaces, the integration is taken along a field line until it closes on itself. For passing particles on irrational surfaces, integration is taken many times along a field until it comes arbitrarily close to its starting point. For trapped particles, integration is taken along a closed orbit between $l_-$ and $l_+$, where $B(\psi,\alpha,l_-) = B(\psi,\alpha,l_+) = B_{\text{crit}}$. The integrals are evaluated holding $E$, $\mu$, $\psi$, and $\alpha$ constant. The quantity $\oint dl/V_{||}$ is often denoted by $\tau$, the periodic transit time along a field line. 

Evaluating \eqref{eq:passing_drift} with the expression for the radial drift \eqref{eq:radial_drift}, we find that the averaged radial drift vanishes, $\langle \dot{\bm{R}}_{\perp} \cdot \nabla \psi \rangle = 0$, if 
\begin{align}
    \partder{}{\alpha} \left( \oint dl \, V_{||} \right) = 0,
    \label{eq:omnigeneity}
\end{align}
due to periodicity in $l$.
This is a consequence of the fact that the integration in \eqref{eq:passing_drift} is performed at constant $\alpha$ such that the derivative can be pulled out of the integral. 

The quantity $J_{||} = \oint dl \, V_{||}$ is often called the parallel adiabatic invariant, which is approximately constant along the averaged drift motion. This can be seen by computing the averaged drifts,
\begin{subequations}
\begin{align}   
    \left \langle \dot{\bm{R}} \cdot \nabla \psi \right \rangle &= \frac{m}{q \tau} \partder{J_{||}}{\alpha} \\
    \left \langle \dot{\bm{R}} \cdot \nabla \alpha \right \rangle &= -\frac{m}{q \tau} \partder{J_{||}}{\psi},
\end{align}
\end{subequations}
to compute the total variation of $J_{||}$ along an averaged guiding center trajectory,
\begin{align}
    \der{J_{||}}{t} &= \partder{J_{||}}{\alpha} \left\langle \dot{\bm{R}} \cdot \nabla \alpha \right \rangle + \partder{J_{||}}{\psi} \left \langle \dot{\bm{R}} \cdot \nabla \psi \right \rangle = 0. 
\end{align}
If the above is satisfied, then $J_{||}$ is conserved along the averaged drift trajectories.
Thus the drift orbits are confined to surfaces of constant $J_{||}$. This implies a condition for good confinement: surfaces of constant $J_{||}$  should align with surfaces of constant $\psi$, or $\partial J_{||}/\partial \alpha = 0$, for all particles (equivalently, all values of $E/\mu$). A given magnetic field satisfying the above property is said to be \textit{omnigeneous} \cite{Hall1975,Cary1997}, implying that the time-averaged magnetic drift off of a magnetic surface vanishes for all particles.


It can be shown that the omnigeneity condition \eqref{eq:omnigeneity} is automatically satisfied by passing particles. In this case, as in \cite{2012Dhaeseleer,Helander2014}, the integrals in \eqref{eq:passing_drift} can be evaluated in the following way,
\begin{align}
       \left\langle \dot{\bm{R}}_{\perp} \cdot \nabla \psi  \right\rangle = \frac{\int_0^{2\pi} d \alpha \int_0^L \frac{dl}{V_{||}} \, \dot{\bm{R}}_{\perp} \cdot \nabla \psi}{\int_0^{2\pi} d \alpha \int_0^L \frac{dl}{V_{||}}}, 
       \label{eq:passing_drift2}
\end{align}
where the $l$ integral is taken along one toroidal loop from 0 to $L$. 
Inserting \eqref{eq:radial_drift} into \eqref{eq:passing_drift2}, we find that the averaged drift vanishes for passing orbits due to periodicity in $\alpha$. This result only relies on the existence of magnetic surfaces and places no additional constraint on magnetic geometry. Thus passing particles are well-confined if magnetic surfaces exist. However, confinement of trapped particles requires an additional restriction on the magnetic geometry.

We can note that quasisymmetry implies omnigeneity. The relationship between quasisymmetry and omnigeneity is discussed in several references \cite{Cary1997,Cary1997b,Landreman2012}. If a magnetic field satisfies \eqref{eq:quasisymmetry2}, then the radial drift can be written as 
\begin{align}
    \dot{\bm{R}}_{\perp} \cdot \nabla \psi &= V_{||} \partder{}{l}\left( \frac{V_{||}}{\Omega} \right) F(\psi).
\end{align}
This implies that $\langle \dot{\bm{R}}_{\perp} \cdot \nabla \psi \rangle$ vanishes from periodicity by evaluating \eqref{eq:passing_drift}.

However, omnigeneity includes a much wider class of magnetic fields than quasisymmetry. It has been shown that an infinitely differentiable omnigeneous magnetic field must be quasisymmetric \cite{Cary1997}. Nonetheless, a magnetic field which is arbitrarily close to omnigeneity can be very far from quasisymmetry. Construction of nearly-omnigeneous magnetic fields has been demonstrated \cite{Plunk2019} based on equilibria near the magnetic axis (see Section \ref{sec:axis_expansion}). The Wendelstein 7-X stellarator is an example of an existing quasi-omnigeneous configuration obtained with numerical optimization techniques \cite{Grieger1989,Grieger1992}.

\subsection{$N_P$ symmetry}
\label{sec:Np_symmetry}

While quasisymmetry is a continuous symmetry and results in an independent coordinate, many physical properties of stellarators can also exhibit discrete symmetries, for example, invariance with respect to reflection or translation.

Specifically, the magnetic field in stellarators often exhibits a (discrete) symmetry with respect to the number of periods, $N_P$, in a full toroidal turn, namely,
\begin{gather}
    \bm{B}(\psi,\theta,\zeta + 2\pi/N_P) = \bm{B}(\psi,\theta,\zeta),
    \label{eq:NP_symmetry}
\end{gather}
when expressed in flux coordinates (see Section \ref{sec:flux_coordinates}), $(\psi,\theta,\zeta)$. The same periodicity holds in the cylindrical coordinate system (see Section \ref{sec:toroidal_geom}),  $(R,\phi,Z)$.

The above notation in \eqref{eq:NP_symmetry} indicates that each of the vector components obeys the symmetry.
As a consequence, other quantities, such as the cylindrical coordinates describing magnetic surfaces, $R$ and $Z$, exhibit the same periodicity when expressed in flux coordinates. As many physical quantities are periodic with respect to $N_P$, this yields computational savings for many problems. In the stellarator literature, this is referred to as $N_P$-symmetry or field period symmetry, referring to a periodicity with respect to a toroidal turn of $2\pi/N_P$.

\begin{figure}
    \centering
    \includegraphics[trim=6cm 5cm 5cm 5cm,clip,width=0.8\textwidth]{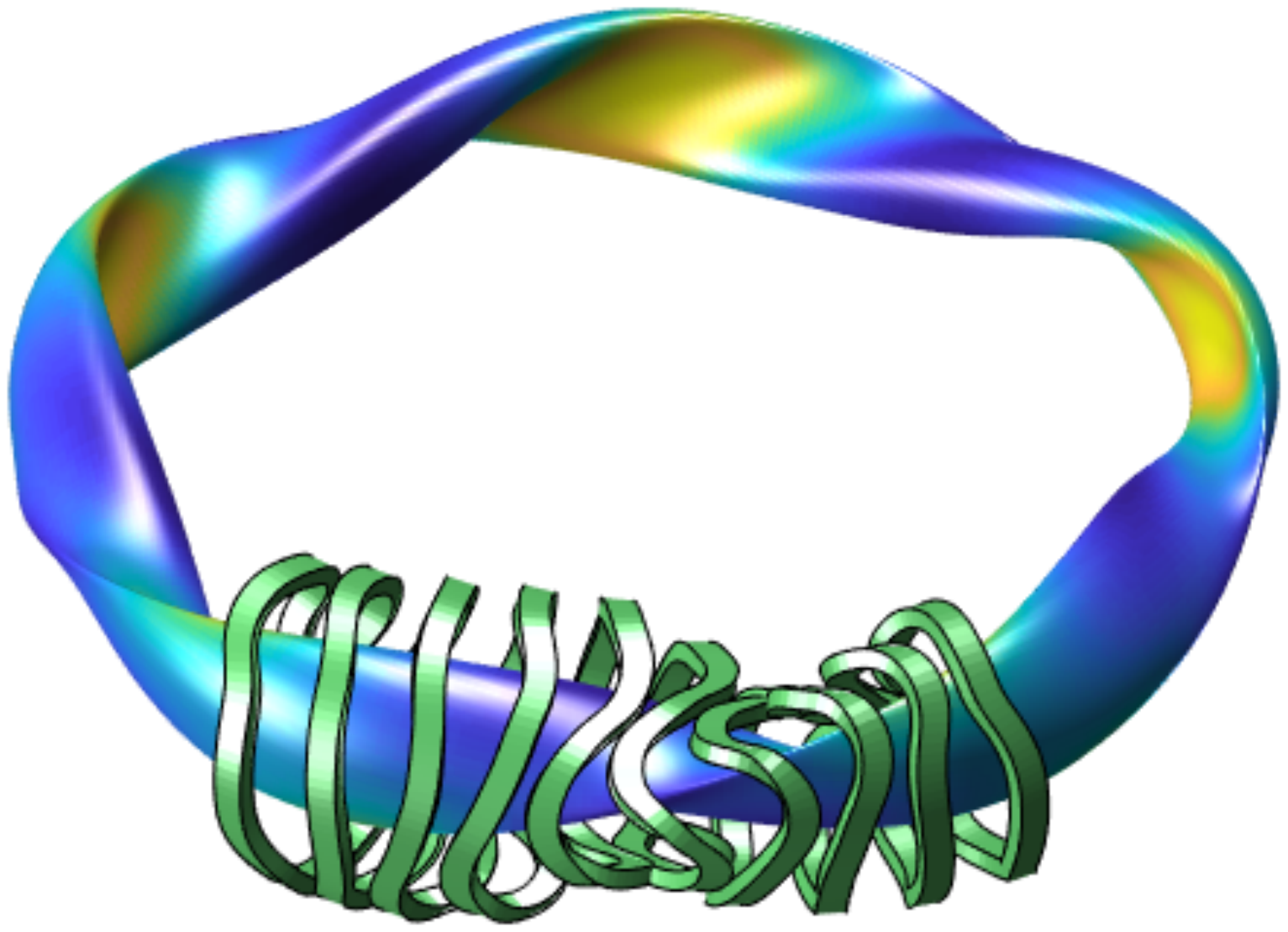}
    \caption{The last closed magnetic surface of the W7-X configuration is shown with the color scale indicating the field strength. The electromagnetic coil shapes are shown for one field period of the device.}
    \label{fig:w7x_periods}
\end{figure}

For example, the W7-X stellarator has 5 field periods, and as a result, the same coil shapes can be used for each field period. 
In Figure \ref{fig:w7x_periods}, a magnetic surface of W7-X is shown, with the color scale indicating the field strength. As can be seen, the field strength exhibits 5 toroidal periods. The electromagnetic coils are shown for one field period.

\subsection{Stellarator symmetry}
\label{sec:stellarator_symmetry}

Stellarator symmetry refers to a discrete symmetry with respect to a specific transformation as defined below. 
Consider the cylindrical coordinates, $(R,\phi,Z)$ (see Section \ref{sec:toroidal_geom}). 

\begin{figure}
    \centering
    \includegraphics[trim=2cm 10cm 2cm 10cm,clip,width=0.7\textwidth]{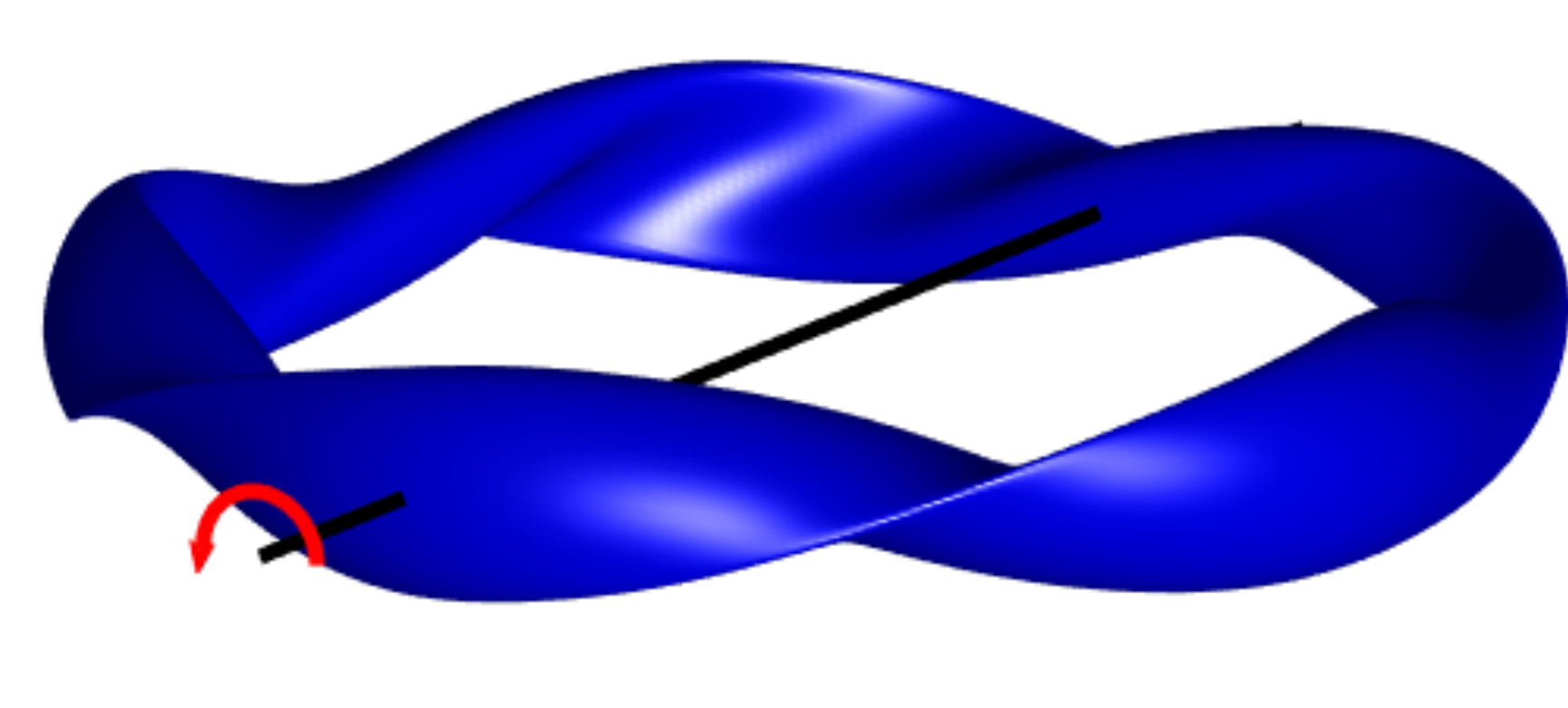}
    \caption{Stellarator symmetry describes an inversion about the line $Z = 0$, $\phi = 0$. The stellarator-symmetric W7-X configuration is shown.}
    \label{fig:stellarator_symmetry}
\end{figure}

We define the transformation $\mathcal T$ by
\begin{align}\label{defT1}
    \mathcal {T}:\bm{F}=F_R \hat{\bm{R}} +F_{\phi} \hat{\bm{\phi}} + F_Z \hat{\bm{Z}} \mapsto  \bm{G}  = G_R \hat{\bm{R}} + G_{\phi} \hat{\bm{\phi}} + G_Z \hat{\bm{Z}}
    \end{align}
    such that 
    \begin{align}\label{defT2}
    \left\{
    \begin{array}{l}
    G_R(R,\phi,Z) = F_R(R,-\phi,-Z)\\
    G_\phi(R,\phi,Z) = F_\phi(R,-\phi,-Z) \\
    G_Z(R,\phi,Z) = F_Z(R,-\phi,-Z)
    \end{array}\right. .
\end{align}
The transformation $\mathcal {T}$ can be thought of as an inversion about the line $\phi = 0, Z = 0$ (see Figure \ref{fig:stellarator_symmetry}). The term stellarator symmetry describes a vector field $\bm{F}$ satisfying the following property
\begin{gather}
    \mathcal{T}[ \bm{F}]= -F_R \hat{\bm{R}} +F_{\phi} \hat{\bm{\phi}} + F_Z \hat{\bm{Z}}.
       \label{eq:stellarator_symmetry}
\end{gather}
A given stellarator configuration is said to possess stellarator symmetry if the equilibrium magnetic field $\bm{B}$ is stellarator symmetric according to \eqref{eq:stellarator_symmetry}. As will be discussed further at the end of this Section, stellarator symmetry provides computational efficiency, as many physical quantities have definite parity (are even or odd). It can be shown that if the magnetic field $\bm{B}$ exhibits this symmetry, then so does the current density $\bm{J}$ upon application of Ampere's law \eqref{eq:ampere_magnetostatic} in cylindrical coordinates. 

Suppose a vector field $\bm{F}$ possesses stellarator symmetry. This implies that the magnitude $F$ of the vector field exhibits the following symmetry,
\begin{gather}
    F(R,-\phi,-Z) = F(R,\phi,Z).
    \label{eq:stellarator_symmetry_scalar}
\end{gather}
Thus the field strength $B$ and magnitude of the current density $J$ exhibit this property for a stellarator symmetric configuration.

Consider now what happens to
the position vector which follows a field line, $\bm{r}$, under the assumption that $\bm{B}$ is stellarator symmetric. The curve $\bm{r}$ satisfies,
\begin{align}
(\hat{\bm{b}} \cdot \nabla)
\bm{r}&= \hat{\bm{b}},
\label{eq:field_line_symmetric}
\end{align}
where $\hat{\bm{b}} = \bm{B}/B$ is a unit vector in the direction of the magnetic field.
Assume that $\bm{r}$ is a curve  parameterized by $\phi$, $\bm{r}(\phi)=R(\phi) \hat{\bm{R}}(\phi) + Z(\phi) \hat{\bm{Z}}$, then in cylindrical coordinates \eqref{eq:field_line_symmetric} reads,
\begin{subequations}
\begin{align}
\der{R(\phi)}{\phi} &= \frac{R(\phi) B_R(R(\phi),\phi, Z(\phi))}{B_{\phi}(R(\phi),\phi,Z(\phi))} \\
\der{Z(\phi)}{\phi} &= \frac{R(\phi) B_{Z}(R(\phi),\phi, Z(\phi))}{B_{\phi}(R(\phi),\phi,Z(\phi))},
\end{align}
\end{subequations}
where the notation $d/d\phi$ indicates the derivative with respect to $\phi$ along a field line and we have used $d \bm{r}/d\phi = d R/d\phi \hat{\bm{R}} + R \hat{\bm{\phi}} + d Z/d\phi \hat{\bm{Z}}$. If $\bm{B}$ is stellarator symmetric, then under the transformation $Z \rightarrow -Z$ and $\phi \rightarrow -\phi$, we see that $ d R/d\phi \rightarrow - dR/d\phi$ and $dZ/d\phi \rightarrow dZ/d\phi$. This implies that if $R(\phi)$ and $Z(\phi)$ parameterize a field line of a stellarator symmetric magnetic field, then so do $R(-\phi)$ and $-Z(-\phi)$. If in addition there exists a flux surface, this results in the same symmetry of the cylindrical components, $R$ and $Z$, of the surface on which field lines lie.

Stellarator symmetry can also be expressed in general flux coordinates $(\psi,\theta,\zeta)$, where $\psi$ is a flux label, $\theta$ is a poloidal angle, and $\zeta$ is a toroidal angle (see Section \ref{sec:flux_coordinates}). We will now see that it can simplify the expression of some physical quantities of a stellarator symmetric field. For a coordinate system which preserves stellarator symmetry, the following property holds for the cylindrical components ($R$,$\phi$,$Z$),
\begin{subequations}
\begin{align}
    R(\psi,-\theta,-\zeta) &= R(\psi,\theta,\zeta) \\
    \phi(\psi,-\theta,-\zeta) &= -\phi(\psi,\theta,\zeta) 
    \\
    Z(\psi,-\theta,-\zeta) &= -Z(\psi,\theta,\zeta)
    .
\end{align}
\end{subequations}
Therefore the transformation $\mathcal T$ defined in \eqref{defT1}-\eqref{defT2} can now be expressed in terms of flux coordinates as follows
\begin{align}
    \mathcal {T}:\bm{F}=F_R \hat{\bm{R}} +F_{\phi} \hat{\bm{\phi}} + F_Z \hat{\bm{Z}} \mapsto  \bm{G}  = G_R \hat{\bm{R}} + G_{\phi} \hat{\bm{\phi}}+ G_Z \hat{\bm{Z}} 
    \end{align}
    defined by 
    \begin{align}
    \left\{
    \begin{array}{l}
    G_R(\psi,\theta,\zeta) = F_R(\psi,-\theta,-\zeta)\\
    G_\phi(\psi,\theta,\zeta) = F_\phi(\psi,-\theta,-\zeta) \\
    G_Z(\psi,\theta,\zeta) = F_Z(\psi,-\theta,-\zeta)
    \end{array}\right. .
\end{align}
We see that for a stellarator symmetric field, $F_R$ is odd with respect to ($\theta$,$\zeta$) while $F_Z$ and $F_{\phi}$ are even. Thus stellarator symmetry implies a definite parity of many physical quantities; these quantities are even or odd with respect to $(\theta,\zeta)$. Quantities with a definite parity can be expressed in terms of only a sine or cosine series rather than a generic Fourier series. For example, in a stellarator symmetric field the field strength can be expressed with just a cosine series from \eqref{eq:stellarator_symmetry_scalar},
\begin{gather}
    B(\psi,\theta,\zeta) = \sum_{m,n} B_{mn}(\psi) \cos(m\theta - n\zeta).
\end{gather}

  Unlike quasisymmetry, stellarator symmetry is not inherently beneficial for confinement. However, the assumption of stellarator symmetry has been made in the design of almost \cite{Landreman2018b} every stellarator configuration to date. This is partly motivated by computational efficiency, as fewer Fourier coefficients are required to represent equilibrium quantities. A detailed discussion of stellarator symmetry is given in \cite{Dewar1998}.
  
\section{Optimization for stellarator design}
\label{sec:optimization}

As we have seen in Section \ref{sec:magnetic_confinement_devices}, one advantage of stellarators compared to tokamaks is that they do not rely on a large plasma current to produce the necessary rotational transform. On the other hand, neither the existence of magnetic surfaces nor single particle confinement are guaranteed in 3D configurations as they are in axisymmetry, as described respectively in Sections \ref{sec:integrability} and \ref{sec:confinement_axisymmetry}. Thus stellarators must be carefully designed with respect to a set of desirable physical properties, presented in Section \ref{ssec:objectives}. We note that optimization of tokamak magnetic fields is also possible \cite{Highcock2018}, though it is much more difficult than stellarator optimization as confinement properties become very sensitive to the plasma current and pressure profiles. These profiles can be determined with modeling on turbulent and transport time scales, which is very computationally intensive. On the other hand, the physical properties of stellarators are relatively insensitive to these profiles, as they primarily rely on the externally produced magnetic field for confinement \cite{Boozer2019}. 

Traditionally, the stellarator design process consists of two steps:
\begin{enumerate}
    \item the magnetic field is designed based on equilibrium models,
    \item coils are designed to be able to generate this desired magnetic field.
\end{enumerate}
 These problems are commonly formulated in terms of numerical optimization: a function defining various physics objectives is minimized, possibly subject to additional constraints. Iterative optimization algorithms are then leveraged to approximate a solution to the numerical problem, either a local or global minimum.
In this section we will describe standard optimization considerations, techniques, and tools for stellarator design.
Based on models presented in Section \ref{sec:equilibrium_fields}, we will describe the two steps of stellarator design: optimization of the plasma boundary with the fixed-boundary approach in Section \ref{sec:fixed_optimization} and optimization of coil shapes in Section \ref{sec:coil_opt}. In Section \ref{sec:device_examples} we present a few examples of optimized configurations to illustrate some successes and challenges.

\subsection{Physics objectives for stellarator optimization}\label{ssec:objectives}
\TBD
Stellarators must be carefully designed to ensure existence of magnetic surfaces within a large volume and confinement of single particle trajectories. As was already discussed in Section \ref{sec:symmetry}, both quasi-omnigeneity and quasisymmetry guarantee single particle confinement under the guiding center approximation. However, these properties alone are not sufficient to actually design a stellarator configuration, and there are additional considerations, both physical and practical, that should be taken into account. Often proxy functions are applied, simplified figures of merit which approximate the behavior of more complex physics objectives. We now outline important physics considerations and the corresponding proxy functions used during optimization.

\subsubsection{Equilibrium properties}

One of the most fundamental objectives in designing a stellarator is obtaining a large volume of continuously nested magnetic flux surfaces. As explained in Chapter \ref{sec:3D_difficulties}, a general 3D magnetic field contains islands and stochastic regions which may degrade confinement properties. The operating space of stellarators, the allowable experimental values of density and temperature, is often restricted due to MHD equilibrium properties. A common measure of the performance of a magnetic confinement device is $\beta := p/(B^2/(2\mu_0))$, the ratio of the plasma pressure, $p$, to the magnetic pressure, $B^2/(2\mu_0)$. One specific way in which the equilibrium properties can limit performance is through the \textit{Shafranov shift}, the outward shift of the magnetic axis in the presence of pressure gradients. When $\beta \sim \epsilon \iota^2/2$ where $\epsilon$ is the inverse aspect ratio and $\iota$ is the rotational transform, the Shafranov shift becomes comparable to the minor radius of the device, which results in flux-surface break-up \cite{Spong2010,Helander2014}. There is a tendency of the edge magnetic field to become stochastic at large beta \cite{Sakakibara2008}, so a design should try to maximize the volume of continuously nested flux surfaces \cite{Hudson2002}. One should also minimize the island width near low-order rational values of the rotational transform, which can be estimated using analytic expressions \cite{Lee1990,Cary1991}, assuming the magnetic field is close to having perfect magnetic surfaces. Such islands can also be minimized by controlling the rotational transform. 
If the magnetic shear, the derivative $\iota'(\psi)$ where $\iota$ is the rotational transform, is sufficiently small,  low-order rational surfaces can be avoided altogether. On the other hand, the width of islands can be minimized if the magnetic shear is large, as the magnetic island width scales as $1/\sqrt{\iota'(\psi)}$ \cite{Boozer2015}. Thus the design of a stellarator often either adopts the low or high-shear strategy. 
     
\subsubsection{Plasma current}

There are several sources of self-driven plasma current \cite{Helander2014}: the parallel bootstrap current arises due to collisions between trapped and passing particles in the presence of density and temperature gradients, and the parallel Pfirsch-Sch\"{u}ter and perpendicular diamagnetic currents occur due to equilibrium pressure gradients. The bootstrap current can cause shifts in the rotational transform toward low-order rational values, which must especially be avoided in low-shear devices. Control of the edge rotational transform is also vital for designs with an island divertor \cite{Geiger2010}. The \textit{divertor} is the device used for removal of heat and unwanted fusion products, such as helium (Section \ref{sec:fusion_reactions}), from the confinement region. The functioning of an island divertor requires that material structures intersect a large island at a specific location; thus an uncontrolled shift in the rotational transform may lead to malfunction of such a divertor system. In the presence of reduced bootstrap current, the magnetic field structure becomes less sensitive to changes in beta. For these reasons, the Wendelstein 7-X (W7-X) configuration was designed for minimal bootstrap current \cite{Grieger1992}. Often optimization is performed with a low-collisionality semi-analytic bootstrap current model \cite{Shaing1989}. 

The Pfirsch-Schl\"{u}ter current, defined as the first term in \eqref{eq:parallel_current_sum}, is a parallel current driven by pressure gradients which varies on a surface. This type of current does not provide any \textit{net} current when  integrated around the torus and therefore does not shift the rotational transform. However, it can give rise to a Shafranov shift and thus affect the equilibrium beta limit \cite{Wagner1998}. The Pfirsch-Schl\"{u}ter current can be reduced by minimizing the magnitude of the geodesic curvature, the component of the magnetic field curvature which is tangent to a flux surface. The net diamagnetic current will only be non-zero in the presence of another source of net current; thus, the reduction of the bootstrap current will automatically reduce the diamagnetic current. 

While the presence of self-driven current can give rise to unfavorable shifts in the rotational transform, there are situations in which significant bootstrap current may be desirable. If the bootstrap current provides a source of rotational transform in addition to the external coils, the coil complexity may be reduced and a more compact device may be possible. Plasma current can also reduce the width of islands in comparison with those in the vacuum configuration \cite{Hegna1998}. For these reasons, the National Compact Stellarator Experiment (NCSX) was designed to be quasi-axisymmetric with a significant fraction of rotational transform provided by the plasma current \cite{Hirshman1999}. 

\subsubsection{Quasisymmetry and omnigeneity}

A general 3D field does not provide confinement of guiding center trajectories. Two strategies to achieve guiding center confinement in an optimized configuration are quasisymmetry and omnigeneity. 

As described in Section \ref{sec:quasisymmetry}, a quasisymmetric magnetic field possesses a symmetry direction of the magnetic field strength when expressed in Boozer coordinates,
\begin{align}
    B(\psi,\vartheta_B,\varphi_B) = B(\psi,M\vartheta_B - N \varphi_B),
\end{align}
for fixed integers $M$ and $N$. If $M = 0$, the contours of the magnetic field strength close poloidally, known as quasi-poloidal symmetry. If $N = 0$, the contours of the magnetic field strength close toroidally, known as quasi-axisymmetry. If both $M$ and $N$ are non-zero, known as quasi-helical symmetry, the contours of the field strength close both toroidally and poloidally. Quasisymmetry implies guiding center confinement \cite{Boozer1983} and neoclassical properties that are comparable to those of an equivalent tokamak \cite{Helander2014}, including the ability to rotate in the direction of quasisymmetry \cite{Helander2008}. Quasisymmetry is typically targeted in optimization by minimizing the symmetry-breaking Fourier harmonics of the magnetic field strength.

As described in Section \ref{sec:omnigeneity}, an omnigeneous magnetic field does not allow any time-averaged drift of particles off of magnetic surfaces,
\begin{align}
    \langle \dot{\bm{R}} \cdot \nabla \psi \rangle = 0,
\end{align}
where $\dot{\bm{R}}$ is the guiding center velocity, and the brackets $\langle \dots \rangle$ indicate an average over the trajectory. A quasisymmetric field is omnigeneous, though the converse is not necessarily true. Omnigeneity can be targeted in optimization by computing properties of the parallel adiabatic invariant introduced in Section \ref{sec:omnigeneity},
\begin{align}
J_{||} = \oint dl \,V_{||}, 
\label{eq:adiabatic_invariant}
\end{align}
 where $V_{||}$ is the velocity parallel to the magnetic field, $l$ measures length along a field line, and the integration is taken along the bounce or transit orbit. If $J_{||}$ is constant on a magnetic surface for all values of the particle energy and magnetic moment, then the collisionless trajectories will experience no net radial drift, or the magnetic field is omnigeneous \cite{Cary1997}. Thus several properties involving $J_{||}$, such as its variation within a flux surface, have been targeted in stellarator optimization \cite{Spong1998,Drevlak2014}.


\subsubsection{Neoclassical transport}

\begin{figure}
    \centering
    \includegraphics[width=0.8\textwidth]{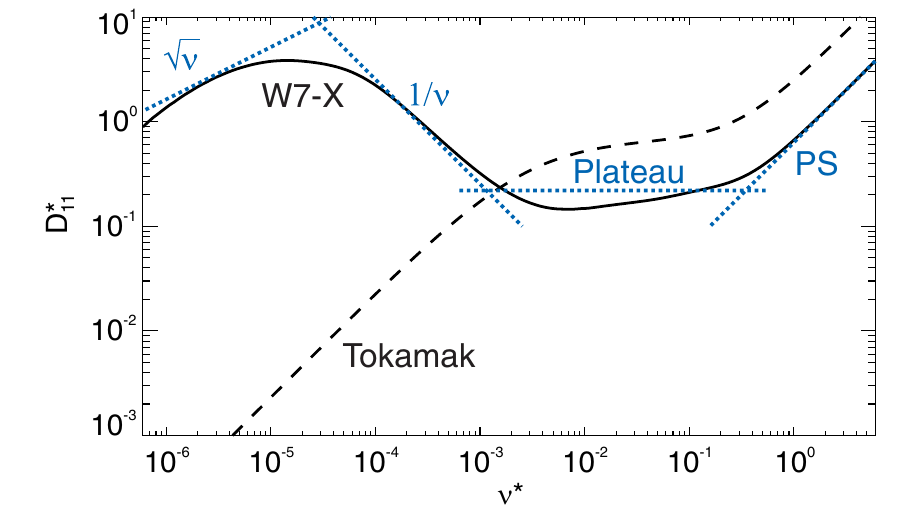}
    \caption{The neoclassical diffusion coefficient, $D_{11}^*$, as a function of the normalized collisionality, $\nu_* = \nu R/(\iota v)$, where $\nu$ is the collision frequency, $\iota$ is the rotational transform, $v$ is the speed, and $R$ is the major radius. An axisymmetric field exhibits a low-collisionality regime in which $D_{11}^* \sim \nu$, while a stellarator exhibits $D_{11}^* \sim 1/\nu$. Thus the neoclassical transport in a general three-dimensional field can be especially deleterious at low collisionality. Figure reproduced from \cite{Helander2012}.}
    \label{fig:collisionality}
\end{figure}

Collisions between charged particles will generally enhance the transport due to guiding center drifts. Known as \textit{neoclassical transport}, this process is amplified in non-omnigeneous configurations. Neoclassical theory predicts the behavior of the bulk population of charged particles, which are described by a Maxwell-Boltzmann distribution (see Appendix \ref{debye}) to lowest order. Neoclassical transport is typically the dominant transport channel, as opposed to turbulent transport due to fluctuations, in classical (unoptimized) stellarators. For a brief overview of neoclassical physics in stellarators, see Section 4 of \cite{Helander2014}. Stellarators experience enhanced neoclassical transport at low collisionality in comparison with tokamaks (Figure \ref{fig:collisionality}). Neoclassical transport is often computed with reduced models \cite{Zarnstorff2001,Ku2008},
such as the effective ripple ($\epsilon_{\text{eff}}^{3/2}$) \cite{Nemov1999}, which quantifies the geometric dependence of transport under the assumption of low collisionality.  A review of neoclassical optimization strategies is given in \cite{Mynick2006}.

\subsubsection{Energetic particles}
In addition to the bulk population, it is critical to consider the \textit{confinement of energetic particles}, such as alpha particles produced through a nuclear fusion reaction (Section \ref{sec:fusion_reactions}). In order for the fusion reactions to be self-sustaining, alpha particles must be confined long enough to deposit their energy in the bulk population. Losses of energetic particles concentrated in a region of space can lead to damage of in-vessel experimental components and must be avoided. The confinement of energetic particles can be studied by following single-particle trajectories in the presence of collisions with other charged particles \cite{Drevlak2009,Mcmillan2014}. Collisional diffusion and deflection are minimal at energies near the birth energy of $3.5$ MeV for a D-T reaction (Chapter 3 in \cite{Helander2005}), so collisionless guiding center orbits are an informative metric of energetic particle confinement. If the collision frequency is small enough that energetic ions can complete their bounce or transit orbits, then the parallel adiabatic invariant \eqref{eq:adiabatic_invariant}
is a conserved quantity. Thus one method of improving energetic particle confinement under the collisionless approximation is by targeting omnigeneity. There is evidence that targeting quasisymmetry near the half-radius may also improve energetic particle confinement \cite{Henneberg2019b}. Equilibria have been numerically designed with good confinement of energetic particles \cite{Drevlak2018,Bader2019} by considering proxy functions related to the variation of the magnetic field strength along a field line  \cite{Drevlak2014,Nemov2008}. 

\subsubsection{Stability}
\label{sec:stability}

The robustness of a plasma configuration to perturbations is referred to as \textit{stability}: perturbations which are damped in time are referred to as stable, while those which grow are referred to as unstable. The stability of magnetically confined fusion plasmas to large-scale perturbations is predicated well by MHD theory, see for instance Chapter 2 in \cite{Freidberg2008}. Given suitable conditions, an MHD unstable plasma will tend to evolve towards a lower energy state. This process can result in significant expulsion of energy from the plasma, lowering confinement, potentially physically damaging device components, and limiting the space of possible operating regimes. Energetic particles can also drive instabilities \cite{Spong2011,Cooper2006}, potentially resulting in enhanced particle losses. Common proxies for MHD stability include the magnetic well parameter \cite{Greene1997} and the Mercier criterion \cite{Mercier1974}. These criteria have been widely applied in stellarator optimization \cite{Yokoyama1997,Sanchez2001,Castejon2012,Henneberg2019}. One can also try to increase magnetic shear, the radial derivative of the rotational transform $\iota'(\psi)$, to improve large toroidal mode number $n$ ballooning stability and Mercier stability \cite{Hegna1998}. There is some experimental evidence that stellarator configurations may be able to operate above the linear MHD stability beta threshold \cite{Weller2006} rather than being terminated by a disruption. The Large Helical Device (LHD) has operated up to a volume-averaged $\beta$ of $5\%$ without any disruptive MHD phenomena, though the heat transport increases due to low-$n$ mode activity \cite{Sakakibara2008}. However, it is desirable to design a stellarator with an increased beta limit to reduce enhanced transport caused by MHD modes.



\subsubsection{Summary}

\begin{table}[]
    \centering
    \begin{tabular}{|c|c|c|}
    \hline 
     \textbf{Physics parameter} & \textbf{General consideration} \\ \hline
        Island width & Equilibrium properties \\ 
      Shafranov shift & Equilibrium properties \\
      Low shear & Equilibrium properties \\
      High shear & Equilibrium properties, stability \\
     Bootstrap current &  Equilibrium properties, stability \\
     Pfirsch-Schluter current &  Equilibrium properties, stability \\
     Geodesic curvature & Equilibrium properties, stability, collisionless particle confinement \\
     Mercier criterion & Stability \\
     Magnetic well & Stability \\
     Quasisymmetry & Collisionless particle confinement \\
     Omnigeneity & Collisionless particle confinement \\
    Effective ripple ($\epsilon_{\text{eff}}^{3/2}$) & Neoclassical particle confinement \\ \hline
    \end{tabular}
    \caption{Summary of common stellarator equilibrium objectives.}
    \label{tab:equilibrium_objectives}
\end{table}

In Table \ref{tab:equilibrium_objectives}, we summarize common stellarator optimization objectives used and their purposes. In practice, additional engineering constraints are also introduced into the design process. Overly-complex coils are costly and may interfere with maintenance and diagnostics of the experiment. The tolerances to which coils must be manufactured and positioned are set by the allowable deviations of physics objectives. Tight tolerances are a large driver of costs for the stellarator program and cannot be ignored in the design process. We reserve discussions related to coils for Section \ref{sec:coil_opt}. 

\subsection{Fixed-boundary MHD optimization}\label{sec:fixed_optimization}
In the fixed-boundary approach to compute an MHD equilibrium on a given domain $\Omega$, see Section \ref{sec:3d_mhd_surf}, the boundary $\partial \Omega$ of the computational domain is the plasma boundary. The MHD equilibrium equations are repeated here for convenience,
\begin{subequations}
\begin{align}
\left.\begin{array}{l}
   \left(\nabla \times \bm{B}\right) \times \bm{B} = \mu_0 \nabla p \\
    \nabla \cdot \bm{B} = 0 \\
    \end{array} \right\} \hspace{1cm} &\text{ in }\Omega \\
    \bm{B} \cdot \hat{\bm{n}} = 0 \hspace{1cm} &\text{ on }\partial \Omega.
\end{align}
\label{eq:equilibrium_fixed}
\end{subequations}
Solutions are computed assuming that a continuously nested set of magnetic surfaces exist, labeled by the toroidal flux $2\pi\psi$ (see Section \ref{sec:flux_function}). The pressure, $p(\psi)$, and one other function of flux, such as the rotational transform $\iota(\psi)$ or enclosed toroidal current $I_T(\psi)$, are prescribed.
We note that many models of 3D equilibria exist (see Section \ref{sec:equilibrium_fields}); however, this model has historically been the standard choice for fixed-boundary optimization using codes such as VMEC \cite{Hirshman1983}, NSTAB \cite{Garabedian2002}, or BETA \cite{Bauer1987}. This is partly for computational efficiency, as the variational principle (see Section \ref{sec:variational_principle}) can be employed. The assumption of continuously nested flux surfaces is also convenient as many physical quantities of interest are based on models which require magnetic surfaces, such as the Boozer coordinate transformation (see Section \ref{sec:boozer_coordinates}).

In this context, the magnetic geometry throughout the volume $\Omega$ is determined by the shape of $\partial \Omega$. As many physical parameters depend on the magnetic geometry, it is thus reasonable to search for the optimal shape of $\partial \Omega$ with respect to a set of physics objectives. This is the fixed-boundary optimization approach, first introduced by N\"{u}hrenburg and Zille \cite{Nuhrenberg1988}. 

The plasma boundary, $\partial \Omega$, is a toroidal surface and can be parameterized by a poloidal angle $\theta$ and toroidal angle $\phi$. The toroidal angle is often chosen to coincide with the cylindrical angle $\phi$. In cylindrical coordinates the plasma boundary can be described by the radius $R(\theta,\phi)$ and the height $Z(\theta,\phi)$. These coordinates can be expressed in Fourier series. In particular, under the assumption of stellarator symmetry, see Section \ref{sec:stellarator_symmetry}, $R(\theta,\phi)$ is even in $(\theta,\phi)$ while $Z(\theta,\phi)$ is odd in $(\theta,\phi)$. Since they are also real-valued, they can then be expressed compactly in terms of real Fourier series as,
\begin{subequations}
\begin{align}
    R(\theta,\phi) &= \sum_{(m,n)\in \mathcal D} R_{m,n} \cos(m \theta - n \phi) \\
    Z(\theta,\phi) &= \sum_{(m,n)\in \mathcal D} Z_{m,n} \sin(m \theta - n \phi),
\end{align}
\label{eq:RZsurf}
\end{subequations}
where $\mathcal D\subset \mathbb Z^2$ is the set of $(m,n)$ pairs such that $n \geq 0$ only for $m = 0$ otherwise $n\in\mathbb Z$ for $m\geq 0$; that is $\mathcal D:=(\mathbb N^*\times \mathbb Z )\cup (\{0\}\times \mathbb N )$ by combination of the real-valued and parity properties of $R$ and $Z$. Equivalently we could choose $\tilde {\mathcal D}:=(\mathbb Z\times \mathbb N^* )\cup  (\mathbb N\times \{ 0 \} )$, however the previous choice is more common.
So the set of parameters $\left\{(R_{m,n},Z_{m,n})\in\mathbb R^2, (m,n)\in(\mathbb N^*\times \mathbb Z) \cup (\{0\}\times \mathbb N)\right\}$ uniquely defines the shape of $\partial\Omega$. In practice, a finite number of these modes are retained. Assuming that we  choose maximum poloidal and toroidal mode numbers, $M$ and $N$ respectively, the corresponding $N+1+M(2N+1)$ parameters $\left\{(R_{m,n},Z_{m,n})\in\mathbb R^2, (m,n)\in(\mathbb N^*\times \mathbb Z) \cup (\{0\}\times \mathbb N), m\leq M, |n|\leq N\right\}$ are the output of the fixed-boundary MHD optimization. This is simply one choice of parameterization for toroidal surfaces, used in the widely-used VMEC code. Another possible choice, used in the NSATB code, is  the Garabedian representation \cite{Garabedian1970,Garabedian2008}, representing the outer surface as in \eqref{eq:RZsurf} with the additional assumption that $R_{m,n}=Z_{m,n}$. 

Recall that the calculation of the MHD equilibrium also requires the prescription of two free functions of flux, such as the pressure, $p(\psi)$, and the toroidal current enclosed by a flux surface, $I_T(\psi)$. These quantities can be prescribed and held fixed throughout the optimization. This is generally the method used for choosing these profiles.
Alternatively, the current can be considered an output of the calculation, determined by self-consistent modeling of the bootstrap current, the parallel current predicted by neoclassical theory, throughout the optimization \cite{Hirshman1999,Strickler2004}. In this approach, at each step in the optimization, the equilibrium is computed with an iterative method: (1) an MHD equilibrium with a prescribed $I_T(\psi)$ is computed according to \eqref{eq:equilibrium_fixed}, and (2) the resulting bootstrap current $I_T^{\text{boots}}(\psi)$ is computed from the equilibrium field resulting from (1). If the two currents do not match, then $I_T(\psi)$ is updated to $I_T^{\text{boots}}(\psi)$ for the next iteration. Steps (1) and (2) are iterated until $I_T(\psi)$ and $I_T^{\text{boots}}(\psi)$ match.

The fixed-boundary optimization problem can be expressed as
\begin{align}
        \min_{R_{m,n},Z_{m,n}} \chi^2,
\end{align}
where $\chi^2$ is an objective function, often taken to be the sum of squares,
\begin{align}
        \chi^2 = \sum_i \frac{(f_i^{\text{target}}-f_i^{\text{equilibrium}})^2}{\sigma_i^2},
        \label{eq:chi2_fixedboundary}
\end{align}
    where $f_i^{\text{target}}$ is the target value of the physics objective and $\sigma_i$ is a scaling factor which sets the relative importance of each of the targets. In general, each of the target functions depends on the optimization parameters, $\{R_{m,n},Z_{m,n}\}$, through the solution of the MHD equilibrium equations \eqref{eq:equilibrium_fixed}.
    The choice of optimization algorithm will determine how the optimization parameters are adjusted to arrive at a local or global minimum within a given tolerance.
    
    Considering that are given:
    \begin{itemize}
        \item two flux functions, $p(\psi)$ and $I_T(\psi)$,
        \item an MHD equilibrium code solving \eqref{eq:equilibrium_fixed},
        \item fixed numbers $N+1+M(2N+1)$  of parameters $R_{m,n},Z_{m,n}$ to describe the boundary,
        \item an optimization algorithm,
        \item values defining the objective function \eqref{eq:chi2_fixedboundary} $\left\{f_i^{\text{target}},\sigma_i\right\}_i$,
    \end{itemize}
    the complete optimization process proceeds as follows:
\begin{enumerate}
    \item Find an initial boundary shape $\partial \Omega^{\text{init}}$. 
    \begin{enumerate}
        \item Compute the initial MHD equilibrium  magnetic field in $\Omega$ from $\partial \Omega^{\text{init}}$, $p(\psi)$, and $I_T(\psi)$.
        \item Evaluate the set of objectives $\left\{f_i^{\text{equilibrium}}\right\}_i$ and $\chi^2$ from the initial magnetic field. 
        \end{enumerate}
    \item Until $\chi^2$ satisfies a stopping criterion, repeat the following steps.
    \begin{enumerate}
        \item Adjust $\partial \Omega$ according to the optimization algorithm.
        \item Compute the MHD equilibrium magnetic field in $\Omega$ from $\partial \Omega$, $p(\psi)$ and $I_T(\psi)$.
        \item Evaluate the set of objectives $\left\{f_i^{\text{equilibrium}}\right\}_i$ and $\chi^2$ from the resulting magnetic field.
    \end{enumerate}
\end{enumerate}

This scheme is implemented in the STELLOPT \cite{Spong2001} and ROSE \cite{Drevlak2018} optimization codes based on VMEC equilibrium calculations. Several optimization algorithms have been applied to the fixed-boundary approach \cite{Spong1998,Mynick2002,Drevlak2018}, including gradient-based methods such as the Levenberg-Marquardt algorithm \cite{More1978} and gradient-free methods such as the Brent algorithm \cite{Brent2013}, particle swarm \cite{Parsopoulos2002}, and differential evolution \cite{Storn1997}.

The first demonstration of this approach by N\"{u}hrenberg and Zille \cite{Nuhrenberg1988} based on the BETA equilibrium code resulted in the Helias class of stellarators. The boundary was optimized to obtain a quasisymmetric magnetic field, in addition to considerations of the rotational transform, magnetic well (a proxy for MHD stability), and bootstrap current. The W7-X stellarator was later designed based on further optimization of one of the Helias configurations \cite{Beidler1990}. Several other examples will be given in Section \ref{sec:device_examples}.

Once the fixed-boundary optimization is complete, a set of electromagnetic coils which are consistent with $\partial \Omega$ must be realized. This is generally a non-trivial task and will be discussed in Section \ref{sec:coil_opt}. In this way, the fixed-boundary approach delays any consideration of engineering constraints related to coil construction. For this reason, the fixed-boundary approach may lead to improved physics properties as engineering considerations can be initially ignored.  It also allows for a wider coil design space, allowing for multiple coil types and topologies. However, it is possible that the resulting configuration requires overly-complex coils that cannot ultimately be constructed. While MHD equilibrium calculations can also be performed with prescribed electromagnetic coils rather than an outer boundary, the fixed-boundary equilibrium calculations tend to be less computationally intensive than free-boundary calculations; thus the fixed-boundary optimization approach has prevailed historically.

\subsection{Coil optimization}\label{sec:coil_opt}

In the traditional optimization approach, electromagnetic coils are optimized as a second step, given a target outer boundary of the plasma, $\partial \Omega$, obtained from the first step. Coils are a set of electro-magnets, or currents, outside the confinement region that produce the necessary confining magnetic field. It is important to keep in mind that this magnetic field, in practice, is  numerically computed from an equilibrium model.

The goal here is thus to find a set of currents, $\bm{J}^{\text{coil}}$, in the vacuum regions such that the total magnetic field is tangent to $\partial \Omega$,
\begin{align}   \label{eq:BCsurf}
\bm{B} \cdot \hat{\bm{n}} \rvert_{\partial \Omega} = 0.
\end{align}
Sometimes coil optimization is instead performed given a prescribed normal field on a reference surface, but for simplicity we will assume that $\partial \Omega$ is a magnetic surface. The following discussion would still hold if we replace \eqref{eq:BCsurf} by a boundary condition imposing a non trivial value of $\bm{B} \cdot \hat{\bm{n}} \rvert_{\partial \Omega}$ on a reference surface.
The total magnetic field consists of the magnetic field due to the plasma current, $\bm{B}^{\text{plasma}}$, and that due to coil current, $\bm{B}^{\text{coil}}$. The magnetic field due to the coils can be computed from the Biot-Savart law \eqref{eq:biot_savart}, while the magnetic field due to the plasma current is an input for this step. We restrict our interest to coils that can be feasibly constructed, thanks to the two following assumptions. We will assume that there is a distance $d_c>0$ between the boundary of the plasma and the coils. This region between the plasma boundary and the coils is needed to allow for several experimental components, such as the first wall and vacuum vessel (see Section \ref{sec:reformulation}). In a reactor, the minimum coil-plasma distance is set by the requirement of having a sufficiently thick blanket for absorption of neutrons and tritium breeding \cite{Ku2008}. We will also assume that the coils are contained in a bounded volume. These two assumptions are summarized by the definition of a volume $\Omega_c$ outside of which the current $\bm{J}^{\text{coil}}$ vanishes. 
Therefore the coil design problem is expressed as: for a fixed plasma volume $\Omega$, given $\bm{B}^{\text{plasma}}(\bm{r})$ on the plasma boundary $\partial\Omega$, find $\bm{J}^{\text{coil}}(\bm{r})$ for all $\bm{r} \in \Omega_c\subset\mathbb{R}^3 \backslash \Omega$ such that
\begin{align}
\bm{B}^{\text{plasma}}(\bm{r}) \cdot \hat{\bm{n}}(\bm{r}) = -\frac{\mu_0}{4\pi} \int_{
\Omega_c 
} \frac{\bm{J}^{\text{coil}}(\bm{r}') \times (\bm{r}-\bm{r}') \cdot \hat{\bm{n}}(\bm{r})}{\rvert{\bm{r}-\bm{r}'\rvert^3}}  \, d\bm{r}',
\label{eq:coil_design}
\end{align}
for all $\bm{r} \in \partial \Omega$. 

Problems of the form \eqref{eq:coil_design} are generally ill-posed, as is discussed in Section \ref{sec:ill_posed}. We then discuss several regularization methods for such problems. As we will see in \ref{sec:regularization}, a common approach is a reformulation as an optimization problem. 
The material in these two sections is independent of the following ones and is intended to provide a theoretical view of the mathematical problem expressed in \eqref{eq:coil_design}.

The problem as stated in \eqref{eq:coil_design} seeks for the unknown $\bm{J}^{\text{coil}}$ with a support in $\Omega_c\subset\mathbb R^3\backslash \Omega$. 
However, from the experimental point of view, the currents will be supported by coils. 
Such stellarator coils have an actual volume, referred to as their finite build, consisting of several layers, each with several turns of the conducting material. As a first approximation, the finite build of coils is often not taken into account during the design process: the support of the unknown function $\bm{J}^{\text{coil}}$ is limited to having no volume. 
There are two most common ways to model such coils: assuming that all currents in the region $\Omega_c$ are restricted to a toroidal surface, as discussed in Section \ref{sec:winding_surface}, or assuming that the currents are restricted to a finite number of filamentary lines, as discussed in Section \ref{sec:filamentary}. In the latter case, the curves will define the shapes of the physical coils, while in the former case the coil shapes will be chosen in a post-processing step following some streamlines of the computed current.

\subsubsection{Ill-posedness}
\label{sec:ill_posed}
This section is dedicated to showing that the coil design problem is ill-posed in the following sense. In the mathematical community, a problem is said to be well-posed in the sense of Haddamard \cite{Hadamard1932} if:  
\begin{enumerate}
    \item it has a unique solution;
    \item small changes in prescribed data result in small changes in the solution.
\end{enumerate}
Otherwise, the problem is said to be ill-posed.  

The coil design problem \eqref{eq:coil_design} is in the form of a Fredholm integral equation of the first kind. Given functions $K$ and $g$, such an equation has the general form:
\begin{equation}
\begin{array}{l}
\displaystyle
\text{ Find } f \text{ such that } 
    g(t) = \int K(t,s) f(s) \, ds.
\end{array}
\label{eq:FirstKindIntegral}
\end{equation}
The function $K$ is called the kernel of the integral term. Each of the functions $f$ and $g$ can be defined in any dimension $d\in\mathbb N$, their domains of definition can be bounded or not, and they may be vector valued.
We will now infer that integral equations of the first kind like \eqref{eq:coil_design} are ill-posed.

For instance the coil design problem \eqref{eq:coil_design} involves a vector-valued unknown $\bm{J}^{\text{coil}}$, the integral is taken over a three-dimensional bounded set $\Omega_c$, and the kernel is vector-valued. More details on the domain of definition  of the unknown $\bm{J}^{\text{coil}}$ will be subject to discussion in following Sections.

The theory of integral operators and integral equations has been widely developed in the literature,  
and standard applications include scattering theory  and inverse problems \cite{CKscat} or linear flow problems \cite{BdryIntFlows}.
The discussion in this Section is based on \cite{Hansen2005,Kress1989,Wing1991}. 
%
Here, to present the mathematical challenges associated with Fredholm integral equations of the first kind, we will consider the following setting:
\begin{itemize}
    \item two domains, $\Omega_1 \subset\mathbb R^{d_1}$ on which the equation holds, and the domain of integration $\Omega_2\subset\mathbb R^{d_2}$, not necessarily bounded;
%
    \item two real Hilbert spaces, $\mathcal{H}_1=L^2(\Omega_1,\mathbb R^{n_1})
    $ 
    and $\mathcal{H}_2=L^2(\Omega_2,\mathbb R^{n_2})
    $, 
    the Lebesgue spaces of functions $G:\Omega_i\rightarrow \mathbb R^{n_i}
    $ such that $\int_{\Omega_{i}}
    G(t)\cdot G(t)\ dt<\infty$, equipped with the standard inner product
\begin{align}
    \langle G,H \rangle &= \int_{\Omega_{i}} G(t) \cdot H(t) \, dt  \quad \text{ for all } G,H\in\mathcal H_i \text{ for } i=1,2,
\end{align} 
    these are also called spaces of measurable square integrable functions, either scalar-valued (if $n_i=1$) or vector-valued (if $n_i>1$);
%
    \item 
    square integrable kernels $K:\Omega_1\times\Omega_2\rightarrow \mathbb R^{n_1\times n_2}$ in the following sense
 \begin{align}
 \int \int_{\Omega_1\times\Omega_2} \left\|K(t,s)\right\|^2_{\mathbb R^{n_1\times n_2}} \, dt ds < \infty,
 \label{eq:kernel_conditions}
 \end{align}
 where $\|\cdot\|_{\mathbb R^{n_1\times n_2}}$ is the matrix norm induced by the Euclidean vector norms, that is to say $\|A\|_{\mathbb R^{n_1\times n_2}}=\sup\{ \|Ax\|_{\mathbb R^{n_1}}: x\in\mathbb R^{n_2}, \|x\|_{\mathbb R^{n_2}}=1\}$.
\end{itemize}
For instance for the coil design problem \eqref{eq:coil_design}, we see that the data is prescribed on the surface defining the plasma boundary, $\Omega_1 = \partial \Omega\in\mathbb R^3$, so $d_1=3$, the volume outside of which the current vanishes is $\Omega_2 = \Omega_c\in\mathbb R^3$, so $d_2=3$,  the data is scalar-valued, $n_1=1$, the unknown $\bm J^{\text{coil}}$ is vector-valued, $n_2=3$, and the kernel is square integrable as it is  bounded on the bounded domain $\partial \Omega\times \Omega_c$.

For all the following arguments we will consider $K$ to be square integrable. Given $f\in\mathcal H_2$, the function defined on $\Omega_1$ by
\begin{equation}
 t\mapsto \int_{\Omega_2}K(t,s) f(s) \, ds
\end{equation}
is an element of $\mathcal H_1$, since as $K$ and $f$ are both square integrable
\begin{equation}
\begin{array}{l}
\displaystyle
    \int_{\Omega_1} \left(\int_{\Omega_2} K(t,s) f(s) \, ds \right)\cdot \left(\int_{\Omega_2} K(t,s) f(s) \, ds \right) \, dt
    \\
\displaystyle
    \le \int_{\Omega_1} \left(\int_{\Omega_2} \left\|K(t,s)\right\|_{\mathbb R^{n_1\times n_2}}^2 ds \right)\left(\int_{\Omega_2} f(s)\cdot f(s) ds \right) \, dt < \infty.
\end{array}
\end{equation}
Hence, we can define a linear operator $F_K:\mathcal H_2 \rightarrow \mathcal H_1$ as the mapping
\begin{align}
f \mapsto \int_{\Omega_2}K(\cdot,s) f(s) \, ds.
    \label{eq:integral_operator}
\end{align}

The problem at stake can then be posed abstractly as
\begin{equation}\label{PB:LiCopState}
\begin{array}{l}
   \text{Assume } g\in \mathcal H_1 \text{ and } K \text{ are given.}\\
   \text{Find }
   f\in\mathcal H_2 \text{ such that } F_K[f]=g.
   \end{array}
\end{equation}

Square integrable kernels give rise to integral operators belonging to a class known as compact operators. Compact operators share many properties with operators in finite-dimensional spaces however, due to the infinite dimensionality of the spaces involved, they are very sensitive to variations in problem data. In a metric space, a compact operator can be characterized by the fact that the image of any bounded sequence by the operator contains a converging subsequence. See for instance Chapter 22 in \cite{lax2002functional}. 
The key property leading to the ill-posedness of 
the coil design problem \eqref{eq:coil_design} 
is that the associated operator $F_K$ is compact. 
 

The operator $F_K$  has an adjoint operator, $F_K^*$, defined by,
\begin{align}
\langle F_K[G],H \rangle = \langle G,F_K^*[H] \rangle \quad \text{ for all } (G,H)\in\mathcal H_2\times\mathcal H_1,
\end{align}
which is also compact. The composition $F_K^*F_K$ is self-adjoint, since
\begin{align}
\text{ for all } G,H\in\mathcal H_2,  \ 
\langle F_K^*F_K[G],H \rangle 
= \langle F_K[G],F_K[H] \rangle
= \langle G,F_K^*F_K[H] \rangle.
\end{align}
 The set of eigenvalues of a self-adjoint operator is countable. For $F_K^*F_K$ they will  be denoted $\{ \alpha_i\}_{i\in\mathbb N}$, repeated according to their multiplicity  while the set of associated eigenfunctions forms a basis of $\mathcal H_2$. It can be shown that $\{ \alpha_i\}_{i\in\mathbb N}$ are non-negative and they either form a finite set or a countably infinite set with $\alpha_i \rightarrow 0$ \cite{Kress1989}. Therefore, we can number the eigenvalues $\{\alpha_i\}$ in decreasing order such that $\alpha_1 \ge \alpha_2 \ge ... \ge 0$. We define $N_F$ to be the largest integer such that $\alpha_{N_F} > 0$ and $\alpha_{i}=0$ for all $i > N_F$ if there is a finite number of eigenvalues, or $N_F=\infty$ if there is a countable infinite set of eigenvalues. 
 
The \textit{singular values} of $F_K$ are then defined as $\mu_i :=  \sqrt{\alpha_i} $ for $i$ from $1$ to $N_F$ while the associated \textit{singular functions} are defined for $i$ from $1$ to $N_F$ by,
\begin{subequations}
\begin{align}
    u_i &= \varphi_i \in\mathcal H_2\\
    v_i &= \frac{F_K[u_i]}{\mu_i}\in\mathcal H_1,
\end{align}
\end{subequations}
which satisfy for $i$ from $1$ to $N_F$,
\begin{subequations}
\begin{align}
    F_K[u_i] &= \mu_i v_i \\
    F_K^*[v_i] &= \mu_i u_i.
\end{align}
\label{eq:singular_functions}
\end{subequations}
Furthermore, from the orthonormality of the $\{\varphi_i\}$ basis, the singular functions satisfy the following orthonormality conditions,
\begin{subequations}
\begin{align}
    \langle  u_i, u_j \rangle &= \delta_{ij} \\
    \langle v_i, v_j \rangle &= \delta_{ij}.
\end{align}
\label{eq:othonormality_svd}
\end{subequations}
Finally, if $N_F<\infty$, we can conveniently define a basis of the null space of $F_K$ by $\{u_i=\varphi_i, i>N_F\}$, and let $\{v_i, i>N_F\}$ be any orthonormal basis of $\mathcal H_1 \backslash \ \text{Span} \{v_i, i\leq N_F \}$.  As a result, independently of the value of $N_F$, the sets $\{u_i, i\in\mathbb N\}$ and $\{v_i, i\in\mathbb N\}$ form orthonormal bases of  $\mathcal H_2$ and $\mathcal H_1$. We can note that \eqref{eq:singular_functions} still holds for $\{u_i\}$ and $\{v_i\}$ with $i \in \mathbb{N}$,
\begin{subequations}
\begin{align}
F_K[u_i] &= 0 = \mu_i v_i \\
    F_K^*[v_i] &= \sum_{j=1}^{\infty} u_j \langle F_K^*[v_i],u_j\rangle =\sum_{j=1}^{N_F} u_j \langle v_i,F_K[u_j] \rangle = \mu_i u_i.
\end{align}
\end{subequations}

From \eqref{eq:singular_functions} and \eqref{eq:othonormality_svd}, the image of the operator $F_K$ can then be described in terms of singular functions as
\begin{align}
\label{eq:RangeF}
\text{ for all } f\in\mathcal H_2,\ 
    F_K[f] &= \sum_{i=1}^{\infty} v_i \langle F_K[f], v_i \rangle =\sum_{i=1}^{\infty} v_i \mu_i \langle f, u_i \rangle
    = \sum_{i=1}^{N_F} v_i \mu_i \langle f, u_i \rangle.
\end{align}
We can now turn back to the problem \eqref{PB:LiCopState}, 
which can now be expressed as follows.
    For a given square integrable kernel $K$, there exists an associated SVD of the corresponding compact integral operator $F_K$, providing singular values and functions $\{(\mu_j,u_j, v_j), j\in\mathbb N \}$;
    any $g\in \mathcal H_1$ is uniquely defined by its coordinates in the $\{v_i\}_{i}$ basis: $\displaystyle g = \sum_{i=1}^\infty \langle v_i,g \rangle v_i$;
so the problem reads
\begin{equation}\label{PB:LiCop}
\begin{array}{l}
\displaystyle
   \text{Assume } g = \sum_{i=1}^\infty \langle v_i,g \rangle v_i \in \mathcal H_1 \text{ is given.}\\
   \displaystyle
   \text{ Find }
   f\in\mathcal H_2 \text{ such that } \sum_{i=1}^{N_F} v_i \mu_i \langle f, u_i \rangle=g.
   \end{array}
\end{equation} 

We first consider the case in which $N_F < + \infty$. In this case, problem \eqref{PB:LiCop}  has no unique solution, as the linear operator \eqref{eq:integral_operator} has a non-trivial null space. To demonstrate this, suppose that $\hat{f}(s)$ satisfies \eqref{PB:LiCop}. Then for any $n > N_F$, $f := \hat{f} + u_n$ will also be a solution, since $u_n$ is in the null space of $F_K$. Thus the problem is not well posed.

In this case, the image of $\mathcal H_2$ by the operator $F_K$ is included in $\mathcal H_1$, but here it is  smaller than $\mathcal H$. From \eqref{eq:RangeF} we see that $F_K[\mathcal H_2] = \text{Span}\{ v_i, 1\leq i\leq N_F \}$, so $F_K[\mathcal H_2]\neq \mathcal H_1$. As a result, although a solution may not exist for all function $g$ in $\mathcal H_1$, there exists a solution as long as $g$ belongs to the image $F_K[\mathcal H_2]$, 
$$
g\in F_K[\mathcal H_2]
\Leftrightarrow
\langle v_i,g \rangle = 0 \, \text{ for all } i > N_F.
$$
Under this condition, then $g$ can be expressed in terms of $v_i$ singular functions as,
\begin{align}
 g &= \sum_{i=1}^{N_F} \langle v_i,g \rangle v_i,
\end{align}
and one particular (non-unique) solution to \eqref{PB:LiCop} can be written as,
\begin{align}
    f = \sum_{i=1}^{N_F} \frac{u_i \langle v_i,g \rangle}{\mu_i},
    \label{eq:solution_non_unique}
\end{align}
while any solution is given, for any set of coefficients $\{\alpha_i\in\mathbb R, i\geq N+1\}$, by
\begin{align}
    f = \sum_{i=1}^{N_F} \frac{u_i \langle v_i,g \rangle}{\mu_i}
    +\sum_{i=N_F+1}^{\infty}
    \alpha_i \varphi_i(s).
\end{align}



We now consider the case  $N_F = \infty$, so that $\mu_i > 0$ for all $i$ with $\mu_i \rightarrow 0$ as $i \rightarrow \infty$. 
Then $g\in\mathcal H_1$ can be expressed in the singular function basis as
\begin{align}
 g &= \sum_{i=1}^{\infty} \langle v_i,g \rangle v_i.
\end{align}
So, from \eqref{eq:RangeF}, if a function $f\in \text{Span}\{u_i, i\in\mathbb N\}$ is a solution to \eqref{PB:LiCop}, it necessarily reads,
\begin{align}
    f = \sum_{i=1}^{\infty} \frac{\langle v_i,g \rangle}{\mu_i}u_i.
    \label{eq:f_N_infty}
\end{align}
Thus for a square-integrable solution to exist, we must have,
\begin{align}
    \langle f,f \rangle<\infty \Leftrightarrow  \sum_{i = 1}^{\infty} \left( \frac{\langle v_i,g \rangle}{\mu_i} \right)^2 < \infty.
    \label{eq:condition_g}
\end{align}
The above condition effectively sets a constraint on the allowable $g\in\mathcal H_1$ that can be prescribed such that \eqref{PB:LiCop} has a solution in $\mathcal{H}_2$. However, under this condition the solution is unique. Note that the constraint on $g$ is stronger than that required for square-integrability, namely
\begin{align}
    \langle g, g \rangle <\infty \Leftrightarrow \sum_{i = 1}^{\infty} \langle v_i,g \rangle^2 < \infty,
\end{align}
since the $\mu_i \rightarrow 0$. Therefore, we see that again the image $F_K[\mathcal{H}_2]$ is a subspace of $\mathcal{H}_1$ that is smaller than $\mathcal{H}_1$. Thus the problem as stated in \eqref{PB:LiCop} is ill-posed, as a solution does not exist unless additional constraints are placed on the prescribed data $K$ and $g$.  

We have now considered both cases of $N_F$ finite and $N_F = + \infty$. In both cases, we find that the problem stated as \eqref{PB:LiCop} does not have a solution unless an additional constraint is placed on $g$. Thus the general problem \eqref{PB:LiCopState} is ill-posed, as a square-integrable solution does not exist for every $g \in \mathcal{H}_1$. In other words, the compact linear operator $F_K$ defined on the infinite-dimensional Hilbert space $\mathcal H_2$ does not have a continuous inverse.


Next, suppose that we consider $g$ such that \eqref{eq:condition_g} is satisfied and a square-integrable solution exists. Now we consider the second condition for a well-posed problem by evaluating the result of perturbation in the prescribed data. Consider a perturbation of the data by $g \rightarrow g + \delta g$. For the coils problem, this could come from any numerical error in the computed normal magnetic field from the equilibrium. In particular, we consider a perturbation $\delta g = \Delta v_{i_0}$ for a scalar $\Delta$ and particular singular vector $v_{i_0}$. This will result in a perturbation of the solution by,
\begin{align}
    \delta f = \frac{u_{i_0} \Delta }{\mu_{i_0}},
\end{align}
which follows from \eqref{eq:f_N_infty}. The condition for square-integrability of $\delta f$ now becomes,
\begin{align}
   \sum_{i=1}^\infty \frac{\langle v_i,\delta g \rangle^2}{\mu_i^2} = \frac{\Delta^2}{\mu_{i_0}^2} < \infty.
\end{align}
The above condition does not hold for an arbitrary perturbation of the data, as the $\mu_i \rightarrow 0$ as $i \rightarrow \infty$, and a finite perturbation $\delta g$ can be made in the direction $v_{i_0}$ for $i_0 \rightarrow \infty$. Thus the perturbation of the data may result in an arbitrarily large perturbation of the solution, and the problem is generally ill-posed. In this way, solving an integral equation of the first kind tends to amplify noise. 

Despite the possible ill-posedness of the problem, there is considerable interest in methods for a stable numerical approximation of a solution, when a solution exists. In general, a numerical approximation of a given equation may be viewed as the solution to the same equation with perturbed data, so the ill-posed nature of our problem has consequences for the numerical approximation of a solution. We will now turn to classical methods for the numerical treatment of integral equations of the first kind.

\subsubsection{Regularization}
\label{sec:regularization}

In order to be able to compute a numerical approximation to a solution of an ill-posed problem, one first needs to formulate a well-posed problem to be solved approximately by a numerical method. In the context of ill-posed problems, regularization methods refer to stable numerical methods to compute an approximate solution to the initial problem. Such methods usually rely on a modified well-posed problem, the so-called {\it regularized} problem, which unlike the original problem is well-posed, so it can be solved thanks to a stable numerical method.
This process introduces an error due to the approximation of the problem, and accuracy has to be balanced with stability. We limit our discussion here to two standard regularization methods, while more details can be found in \cite{Kress1989}.

One regularization method is based on a truncated singular value decomposition. Given the singular vectors $\{v_i\}$ and $\{u_i\}$ and singular values $\{\mu_i\}$ of the integral operator $F_K$, for a parameter $k$, the regularized problem then reads
\begin{equation}\label{PB:LiCop_reg}
\begin{array}{l}
\displaystyle
   \text{Given } g_k = \sum_{i=1}^k \langle v_i,g \rangle v_i \in \mathcal H_1\\
   \displaystyle
   \text{ find }
   f_k\in\mathcal H_2 \text{ such that } \sum_{i=1}^k v_i \mu_i \langle f_k, u_i \rangle=g_k.
   \end{array}
\end{equation} 
This naturally has a unique solution given by
\begin{align}
\label{solSVD}
    f_k = \sum_{i=1}^k \frac{u_i \langle g, v_i \rangle}{\mu_i}.
\end{align}
By construction, $f_k$ is square-integrable as it is a finite sum: the effect of the small singular values as $i \rightarrow \infty$ are not retained in comparison with the solution of the ill-posed problem \eqref{eq:f_N_infty}. 
The application of this truncation also improves stability of the solution with respect to noise in the prescribed data. Indeed, consider a perturbation of the data $g \rightarrow g + \delta g$, resulting in a perturbation of the solution,
\begin{align}
    \delta f_k = \sum_{i = 1}^k \frac{u_i}{\mu_i} \langle \delta g, v_i \rangle.
    \label{eq:solution_perturbation_tsvd}
\end{align}
Thanks to the truncation, the noise associated with large $i$ is removed; thus $\delta f_k$ is square-integrable. If $\langle g, v_i \rangle/\mu_i$ decays sufficiently fast as $i \rightarrow \infty$, then $f_k$ will yield a good approximation for $f$ as $k\rightarrow \infty$.  In this way, the regularized problem \eqref{PB:LiCop_reg} is well-posed according to the definition presented in Section \ref{sec:ill_posed}.

This then provides a natural approach to construct a numerical method which approximates a solution of \eqref{PB:LiCopState}, by solving numerically the well-posed regularized problem \eqref{PB:LiCop_reg}. It is however important to understand the role of the regularization parameter $k$.  For the regularized solution \eqref{solSVD}, we can compute the residual $R[f_k]:= g- F_K[f_k] $ for the initial problem, 
 \begin{align}
     R[f_k]= \sum_{i={k+1}}^{\infty} v_i \langle v_i,g \rangle.
 \end{align}
As the regularization parameter $k$ goes to infinity, the residual $R[f_k]$ goes to zero since it contains fewer non-zero components,
while the  numerical stability decreases. Indeed, although $\delta f_k$ will remain square-integrable as long as $k$ is finite, a given perturbation of the data, $\delta g$, in the direction of one of the singular vectors, $v_i$, will result in a larger perturbation of the solution as in \eqref{eq:solution_perturbation_tsvd} due to its multiplication by $1/\mu_i$.  On the other hand, for $k\rightarrow 1$, the  numerical stability increases while the residual $R[f_k]\rightarrow g$, so $f_k$ is a poor approximation to the initial problem. As a consequence, the role of the regularization parameter $k$ is to balance accuracy and stability of the regularized problem.

As the mere definition of the regularized problem relies here on the singular values of the operator $F_K$, it is crucial to keep in mind that any numerical method relying on the truncated SVD regularization will require some approximation of these singular values. The truncated SVD technique has been applied in several stellarator coil design codes \cite{Pomphrey2001,Landreman2016} which compute approximate solutions of \eqref{eq:coil_design}. In these two references, both the right hand side $g$ and the unknown $f$ are supported on toroidal surfaces. The problem is then naturally approximated in terms of truncated Fourier series, and the approximated truncated SVD is computed as the SVD of the discretized linear integral operator. The minimum approximated singular value retained in the SVD computation must be chosen carefully in order to improve numerical stability while maintaining accuracy.  Another application of a similar truncated SVD technique to stellarator design can be found in \cite{Ku2010}.

 A second common regularization method is Tikhonov regularization \cite{Tikhonov1963}, which introduces additional information about the nature of the solution in the regularized problem. 
 One might imagine formulating the first kind integral problem \eqref{PB:LiCopState} as an optimization problem by seeking $W$ which minimizes the residual $g - F_K[W]$. In the Tikhonov approach, one seeks to minimize the sum of the residual, $g - F_K[W]$, and a regularization (or penalty) term. The additional term can be chosen appropriately depending on the desired or expected features of the solution, the most classical choice being related to the norm of the solution. The corresponding regularized problem
 then reads
\begin{align}
    \min_{W\in\mathcal H} \chi^2(W) \text{ where } \chi^2(W) := \left\langle g - F_K[W],g-F_K[W] \right\rangle + \lambda \left\langle W,W \right\rangle,
    \label{eq:regularized_integral}
\end{align}
where $\lambda$ is a parameter to be fixed.
We will now show that this regularized problem is well-posed for $\lambda > 0$. The above functional, $\chi^2$, is strongly convex (Chapter 3 in \cite{Boyd2004}), as its second variation with respect to $W$ is positive definite,
\begin{align}
    \delta^2  \chi^2(W;\delta W,\delta W) = 2 \left \langle \delta W, \left(F_K^* F_K  + \lambda \right)\delta W \right \rangle > 0,
   \quad
   \text{ for all }\ \ \delta W\in\mathcal H,
\end{align}
for any $\lambda>0$, as $F_K^* F_K$ is a symmetric positive semi-definite operator. Thus $\chi^2$ has a  unique global minimum. The functional reaches its minimum value at its unique stationary point $W_m$, defined by
\begin{subequations}
\begin{align}
    &\delta \chi^2 (W_m;\delta W)=0,\quad
   \text{ for all }\ \ \delta W\in\mathcal H,
    \\
    &\Leftrightarrow 2 \left \langle \delta W, F_K^* \left(F_K W_m-g \right) + \lambda W_m \right \rangle = 0,\quad
   \text{ for all }\ \ \delta W\in\mathcal H,
   \\
   &\Leftrightarrow \left(F_K^* F_K + \lambda \right) W_m = F_K^* g.
\end{align}
\end{subequations}
Solving the regularized problem is therefore equivalent to solving the linear problem,
\begin{equation}
\text{Find }W\in\mathcal H_2\text{ such that }G_K [W] = F_K^* g,
\label{eq:TikhonovProb}
\end{equation}
where the operator $G_K : \mathcal{H}_2 \rightarrow \mathcal{H}_2$ is defined as $G_K =  F_K^*F_K + \lambda I$, $I$ being the identity operator on $\mathcal H$. Due to the strong convexity of $\chi^2$, a unique solution to $G_K[W] = F_K^*g$ exists. Since it is equivalent to the regularized problem, then the regularized problem itself has a unique solution. As the equation is linear, uniqueness can be also be proved as follows: the uniqueness of the solution is equivalent to the null space of the operator $G_K$ being trivial, which we can prove for $\lambda>0$. Indeed, for any $W$ such that $G_K[W] = 0$, then $\langle G_K[W],W\rangle=0$, hence $\lambda \langle W,W \rangle + \langle F_K[W],F_K[W] \rangle = 0$, which implies that $W = 0$ if (and only if) $\lambda >0$. 

 To underscore the benefit of Tikhonov regularization, we consider the eigenspectrum of $G_K$. We note that the eigenfunctions of $F_K^* F_K$, denoted by $\{\varphi_i\}_{i \in \mathbb{N}}$, will also be eigenfunctions of $G_K$ with corresponding eigenvalues $\{\alpha_i + \lambda\}_{i \in \mathbb{N}}$, where $\{\alpha_i\}_{i \in \mathbb{N}}$ are the eigenvalues of $F_K^* F_K$. The image of the operator $G_K$ can then be expressed in terms of its eigen-spectrum as
\begin{align}
    \text{ for all } W \in \mathcal{H}_2, \quad G_K[W] = G_K\left[\sum_{i = 1}^{\infty} \varphi_i \langle W,\varphi_i \rangle\right]= \sum_{i = 1}^{\infty} G_K[\varphi_i] \langle W,\varphi_i \rangle = \sum_{i=1}^{\infty} ( \alpha_i + \lambda)\varphi_i \langle W, \varphi_i \rangle.
\end{align}
Thanks to the properties of the singular functions \eqref{eq:singular_functions}, the quantity $F_K^*g$ can be expanded in the basis of the singular functions of $F_K$ as,
\begin{align}
    F_K^*g = \sum_{i=1}^{\infty} \langle u_i, F_K^* g \rangle u_i = \sum_{i=1}^{\infty} \langle F_K[u_i], g \rangle u_i = \sum_{i = 1}^{\infty} \mu_i \langle v_i, g \rangle u_i.
\end{align}
 This shows again that the  problem \eqref{eq:TikhonovProb} has a unique solution given by
\begin{align}
    W_m = \sum_{i=1}^{\infty} u_i \frac{\mu_i}{\mu_i^2 + \lambda} \langle v_i ,g \rangle,
    \label{eq:W_m}
\end{align}
 which is the unique solution to the regularized problem.
Indeed we can easily verify that this function is indeed square-integrable for any $\lambda > 0$ and $g \in \mathcal{H}_1$, as
\begin{align}
    \langle W_m, W_m \rangle = \sum_{i=1}^{\infty} \frac{\mu_i^2}{(\mu_i^2 + \lambda)^2} \langle v_i, g \rangle^2 \le \frac{1}{4 \lambda} \sum_{i=1}^{\infty} \langle v_i, g \rangle^2 =\frac{1}{4 \lambda}\langle g, g \rangle  < \infty.
    \label{eq:L2W}
\end{align}
The above holds due to the inequality of arithmetic and geometric means, $x + y \ge 2\sqrt{xy}$ for all non-negative $x$ and $y$.
We can now show that Tikhonov regularization improves the stability of the solution to perturbation of the data. Consider a perturbation $g \rightarrow g + \delta g$, which from \eqref{eq:W_m}-\eqref{eq:L2W} results in a perturbation of the solution given by
\begin{align}
\label{regest}
    \delta W_m = \sum_{i = 1}^{\infty} u_i \frac{\mu_i}{\mu_i^2 + \lambda} \langle v_i, \delta g \rangle\quad \text{ with }
    \langle \delta W_m, \delta W_m \rangle \le \frac{\langle \delta g,\delta g \rangle}{4 \lambda}. 
\end{align}
Thus a square-integrable change in the prescribed data, $\delta g$, will result in a square-integrable change to the solution, $\delta W_m$. In this way, the regularized problem \eqref{eq:TikhonovProb} is well-posed according to the definition presented in Section \ref{sec:ill_posed}. The solution of the regularized problem now depends continuously on the prescribed data, $g$, for any $\lambda >0$.

  This provides another natural approach to construct a numerical method which approximate solutions of \eqref{PB:LiCopState}, by solving numerically the well-posed regularized problem \eqref{eq:TikhonovProb}.
 It is however important to understand the role of the regularization parameter $\lambda$. For the regularized solution \eqref{eq:W_m}, we can compute the residual $R[W_m]:=g-F_K[W_m]$ for the initial problem, 
 \begin{align}
 \label{regres}
      R[W_m]  = \sum_i \frac{\lambda}{\mu_i^2 + \lambda} \langle v_i, g \rangle  v_i.
 \end{align}
 So for $\lambda \rightarrow \infty$, that is for a large regularization term, the constant in the stability estimate \eqref{regest} is $\frac{1}{4\lambda} \rightarrow 0$,  while the residual $R[W_m]\rightarrow g$, so $W_m$ is a poor approximation to the initial problem. Large values of $\lambda$ are associated with low accuracy. In contrast, for $\lambda \rightarrow 0$, that is for a small regularization term,  the residual $R[W_m]\rightarrow 0$ independently of the data $g$,  while the constant in the stability estimate \eqref{regest} is $\frac{1}{4\lambda} \rightarrow \infty$, so the solution becomes more sensitive to noise in the data. Small values of $\lambda$ are associated with poor stability. As a consequence, again, the role of the regularization parameter $\lambda$ is to balance accuracy and stability of the regularized problem.

The regularized problem \eqref{eq:regularized_integral} has in this case been formulated as an optimization problem on the infinite dimensional space $\mathcal H_2$. 
The construction of an approximated solution to our first kind integral equation \eqref{eq:coil_design} via numerical treatment of Tikhonov regularized problems will be addressed in Sections \ref{sec:winding_surface}-\ref{sec:filamentary}, after additional assumptions guided by practical considerations lead to a simplified problem.

Now that we have discussed properties of integral equations of the first kind \eqref{eq:FirstKindIntegral}, including its ill-posedness and methods for regularization, we return to the formulation of the coil design problem. Further common assumptions imposed on the coils problem are discussed in the following Sections, including assumptions on the support of the currents along with desirable properties of coils. 

\subsubsection{Reformulation of the coils problem}
\label{sec:reformulation}
\begin{figure}
    \centering
    \includegraphics[width=0.8\textwidth]{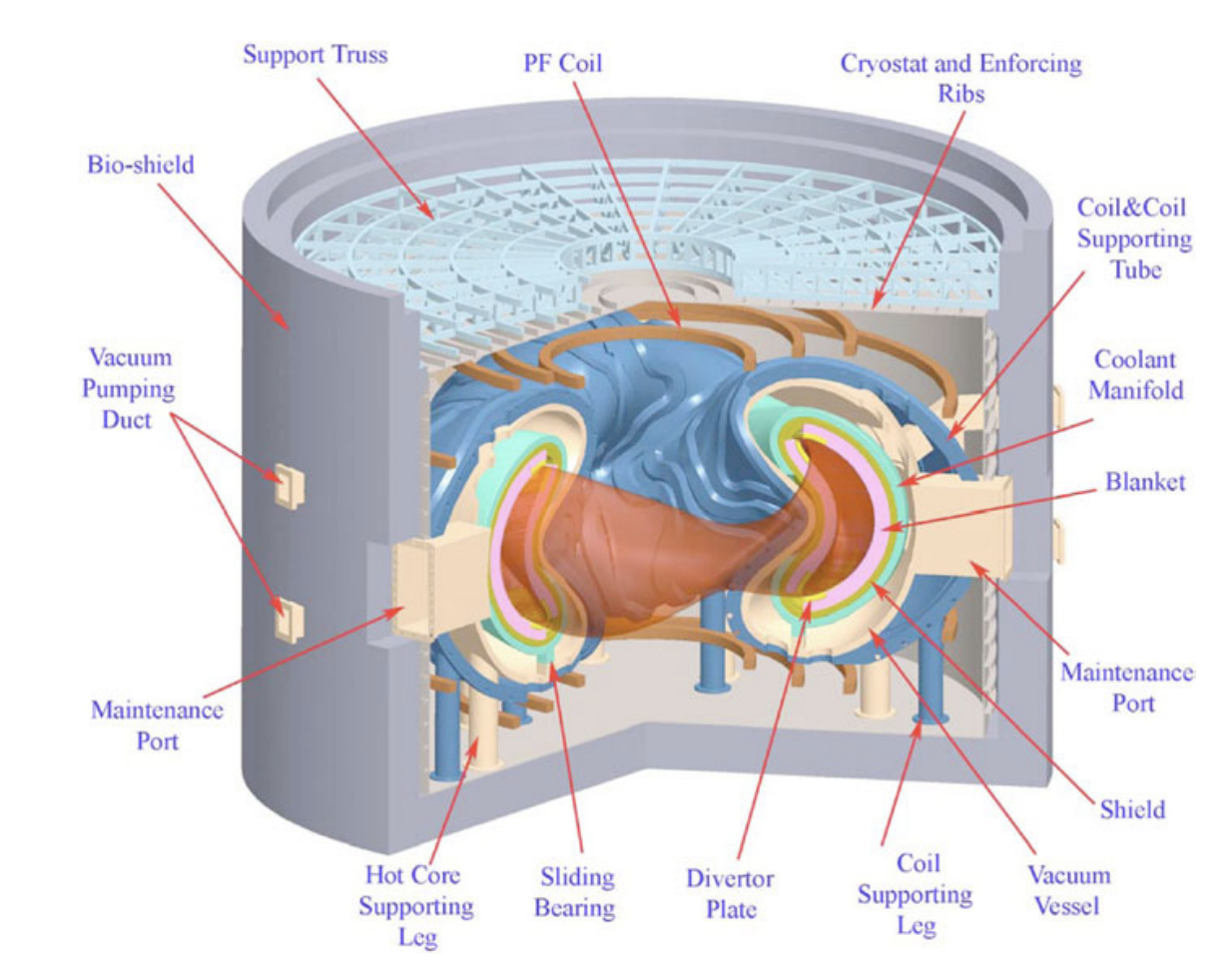}
    \caption{A schematic of the ARIES-CS stellarator power plant concept. The modular coils lie on a supporting surface (blue). There must be sufficient space between coils to accommodate maintenance ports and vacuum pumping ducts. There must be sufficient space between the coils and the plasma (red) to allow for the blanket (pink), divertor plates (yellow), neutron shielding (dark yellow), coolant manifold (green), and vacuum vessel (beige). Figure reproduced from \cite{Najmabadi2008}.}
    \label{fig:aries_CS}
\end{figure}

The ill-posedness of the coils problem may actually be beneficial, as the existence of many possible solutions which reproduce the desired plasma boundary within a given tolerance allows us to choose the solution which possesses the best engineering properties. Given the freedom inherent to the coil design process, desirable properties of coils can be taken into account in formulating a problem for numerical approximation of a solution. 
A critical engineering aspect is the manufacturability of a coil set, as coil shapes that are very complex are more difficult to build and install. The coil complexity can be quantified in several ways, such as the length, torsion, and curvature. In practice, coils have a minimum allowable radius of curvature due to their actual volume. There should be sufficient width between coils to allow for diagnostic and maintenance ports. There should also be adequate room between the plasma surface and coils to account for the vacuum vessel, first wall, and coil casing. The coil-plasma spacing becomes even more critical in a reactor, as a tritium breeding blanket and neutron shielding are required. To illustrate how stellarator coils must accommodate for various structures in an experiment or reactor, we present a schematic diagram of the ARIES-CS stellarator power plant concept in Figure \ref{fig:aries_CS}.

%



Coil design is often regularized in a manner similar to Tikhonov regularization: by formulation in terms of an optimization problem,
\begin{align}
    \min_{\bm{J}^{\text{coil}}} \chi^2 \text{ where } 
    \left\{
    {\renewcommand{\arraystretch}{2.0}
    \begin{array}{l}
    \displaystyle
    \chi^2 := \int_{\Omega} f_{\text{equilibrium}}(\bm{B}) \, d^3 x + \lambda \int_{\Omega_c} f_{\text{coils}}(\bm{J}^{\text{coil}}) \, d^3 x,\\
    \displaystyle
    \bm{B} = \bm{B}^{\text{plasma}} + \bm{B}^{\text{coil}}
    \\
    \displaystyle
   \bm{B}^{\text{plasma}} \text{ given}
    \\
    \displaystyle
   \bm{B}^{\text{coil}}(\bm{r}) = \frac{\mu_0}{4\pi} \int_{\Omega_c} \frac{\bm{J}^{\text{coil}}(\bm{r}') \times (\bm{r}-\bm{r}')}{\rvert{\bm{r}-\bm{r}'\rvert^3}} \, d\bm{r}'.
    \end{array}
    }
    \right.
    \label{eq:regularized_coil_optimization}
\end{align}  
Here the first term in $\chi^2$ accounts for the equilibrium properties of the coils. The most common choice for this term is,
\begin{equation}
    \chi_B^2 = \int_{\partial \Omega} \left(\bm{B} \cdot \hat{\bm{n}}\right)^2 \, d^2 x, 
    \label{eq:chi2_B}
\end{equation}
Other critical physical properties of the equilibrium may also be included, such as the rotational transform (Section \ref{sec:rotation_transform}) or magnetic ripple on axis (a proxy for quasisymmetry (Section \ref{sec:quasisymmetry})). The second term in $\chi^2$ accounts for properties of the coils and introduces regularization. While in \eqref{eq:regularized_integral} the addition of the regularization term favors solutions with a small norm, many other choices for regularization are possible. In the above expressions, $f_{\text{coils}}$ is a regularization function chosen to characterize the desired or expected properties of the coil shapes. The specific choice of regularization term, $f_{\text{coils}}$, will depend on the model used. Thus we will discuss separately specifics of regularization under each assumption.


\subsubsection{Winding surface methods}
\label{sec:winding_surface}

The first stellarator coil optimization tools were based on a winding surface approximation. Rather than seeking a current supported in a volume $\Omega_c$, as in \eqref{eq:coil_design}, the support of the unknown current is assumed to be restricted to a given surface called the winding surface. Furthermore, the current must be tangent to this surface. This leads to a considerably simplified optimization problem, and the individual coil shapes are obtained as a post-processing step. Compared to the individual coils that are to be built in the experiment, this winding surface approximation is often thought of as a limit of a very large number of individual coils.


For a smooth magnetic field $\bm{B}$ and current density $\bm{J}$, we can use the differential form of Ampere's law,
\begin{align}
    \frac{\nabla \times \bm{B}}{\mu_0} = \bm{J},
\end{align}
to conclude that $\nabla \cdot \bm{J} = 0$. We will need to derive a similar condition for a current density supported on a surface. This will allow us to simplify the optimization problem by expressing the current in terms of a scalar potential supported on the surface.

\begin{figure}
    \centering
    \begin{subfigure}{0.49\textwidth}
    \includegraphics[trim=5cm 10cm 3cm 7cm,clip,width=\textwidth]{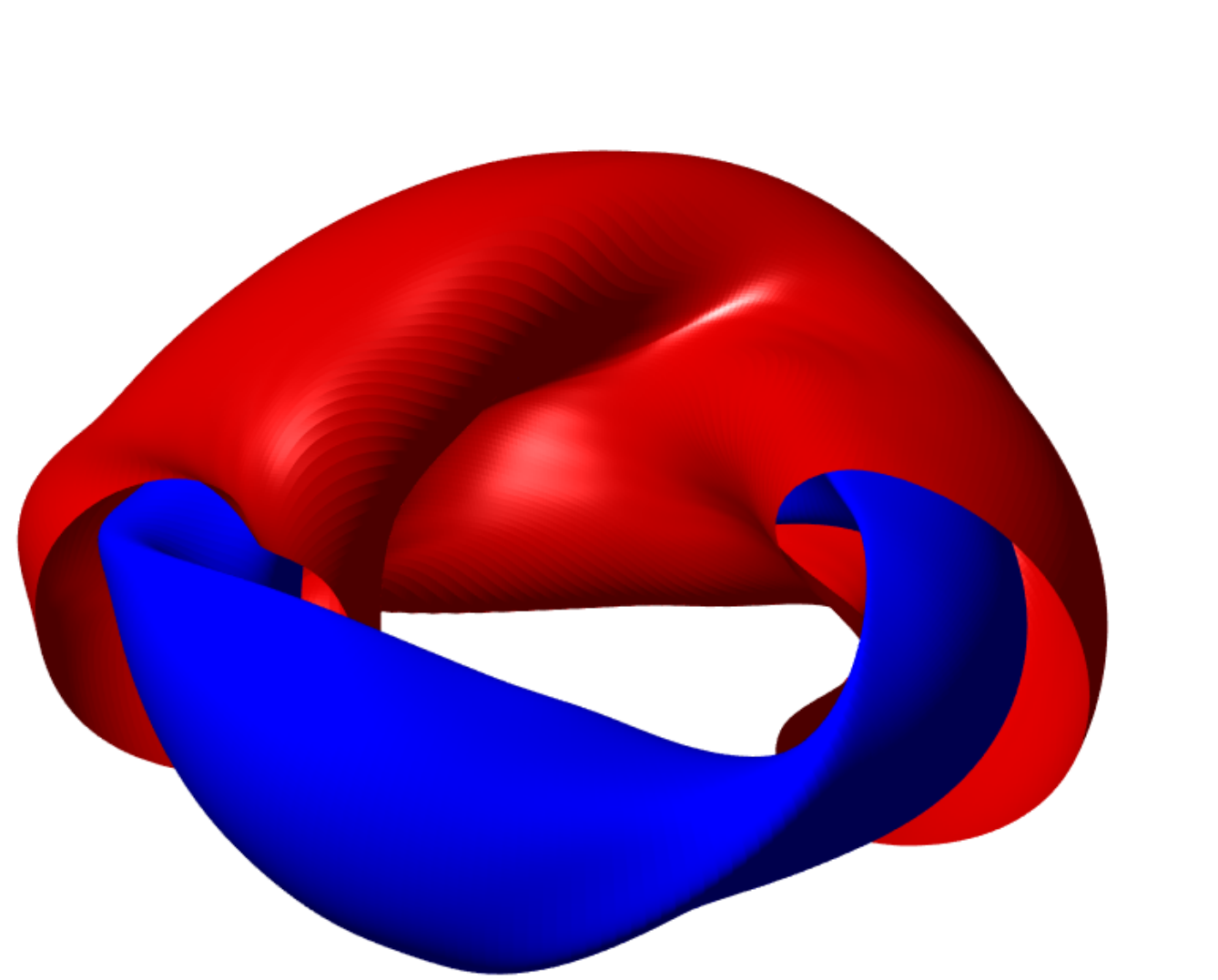}
    \caption{}
    \end{subfigure}
    \begin{subfigure}{0.49\textwidth}
    \includegraphics[trim=6cm 10cm 5cm 9cm,clip,width=\textwidth]{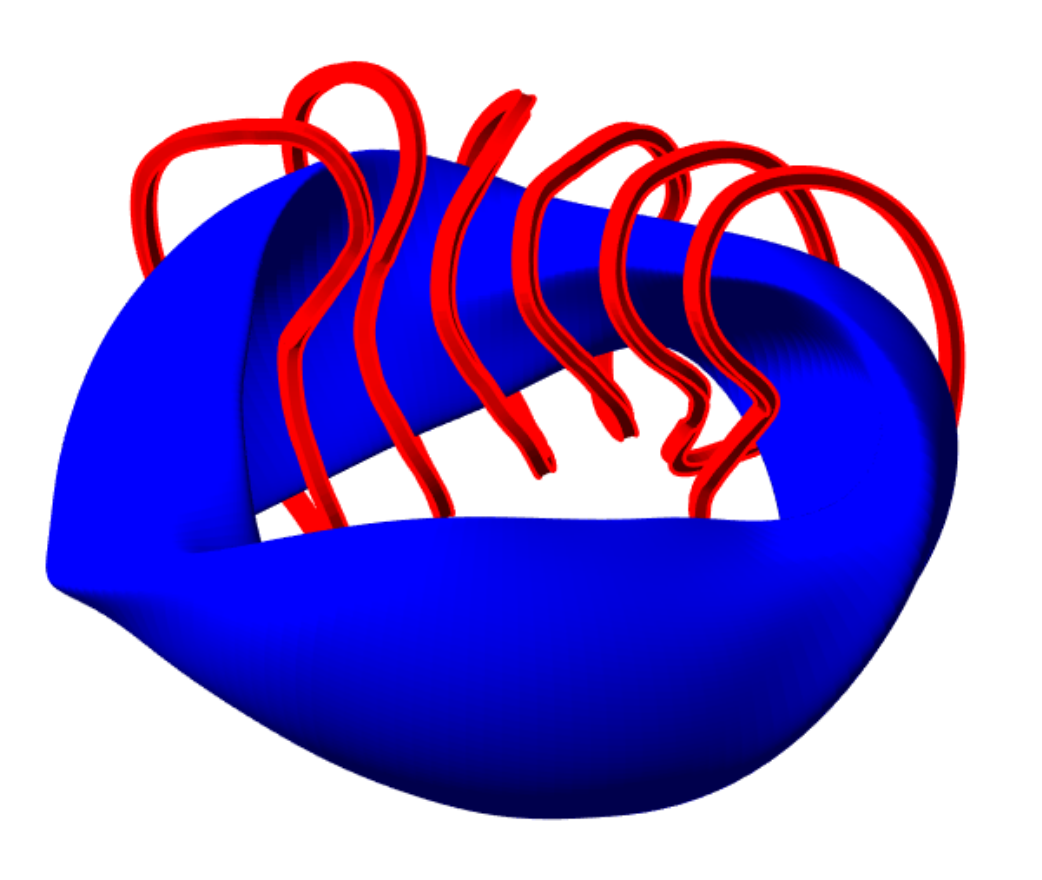}
    \caption{}
    \end{subfigure}
    \begin{subfigure}{0.49\textwidth}
    \includegraphics[trim=1cm 6cm 1cm 6cm,clip,width=1.0\textwidth]{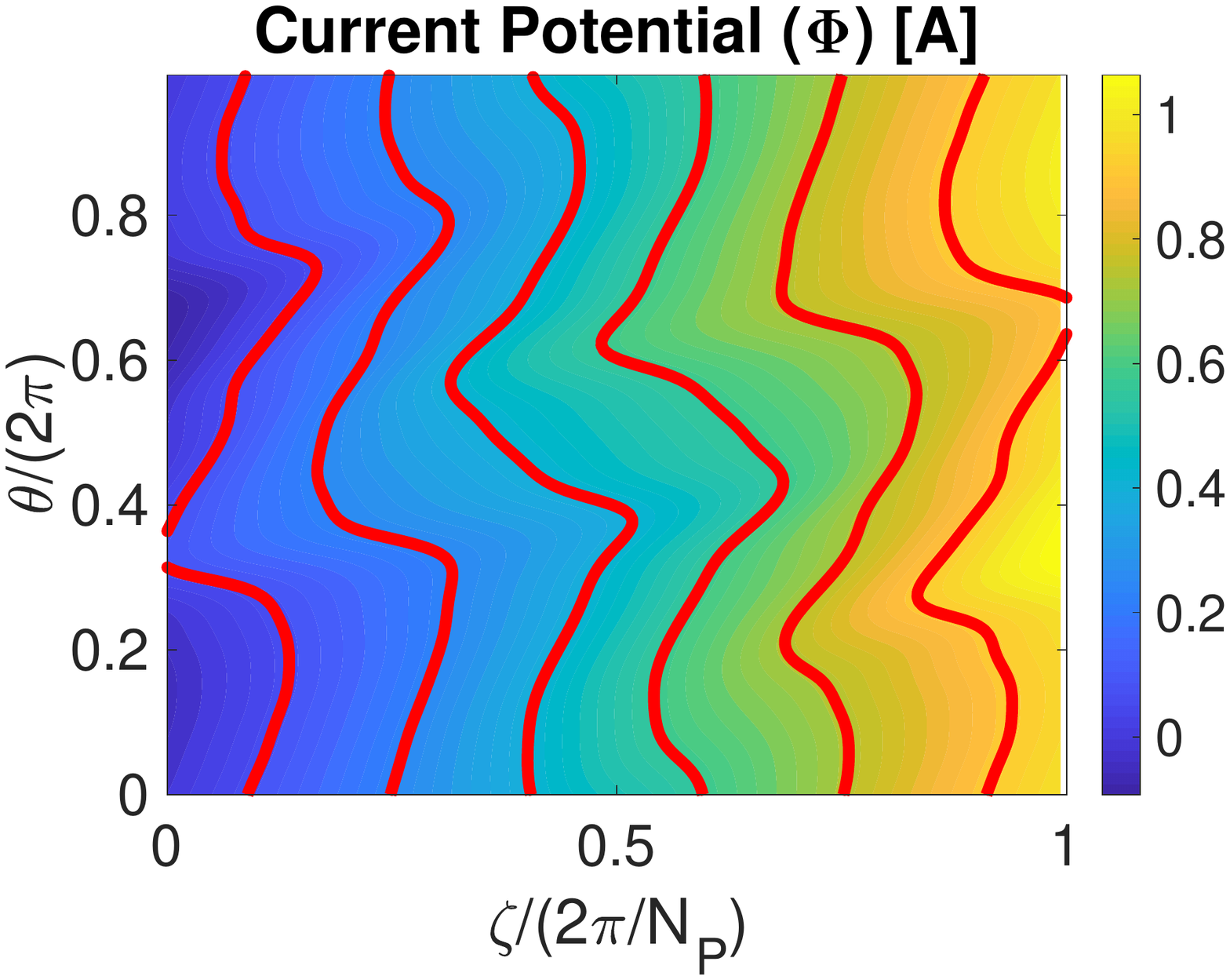}     
    \end{subfigure}
    \caption{(a) The outer boundary of the NCSX LI383 equilibrium \cite{Zarnstorff2001} is shown (blue) along with a winding surface (red) which is chosen to be uniformly offset from the plasma boundary. (b) Coils (red) computed from REGCOIL using the winding surface and target plasma boundary are shown for one period of the device. (c) The normalized current potential solution computed with REGCOIL. The 6 contours corresponding to the coil shapes in (b) are shown. }
    \label{fig:regcoil_NCSX}
\end{figure}

 Assume a genus one surface $S_{\text{coil}}$ is given, parametrized by a poloidal angle $\theta$ and toroidal angle $\zeta$, and that the plasma boundary $\partial\Omega$ is nested inside $S_{\text{coil}}$, see Figure \ref{fig:regcoil_NCSX}. We will here seek the current $\bm{J}^{\text{coil}}$ supported by and tangent to $S_{\text{coil}}$.
 As $\bm{J}^{\text{coil}}$ is a tangential vector field on $S_{\text{coil}}$, we express it as,
 \begin{align}
     \bm{J}^{\text{coil}}(\bm{r}) = \left\{ \begin{array}{l} \bm{K}(\theta(\bm{r}),\zeta(\bm{r})) \hspace{1cm} \bm{r} \in S_{\text{coil}} \\
     0 \hspace{2.9cm} \text{otherwise}
     \end{array}
     \right. ,
     \label{eq:J_delta_function}
 \end{align}
where  $\hat{\bm{n}} \cdot \bm{K} = 0$ on the surface. In order to reformulate the optimization problem in a simpler  manner under this assumption, we will now take the following steps:
 \begin{enumerate}
     \item  Derive the equivalent of the zero divergence condition for a non smooth vector field.
     \item Define and derive an explicit formula for the surface divergence.
     \item Perform a change of unknowns,
     taking advantage of the zero divergence property.
     \item Formulate a simplified optimization problem.
 \end{enumerate} 

As the magnetic field and current density are no longer smooth, we will simply assume that they are integrable. Instead of its differential form, we will then apply the integral form of Ampere's law, 
\begin{align}
    \frac{\int_{\partial \mathcal{S}} \bm{B} \cdot d \bm{l}}{\mu_0} =\int_{\mathcal{S}} \bm{J} \cdot \hat{\bm{n}} \, d^2 x ,
    \label{eq:integral_ampere}
\end{align}
for any surface $\mathcal{S}\in\mathbb R^3$, where  $\partial \mathcal{S}$ is the boundary of the surface if the surface is open while it is the empty set if the surface is closed. Hence for any closed surface, the left-hand-side of the above vanishes. As a result, any integrable field $\bm{J}$ satisfying Ampere's law will necessarily fulfil the condition:
\begin{equation}
    \int_{\mathcal{S}} \bm{J} \cdot \hat{\bm{n}} \, d^2 x = 0\quad \text{for any closed surface } \mathcal S.
\end{equation}
This is the integral formulation of the zero divergence condition, and we will now turn to the derivation of an explicit expression of this condition for a tangential vector field on $S_{\text{coil}}$.

The definition of the divergence of $\bm J$ at a point $\bm r$ is defined as the limit of the ratio between the flux of $\bm J$ through the boundary of a closed volume $V(\bm r)$ enclosing $\bm r$ to the area of $\partial V$, as $V$ shrinks to zero at $\bm{r}$:
\begin{equation}
\label{eq:defdiv}
    \left[ \nabla\cdot \bm{J} \right] (\bm r) := \lim_{V(\bm r)\rightarrow 0}\frac{\int_{\partial V(\bm r)} \bm{J} \cdot \hat{\bm{n}} \, d^2 x}{\int_{\partial V(\bm r)}  \, d^2 x}.
\end{equation}
It can be shown that the limit is independent of how the volume goes to zero. In the case of a smooth vector field, the divergence can equivalently be defined in curvilinear coordinate as in Table \ref{table:non_orthogonal}. This formula does not apply for a vector field supported on a surface, as such a field is not differentiable. 
In the case of a tangential vector field supported on a surface, as $\bm J^{\text{coil}}$ described in \eqref{eq:J_delta_function}, and smooth \textit{along} the surface, meaning that the field $\bm K$ is smooth, there is a differential formula, 
which defines the surface divergence operator of $\bm K$, denoted $\nabla_{S_{\text{coil}}} \cdot \bm K$. In what follows we derive this formula, namely \eqref{eq:surface_divergence_condition}.

The differential definition of the surface divergence will be obtained from the definition of the divergence \eqref{eq:defdiv}, for a particular choice of volumes $V_0\rightarrow 0$. To do so, we will use the signed distance function to the surface $S_{\text{coil}}$, $b: \mathbb{R}^3\rightarrow \mathbb{R}$, defined by
\begin{align}
     b(\bm{r}) := \left \{ \begin{array}{ll} \text{dist}(\bm{r},S_{\text{coil}}),& \bm{r} \in \mathbb{R}^3 \setminus \overline{V_{\text{coil}}}, \\
     0, &  \bm{r} \in S_{\text{coil}}, \\
     - \text{dist}(\bm{r},S_{\text{coil}}), & \bm{r} \in V_{\text{coil}}.
     \end{array} \right . 
     \label{eq:signed_distance_function}
 \end{align}
 Here $V_{\text{coil}}$ is the volume enclosed by $S_{\text{coil}}$ and $\text{dist}(\bm{r},S_{\text{coil}})$ is the minimum distance from $\bm{r}$ to a point on $S_{\text{coil}}$. An important property of the signed distance function is that its gradient restricted to the surface $S_{\text{coil}}$ is the unit normal on the surface,
 \begin{align}
 \label{eq:gradbnorm}
     \nabla b \rvert_{S_{\text{coil}}} = \hat{\bm{n}}.
 \end{align}
 Therefore, $\nabla b$ is an extension of the normal vector in a neighborhood of $S_{\text{coil}}$. Using the projection from a point $\bm r \in\mathbb R^3$ onto the surface $S_{\text{coil}}$ along the $\nabla b$ direction, namely $\bm{r}-b(\bm{r}) \nabla b(\bm{r})$, we also define extensions $\left(\widetilde\theta,\widetilde\zeta\right)$ of the angles $(\theta,\zeta)$ away from $S_{\text{coil}}$, 
 \begin{subequations}
 \begin{align}
     \widetilde\theta(\bm{r}) &= \theta(\bm{r}-b(\bm{r}) \nabla b(\bm{r})) \\
     \widetilde\zeta(\bm{r}) &= \zeta(\bm{r}-b(\bm{r}) \nabla b(\bm{r})).
 \end{align}
 \end{subequations}
In this way, the position vector satisfies,
 \begin{align}
     \bm{r}\left(b,\widetilde{\theta},\widetilde{\zeta}\right) = \bm{r}_s(\widetilde{\theta},\widetilde{\zeta}) + b \nabla b,
 \end{align}
 where $\bm{r}_s$ is the position vector on $S_{\text{coil}}$. As a consequence, we find that
     \begin{align}
        \partder{\bm{r}\left(b,\widetilde{\theta},\widetilde{\zeta}\right)}{b} &= \nabla b + b \partder{\left(\nabla b\right)}{b}.
     \end{align}
 This implies that $\partial \bm{r}\left(b,\widetilde{\theta},\widetilde{\zeta}\right)/\partial b \rvert_{S_{\text{coil}}} = \nabla b\rvert_{S_{\text{coil}}}$, and in turn from \eqref{eq:gradbnorm} it implies that this is equal to the normal, $\partial \bm{r}\left(b,\widetilde{\theta},\widetilde{\zeta}\right)/\partial b \rvert_{S_{\text{coil}}} = \hat{\bm{n}}$. As a result we see that on the surface we have that the Jacobian $\partial \bm{r}\left(b,\widetilde{\theta},\widetilde{\zeta}\right)/\partial b  \times \partial \bm{r}\left(b,\widetilde{\theta},\widetilde{\zeta}\right)/\partial\zeta \cdot \partial \bm{r}\left(b,\widetilde{\theta},\widetilde{\zeta}\right)/\partial\theta \neq 0$, and therefore in a neighborhood of the surface $S_{\text{coil}}$, denoted $\mathcal N$, the triplet $\left(b,\widetilde\theta,\widetilde\zeta\right)$ defines a coordinate system.
 
\begin{figure}
    \centering
    \includegraphics[width=0.8\textwidth]{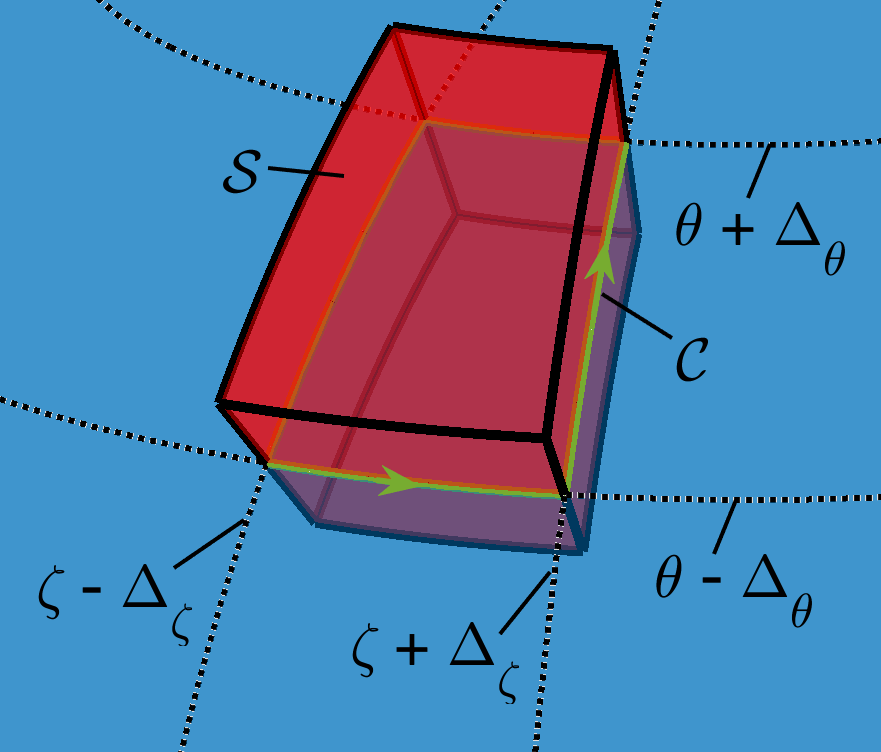}
    \caption{The surface, $\mathcal{S}$, used for evaluating the integral form of Ampere's law \eqref{eq:integral_ampere}. This integral involves a line integral along $\mathcal{C}$, the intersection of $\mathcal{S}$ with $S_{\text{coil}}$.}
    \label{fig:surface_div}
\end{figure}

 So around any point
 $\left(b,\widetilde{\theta},\widetilde{\zeta}\right) = (0,\theta_0,\zeta_0)\in S_{\text{coil}}$, there exists $\Delta_b>0$, $\Delta_\theta>0$, and $\Delta_\zeta>0$ such that
 \begin{align}
V_0=\left\{ \left(b,\widetilde{\theta},\widetilde{\zeta}\right)\in\mathbb R^3; |b| \leq \Delta_b, \left|\theta_0-\widetilde{\theta}\right|\leq \Delta_\theta,\left|\zeta_0-\widetilde{\zeta}\right|\leq \Delta_\zeta\right\}
 \end{align}
 defines a volume intersecting the surface $S_{\text{coil}}$ while $V_0\in \mathcal N$. 
In the $\left(b,\widetilde{\theta},\widetilde{\zeta}\right)$ coordinate system, $V_0$ is the cube centered at $(0,\theta_0,\zeta_0)$, and its boundary $\mathcal S$ is represented on Figure \ref{fig:surface_div}. In other words $\mathcal{S}$ is the union of six faces that can be defined as,
\begin{align}
 \left\{
 \def\arraystretch{1.7}
\begin{array}{c}
\{ \bm{r}(-\Delta_b,\widetilde\theta,\widetilde\zeta); \left|\theta_0-\widetilde{\theta}\right|\leq \Delta_\theta, \left|\zeta_0-\widetilde{\zeta}\right|\leq \Delta_\zeta \}\\
\{\bm{r}(+\Delta_b,\widetilde\theta,\widetilde\zeta); \left|\theta_0-\widetilde{\theta}\right|\leq \Delta_\theta, \left|\zeta_0-\widetilde{\zeta}\right|\leq \Delta_\zeta \}\\
\{ \bm{r}(b,\theta_0-\Delta_{\theta},\widetilde\zeta); |b| \leq \Delta_b, \left|\zeta_0-\widetilde{\zeta}\right|\leq \Delta_\zeta \}\\
\{ \bm{r}(b,\theta_0+\Delta_{\theta},\widetilde\zeta); |b| \leq \Delta_b, \left|\zeta_0-\widetilde{\zeta}\right|\leq \Delta_\zeta \}\\
\{ \bm{r}(b,\widetilde\theta,\zeta_0-\Delta_{\zeta}); |b| \leq \Delta_b, \left|\theta_0-\widetilde{\theta}\right|\leq \Delta_\theta\}\\
\{ \bm{r}(b,\widetilde\theta,\zeta_0+\Delta_{\zeta}); |b| \leq \Delta_b, \left|\theta_0-\widetilde{\theta}\right|\leq \Delta_\theta\}.
\end{array} \right.
\end{align}
As a result, integrating over this surface $\mathcal S$, for any vector field $\bm{J}$, we can easily see that
\begin{multline}
\label{eq:intJdotn}
    \int_{\mathcal{S}} \bm{J} \cdot \hat{\bm{n}}_{\mathcal{S}} \, d^2 x =  \int_{\theta_0-\Delta_{\theta}}^{\theta_0+\Delta_{\theta}} \int_{\zeta_0-\Delta_{\zeta}}^{\zeta_0+\Delta_{\zeta}} \bm{J} \cdot \nabla b \sqrt{g}  \, d\widetilde{\zeta} d \widetilde{\theta} \bigg\rvert_{b=-\Delta_b}^{b=+\Delta_b} \\
    +\int_{-\Delta_b}^{+\Delta_b} \int_{\zeta_0-\Delta_{\zeta}}^{\zeta_0+\Delta_{\zeta}} \bm{J} \cdot \nabla \widetilde{\theta} \sqrt{g}  \, d\widetilde{\zeta} d b \bigg\rvert_{\widetilde{\theta}=\theta_0-\Delta_{\theta}}^{\widetilde{\theta}=\theta_0+\Delta_{\theta}} + \int_{-\Delta_b}^{+\Delta_b} \int_{\theta_0-\Delta_{\theta}}^{\theta_0+\Delta_{\theta}} \bm{J} \cdot \nabla \widetilde{\zeta} \sqrt{g} \, d\widetilde{\theta} db  \bigg \rvert_{\widetilde{\zeta} = \zeta_0 - \Delta_{\zeta}}^{\widetilde{\zeta} = \zeta_0 + \Delta_{\zeta}},
\end{multline}
where the Jacobian in the coordinate system $(b,\widetilde \theta,\widetilde \zeta)$ is expressed in terms of covariant vectors as $\sqrt{g} = \left(\nabla b \times \nabla \widetilde \theta \cdot \nabla \widetilde \zeta\right)^{-1}$. In the particular case of our vector field $\bm{J}^{\text{coil}}$ supported on $S^{\text{coil}}$, as expressed in \eqref{eq:J_delta_function}, it is clear that $\bm J^{\text{coil}}(\pm\Delta_b,\widetilde\theta,\widetilde\zeta)=0$, and that $\bm J^{\text{coil}}(b,\widetilde\theta,\widetilde\zeta)=0$ unless $b=0$. Hence the first contribution from \eqref{eq:intJdotn} to $\int_{\mathcal{S}} \bm{J}^{\text{coil}}\cdot \hat{\bm{n}}_{\mathcal{S}} \, d^2 x$ vanishes while the last two contributions reduce to line integrals, 
\begin{subequations}
\begin{align}
\label{eq:Jcinttmp}
    \int_{\mathcal{S}} \bm{J}^{\text{coil}} \cdot \hat{\bm{n}}_{\mathcal{S}} \, d^2 x &= 
     \int_{\zeta_0-\Delta_{\zeta}}^{\zeta_0+\Delta_{\zeta}} \bm{K} \cdot \nabla \theta \sqrt{g}  \, d\zeta \bigg\rvert_{\theta=\theta_0-\Delta_{\theta}}^{\theta=\theta_0+\Delta_{\theta}} 
     +  \int_{\theta_0-\Delta_{\theta}}^{\theta_0+\Delta_{\theta}} \bm{K} \cdot \nabla \zeta \sqrt{g} \, d\theta   \bigg \rvert_{\zeta = \zeta_0 - \Delta_{\zeta}}^{\zeta = \zeta_0 + \Delta_{\zeta}} \\
     &= \int_{\theta_0-\Delta_{\theta}}^{\theta_0+\Delta_{\theta}} \int_{\zeta_0-\Delta_{\zeta}}^{\zeta_0+\Delta_{\zeta}} \left[ \partder{\left(\sqrt{g} \bm{K} \cdot \nabla \theta \right)}{\theta} +  \partder{\left(\sqrt{g} \bm{K} \cdot \nabla \zeta \right)}{\zeta}\right] \, d \zeta d \theta.
\end{align}
\end{subequations}
Now taking the ratio with the surface area and taking the limit when $\Delta_{\theta}\rightarrow 0$ and $\Delta_{\zeta} \rightarrow 0$ in order to evaluate \eqref{eq:defdiv},
\begin{align}
   \lim_{\Delta_{\zeta} \rightarrow 0,\Delta_{\theta} \rightarrow 0} \frac{\int_{\mathcal{S}} \bm{J}^{\text{coil}} \cdot \hat{\bm{n}}_{\mathcal{S}} \, d^2 x}{\int_{\mathcal{S}}\, d^2 x} = \lim_{\Delta_{\zeta} \rightarrow 0,\Delta_{\theta} \rightarrow 0} \frac{\int_{\theta_0-\Delta_{\theta}}^{\theta_0+\Delta_{\theta}} \int_{\zeta_0-\Delta_{\zeta}}^{\zeta_0+\Delta_{\zeta}} \left[ \partder{\left(\sqrt{g} \bm{K} \cdot \nabla \theta \right)}{\theta} +  \partder{\left(\sqrt{g} \bm{K} \cdot \nabla \zeta \right)}{\zeta}\right] \, d \zeta d \theta}{\int_{\theta_0-\Delta_{\theta}}^{\theta_0+\Delta_{\theta}} \int_{\zeta_0-\Delta_{\zeta}}^{\zeta_0+\Delta_{\zeta}} \sqrt{g} \,  d \zeta d \theta}.
\end{align}%

Keeping in mind that on the surface the differential surface area is $d^2 x = \sqrt{g} \,  d \zeta d \theta$, the above quantity defines the \textit{surface divergence} (Appendix 3 in \cite{Van2007}),
\begin{align}
 \nabla_{S_{\text{coil}}} \cdot \bm{K} := \frac{1}{\sqrt{g}} \left[ \partder{\left(\sqrt{g} \bm{K} \cdot \nabla \theta \right)}{\theta} +  \partder{\left(\sqrt{g} \bm{K} \cdot \nabla \zeta \right)}{\zeta}\right].
 \label{eq:surface_divergence_condition}
\end{align}
In the literature, the surface divergence operator is often written as $[\nabla_{\Gamma} \cdot ]$.
The zero divergence condition for a current supported on the surface $S_{\text{coil}}$ can finally be expressed in a differential form by $\nabla_{S_{\text{coil}}} \cdot \bm{K} = 0$ \cite{Arnoldus2006},
%
which is similar to the zero-divergence condition on the magnetic field \eqref{eq:div_B_magnetic}, except for the fact that here the field $\bm K$ is a tangential surface field. As a direct consequence, and similarly to  \eqref{eq:Bphitheta}-\eqref{eq:AClambda}, we can define a scalar unknown on the surface, called the current potential,
\begin{subequations}
\begin{align}
   \Phi (\theta,\zeta) = \int_0^\theta \bm{K} \left(\theta',\zeta \right) \cdot \nabla \zeta \left(\theta',\zeta\right)  \sqrt{g}(\theta',\zeta) d \theta' ,
\end{align}
so that we have
\begin{align}
\bm{K} \cdot \nabla \theta &= \frac{1}{\sqrt{g}} \partder{\Phi}{\zeta} \\
\bm{K} \cdot \nabla \zeta &= -\frac{1}{\sqrt{g}} \partder{\Phi}{\theta}.
\end{align}
\end{subequations}
It immediately follows that the vector field $\bm K$ can then be expressed in terms of the current potential as,
\begin{align}
    \bm{K} = \hat{\bm{n}} \times \nabla \Phi.
    \label{eq:K_Phi}
\end{align}
We note that \eqref{eq:K_Phi} implies that $\bm{K} \cdot \nabla \Phi = 0$, or current flows along streamlines of the current potential. Thus, in the postprocessing, it will be reasonable to search for filamentary coils as a set of the contours of the current potential. 
 
As $\bm{K}$ must be a single-valued function of $\theta$ and $\zeta$, the current potential can be decomposed into a single-valued term and the sum of two periodic terms as follows,
\begin{align}
\label{eq:phiIGsv}
    \Phi(\theta,\zeta) = \Phi_{\text{sv}}(\theta,\zeta) + \frac{G \theta}{2\pi} + \frac{I \zeta}{2\pi}, 
\end{align}
where $G$ is the net current linking the winding surface toroidally and $I$ is the net current linking the winding surface poloidally, or in other words,
\begin{subequations}
\begin{align}
    2\pi G &= \int_{S_{\text{coil}}} \, \bm{K} \cdot \nabla \zeta \,  d^2 x \\
    2\pi I &= -\int_{S_{\text{coil}}} \bm{K} \cdot \nabla \theta \, d^2 x.
\end{align}
\end{subequations}
In order to find $\Phi$, we now discuss each of the three terms contributing to \eqref{eq:phiIGsv}.

First, the constant $I$ can be derived from known properties of the target equilibrium as follows. Consider the outer plasma surface, $\partial \Omega$, parameterized by two angles, $\theta_p$ and $\zeta_p$. The integral form of Ampere's law applied to a surface $S_P$  enclosed by $\partial S_P$, a curve at constant $\theta_p$ on $\partial \Omega$, as represented on Figure \ref{fig:poloidal_current}, reads
\begin{align}
    \oint_{\partial S_P} \bm{B} \cdot \partder{\bm{r}}{\zeta_p} \, d \zeta_p = \mu_0 \int_{S_P} \bm{J} \cdot \hat{\bm{n}} \, d^2 x,
\end{align}
where $S_P$ is a surface enclosed by $\partial S_P$, a curve at constant $\theta_p$ on $\partial \Omega$ (Figure \ref{fig:poloidal_current}). While the left hand side is fixed by the desired equilibrium, the right hand side can now be expressed in terms of current potential. 
Indeed, using the local coordinate system $(b,\widetilde{\theta},\widetilde{\zeta})$ and the dual relations between covariant and contravariant basis vectors from Table \ref{table:non_orthogonal}, for a tangential surface vector field we see that 
\begin{align}
    \mu_0 \int_{S_P} \bm{J}^{\text{coil}} \cdot \hat{\bm{n}} \, d^2 x &= \mu_0 \int_{b_-}^{b_+} \int_0^{2\pi} \frac{\bm{J}^{\text{coil}} \cdot \nabla \widetilde{\theta} }{|\nabla \widetilde{\theta}|} \bigg\rvert \partder{\bm{r}}{\widetilde{\zeta}} \times \partder{\bm{r}}{b} \bigg\rvert \, d \widetilde{\zeta} d b .
\end{align}
We will use the dual relation $\nabla \widetilde{\theta}=\left(\partial \bm{r}/\partial b \times \partial \bm{r}/\partial \widetilde \zeta \right)\sqrt{g}$, combined with the fact that the Jacobian in the coordinate system $(b,\widetilde \theta,\widetilde \zeta)$ is expressed in terms of contravariant vectors as $\sqrt{g} = \partial \bm{r}/\partial \widetilde\zeta \times \partial \bm{r}/\partial \widetilde\theta \cdot \partial \bm{r}/\partial b$, where $\partial \bm{r}/\partial b\rvert_{S_P}= \hat{\bm{n}}$ so that $|\sqrt{g}| = 
\left|  \partial \bm{r}/\partial \widetilde\zeta \times \partial \bm{r}/\partial \widetilde\theta
\right|$. Moreover, given that $\widetilde{\theta}\rvert_{S_{\text{coil}}} = \theta$, $\widetilde{\zeta}\rvert_{S_{\text{coil}}} = \zeta$ and the derivatives of $\bm{r}$ are evaluated at fixed $b$, $\partial \bm{r}/\partial \widetilde{\theta} \rvert_{S_{\text{coil}}} = \partial \bm{r}/\partial \theta$ and $\partial \bm{r}/\partial \widetilde{\zeta} \rvert_{S_{\text{coil}}} = \partial \bm{r}/\partial \zeta$. Finally we get
\begin{align}
   \mu_0 \int_{S_P} \bm{J}^{\text{coil}} \cdot \hat{\bm{n}} \, d^2 x  &= \mu_0 \int_0^{2\pi} \bm{K} \cdot \nabla \theta  \bigg \rvert \partder{\bm{r}}{\theta} \times \partder{\bm{r}}{\zeta} \bigg\rvert \, d \zeta = \mu_0 I.
\end{align}  
So the constant can  be computed from the desired equilibrium as
\begin{equation}
\label{eq:condition_I}
    I = \frac{1}{\mu_0} \oint_{\partial S_P} \bm{B} \cdot \partder{\bm{r}}{\zeta_p} \, d \zeta_p.
\end{equation}

As a side note, the integral form of Ampere's law applied to a surface $S_T$ enclosed by $\partial S_T$, a curve at constant $\zeta_p$ on $\partial \Omega$ (Figure \ref{fig:toroidal_current}), reads,
\begin{align}
    \oint_{S_T} \bm{B} \cdot \partder{\bm{r}}{\theta_p} \, d \theta_p = \mu_0 \int_{S_T} \bm{J} \cdot \hat{\bm{n}} \, d^2 x .
\end{align}
The current density that appears on the right hand side has no contribution from the currents on the winding surface, as it is an integral over a toroidal cross-section of the plasma. As a consequence, this identity is not related to $G$, and therefore, unlike $I$, the constant $G$ is not fixed by the equilibrium.

Furthermore note that $G$ and $I$ are related to the winding of streamlines of the current density around the winding surface, respectively with respect to the toroidal and poloidal directions. In order to choose the value of $G$, we will take into account the post-processing step. Specifically, we will want to choose a set of discrete coils to be the streamlines of $\bm{K}$, thus we will choose $G$ such that the streamlines are closed. This corresponds to the additional assumption that $G/I$ takes a rational value. In this case, consider any closed streamline $C_{\Phi_0}$ corresponding to a particular value of the current potential, $\Phi_0$, and parameterized by $(\theta(l),\zeta(l))$, where $l$ is a coordinate that measures length along the streamline. The number of poloidal and toroidal turns of the streamline are along the surface $S_{\text{coil}}$,
\begin{subequations}
\begin{align}
    M(\Phi_0) &= \frac{1}{2\pi} \oint_{C_{\Phi_0}} \der{\theta(l)}{l} \, dl \\
    N(\Phi_0) &= \frac{1}{2\pi} \oint_{C_{\Phi_0}} \der{\zeta(l)}{l} \, dl .
\end{align}
\end{subequations}
 We note that by definition of a streamline,
\begin{equation}
\left\{
    \def\arraystretch{2}
    \begin{array}{l}
    \displaystyle
    \frac{d \theta(l)}{dl} = \left[\frac{\bm{K} \cdot \nabla \theta}{|\bm{K}|} \right] (\theta(l),\zeta(l))
    \\
    \displaystyle
    \frac{d \zeta(l)}{dl} = \left[\frac{\bm{K} \cdot \nabla \zeta}{|\bm{K}|} \right] (\theta(l),\zeta(l)).
    \end{array}
\right.
\end{equation}
We now average over all possible streamlines by integrating over $\Phi$,
\begin{subequations}
\begin{align}
  \int M(\Phi) \, d \Phi &= \frac{1}{2\pi} \int \oint_{C_{\Phi}} \frac{\bm{K} \cdot \nabla \theta}{|\bm{K}|}  \, dl d \Phi = \frac{1}{2\pi} \int_{S_{\text{coil}}} \bm{K} \cdot \nabla \theta \, d^2 x  \\
  \int N(\Phi) \, d \Phi &= \frac{1}{2\pi} \int \oint_{C_{\Phi}} \frac{\bm{K} \cdot \nabla \zeta}{|\bm{K}|} \, dl d \Phi =  \frac{1}{2\pi} \int_{S_{\text{coil}}} \bm{K} \cdot \nabla \zeta \, d^2 x ,
\end{align}
\end{subequations}
where in the second step we have changed coordinates, noting that $dl d \Phi = d\theta d \zeta |\bm{K}| \rvert \partial \bm{r}/\partial \theta \times \partial \bm{r}/\partial \zeta \rvert $. 
Thus in particular we see that $G$ quantifies the average winding of streamlines of the current density in the toroidal direction, while $I$ quantifies the average winding of the streamlines in the poloidal direction. Here the ratio of $I/G$ is analogous to the rotational transform \eqref{eq:iota}.  The value of $G$ is 
chosen based on the desired topology of the coil set. If one desires \textit{modular} coils, which link the plasma poloidally but not toroidally, then 
$G=0$ is chosen. If one desires \textit{helical} coils, which link the plasma both poloidally and toroidally, then 
$G$ is chosen based on the desired relative number of toroidal turns of the coils.

The case of \textit{saddle} coils, which do not link the plasma toroidally or poloidally, might also be of interest. However, in this case then both $I$ and $G$ are chosen to vanish, therefore such a coil set cannot be consistent with the equilibrium according to \eqref{eq:condition_I}. In order to satisfy Ampere's law, saddle coils must be supplemented by a coil set which links the plasma poloidally, so the resulting set of coils cannot be obtained under the assumption that the current is supported by a single winding surface. In this case, one would assume either two disjoint winding surfaces, or a winding surface with extra filamentary coils.

Finally, the single-valued part of the current potential, $\Phi_{\text{sv}}$, is expressed as a Fourier series,
\begin{align}
    \Phi_{\text{sv}}(\theta,\zeta) &= \sum_{m,n} \Phi_{m,n} \sin(m\theta - n \zeta).
\end{align}
Under the assumption of stellarator symmetry (see Section \ref{sec:stellarator_symmetry}), only a sine series is needed, as $\bm{K}$ must be even with respect to $\theta$ and $\zeta$. The Fourier harmonics are then chosen to minimize a certain objective function as announced in \eqref{eq:regularized_coil_optimization}.

As a conclusion, our simplified optimization can be summarized as follows:
\begin{enumerate}
    \item Fix $ I$ according to \eqref{eq:condition_I}.
    \item Fix $G$ according to desired topology of the coils.
    \item Solve 
    \begin{equation}
        \min_{\{\Phi_{m,n}\}} \chi^2 \text{ where } \chi^2 := \chi^2_B + \lambda \int_{S_{\text{coil}}} f(\bm{K}) \, d^2 x.
        \label{eq:chi2}
    \end{equation}
\end{enumerate}
The first term, $\chi^2_B$, quantifies how close the target magnetic surface is to being an actual magnetic surface while the second term quantifies coil complexity. The regularization term is most often chosen such that \eqref{eq:chi2} is a convex optimization problem. One particular choice for quantifying coil complexity is,
\begin{align}
    \chi^2_K =  \int_{S_{\text{coil}}}  |\bm{K}|^2 \, d^2 x . 
\end{align}
Larger values of $|\bm{K}|^2$ indicate that contours of the current potential are closer to each other, implying a smaller coil-coil distance (a feature that is difficult to engineer and that makes experimental access difficult). Other choices for the regularization term are possible, depending on the desired features of the current potential.  Table \ref{tab:coil_metrics_winding} summarizes common figures of merit to quantify the complexity of the current potential.

\begin{table}[]
    \centering
    {\renewcommand{\arraystretch}{1.7}
    \begin{tabular}{|c|c|}
    \hline \hline 
    RMS coil-coil spacing & $||\bm{K}||_2 =  \sqrt{\int_{S_{\text{coil}}}  |\bm{K}|^2 \, d^2 x/\int_{S_{\text{coil}}} d^2 x } $ \\ \hline
    Maximum coil-coil spacing & $K_{\text{max}} = \max_{\theta,\zeta} |\bm{K}(\theta,\zeta)|$ \\ \hline
    Harmonic content & $\mathcal{H} = \sum_{m,n} \Phi_{m,n}^2(m^2 + n^2)$ \\ \hline
   Curvature & $f_{\kappa} = \int_{S_{\text{coil}}} \kappa^2 |\bm{K}|^2 \, d^2 x$ \\ \hline
   Coil-plasma spacing & $\min_{\theta,\zeta,\theta_p,\zeta_p} \text{dist}\left(\bm{r}(\theta,\zeta),\bm{r}(\theta_p,\zeta_p) \right)$ \\
     \hline \hline 
    \end{tabular}}
    \caption{Properties of the continuous current, $\bm{K}(\theta,\zeta)$, on a winding surface, $S_{\text{coil}}$. These properties allow one to approximately quantify the complexity of filamentary coils which are taken to be streamlines of $\Phi$. In the above expression, $\bm{r}(\theta_p,\zeta_p)$ denotes the position on the desired plasma boundary $\partial \Omega$ parameterized by angles $(\theta_p,\zeta_p)$. We note that $|\bm{K}|$ is a proxy for coil-coil spacing, as contours of the spacing between contours of the current potential will be reduced in regions of small current density. The harmonic content is a measure of the high mode content required to produce an equilibrium. The curvature metric quantifies the average curvature of the streamlines. Here a streamline of the current potential is expressed as a curve parameterized by the length, $\bm{r}(l)$, and the curvature is defined as $|\bm{r}''(l)|$. These metrics are described in \cite{Landreman2017,Paul2018,Drevlak2018}.}
    \label{tab:coil_metrics_winding}
\end{table}



We can consider the formulation in \eqref{eq:chi2} to be a form of Tikhonov regularization. As discussed in Section \ref{sec:regularization}, there is a unique global minimum of $\chi^2$ due to its strict convexity. This global minimum of $\chi^2$ can be found by solving a linear least-squares system for $\{\Phi_{m,n}\}$. Thus for a given winding surface and target plasma boundary, winding surface methods are guaranteed to obtain the \textit{global} minimum of $\chi^2$ by solving a single linear system. Once the current potential is obtained, a set of filamentary coils can be taken to be a set of the contours of the current potential. For a fixed target plasma surface and winding surface, the current potential method allows one to very efficiently compute optimized coil shapes. However, this method is based on very crude assumptions. Certain coil topologies, such as inter-linked coils, cannot be treated with this method, and effects due to the finite nature of filamentary coils are not included. It remains an important tool in the stellarator design community, as the result from a current potential calculation provides a reasonable initial guess for filamentary methods described in the following Section. 

The first implementation of this method was the NESCOIL code \cite{Merkel1987}, which solves the problem without regularizaion ($\lambda = 0$), or $\chi^2 = \chi^2_B$. Due to its ill-posedness, a truncated SVD approach was implemented during the NCSX design \cite{Pomphrey2001}. Tikhonov regularization was introduced with the REGCOIL code \cite{Landreman2017} by adding the second term to \eqref{eq:chi2}. An example REGCOIL calculation for the NCSX equilibrium is shown in Figure \ref{fig:regcoil_NCSX}. 

The calculations described in this Section have assumed a fixed winding surface. Often an initial guess for such a surface is taken to be a surface uniformly offset from $\partial \Omega$. In reality, $S_{\text{coil}}$ itself can be optimized \cite{Paul2018} to improve properties of the resulting coils. For example, such an optimization was performed during the design of the W7-X coils \cite{Grieger1992}.

\subsubsection{Filamentary methods}
\label{sec:filamentary}

\begin{figure}
    \centering
    \includegraphics[width=0.8\textwidth]{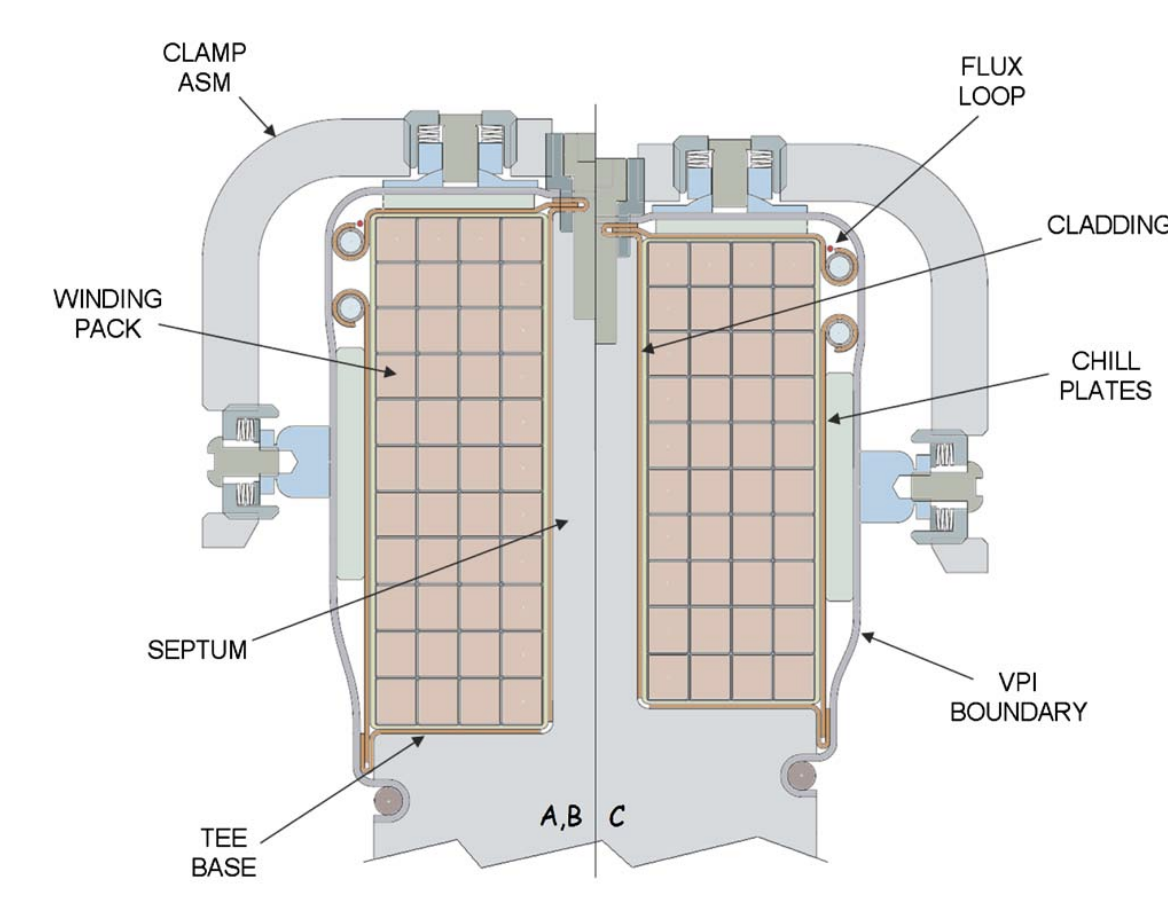}
    \caption{A cross-section of a modular coil of the NCSX stellarator, displaying the conducing material which comprises the winding pack in addition to support and cooling structures. Figure reproduced from \cite{Stratton2006}.}
    \label{fig:ncsx_winding}
\end{figure}

As opposed to the winding surface methods, filamentary methods assume that coils can be approximated by a closed set of curves. Each curve will then be the center of the winding pack, the material making up the physical coil. For context, the winding pack of a NCSX modular coil is provided in Figure \ref{fig:ncsx_winding}.

In the optimization problem, the current is then assumed to be supported by a finite number of lines, and the unknowns are the shape of these lines $\{C_k\}$, and the magnitude of current through each, $\{I_k\}$. The parameterized curves, $\{\bm{x}_k(s)\}$, are often represented by a set of Fourier coefficients,
\begin{align}
    \bm{x}_k(s) = \bm{x}_0^{k,c} + \sum_{m>0} \bm{x}_m^{k,c} \cos(m s) + \bm{x}_m^{k,s} \sin(m s).
\end{align}
Here $s \in[0,2\pi)$ parameterizes the length along the coil. Thus the optimization parameters are the sets of Fourier coefficients together with the set of currents. The objective function is often taken to be the sum of squares as in \eqref{eq:chi2_fixedboundary}, with at least one term that quantifies the departure from the desired equilibrium, such as,
\begin{align}
    \chi^2_B = \int_{\partial \Omega} \left(\bm{B} \cdot \hat{\bm{n}}\right)^2 \, d^2 x ,
\end{align}
in addition to other terms which quantify the desired engineering properties of the coils. The optimization problem then reads: 
\begin{equation}
        \min_{\{ \bm{x}^{k,c}, \bm{x}^{k,s},I_k\}} \chi^2 \text{ where } \chi^2 := \chi^2_E + \chi^2_C, 
        \label{eq:chi2fil}
\end{equation}
where $\chi^2_E$ accounts for properties of the equilibrium and $\chi^2_C$ accounts for properties of the coil geometry. There are many choices for the objective function. For example, the FOCUS code employs the coil length as a form of regularization, as coils that are lengthy are more expensive to build and may develop large curvature. The ONSET \cite{Drevlak1998} code includes terms that penalize the curvature and minimum distance between coils. The COILOPT \cite{Strickler2002} code, developed during the design of NCSX, includes terms that penalize the coil-plasma separation, and an updated version of the code \cite{Brown2015} enables a constraint that imposes for the filaments to be straight on the outboard side of the device. Table \ref{tab:coil_metrics} provides a summary of common metrics employed during filamentary coil optimization. 

As a side note, the set of filamentary coils can also be modeled by a set of closed curves on a winding surface, which itself is allowed to evolve throughout the optimization \cite{Drevlak1998,Strickler2002}. The winding surface may be constrained to lie within two bounding toroidal surfaces. 

While for the winding surface approximation the objective function can be chosen to lead to a convex optimization problem that can be solved with a linear least-squares method, the filamentary methods require the solution of a non-convex, non-linear optimization problem. Therefore, several local minima may exist, and the result of a configuration optimization will generally depend on the initial guess. The output of a winding surface calculation can therefore provide a reasonable initial solution for such non-linear optimization methods.

\begin{table}[]
    \centering
    {\renewcommand{\arraystretch}{1.7}
    \begin{tabular}{|c|c|}
    \hline \hline 
     Length  &  $L_i = \oint_{C_i} dl_i$  \\  \hline
     Curvature  & $\kappa = |\bm{r}_i''(l)|$ \\  \hline
     Torsion & $\tau_i = \frac{|\bm{r}_i'(l) \cdot ( \bm{r}_i''(l) \times \bm{r}_i'''(l))|}{|\bm{r}_i''(l)|^2}$ \\ \hline
     Coil-coil separation & $\min_{i,j,l} \text{dist}(\bm{r}_i(l),\bm{r}_{j \ne i}(l))$ \\ \hline
     Coil-plasma separation & $\min_{i,\theta_p,\zeta_p,l} \text{dist}(\bm{r}_i(l),\bm{r}(\theta_p,\zeta_p))$\\
     \hline \hline 
    \end{tabular}}
    \caption{Properties of a set of coils approximated by  filamentary curves, $\{C\}_i$, parameterized by the length along the curve $l$. In the above expressions, $\bm{r}(\theta_p,\zeta_p)$ denotes the position on the desired plasma boundary, $\partial \Omega$, parameterized by angles $\{\theta_p,\zeta_p\}$. In practice, the maximum of the curvature is especially important, as it sets the allowable thickness of finite-build coils. These metrics are discussed in \cite{Zhu2018,Drevlak2018,Hudson2018}.}
    \label{tab:coil_metrics}
\end{table}



\subsection{Examples of optimized configurations}
\label{sec:device_examples}

Now that we have discussed numerical methods and challenges associated with stellarator design, we provide a brief history of some optimized stellarators. While early stellarator experiments were not optimized, we present an overview of these efforts to provide historical context in Section \ref{sec:early_stellarators}. 


\subsubsection{Early stellarator experiments}
\label{sec:early_stellarators}

\begin{figure}
    \centering
    \includegraphics[width=0.8\textwidth]{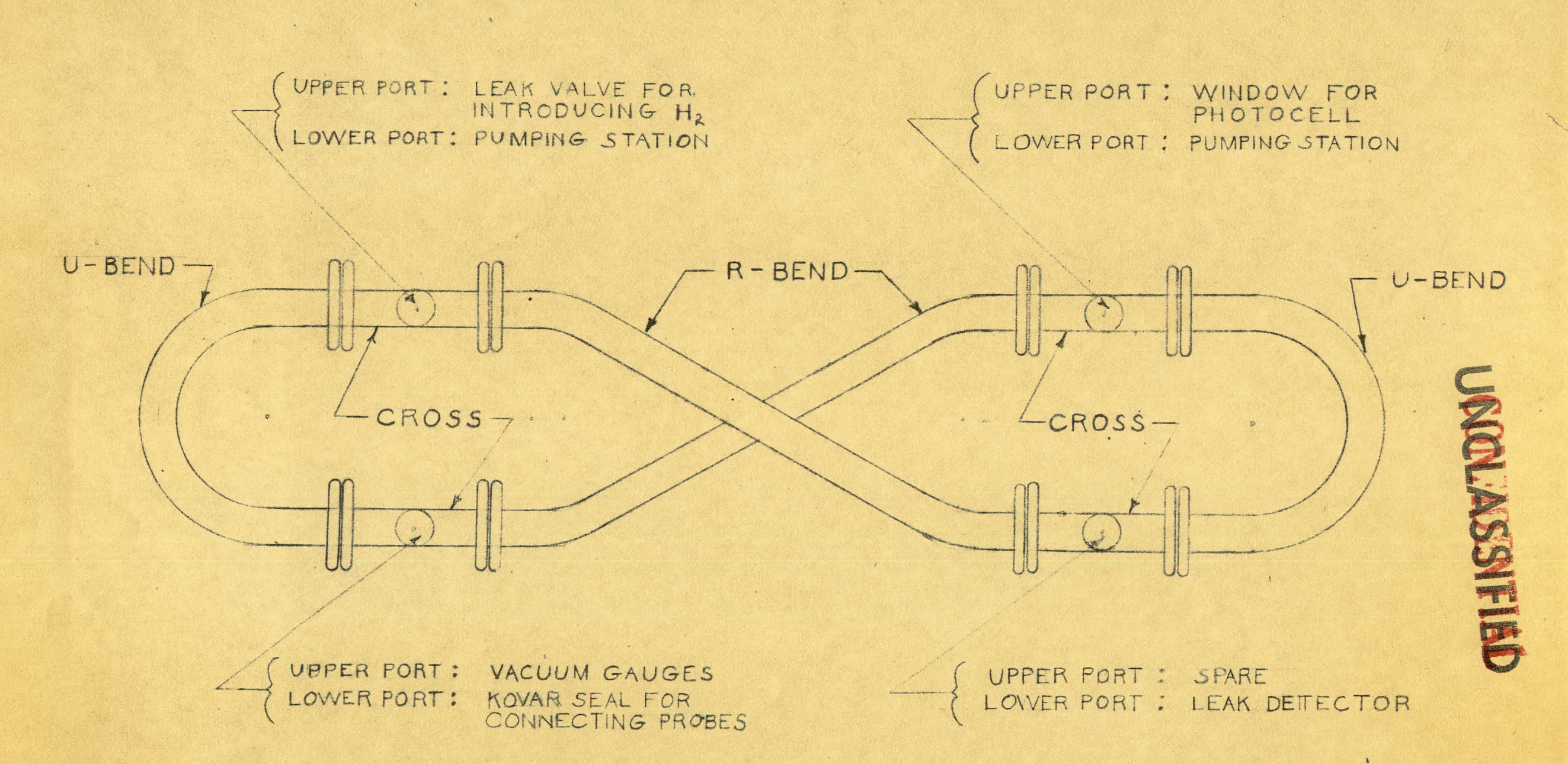}
    \caption{A diagram of the Princeton Model A stellarator based on Lyman Spitzer's figure-eight design. Figure reproduced from \cite{Willis1953}.}
    \label{fig:modelA}
\end{figure}

As discussed in Section \ref{sec:producing_rotational_transform}, rotational transform can be produced with three physical mechanisms: by current in the plasma, twisting ellipticity of magnetic flux surfaces, and non-planarity (torsion) of the magnetic axis. The first stellarators, based on Lyman Spitzer's ``figure-eight" design \cite{Spitzer1953}, used torsion of the magnetic axis to produce rotational transform. Such a configuration is known as a \textit{heliac}, a stellarator whose rotational transform is provided by a set of planar toroidal field coils arranged such that their centers follow a toroidal helix. Such a configuration produces a helical magnetic axis. The confinement was provided by a set of circular, planar toroidal field coils, and the plasma was contained in a twisted glass tube of circular cross-section (Figure \ref{fig:modelA}). The figure-eight design was the basis for several early stellarator experiments at Princeton, the Model A and B devices. 

While the figure-eight design provided rotational transform, Spitzer discovered that stability properties could be improved by introducing magnetic shear, a non-zero derivative in the rotational transform, $\iota'(\psi)$. This shear was introduced experimentally in the Model C stellarator with helical coils, which link the plasma both poloidally and toroidally \cite{Stix1998}. Such a configuration is known as a \textit{heliotron}\footnote{ Sometimes a distinction is made between a heliotron and a similar configuration known as a \textit{torsatron}. We will instead use the term heliotron to refer to a stellarator with any helical coil.}. While the Model C team was able to experimentally demonstrate the existence of magnetic surfaces \cite{Sinclair1970}, the device was plagued by poor particle confinement \cite{Yoshikawa1985}. These early stellarator experiments at Princeton operated until the late 1960s when promising results from the Soviet T-3 tokamak became available, and it was decided that the Model C would be converted to a tokamak \cite{PPPLTimeline}. 

\begin{figure}
    \centering
    \includegraphics[width=0.8\textwidth]{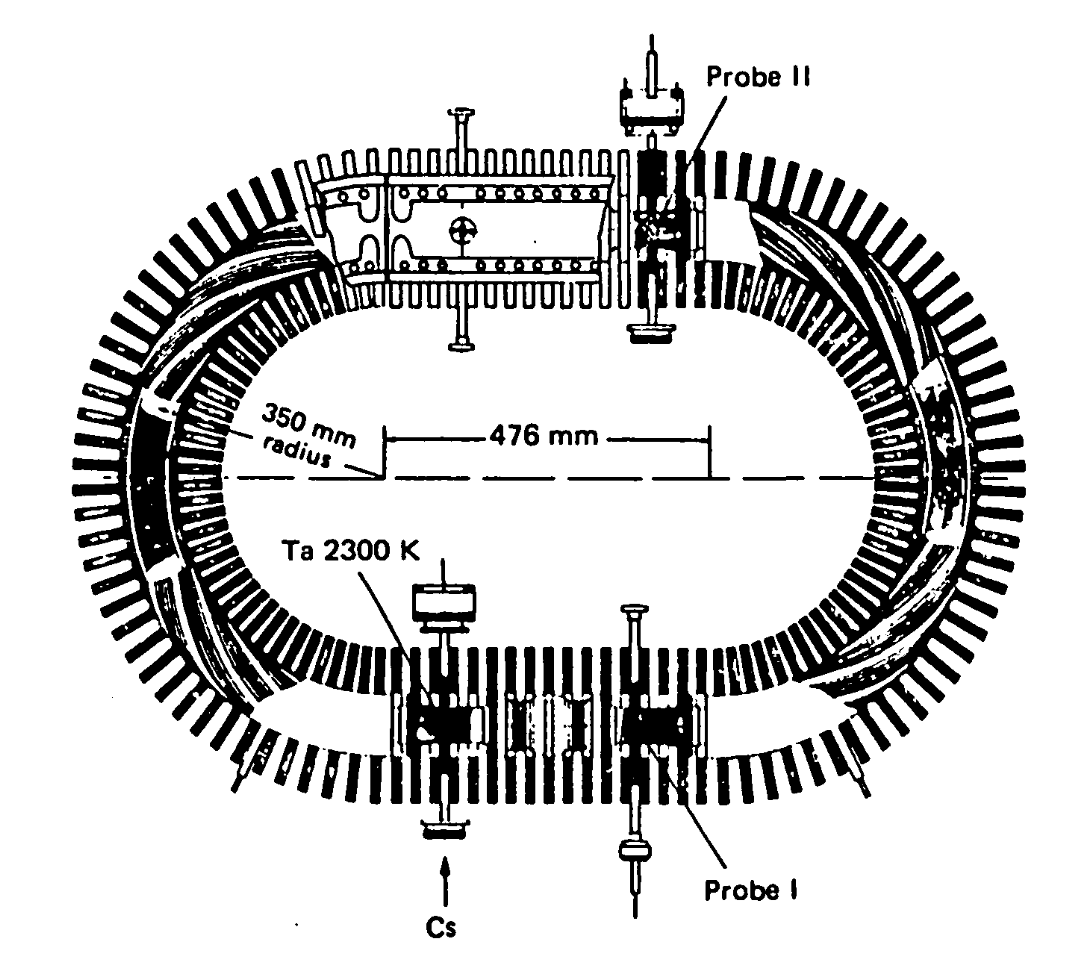}
    \caption{A schematic diagram of the Wendelstein-I (WI) experiment, which adopted the racetrack design and helical coils of the Princeton Model C device.}
    \label{fig:WI}
\end{figure}

Meanwhile, the Wendelstein line of stellarators was active at IPP Garching, initially adopting designs similar to Princeton's Model C with the WI-A and WI-B devices (Figure \ref{fig:WI}). Experiments on WII-A, which operated from 1968-1974, provided insight into the benefits of low magnetic shear and accurate construction of the coil system for avoiding magnetic islands \cite{Berkl1968}. The performance continued, however, to be limited by poor transport properties and low equilibrium beta limits due to the Shafranov shift \cite{Hirsch2008}, an outward shift of the magnetic axis. 

\begin{figure}
    \centering
    \includegraphics[trim=1cm 1cm 0cm 0cm,clip,width=0.8\textwidth]{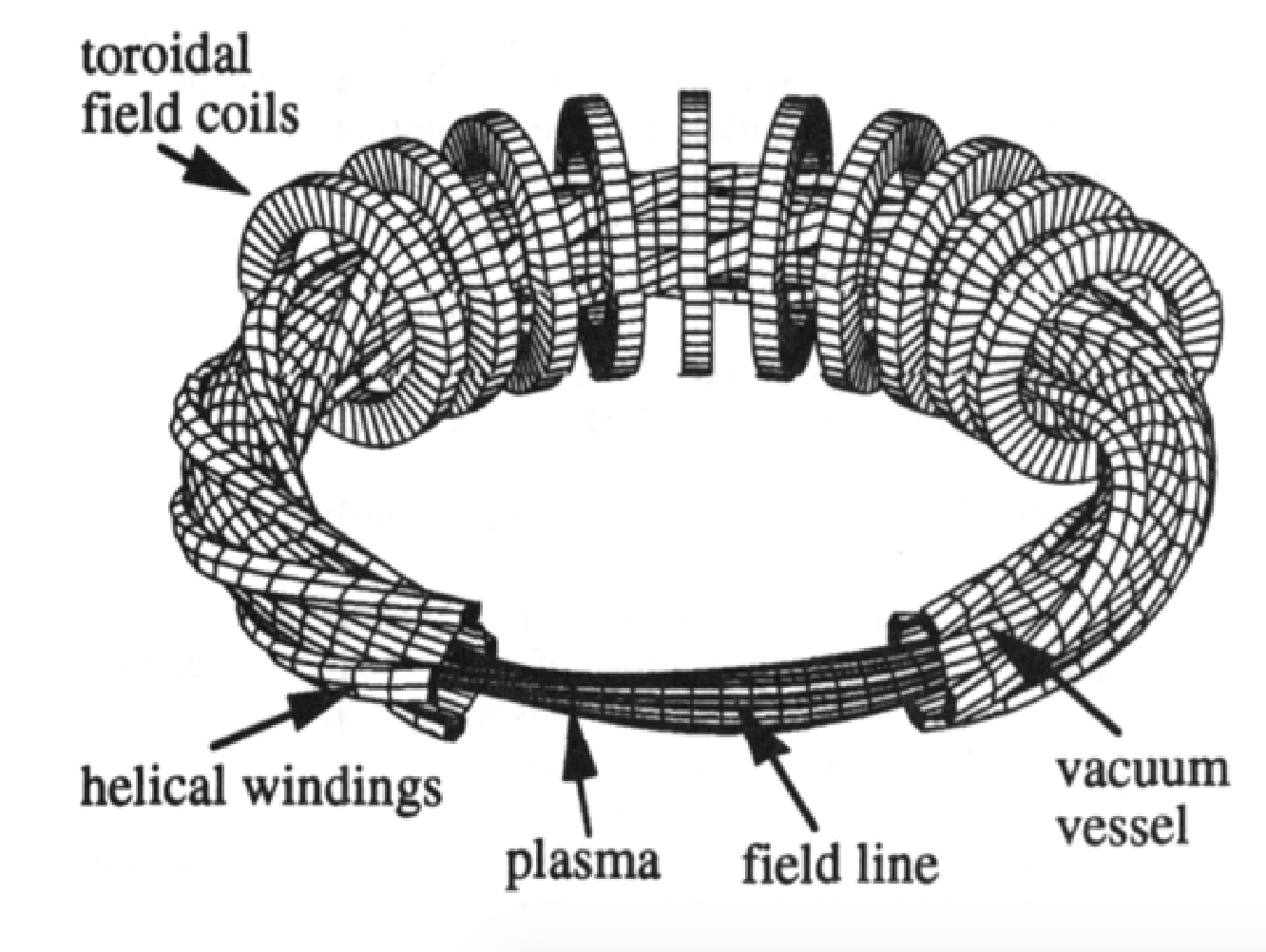}
    \caption{The coil system of the Wendelstein 7-A stellarator, a classical stellarator whose confinement is provided by a set of helical and planar toroidal field coils. Figure reproduced from \cite{Wagner1998}.}
    \label{fig:W7A}
\end{figure}

\subsubsection{Wendelstein 7-Advanced Stellarator (W7-AS)}

The first experiment designed using optimization techniques 
was the Wendelstein 7-Advanced Stellarator (W7-AS). The principal objective of the design was minimization of equilibrium Pfirsch-Schl\"{u}ter currents \cite{Hofmann1996}. Here the Pfirsch-Schlu\"{u}ter current is defined as the first term in \eqref{eq:parallel_current_sum}, a parallel current driven by pressure gradients which varies on a surface. As discussed in Section \ref{ssec:objectives}, this particular type of current gives rise to a Shafranov shift; thus a reduction of these parallel currents can increase the maximum pressure at which good flux surfaces exist.

\begin{figure}
    \centering
    \includegraphics[width=0.8\textwidth]{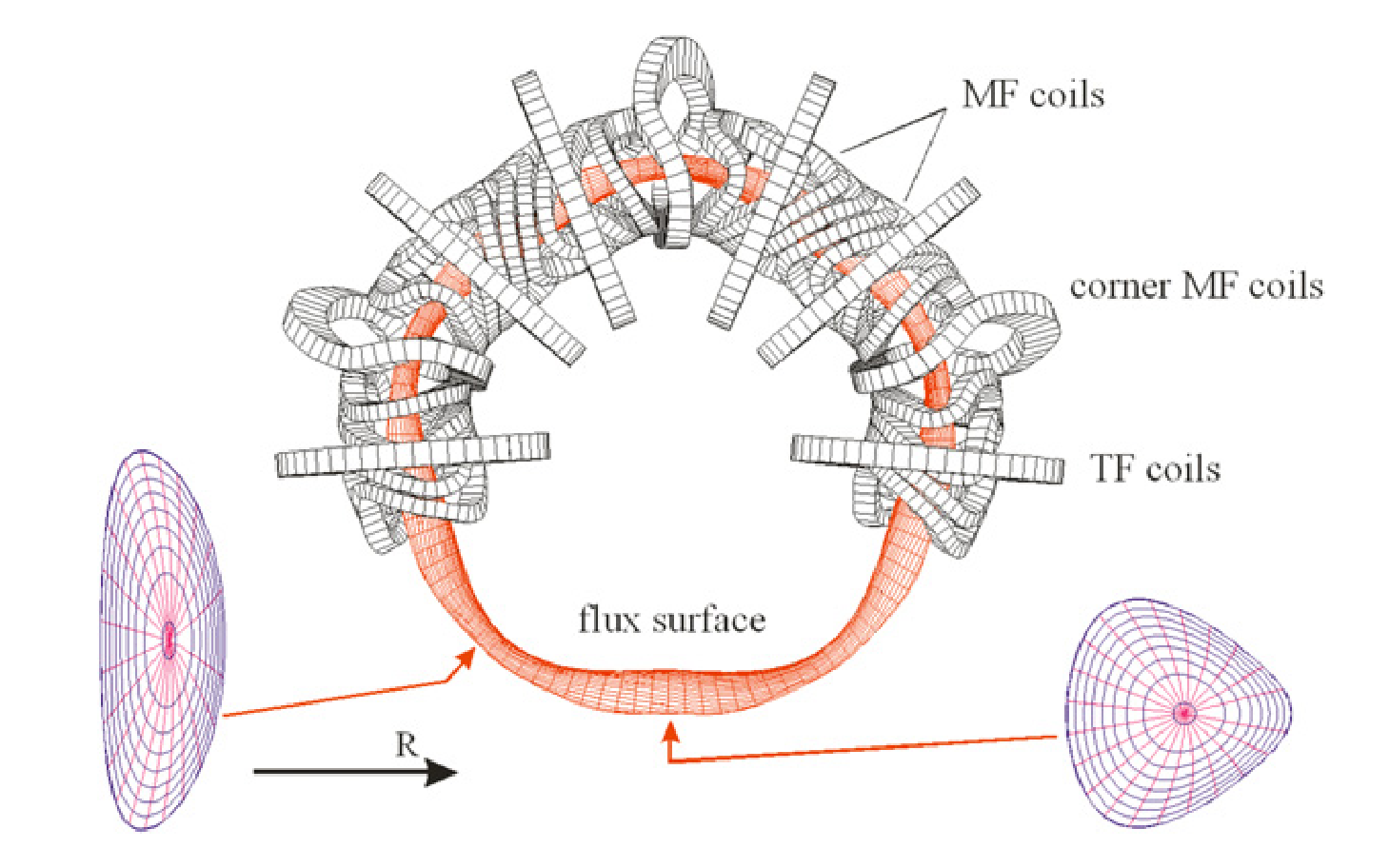}
    \caption{The modular field coils, planar toroidal field coils, and flux surface of the Wendelstein 7-Advanced Stellarator (W7-AS). Figure reproduced from \cite{Hirsch2008}.}
    \label{fig:W7AS}
\end{figure}

To achieve this, the geodesic curvature, the component of the magnetic field curvature which is tangent to a flux surface, defined as
\begin{align}
    \kappa_g = \left(\nabla \psi \times \hat{\bm{b}} \right)\cdot  \left(\hat{\bm{b}} \cdot \nabla \hat{\bm{b}} \right),
\end{align}
was minimized. In comparison with the unoptimized W7-A configuration, see Figure \ref{fig:W7A}, W7-AS achieved a reduction of the parallel currents by about a factor of 2. In addition to reduction of the plasma current, this resulted in reduced neoclassical transport in the higher collisionality plateau and Pfirsch-Schl\"{u}ter regimes and improved guiding center confinement \cite{Hirsch2008}. The magnetic field was produced by modular coils, which link the plasma poloidally, rather than helical coils, which link the plasma both poloidally and toroidally. This  coil topology is represented in Figure \ref{fig:W7AS}. It makes the careful design of the magnetic field more tractable and reduces electromagnetic forces between interlinked coils that were present in early stellarator devices. The coils of W7-AS were designed using a current potential method similar to that described in Section \ref{sec:winding_surface}, with the desired plasma configuration defined by an analytic solution of Laplace's equation, known as Dommaschk potentials \cite{Dommaschk1986}. W7-AS, which operated from 1988-2002, demonstrated the success of the stellarator optimization technique, including experimental verification of a reduction of neoclassical transport in higher collisionality regimes and improved  stability properties \cite{Hofmann1996}. 

\subsubsection{Wendelstein 7-X (W7-X)}

The resounding scientific success of W7-AS served as validation of the fixed-boundary optimization technique, wherein plasma properties are first optimized with the requisite coils subsequently determined. In particular, the W7-AS experiment was able to demonstrate that modular coils could provide a large volume of good magnetic surfaces, with a few islands that are sufficiently small to not inhibit confinement properties. Furthermore, W7-AS demonstrated that parallel currents could be reduced with equilibrium optimization. Experimental verification of neoclassical theory was also performed \cite{Erckmann1994,Brakel1997}, providing confidence in optimization of the bootstrap current and neoclassical transport for the later design of W7-X. There were also significant theoretical developments after the design of W7-AS, in particular the discovery of quasisymmetry by Boozer in 1983 \cite{Boozer1983b}. Soon after, in  Garching, N\"{u}hrenberg and Zille demonstrated that quasisymmetric configurations could be obtained through equilibrium optimization \cite{Nuhrenberg1988}. The W7-X configuration was designed using these equilibrium optimization techniques
to achieve additional objectives: nested magnetic surfaces, fast-particle confinement, reduced parallel currents, minimal neoclassical transport at low collisionality, and MHD stability up to an average beta of $5\%$ \cite{Beidler1990}. The resulting configuration was quasi-isodynamic, a quasi-omnigeneous magnetic field with poloidally closed econtours of the magnetic field strength \cite{Nuhrenberg1994,Helander2009}. 

\begin{figure}
    \centering
    \includegraphics[width=0.8\textwidth]{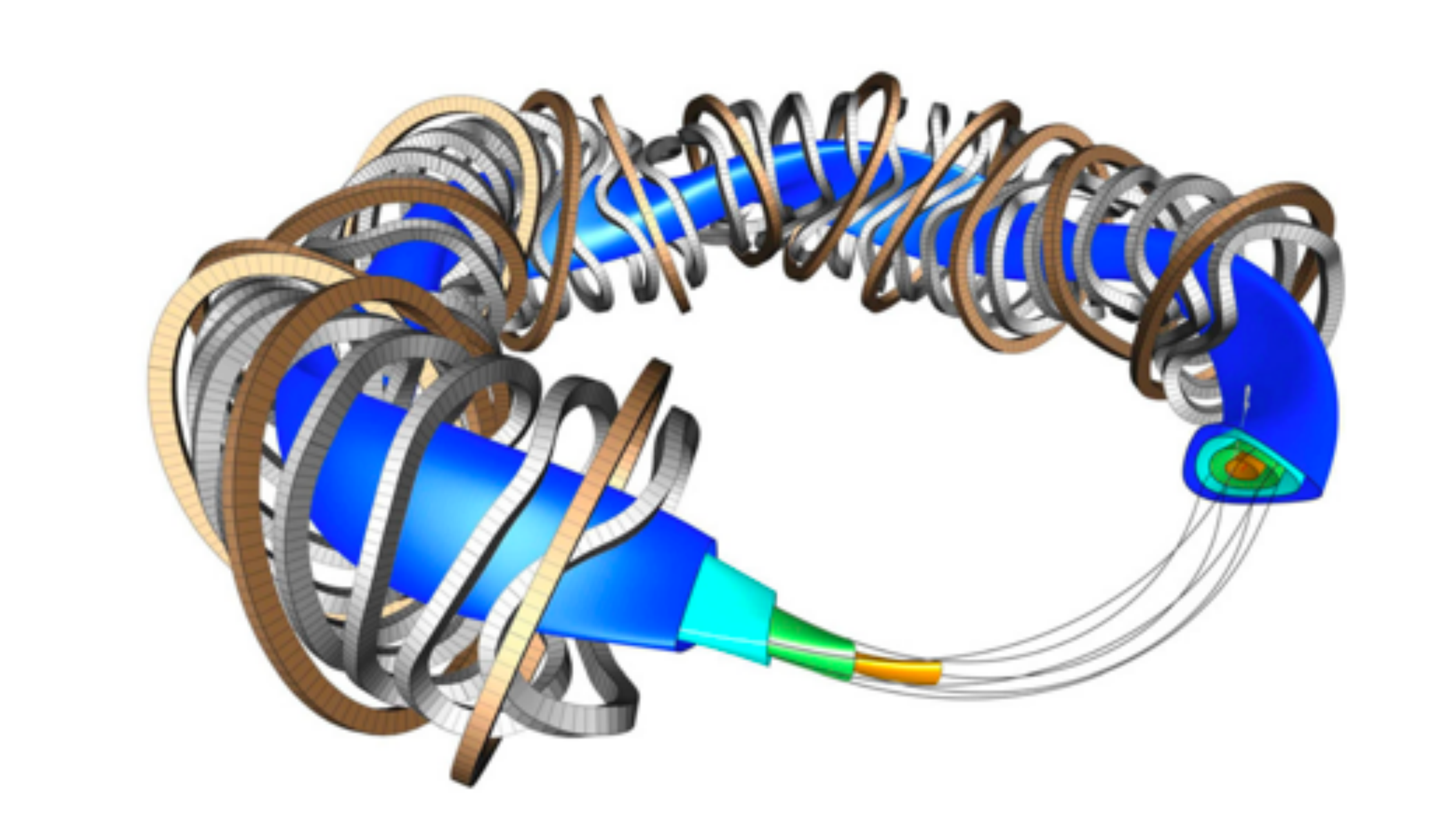}
    \caption{Modular field coils (silver), toroidal field coils (bronze), and magnetic surfaces of the W7-X stellarator. Figure reproduced from \cite{Pedersen2017}.}
    \label{fig:w7x}
\end{figure}

As can be seen in Figure \ref{fig:w7x}, the device's coils are geometrically complex, with significant non-planarity. The construction of W7-X took more than ten years and the project faced numerous challenges associated with quality assurance and limitations in engineering capacity, resulting in delays and significant project recovery efforts. These issues have been attributed to insufficient design tolerances and very strict margins and clearances, which considerably impacted project delivery \cite{klinger2013}. Due to tight tolerances, the construction of the 50 modular coils alone took $10^6$ man-hours \cite{Bosch2017}. Nonetheless, W7-X was ultimately completed successfully, and the first experimental campaign began in 2015. 

Experiments from the initial campaigns of W7-X have demonstrated the success of the stellarator equilibrium optimization concept, confirming the desired magnetic geometry to within a tolerance of $10^{-5}$ \cite{Pedersen2016}. High-beta operation will not be demonstrated until an actively-cooled divertor is installed for the next operating campaign. However, there is initial evidence that recent high-performance shots could not have been achieved without neoclassical optimization \cite{Wolf2019}.

\subsubsection{Large Helical Device (LHD)}

\begin{figure}
    \centering
    \includegraphics[width=0.8\textwidth]{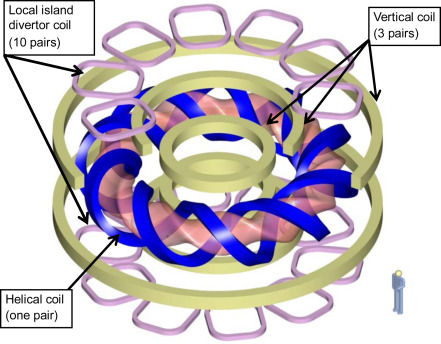}
    \caption{The plasma (pink) and coils of the Large Helical Device (LHD). Figure reproduced from \cite{Kaneko2016}.}
    \label{fig:LHD}
\end{figure}

The success of the W7-A experiment and other early stellarators \cite{Sudo1990}, which demonstrated energy confinement comparable to some tokamak operating regimes, inspired physicists at the National Institute for Fusion Studies in Japan to probe the performance of stellarators at a larger size and higher plasma temperature. As opposed to W7-AS and W7-X, which utilized modular coil systems, LHD was designed with a helical coil system. This coil type may provide certain advantages, including the presence of large magnetic shear at the edge, which improves stability and equilibrium properties, as well as a simplified divertor system \cite{Yamazaki1992}. The design of LHD aimed to achieve $\beta = 5\%$, maximal confinement properties, and sufficient distance between coils and plasma to enable simple installation. Rather than performing equilibrium optimization, the winding of the helical coil was chosen based on simplified scaling relations of physics and engineering properties on coil parameters. These models were constructed from a database of simulations, empirical scalings based on experimental measurements, and theoretical scalings of confinement \cite{Yamazaki1992}. In particular, the device parameters were chosen to maximize the triple product, $n T \tau_E$, where $n$ is the plasma density, $T$ is the temperature, and $\tau_E$ is the energy confinement time, a parameter which must be sufficiently large in order for the fusion reaction to be self-sustaining, as described in Section \ref{sec:magnetic_confinement_fusion}. The resulting design is shown in Figure \ref{fig:LHD}.

The LHD experiment enables great flexibility in the magnetic configuration space by adjusting the coil currents. The standard configuration features good MHD stability properties, but relatively poor neoclassical confinement properties. Another configuration is also available in which the magnetic axis is shifted inward. In this inward-shifted configuration, the MHD stability properties are predicted to be deteriorated, but the neoclassical diffusion is reduced by about a factor of ten \cite{Murakami2002}. Therefore, although LHD was not designed based on equilibrium optimization studies as W7-AS and W7-X were, the inward-shifted configuration is sometimes considered to be an optimized configuration, and improved energetic particle confinement \cite{Murakami2004} and energy confinement \cite{Yamada2001} have been confirmed in comparison with the standard configuration. 

\subsubsection{Helically Symmetric eXperiment (HSX)}

\begin{figure}
    \centering
    \includegraphics[width=0.8\textwidth]{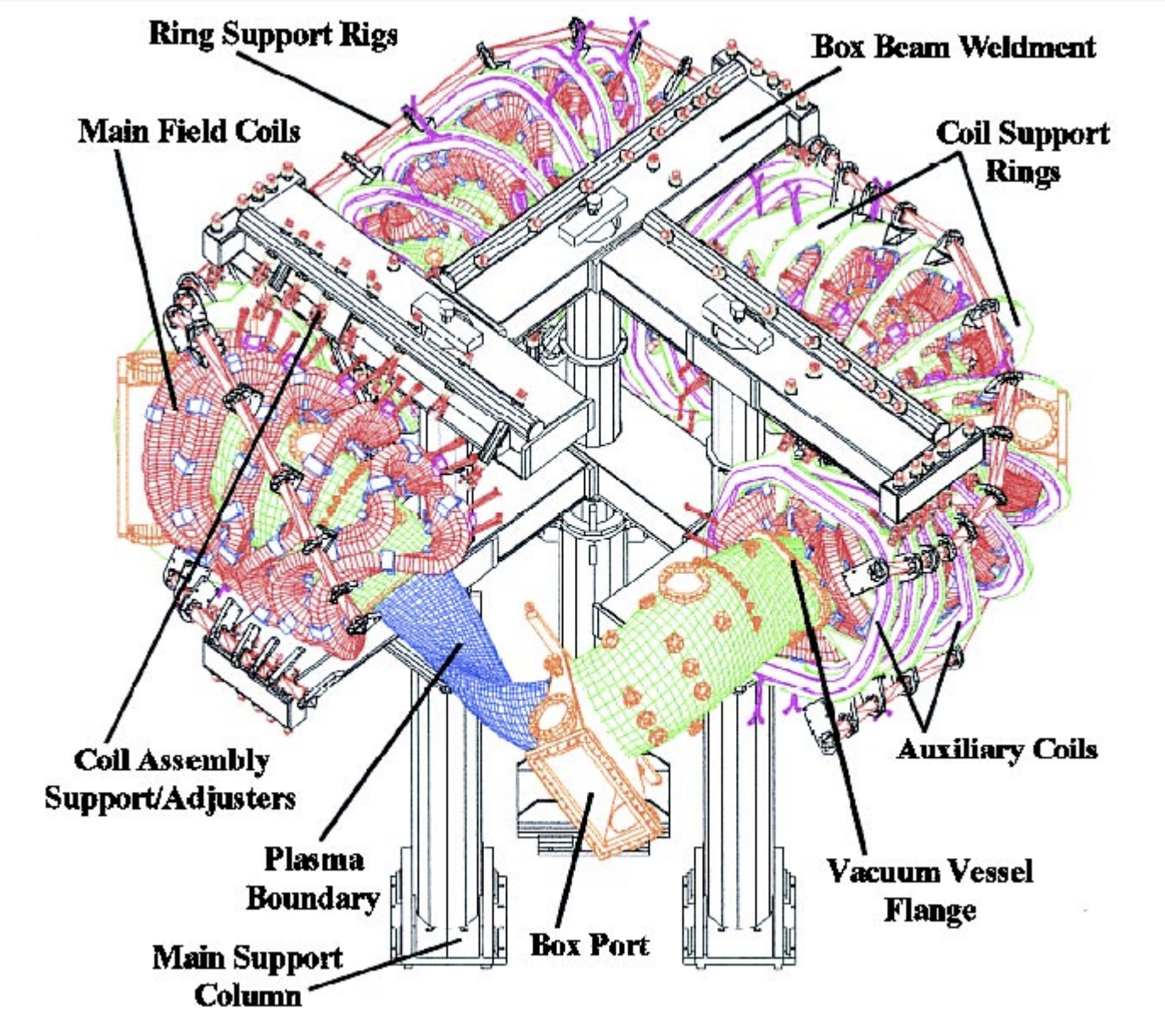}
    \caption{A schematic diagram of the Helically Symmetric Experiment (HSX). Figure reproduced from \cite{Almagri1999}.}
    \label{fig:HSX}
\end{figure}

The Helically Symmetric Experiment (HSX) was the first, and is still  presently the only, experimental validation of the quasisymmetry concept. The design of the device was based on the equilibrium optimization concept pioneered by Garching scientists \cite{Nuhrenberg1994}. The magnetic equilibrium was designed to have quasi-helical symmetry, Mercier stability, and low magnetic shear \cite{Anderson1995} using the equilibrium optimization tools developed by the Wendelstein team \cite{Anderson2019}. The coils were then designed using a current potential method, as described in Section \ref{sec:winding_surface}. The resulting design is shown in Figure \ref{fig:HSX}. HSX has demonstrated a reduction of electron heat transport \cite{Canik2007} and a reduction of flow damping in the symmetry direction \cite{Gerhardt2005}, both of which are predicted to occur in quasisymmetric devices.

\subsubsection{National Compact Stellarator eXperiment (NCSX)}

\begin{figure}
\centering
\includegraphics[width=0.8\textwidth]{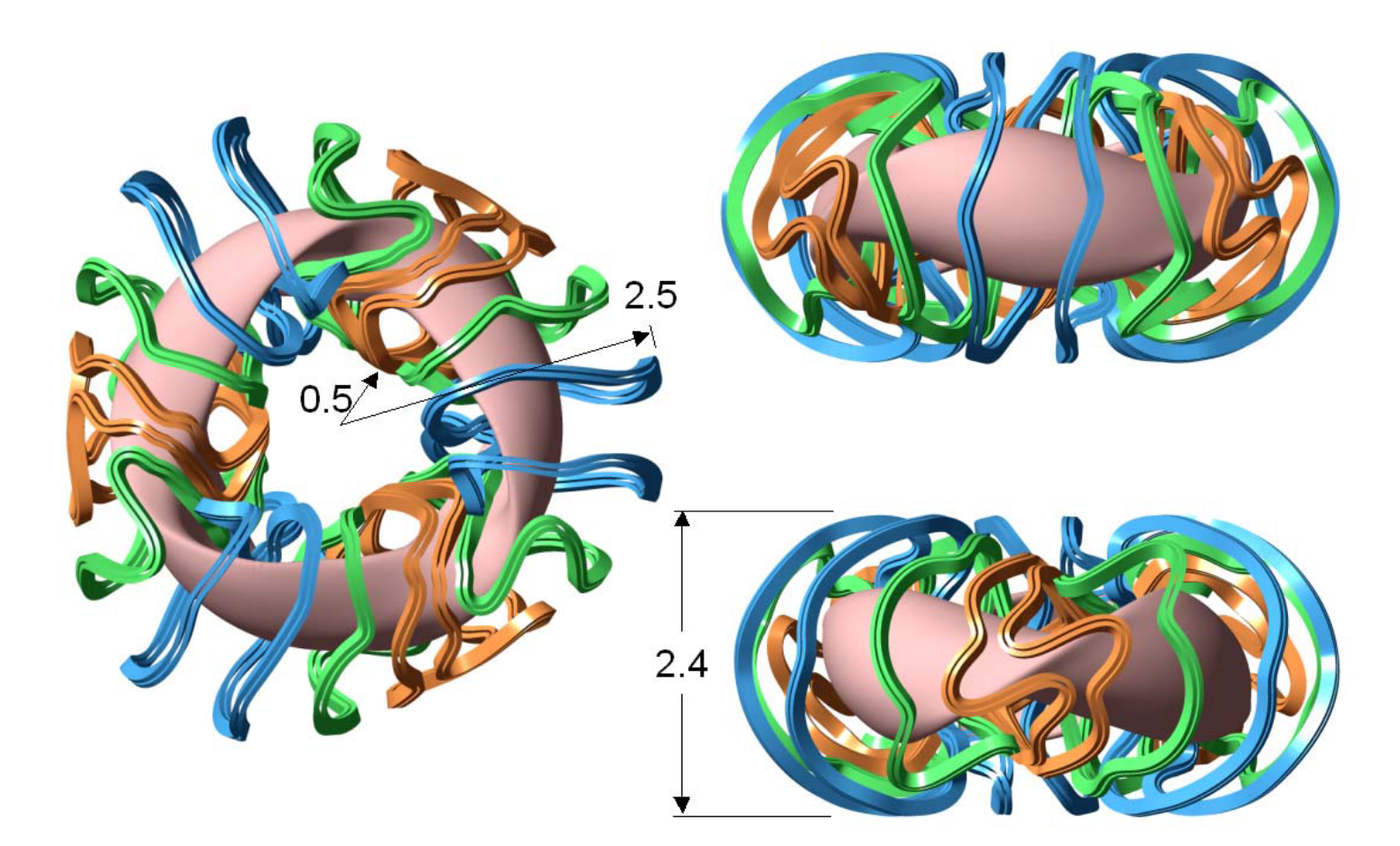}
\caption{The NCSX modular coils obtained through integrated coil-plasma optimization. Figure reproduced from \cite{Strickler2003}.}
\label{fig:ncsx}
\end{figure}

Following completion of the construction of the Helically Symmetric Experiment, a team based at the Princeton Plasma Physics Laboratory (PPPL) began the design of a quasi-axisymmetric stellarator, the National Compact Stellarator eXperiment (NCSX). An advantage of quasi-axisymmetric (QA) over quasi-helical symmetry (QH) is an increased magnitude of the bootstrap current. If the bootstrap current provides a source of rotational transform in addition to the external coils, the coil complexity may be reduced and a more compact device may be possible. Plasma current can also provide island healing \cite{Hegna1998}, reducing the width of islands in comparison with those in the vacuum configuration. NCSX was designed using the fixed-boundary equilibrium optimization technique to achieve several physics objectives: demonstration of MHD stability at $\beta=4.5\%$, a monotonically increasing $\iota$ profile (which is predicted to reduce island widths and provide stabilizing properties \cite{Yokoyama2001}), good flux surface quality, and good confinement properties provided by a quasi-axisymmetric magnetic field. The STELLOPT code \cite{Spong1998,Reiman1999} was written and used for the design of the NCSX equilibrium and remains in use today. The coils were then designed using the COILOPT code  \cite{Strickler2002}, which models the currents as filamentary lines discussed in Section \ref{sec:filamentary}. In the final stages of the NCSX design, the coils and plasma were optimized simultaneously \cite{Hudson2002,Strickler2003}; that is, rather than optimize the plasma boundary for equilibrium properties and design the coils as a second step, the coils are optimized directly for equilibrium properties. This enabled the NCSX team to obtain a set of coils, represented in Figure \ref{fig:ncsx}, which simultaneously meet the physics objectives and coil engineering requirements, such as increased coil-coil spacing, increased coil-plasma spacing, sufficient access for neutral beams, and a sufficiently large curvature radius.

The construction of NCSX began in 2004. Like W7-X, the NCSX project faced significant implementation challenges due to the complexity of the device design. The inability to meet the stringent tolerances led to a redesign of critical components after construction of the device was already underway. 
The revisions resulted in a 50\% increase in the projected final cost, and a four-year delay in expected completion. It ultimately resulted in cancellation of the project by the funding body \cite{dudek2009}. An analysis of the project indicated that the majority of the cost growth arose due to the tight tolerances required on the modular coil assembly \cite{Strykowsky2009}, as calculations of the sensitivity of the magnetic island width indicated that an engineering tolerance of $\approx 1.5$ mm on coil positions was required \cite{Brooks2003}. Lessons learned from both the W7-X and NCSX projects emphasize the importance of coil optimization which accounts for practical constraints, most importantly including reasonably achievable tolerances and clearances.

\subsubsection{Summary of major stellarator experiments}

In the Table \ref{tab:stellarator_experiments}, we provide a summary of major stellarator experiments. We indicate the rough categorization of each experiment in the third column. Here {\it heliotron} indicates a device whose confinement is provided by a helically-wound coil. A {\it heliac} is a stellarator for which rotational transform is provided by a helical magnetic axis produced by circular toroidal field coils centered around a helical axis. The confinement of CNT is provided by two coils that are linked together, which we denote by an inter-locking type stellarator, while QH indicates quasi-helical symmetry, QA indicates quasi-axisymmetry, and optimized indicates a device that was designed using equilibrium optimization techniques. 

\begin{table}[]
\small
    \centering
    {\renewcommand{\tabcolsep}{3pt}
    \begin{tabular}{|c|c|c|c|c|c|c|c|c|c|}
    \hline 
        \textbf{Device} & \textbf{Location} & \textbf{Type} & $\bm{N_P}$ & $\bm{R}$ \textbf{[m]} & $\bm{a}$ \textbf{[m]} & $\bm{B}$ \textbf{[T]} & \textbf{Status}  \\ \hline 
        Heliotron E \cite{Wakatani1996} & Kyoto & heliotron & 19 & 2.2 & 0.2 & 1.9 
        & 1980-1997 \\
        ATF \cite{Thompson1985} & Oak Ridge & heliotron & 12 & 2.1 & 0.27 & 2 
        & 1988-1994 \\
        W7-AS \cite{Hirsch2008} & Garching & optimized & 5 & 2.0 & 0.18 & 3 
        & 1988-2002 \\
        CHS \cite{Nishimura1990} & Nagoya/Toki & heliotron & 8 & 1.0 & 0.2 & 2 
        & 1989-2006 \\
        H1 \cite{Hamberger1990} & Canberra & heliac & 3 & 1.0 & 0.2 & 0.5 & 1992-2017 \\
        TJ-II \cite{Hidalgo2005} & Madrid & heliac & 4 & 1.5 & 0.2 & 1 
        & 1998-running \\
        LHD \cite{Iiyoshi1999} & Toki & heliotron & 10 & 3.9 & 0.6 & 3 
        & 1998-running \\
        HSX \cite{Almagri1999} & Madison & QH, optimized & 4 & 1.2 & 0.15 & 1.4 
        & 1999-running \\ 
        Heliotron J \cite{Sano2000} & Kyoto & heliotron & 4 & 1.2 & 0.2 & 1.5 
        & 2000-running \\ 
        CNT \cite{Brenner2012} & Columbia & inter-locking, optimized & 2 & 0.30 & 0.15 & 0.2 & 2004-running \\
        CTH \cite{Hartwell2017} & Auburn & heliotron & 2 & 0.75 & 0.29 & 0.7 & 2007-running\\
        NCSX \cite{Zarnstorff2001} & Princeton & QA, optimized & 3 & 1.4 & 0.33 & 1.7 
        & terminated \\
        W7-X \cite{Wurden2017,Wolf2019} & Greifswald & optimized & 5 & 5.5 & 0.53 & 3 & 2015-running 
        \\
        CFQS \cite{Isobe2019} & Chengdu & QA, optimized & 2 & 1 & 0.25 & 1
        & under construction 
        \\
        \hline
    \end{tabular}}
    \caption{A summary of some major stellarator devices \cite{ippsummerschool}, approximately
    in chronological order. Here $N_P$ is the number of field periods (Section \ref{sec:Np_symmetry}), $R$ is the average major radius, $a$ is the average minor radius, and $B$ is the maximum magnetic field strength. All of the above devices were constructed and conducted experiments except for NCSX, whose construction was terminated before its completion.}
    \label{tab:stellarator_experiments}
\end{table}

\section{New frontiers of optimization}
\label{sec:new_frontiers}

In this document, we have aimed to provide the necessary introductory material in order to understand the stellarator concept, the necessity of optimization, and principles of stellarator design. We hope that this content will prove useful to researchers who are new to the field of stellarator theory. In closing, we now discuss recent advances and open problems in stellarator optimization to provide a taste of the current state of the field.
\begin{itemize}
    \item \textbf{Turbulent transport} Experiments such as W7-X and HSX have demonstrated the ability to reduce neoclassical transport with equilibrium optimization methods. However, collisions are not the only process that leads to transport of particles and heat. Transport also results from small-scale fluctuations of electric and magnetic fields, processes that are described by gyrokinetic theory. In fact, recent experiments on W7-X indicate that with neoclassical transport effectively reduced with optimization, turbulence is the dominant transport mechanism \cite{Geiger2019}. Thus an important new frontier in the design of stellarators is the optimization of the magnetic field to reduce turbulent transport. While stellarators have historically been designed for stability with respect to large-scale MHD modes, optimization for microstability is much more challenging due to the computational expense of running non-linear gyrokinetic simulations. There have been efforts to construct proxy functions which faithfully capture the turbulent transport properties of stellarator configurations, such as reducing the overlap between bad curvature and trapping regions \cite{Xanthopoulos2014} or increasing non-linear energy transfer between unstable and damped modes \cite{Hegna2018}.
        \item \textbf{Nonlinear MHD stability} As described in Section \ref{sec:stability}, it is important to ensure that an equilibrium is stable with respect to macroscopic MHD modes, since these can lead to large-scale changes in the plasma possibly degrading confinement. 
        Typically linear ideal MHD stability has been a criterion for stellarator MHD equilibria. However, LHD has been shown to operate in regimes which exceed the pressure limits prescribed by linear stability analysis without substantially degrading plasma performance \cite{Komori2003,Weller2006}. This suggests that the nonlinear operational parameter space may differ significantly compared to limits identified by linear analysis and that the latter may be overly stringent. Given the strong dependence of stellarator confinement on shaping by external coils, developing further understanding of nonlinear MHD stability may expand the space of optimized stellarator configurations, including some with more desirable coil properties.
        
    \item \textbf{Integrated coil design} As described in Section \ref{sec:coil_opt}, stellarator design has traditionally decoupled coil optimization from optimization of the MHD equilibrium. However, such a decoupling may result in an equilibrium which can only be produced with overly-complex coils. As the cost of constructing a stellarator experiment is closely related to coil complexity, it is important to consider how to couple coil design with equilibrium optimization. There are several approaches to address this issue. One option is to directly optimize coils based on free-boundary solutions of the MHD equilibrium equations, eliminating the need to design coils as a second step. For example, this approach was used in the final stages of the NCSX design \cite{Hudson2002,Strickler2003}. The direct optimization approach is somewhat more challenging for several reasons. Free-boundary equilibrium calculations tend to be more expensive than fixed-boundary calculations, as they require iterations between an equilibrium solve and vacuum field calculations. This iterative scheme will not always converge in practice, hence the historical use of the more robust fixed-boundary method. It has also been suggested that fixed-boundary optimization may yield better equilibrium properties, as the model assumes the existence of at least one magnetic surface. An alternative approach is to perform fixed-boundary equilibrium optimization in tandem with the optimization of a set of coils to be consistent with the fixed-boundary \cite{Hudson2018}. Finally, fixed-boundary optimization can be performed with the inclusion of figures-of-merit which quantify the complexity of coils required to produce a given plasma boundary. Examples of proxy functions include properties of the surface current computed with winding surface methods (Section \ref{sec:winding_surface}) and the principal curvatures of the plasma boundary \cite{Drevlak2018,Paul2018}.
    \item \textbf{Advances in numerical optimization} 
    The design of modern stellarators require numerical optimization tools; thus synergy between optimization experts and stellarator physicists can be very fruitful. An example is the application of \textit{stochastic optimization} techniques for the design of stellarator coils. Rather than optimize a single set of design variables, one can optimize the expected value of an objective function by performing a sample average over a distribution of possible deviations of the design variables. These techniques can improve the robustness of the optimum by avoiding small local minima and obtaining solutions with reduced risk. This technique has proven effective for the optimization of coil shapes with increased tolerances \cite{Lobsien2018,Lobsien2020}. Another example is the application of \textit{adjoint methods} \cite{Paul2018,Paul2019}, which allow analytic gradient information to be computed very efficiently. Gradient information can also be obtained through \textit{automatic differentiation} tools, which numerically compute the derivative of an arbitrary function defined by a computer program. This approach is currently being applied to the optimization of coil shapes \cite{simonshour}. The availability of gradient information allows one to more efficiently converge toward an optimum configuration, especially in the high-dimensional spaces of relevance for stellarator design. \textit{Surrogate methods} may also prove useful for stellarator optimization \cite{simonshour}, by replacing an expensive objective function with a relatively cheap model of the data, such as through Gaussian process regression.
    \item \textbf{Direct construction of equilibria} As an alternative to optimizing the plasma boundary or a set of coils to obtain an omnigeneous or quasi-symmetric equilibrium, one can directly construct a configuration by obtaining numerical solutions to the MHD equilibrium equations upon an asymptotic expansion. The techniques used in this approach are similar to those introduced in Section \ref{sec:axis_expansion}, in which a power series expansion of the vacuum magnetic field equations near the magnetic axis was performed in order to evaluate the rotational transform. In particular, this approach has been used to construct quasisymmetric \cite{Landreman2018a,Landreman2018b,Landreman2019,Jorge2020} and quasi-omnigeneous \cite{Plunk2019} configurations near the magnetic axis. This method is several orders of magnitude more computationally efficient than the equilibrium optimization approach. A similar approach is based on an asymptotic expansion in the breaking of axisymmetry to obtain quasi-axisymmetric equilibria \cite{Plunk2018,Plunk2020}. 
    \item \textbf{Permanent magnets} 
    There has been a recent effort to design permanent magnets rather than electromagnetic coils to confine stellarator plasmas. This became possible thanks to advances in materials science, allowing magnetic materials to produce large enough magnetic fields to be of relevance for stellarator confinement. Numerical simulations indicate that using permanent magnets in conjunction with toroidal field coils may significantly reduce coil complexity of a design and improve experimental access to the plasma chamber. Several numerical methods have been developed to optimize the distribution and orientation of permanent magnets, some of which are based on a current potential solution \cite{Helander2020,Zhu2020a}, and some of which require non-linear optimization techniques \cite{Zhu2020b,Hammond2020}.
    \item \textbf{Finite-build coils}
    As was mentioned in Section \ref{sec:coil_opt}, electromagnetic coils have traditionally been optimized with simplifying assumptions by approximating them as a sheet currents or filamentary lines of current. Recently, more realistic modeling of coils has been performed with codes which take into account the finite build of coils \cite{Singh2019}. This requires an optimization over not only the curve defining the center of the winding pack, but of the orientation of the rectangular winding pack with respect to a fixed axis. Such calculations are expected to be more important for smaller devices, where the size of the winding pack relative to the plasma-coil spacing may be significant.
    \item \textbf{Theoretical understanding of quasisymmetry} 
    While several decades have passed since the discovery of quasisymmetry \cite{Boozer1983b}, there continues to be activity in the theoretical understanding of quasisymmetry. This includes a reformulation of the conditions required for quasisymmetry without assumptions of ideal MHD equilibrium \cite{Burby2019,Rodriguez2020} and the derivation of a Grad-Shafranov equation for quasisymmetric magnetic fields \cite{Burby2019}. There remain open problems in this area. For example, the near-axis expansion of quasisymmetric equilibria results in an overdetermined system of equations at sufficiently high order \cite{Garren1991}. It remains to be seen if it is possible to circumvent the overdetermined nature with a modified formulation or if quasisymmetry can be obtained globally within a volume.
    \item \textbf{Divertor optimization} 
    The divertor systems of magnetic confinement devices must be designed such that the incident heat flux does not exceed the material limits of the plasma-facing components. The interaction of the material structures with the plasma can also lead to the presence of \textit{impurities} in the device, higher atomic number species which can radiate and decrease the plasma energy.
    Stellarator divertor systems have not historically been designed using numerical optimization techniques. Rather, the shape and position of divertor plates are modified manually to achieve an acceptable heat flux and consistency with engineering constraints after the equilibrium and coil shapes have been designed. With the development of simulation tools for modeling of heat flux on a 3D material surface \cite{Lore2014}, numerical optimization of the divertor is now possible \cite{Gates2017}, which may enable improved performance.
\end{itemize}

\appendixpage
\appendix 

\section{Debye shielding}
\label{debye}

As an example of the separation of scales that arise in plasma physics, we discuss the concept of Debye shielding. 

Plasmas are dielectric, in that they tend to shield out electrostatic potentials. The Debye length, $\lambda_D$, is the characteristic length scale on which the electric field produced by a charge is screened by other mobile  charges. It is defined, in the presence of species with number density $n_s$, temperature $T_s$ and charge $q_s$, by
\begin{gather}
\label{eq:lambdad}
\lambda_D = \left( \sum_s \frac{n_s q_s^2}{\epsilon_0 T_s} \right)^{-1/2}.
\end{gather}
If the plasma is cold $(T_s \rightarrow 0)$, then the shielding is perfect and $\lambda_D \rightarrow 0$. In a hot plasma, some charges will have enough thermal energy to escape the electrostatic potential, resulting in a finite $\lambda_D$.

We can find this length scale by considering Poisson's equation (see Section \ref{sec:electrostatics}) for a test charge $q_T$ in the presence of species with number density $n_1^s$ and charge $q_s$
\begin{gather}
\Delta \Phi = -\frac{1}{\epsilon_0} \left(\sum_s n_1^s q_s \right) - \frac{q_T}{\epsilon_0} \delta( \bm{r}). 
\end{gather}
Here $\delta (\bm{r})$ is the Dirac delta function. Note that $n_1^s$ will differ from the density $n_s$ that was introduced previously (as in (\ref{eq:quasineutrality}) and (\ref{eq:lambdad})), as it was assumed then that one is looking at length scales longer than $\lambda_D$, such that the electrostatic potential is screened out. Here we will be considering the density at shorter length scales, so this assumption does not hold.

The densities $n_1^s$ can be expressed in terms of distributions functions $f_s$ as $n_1^s = \int d^3 v \, f_s(\bm{v}) $. The distribution function is normalized such that $f(\bm{v}) d^3v$ is the number of particles per unit volume, so $f$ describes the density of particles in phase space (position and velocity). The Maxwell-Boltzmann distribution can be applied to describe a plasma which has reached thermodynamic equilibrium, often through collisions, and we will make the assumption that both the distribution functions can be described by a Maxwell-Boltzmann distribution,
\begin{align}
f_s(\bm{v}) = A \exp \left(\frac{m_s v^2}{2T_s} - \frac{q_s \Phi}{T_s}\right), 
\end{align}
for normalization constant $A$. So, the densities can be written as $n_1^s =   n_s e^{- q_s \Phi/T_s}$ where $n_s=A \int d^3 v \, \exp \left(\frac{m_s v^2}{2T_s}\right)$ is the number density of species $s$ in the limit $\Phi \rightarrow 0$. In other words, far away from the test charge the potential from the charge is shielded out so that quasineutrality holds, that is to say $\sum_s n_s q_s = 0$. 

Assuming $q_s\Phi/T_s \ll 1$, we can Taylor expand the exponents, 
\begin{align}
\Delta \Phi &= -\frac{1}{\epsilon_0} \sum_s n_s q_s \left(1 - \frac{q_s \Phi}{T_s} \right) - \frac{q_T}{\epsilon_0} \delta(\bm{r}).
\end{align}
Since $\sum_{s} n_s q_s = 0$,
\begin{align}
\frac{1}{r^2} \der{}{r} \left( r^2 \der{\Phi}{r} \right) &= \lambda_D^{-2}  \Phi - \frac{q_T}{\epsilon_0} \delta(\bm{r}).
\end{align}
The solution for electrostatic potential is
\begin{gather}
\Phi = \frac{q_T}{r} \exp \left( - r/\lambda_D \right).
\end{gather}
So on length scales longer than $\lambda_D$, the plasma is quasineutral. In typical fusion plasmas the number of particles in a sphere of radius $\lambda_D$ is very large,
\begin{gather}
\frac{4\pi n \lambda_D^{3}}{3} \gg 1. 
\end{gather}
This allows one to consider the weak coupling assumption, such that the collective motion of the plasma dominates over the short-range electrostatic interaction between charges.

\FloatBarrier
\section*{Acknowledgements}
This work was supported by the Simons Foundation (Grant \# 560651, LMIG).
 The \href{https://hiddensymmetries.princeton.edu/}{Simons Collaboration on Hidden Symmetries and Fusion Energy} has been formed to foster collaboration between experts in numerical optimization, dynamical systems, PDE analysis, and plasma physics in order to find stellarator configurations with hidden symmetries. This work was also supported by the Center for Scientific Computing and Mathematical modeling from the University of Maryland.

The authors would like to thank Matt Landreman, Per Helander, Stuart Hudson, Antoine Cerfon, Joaquim Loizu, Georg Stadler, Rogerio Jorge, Wrick Sengupta, Chris Smiet, Allan Reiman, Tonatiuh S\'anchez Vizuet, Caoxiang Zhu, Nick McGreivy, Sophia Henneberg, Silke Glas, Craig Beidler, Harold Weitzner, Nate Ferraro, Yao Zhou, and Craig Beidler for their valuable comments on the document.

\bibliographystyle{siam}
\bibliography{PEqtionary}

\begin{thebibliography}{100}

\bibitem{simonshour}
\url{https://hiddensymmetries.princeton.edu/meetings/simons-hour-talks}.

\bibitem{NIF}
{\em How {NIF} works}.
\newblock \url{https://lasers.llnl.gov/about/how-nif-works}.
\newblock Accessed: 2019-08-14.

\bibitem{PPPLTimeline}
{\em {Princeton Plasma Physics Laboratory - Timeline}}.
\newblock date accessed: 01/03/2019.

\bibitem{ITER}
{\em What will {ITER} do?}
\newblock \url{https://www.iter.org/sci/Goals}.
\newblock Accessed: 2019-08-14.

\bibitem{ippsummerschool}
{\em {IPP Summer University for Plasma Physics}}.
\newblock
  \url{https://pure.mpg.de/rest/items/item_2146333/component/file_2146332/content#page=197},
  2012.
\newblock Accessed: 2020-06-15.

\bibitem{2015Tongb}
{\em The {Hamiltonian} formalism}.
\newblock \url{http://www.damtp.cam.ac.uk/user/tong/dynamics/four.pdf}, 2015.
\newblock Accessed: 2018-03-05.

\bibitem{2015Tong}
{\em The {Lagrangian} formalism}.
\newblock \url{http://www.damtp.cam.ac.uk/user/tong/dynamics/two.pdf}, 2015.
\newblock Accessed: 2018-03-05.

\bibitem{ippweb2}
{\em {Magnetic Coils and Plasma from Wendelstein 7-X}}.
\newblock \url{https://www.ipp.mpg.de/2523775/konzeptentwicklung}, 2018.
\newblock Accessed: 2018-10-04.

\bibitem{ippweb}
{\em Tokamak}.
\newblock \url{http://www.ipp.mpg.de/14869/tokamak}, 2018.
\newblock Accessed: 2018-10-04.

\bibitem{Almagri1999}
{\sc A.~F. Almagri, D.~T. Anderson, F.~S.~B. Anderson, P.~H. Probert, J.~L.
  Shohet, and J.~N. Talmadge}, {\em A {Helically Symmetric Stellarator (HSX)}},
  IEEE Transactions on Plasma Science, 27 (1999), pp.~114--115.

\bibitem{Anderson2019}
{\sc D.~Anderson}.
\newblock Personal communication, 9 2019.

\bibitem{Anderson1995}
{\sc F.~S.~B. Anderson, A.~F. Almagri, D.~T. Anderson, P.~G. Matthews, J.~N.
  Talmadge, and J.~L. Shohet}, {\em The {Helically Symmetric Experiment}
  ({HSX}): goals, design and status}, Fusion Technology, 27 (1995),
  pp.~273--277.

\bibitem{Angelo2012}
{\sc R.~Angelo, E.~Duzzioni, and A.~Ribeiro}, {\em Integrability in
  time-dependent systems with one degree of freedom}, Journal of Physics A:
  Mathematical and Theoretical, 45 (2012), p.~055101.

\bibitem{Arnold2009}
{\sc V.~I. Arnold}, {\em Proof of a theorem of {AN Kolmogorov} on the
  invariance of quasi-periodic motions under small perturbations of the
  {Hamiltonian}}, Collected Works: Representations of Functions, Celestial
  Mechanics and KAM Theory, 1957--1965,  (2009), pp.~267--294.

\bibitem{Arnoldus2006}
{\sc H.~F. Arnoldus}, {\em Conservation of charge at an interface}, Optics
  Communications, 265 (2006), pp.~52--59.

\bibitem{Bader2019}
{\sc A.~Bader, M.~Drevlak, D.~T. Anderson, B.~J. Faber, C.~C. Hegna, K.~M.
  Likin, J.~C. Schmitt, and J.~N. Talmadge}, {\em Stellarator equilibria with
  reactor relevant energetic particle losses}, Journal of Plasma Physics, 85
  (2019), p.~905850508.

\bibitem{Bauer1987}
{\sc F.~Bauer}, {\em The Beta Equilibrium, Stability, and Transport Codes:
  Applications of the Design of Stellarators}, Elsevier, 1987.

\bibitem{Beidler1990}
{\sc C.~Beidler, G.~Grieger, F.~Herrnegger, E.~Harmeyer, J.~Kisslinger,
  W.~Lotz, H.~Maassberg, P.~Merkel, J.~N{\"u}hrenberg, F.~Rau, et~al.}, {\em
  Physics and engineering design for {Wendelstein VII-X}}, Fusion Technology,
  17 (1990), pp.~148--168.

\bibitem{Berger1999}
{\sc M.~A. Berger}, {\em Introduction to magnetic helicity}, Plasma Physics and
  Controlled Fusion, 41 (1999), p.~B167.

\bibitem{Berger1984}
{\sc M.~A. Berger and G.~B. Field}, {\em The topological properties of magnetic
  helicity}, Journal of Fluid Mechanics, 147 (1984), pp.~133--148.

\bibitem{Berkl1968}
{\sc E.~Berkl et~al.}, {\em Plasma physics and controlled nuclear fusion
  research}, in Proceedings of the 3rd International Conference Novosibirsk,
  vol.~1, 1968.

\bibitem{Bishop1958}
{\sc A.~S. Bishop}, {\em {Project Sherwood} - the {US} program in controlled
  fusion}, 1958.

\bibitem{Boozer1983}
{\sc A.~H. Boozer}, {\em Evaluation of the structure of ergodic fields}, The
  Physics of Fluids, 26 (1983), pp.~1288--1291.

\bibitem{Boozer1983b}
\leavevmode\vrule height 2pt depth -1.6pt width 23pt, {\em Transport and
  isomorphic equilibria}, The Physics of Fluids, 26 (1983), p.~496.

\bibitem{Boozer1995}
\leavevmode\vrule height 2pt depth -1.6pt width 23pt, {\em Quasi-helical
  symmetry in stellarators}, Plasma Physics and Controlled Fusion, 37 (1995),
  p.~A103.

\bibitem{Boozer2015}
\leavevmode\vrule height 2pt depth -1.6pt width 23pt, {\em Non-axisymmetric
  magnetic fields and toroidal plasma confinement}, Nuclear Fusion, 55 (2015),
  p.~025001.

\bibitem{Boozer2019}
\leavevmode\vrule height 2pt depth -1.6pt width 23pt, {\em Stellarators as a
  fast path to fusion energy}, arXiv preprint arXiv:1912.06289,  (2019).

\bibitem{Bosch2017}
{\sc H.-S. Bosch, R.~Brakel, T.~Braeuer, V.~Bykov, P.~van Eeten, J.-H. Feist,
  F.~F{\"u}llenbach, M.~Gasparotto, H.~Grote, T.~Klinger, et~al.}, {\em Final
  integration, commissioning and start of the {Wendelstein 7-X} stellarator
  operation}, Nuclear Fusion, 57 (2017), p.~116015.

\bibitem{Bouquet1998}
{\sc S.~Bouquet and A.~Bourdier}, {\em Notion of integrability for
  time-dependent {Hamiltonian} systems: illustrations from the relativistic
  motion of a charged particle}, Physical Review E, 57 (1998), p.~1273.

\bibitem{Boyd2004}
{\sc S.~Boyd, S.~P. Boyd, and L.~Vandenberghe}, {\em Convex Optimization},
  Cambridge University Press, 2004.

\bibitem{Brakel1997}
{\sc R.~Brakel, M.~Anton, J.~Baldzuhn, R.~Burhenn, V.~Erckmann, S.~Fiedler,
  J.~Geiger, H.~Hartfuss, O.~Heinrich, M.~Hirsch, et~al.}, {\em Confinement in
  {W7-AS} and the role of radial electric field and magnetic shear}, Plasma
  Physics and Controlled Fusion, 39 (1997), p.~B273.

\bibitem{Brenner2012}
{\sc P.~Brenner and T.~Sunn~Pedersen}, {\em Pure electron plasmas confined for
  90 ms in a stellarator without electron sources or internal objects}, Physics
  of Plasmas, 19 (2012), p.~050701.

\bibitem{Brent2013}
{\sc R.~P. Brent}, {\em Algorithms for minimization without derivatives},
  Courier Corporation, 2013.

\bibitem{Brooks2003}
{\sc A.~Brooks and W.~Reiersen}, {\em Coil tolerance impact on plasma surface
  quality for {NCSX}}, in 20th IEEE/NPSS Symposium on Fusion Engineering, IEEE,
  2003, pp.~553--556.

\bibitem{Brown2015}
{\sc T.~Brown, J.~Breslau, D.~Gates, N.~Pomphrey, and A.~Zolfaghari}, {\em
  Engineering optimization of stellarator coils lead to improvements in device
  maintenance}, in IEEE 26th Symposium on Fusion Engineering (SOFE), Austin,
  Texas, 2015.

\bibitem{Bruno1996}
{\sc O.~P. Bruno and P.~Laurence}, {\em Existence of three-dimensional toroidal
  {MHD} equilibria with nonconstant pressure}, Communications on Pure and
  Applied Mathematics, 49 (1996), pp.~717--764.

\bibitem{Burby2019}
{\sc J.~W. Burby, N.~Kallinikos, and R.~S. MacKay}, {\em Some mathematics for
  quasi-symmetry}, arXiv preprint arXiv:1912.06468,  (2019).

\bibitem{Canik2007}
{\sc J.~Canik, D.~Anderson, F.~Anderson, C.~Clark, K.~Likin, J.~Talmadge, and
  K.~Zhai}, {\em Reduced particle and heat transport with quasisymmetry in the
  {Helically Symmetric Experiment}}, Physics of Plasmas, 14 (2007), p.~056107.

\bibitem{Cary1991}
{\sc J.~R. Cary and J.~D. Hanson}, {\em Simple method for calculating island
  widths}, Physics of Fluids B: Plasma Physics, 3 (1991), p.~1006.

\bibitem{Cary1983}
{\sc J.~R. Cary and R.~G. Littlejohn}, {\em Noncanonical {Hamiltonian}
  mechanics and its application to magnetic field line flow}, Annals of
  Physics, 151 (1983), pp.~1--34.

\bibitem{Cary1997b}
{\sc J.~R. Cary and S.~G. Shasharina}, {\em Helical plasma confinement devices
  with good confinement properties}, Physical Review Letters, 78 (1997),
  p.~674.

\bibitem{Cary1997}
\leavevmode\vrule height 2pt depth -1.6pt width 23pt, {\em Omnigenity and
  quasihelicity in helical plasma confinement systems}, Physics of Plasmas, 4
  (1997), pp.~3323--3333.

\bibitem{Castejon2012}
{\sc F.~Castej{\'o}n, A.~G{\'o}mez-Iglesias, M.~Vega-Rodr{\'\i}guez,
  J.~Jim{\'e}nez, J.~Velasco, and J.~Romero}, {\em Stellarator optimization
  under several criteria using metaheuristics}, Plasma Physics and Controlled
  Fusion, 55 (2012), p.~014003.

\bibitem{catania2011}
{\sc D.~Catania and P.~Secchi}, {\em Global existence for two regularized mhd
  models in three space-dimension}, Portugaliae Mathematica, 68 (2011), p.~41.

\bibitem{CKscat}
{\sc D.~Colton and R.~Kress}, {\em Integral Equation Methods in Scattering
  Theory}, Society for Industrial and Applied Mathematics, Philadelphia, PA,
  2013.

\bibitem{Cooper2006}
{\sc W.~Cooper, J.~Graves, T.~Tran, R.~Gruber, T.~Yamaguchi, Y.~Narushima,
  S.~Okamura, S.~Sakakibara, C.~Suzuki, K.~Watanabe, et~al.}, {\em Stability
  properties of anisotropic pressure stellarator plasmas with fluid and
  noninteractive energetic particles}, Fusion Science and Technology, 50
  (2006), pp.~245--257.

\bibitem{czarny2008}
{\sc O.~Czarny and G.~Huysmans}, {\em B\'{e}zier surfaces and finite elements
  for {MHD} simulations}, Journal of Computational Physics, 227 (2008),
  pp.~7423 -- 7445.

\bibitem{Del1996}
{\sc D.~del Castillo-Negrete, J.~Greene, and P.~Morrison}, {\em Area preserving
  nontwist maps: periodic orbits and transition to chaos}, Physica D: Nonlinear
  Phenomena, 91 (1996), pp.~1--23.

\bibitem{dewar2013}
{\sc R.~Dewar, A.~Bhattacharjee, R.~Kulsrud, and A.~Wright}, {\em Plasmoid
  solutions of the hahm--kulsrud--taylor equilibrium model}, Physics of
  Plasmas, 20 (2013), p.~082103.

\bibitem{Dewar1998}
{\sc R.~Dewar and S.~Hudson}, {\em Stellarator symmetry}, Physica D: Nonlinear
  Phenomena, 112 (1998), pp.~275--280.

\bibitem{Dewar2015}
{\sc R.~L. Dewar, Z.~Yoshida, A.~Bhattacharjee, and S.~R. Hudson}, {\em
  Variational formulation of relaxed and multi-region relaxed
  magnetohydrodynamics}, Journal of Plasma Physics, 81 (2015).

\bibitem{2012Dhaeseleer}
{\sc W.~D. D'haeseleer, W.~N. Hitchon, J.~D. Callen, and J.~L. Shohet}, {\em
  Flux Coordinates and Magnetic Field Structure: A Guide to a Fundamental Tool
  of Plasma Theory}, Springer, 1991.

\bibitem{Dodson1997}
{\sc C.~T. Dodson, P.~E. Parker, and P.~Parker}, {\em A user's guide to
  algebraic topology}, vol.~387, Springer Science \& Business Media, 1997.

\bibitem{Dommaschk1986}
{\sc W.~Dommaschk}, {\em Representations for vacuum potentials in
  stellarators}, Computer Physics Communications, 40 (1986), pp.~203--218.

\bibitem{Drevlak1998}
{\sc M.~Drevlak}, {\em Automated optimization of stellarator coils}, Fusion
  Technology, 33 (1998), pp.~106--117.

\bibitem{Drevlak2009}
{\sc M.~Drevlak}, {\em Thermal load on the {W7-X} vessel from {NBI} losses}, in
  36th EPS Conference on Plasma Physics, European Physical Society, 2009.

\bibitem{Drevlak2018}
{\sc M.~Drevlak, C.~Beidler, J.~Geiger, P.~Helander, and Y.~Turkin}, {\em
  Optimisation of stellarator equilibria with {ROSE}}, Nuclear Fusion, 59
  (2018), p.~016010.

\bibitem{Drevlak2013}
{\sc M.~Drevlak, F.~Brochard, P.~Helander, J.~Kisslinger, M.~Mikhailov,
  C.~N{\"u}hrenberg, J.~N{\"u}hrenberg, and Y.~Turkin}, {\em {ESTELL}: A
  quasi-toroidally symmetric stellarator}, Contributions to Plasma Physics, 53
  (2013), pp.~459--468.

\bibitem{Drevlak2014}
{\sc M.~Drevlak, J.~Geiger, P.~Helander, and Y.~Turkin}, {\em Fast particle
  confinement with optimized coil currents in the {W7-X} stellarator}, Nuclear
  Fusion, 54 (2014), p.~073002.

\bibitem{dudek2009}
{\sc L.~E. Dudek, J.~H. Chrzanowski, P.~J. Heitzenroeder, S.~Raftopoulos, M.~E.
  Viola, G.~H. Neilson, D.~Rej, M.~J. Cole, P.~Goranson, and K.~Freudenberg},
  {\em Status of the {NCSX} construction}, Fusion Engineering and Design, 84
  (2009), pp.~351--354.

\bibitem{Elsasser1986}
{\sc K.~Elsasser}, {\em Magnetic field line flow as a {Hamiltonian} problem},
  Plasma Physics and Controlled Fusion, 28 (1986), p.~1743.

\bibitem{Erckmann1994}
{\sc V.~Erckmann and U.~Gasparino}, {\em Electron cyclotron resonance heating
  and current drive in toroidal fusion plasmas}, Plasma Physics and Controlled
  Fusion, 36 (1994), p.~1869.

\bibitem{Ferrell1971}
{\sc T.~L. Ferrell}, {\em {Hamilton-Jacobi} perturbation theory}, American
  Journal of Physics, 39 (1971), pp.~622--627.

\bibitem{Thompson1985}
{\sc T.~for~the ATF Team~PB}, {\em The {Advanced Toroidal Facility (ATF)}},
  Fusion Technology, 8 (1985), pp.~450--455.

\bibitem{Freidberg2014}
{\sc J.~Freidberg}, {\em Ideal {MHD}}, Cambridge University Press, 2014.

\bibitem{Freidberg2008}
{\sc J.~P. Freidberg}, {\em Plasma Physics and Fusion Energy}, Cambridge
  University Press, 2008.

\bibitem{furth1963}
{\sc H.~P. Furth, J.~Killeen, and M.~N. Rosenbluth}, {\em Finite-resistivity
  instabilities of a sheet pinch}, The Physics of Fluids, 6 (1963),
  pp.~459--484.

\bibitem{Garabedian2002}
{\sc P.~Garabedian}, {\em Three-dimensional stellarator codes}, Proceedings of
  the National Academy of Sciences, 99 (2002), pp.~10257--10259.

\bibitem{Garabedian1970}
{\sc P.~R. Garabedian}, {\em A method of canonical coordinates for flow
  computations}, Communications on Pure and Applied Mathematics, 23 (1970),
  pp.~313--327.

\bibitem{Garabedian2008}
\leavevmode\vrule height 2pt depth -1.6pt width 23pt, {\em Three-dimensional
  analysis of tokamaks and stellarators}, Proceedings of the National Academy
  of Sciences, 105 (2008), pp.~13716--13719.

\bibitem{Garren1991}
{\sc D.~Garren and A.~H. Boozer}, {\em Existence of quasihelically symmetric
  stellarators}, Physics of Fluids B: Plasma Physics, 3 (1991), pp.~2822--2834.

\bibitem{Gates2017}
{\sc D.~Gates, A.~Boozer, T.~Brown, J.~Breslau, D.~Curreli, M.~Landreman,
  S.~Lazerson, J.~Lore, H.~Mynick, G.~Neilson, et~al.}, {\em Recent advances in
  stellarator optimization}, Nuclear Fusion, 57 (2017), p.~126064.

\bibitem{Geiger2019}
{\sc B.~Geiger, T.~Wegner, C.~Beidler, R.~Burhenn, B.~Buttensch{\"o}n, R.~Dux,
  A.~Langenberg, N.~Pablant, T.~P{\"u}tterich, Y.~Turkin, et~al.}, {\em
  Observation of anomalous impurity transport during low-density experiments in
  {W7-X} with laser blow-off injections of iron}, Nuclear Fusion, 59 (2019),
  p.~046009.

\bibitem{Geiger2010}
{\sc J.~Geiger, C.~Beidler, M.~Drevlak, H.~Maassberg, C.~N{\"u}hrenberg,
  Y.~Suzuki, and Y.~Turkin}, {\em Effects of net currents on the magnetic
  configuration of {W7-X}}, Contributions to Plasma Physics, 50 (2010), p.~770.

\bibitem{Gerhardt2005}
{\sc S.~P. Gerhardt, J.~N. Talmadge, J.~M. Canik, and D.~T. Anderson}, {\em
  Measurements and modeling of plasma flow damping in the {Helically Symmetric
  eXperiment}}, Physics of Plasmas, 12 (2005), p.~056116.

\bibitem{Goedbloed2004}
{\sc J.~Goedbloed and S.~Poedts}, {\em Principles of Magnetohydrodynamics: with
  Applications to Laboratory and Astrophysical Plasmas}, Cambridge University
  Press, 2004.

\bibitem{Goldstein2002}
{\sc H.~Goldstein, C.~Poole, and J.~Safko}, {\em {Classical Mechanics}},
  Addison-Wesley, 2002.

\bibitem{Goldston1995}
{\sc R.~J. Goldston and P.~H. Rutherford}, {\em Introduction to Plasma
  Physics}, CRC Press, 1995.

\bibitem{grad1967}
{\sc H.~Grad}, {\em Toroidal containment of a plasma}, The Physics of Fluids,
  10 (1967), pp.~137--154.

\bibitem{Greene1997}
{\sc J.~M. Greene}, {\em A brief review of magnetic wells}, Comments on Plasma
  Physics and Controlled Fusion, 17 (1997), pp.~389--402.

\bibitem{Grieger1989}
{\sc G.~Grieger, C.~Beidler, and E.~Harmeyer}, {\em Physics studies for
  helical-axis advanced stellarators}, in Plasma Physics and Controlled Nuclear
  Fusion Research, vol.~2, 1988.

\bibitem{Grieger1992}
{\sc G.~Grieger, W.~Lotz, P.~Merkel, J.~N{\"u}hrenberg, J.~Sapper,
  E.~Strumberger, H.~Wobig, R.~Burhenn, V.~Erckmann, U.~Gasparino, et~al.},
  {\em Physics optimization of stellarators}, Physics of Fluids B: Plasma
  Physics, 4 (1992), pp.~2081--2091.

\bibitem{Hadamard1932}
{\sc J.~Hadamard}, {\em Le probleme de Cauchy et les {\'e}quations aux
  d{\'e}riv{\'e}es partielles lin{\'e}aires hyperboliques}, vol.~220, 1932.

\bibitem{hahm1985}
{\sc T.~Hahm and R.~Kulsrud}, {\em Forced magnetic reconnection}, The Physics
  of Fluids, 28 (1985), pp.~2412--2418.

\bibitem{Hall1975}
{\sc L.~S. Hall and B.~McNamara}, {\em Three-dimensional equilibrium of the
  anisotropic, finite-pressure guiding-center plasma: Theory of the magnetic
  plasma}, The Physics of Fluids, 18 (1975), pp.~552--565.

\bibitem{Hamberger1990}
{\sc S.~M. Hamberger, B.~D. Blackwell, L.~E. Sharp, and D.~Shenton}, {\em H-1
  design and construction}, Fusion Technology, 17 (1990), pp.~123--130.

\bibitem{Hammond2020}
{\sc K.~Hammond, C.~Zhu, T.~Brown, K.~Corrigan, D.~Gates, and M.~Sibilia}, {\em
  Geometric concepts for stellarator permanent magnet arrays}, arXiv preprint
  arXiv:2006.00091,  (2020).

\bibitem{Hansen2005}
{\sc P.~C. Hansen}, {\em Rank-deficient and Discrete Ill-posed Problems:
  Numerical Aspects of Linear Inversion}, vol.~4, SIAM, 2005.

\bibitem{Harafuji1989}
{\sc K.~Harafuji, T.~Hayashi, and T.~Sato}, {\em Computational study of
  three-dimensional magnetohydrodynamic equilibria in toroidal helical
  systems}, Journal of Computational Physics, 81 (1989), pp.~169--192.

\bibitem{Hartwell2017}
{\sc G.~Hartwell, S.~Knowlton, J.~Hanson, D.~Ennis, and D.~Maurer}, {\em
  Design, construction, and operation of the {Compact Toroidal Hybrid}}, Fusion
  Science and Technology, 72 (2017), pp.~76--90.

\bibitem{Hazeltine2003}
{\sc R.~D. Hazeltine and J.~D. Meiss}, {\em Plasma Confinement}, Courier
  Corporation, 2003.

\bibitem{Hegna1991}
{\sc C.~Hegna and A.~Bhattacharjee}, {\em Islands in three-dimensional steady
  flows}, Journal of Fluid Mechanics, 227 (1991), pp.~527--542.

\bibitem{Hegna1998}
{\sc C.~C. Hegna and N.~Nakajima}, {\em On the stability of {Mercier} and
  ballooning modes in stellarator configurations}, Physics of Plasmas, 5
  (1998), p.~1336.

\bibitem{Hegna2018}
{\sc C.~C. Hegna, P.~W. Terry, and B.~J. Faber}, {\em Theory of {ITG} turbulent
  saturation in stellarators: identifying mechanisms to reduce turbulent
  transport}, Physics of Plasmas, 25 (2018), p.~022511.

\bibitem{Helander2014}
{\sc P.~Helander}, {\em Theory of plasma confinement in non-axisymmetric
  magnetic field}, Reports of Progress in Physics, 77 (2014), p.~087001.

\bibitem{Helander2012}
{\sc P.~Helander, C.~Beidler, T.~Bird, M.~Drevlak, Y.~Feng, R.~Hatzky,
  F.~Jenko, R.~Kleiber, J.~Proll, Y.~Turkin, et~al.}, {\em Stellarator and
  tokamak plasmas: a comparison}, Plasma Physics and Controlled Fusion, 54
  (2012), p.~124009.

\bibitem{Helander2020}
{\sc P.~Helander, M.~Drevlak, M.~Zarnstorff, and S.~Cowley}, {\em Stellarators
  with permanent magnets}, Physical Review Letters, 124 (2020), p.~095001.

\bibitem{Helander2009}
{\sc P.~Helander and J.~N{\"u}hrenberg}, {\em Bootstrap current and
  neoclassical transport in quasi-isodynamic stellarators}, Plasma Physics and
  Controlled Fusion, 51 (2009), p.~055004.

\bibitem{Helander2005}
{\sc P.~Helander and D.~J. Sigmar}, {\em Collisional Transport in Magnetized
  Plasmas}, vol.~4, Cambridge University Press, 2005.

\bibitem{Helander2008}
{\sc P.~Helander and A.~Simakov}, {\em Intrinsic ambipolarity and rotation in
  stellarators}, Physical Review Letters, 101 (2008), p.~145003.

\bibitem{Henneberg2019b}
{\sc S.~Henneberg, M.~Drevlak, and P.~Helander}, {\em Improving fast-particle
  confinement in quasi-axisymmetric stellarator optimization}, Plasma Physics
  and Controlled Fusion, 62 (2019), p.~014023.

\bibitem{Henneberg2019}
{\sc S.~Henneberg, M.~Drevlak, C.~N{\"u}hrenberg, C.~Beidler, Y.~Turkin,
  J.~Loizu, and P.~Helander}, {\em Properties of a new quasi-axisymmetric
  configuration}, Nuclear Fusion, 59 (2019), p.~026014.

\bibitem{heywood}
{\sc J.~G. Heywood}, {\em Auxiliary flux and pressure conditions for
  {Navier-Stokes} problems}, in Approximation Methods for Navier-Stokes
  Problems, R.~Rautmann, ed., Berlin, Heidelberg, 1980, Springer Berlin
  Heidelberg, pp.~223--234.

\bibitem{turek}
{\sc J.~G. Heywood, R.~Rannacher, and S.~Turek}, {\em Artificial boundaries and
  flux and pressure conditions for the incompressible {Navier--Stokes}
  equations}, International Journal for Numerical Methods in Fluids, 22 (1996),
  pp.~325--352.

\bibitem{Hidalgo2005}
{\sc C.~Hidalgo, C.~Alejaldre, A.~Alonso, J.~Alonso, L.~Almoguera,
  F.~de~Arag{\'o}n, E.~Ascas{\'\i}bar, A.~Baciero, R.~Balb{\'\i}n, E.~Blanco,
  et~al.}, {\em Overview of {TJ-II} experiments}, Nuclear Fusion, 45 (2005),
  p.~S266.

\bibitem{Highcock2018}
{\sc E.~Highcock, N.~Mandell, M.~Barnes, and W.~Dorland}, {\em Optimisation of
  confinement in a fusion reactor using a nonlinear turbulence model}, Journal
  of Plasma Physics, 84 (2018).

\bibitem{Hirsch2008}
{\sc M.~Hirsch, J.~Baldzuhn, C.~Beidler, R.~Brakel, R.~Burhenn, A.~Dinklage,
  H.~Ehmler, M.~Endler, V.~Erckmann, Y.~Feng, et~al.}, {\em Major results from
  the stellarator {Wendelstein 7-AS}}, Plasma Physics and Controlled Fusion, 50
  (2008), p.~053001.

\bibitem{Hirshman1986}
{\sc S.~Hirshman, P.~Merkel, et~al.}, {\em Three-dimensional free boundary
  calculations using a spectral {Green's} function method}, Computer Physics
  Communications, 43 (1986), pp.~143--155.

\bibitem{Hirshman2011}
{\sc S.~Hirshman, R.~Sanchez, and C.~Cook}, {\em {SIESTA}: A scalable iterative
  equilibrium solver for toroidal applications}, Physics of Plasmas, 18 (2011),
  p.~062504.

\bibitem{Hirshman1999}
{\sc S.~Hirshman, D.~Spong, J.~Whitson, B.~Nelson, D.~Batchelor, J.~Lyon,
  R.~Sanchez, A.~Brooks, G.~Y.-Fu, R.~Goldston, et~al.}, {\em Physics of
  compact stellarators}, Physics of Plasmas, 6 (1999), pp.~1858--1864.

\bibitem{Hirshman1983}
{\sc S.~P. Hirshman and J.~C. Whitson}, {\em Steepest descent moment method for
  three-dimensional magnetohydrodynamic equilibria}, The Physics of Fluids, 26
  (1983), p.~3553.

\bibitem{Hofmann1996}
{\sc J.~Hofmann, J.~Baldzuhn, R.~Brakel, Y.~Feng, S.~Fiedler, J.~Geiger,
  P.~Grigull, G.~Herre, R.~Jaenicke, M.~Kick, et~al.}, {\em Stellarator
  optimization studies in {W7-AS}}, Plasma Physics and Controlled Fusion, 38
  (1996), p.~A193.

\bibitem{Hole2006}
{\sc M.~Hole, S.~R. Hudson, and R.~Dewar}, {\em Stepped pressure profile
  equilibria in cylindrical plasmas via partial {Taylor} relaxation}, Journal
  of Plasma Physics, 72 (2006), pp.~1167--1171.

\bibitem{hosking2016}
{\sc R.~J. Hosking and R.~L. Dewar}, {\em Fundamental fluid mechanics and
  magnetohydrodynamics}, Springer, 2016.

\bibitem{Huba2006}
{\sc J.~D. Huba}, {\em {NRL Plasma Formulary}}, tech. rep., Naval Research
  Laboratory, 2006.

\bibitem{Hudson2012}
{\sc S.~Hudson, R.~Dewar, G.~Dennis, M.~Hole, M.~McGann, G.~Von~Nessi, and
  S.~Lazerson}, {\em Computation of multi-region relaxed magnetohydrodynamic
  equilibria}, Physics of Plasmas, 19 (2012), p.~112502.

\bibitem{Hudson2010a}
{\sc S.~Hudson and N.~Nakajima}, {\em Pressure, chaotic magnetic fields, and
  magnetohydrodynamic equilibria}, Physics of Plasmas, 17 (2010), p.~052511.

\bibitem{Hudson2018}
{\sc S.~Hudson, C.~Zhu, D.~Pfefferl{\'e}, and L.~Gunderson}, {\em
  Differentiating the shape of stellarator coils with respect to the plasma
  boundary}, Physics Letters A, 382 (2018), pp.~2732--2737.

\bibitem{Hudson2011}
{\sc S.~R. Hudson, R.~Dewar, M.~Hole, and M.~McGann}, {\em Non-axisymmetric,
  multi-region relaxed magnetohydrodynamic equilibrium solutions}, Plasma
  Physics and Controlled Fusion, 54 (2011), p.~014005.

\bibitem{Hudson2002}
{\sc S.~R. Hudson, D.~Monticello, A.~Reiman, A.~Boozer, D.~Strickler,
  S.~Hirshman, and M.~Zarnstorff}, {\em Eliminating islands in high-pressure
  free-boundary stellarator magnetohydrodynamic equilibrium solutions},
  Physical Review Letters, 89 (2002), p.~275003.

\bibitem{turekBenchmark}
{\sc S.~Hysing, S.~Turek, D.~Kuzmin, N.~Parolini, E.~Burman, S.~Ganesan, and
  L.~Tobiska}, {\em Quantitative benchmark computations of two-dimensional
  bubble dynamics}, International Journal for Numerical Methods in Fluids, 60
  (2009), pp.~1259--1288.

\bibitem{Iiyoshi1999}
{\sc A.~Iiyoshi, A.~Komori, A.~Ejiri, M.~Emoto, H.~Funaba, M.~Goto, K.~Ida,
  H.~Idei, S.~Inagaki, S.~Kado, et~al.}, {\em Overview of the {Large Helical
  Device} project}, Nuclear Fusion, 39 (1999), p.~1245.

\bibitem{Isobe2019}
{\sc M.~Isobe, A.~Shimizu, H.~Liu, H.~Liu, G.~Xiong, D.~Yin, K.~Ogawa,
  Y.~Yoshimura, M.~Nakata, S.~Kinoshita, et~al.}, {\em Current status of
  {NIFS-SWJTU} joint project for quasi-axisymmetric stellarator {CFQS}}, Plasma
  and Fusion Research, 14 (2019), pp.~3402074--3402074.

\bibitem{Jardin2010}
{\sc S.~Jardin}, {\em Computational Methods in Plasma Physics}, CRC Press,
  2010.

\bibitem{Jardin2007}
{\sc S.~C. Jardin, J.~Breslau, and N.~Ferraro}, {\em A high-order implicit
  finite element method for integrating the two-fluid magnetohydrodynamic
  equations in two dimensions}, Journal of Computational Physics, 226 (2007),
  pp.~2146--2174.

\bibitem{Jorge2017}
{\sc R.~Jorge, P.~Ricci, and N.~Loureiro}, {\em A drift-kinetic analytical
  model for scrape-off layer plasma dynamics at arbitrary collisionality},
  Journal of Plasma Physics, 83 (2017).

\bibitem{Jorge2020}
{\sc R.~Jorge, W.~Sengupta, and M.~Landreman}, {\em Construction of
  quasisymmetric stellarators using a direct coordinate approach}, Nuclear
  Fusion,  (2020).

\bibitem{Jorge2020a}
{\sc R.~Jorge, W.~Sengupta, and M.~Landreman}, {\em Near-axis expansion of
  stellarator equilibrium at arbitrary order in the distance to the axis},
  Journal of Plasma Physics, 86 (2020).

\bibitem{1998classical}
{\sc V.~Jose, J.~Jos{\'e}, and E.~Saletan}, {\em Classical Dynamics: A
  Contemporary Approach}, Classical Dynamics: A Contemporary Approach,
  Cambridge University Press, 1998.

\bibitem{Kaneko2016}
{\sc O.~Kaneko}, {\em {Large Helical Device}}, in Magnetic Fusion Energy,
  Elsevier, 2016, pp.~469--491.

\bibitem{klinger2013}
{\sc T.~Klinger, C.~Baylard, C.~Beidler, J.~Boscary, H.~Bosch, A.~Dinklage,
  D.~Hartmann, P.~Helander, H.~Massberg, A.~Peacock, T.~Pedersen, T.~Rummel,
  F.~Schauer, L.~Wegener, and R.~Wolf}, {\em Towards assembly completion and
  preparation of experimental campaigns of {Wendelstein 7-X} in the perspective
  of a path to a stellarator fusion power plant}, Fusion Engineering and
  Design, 88 (2013), pp.~461 -- 465.
\newblock Proceedings of the 27th Symposium On Fusion Technology (SOFT-27);
  Li\'{e}ge, Belgium, September 24-28, 2012.

\bibitem{Kolmogorov1954}
{\sc A.~N. Kolmogorov}, {\em On conservation of conditionally periodic motions
  for a small change in {Hamilton}'s function}, in Dokl. Akad. Nauk SSSR,
  vol.~98, 1954, pp.~527--530.

\bibitem{Komori2006}
{\sc A.~Komori, T.~Morisaki, T.~Mutoh, S.~Sakakibara, Y.~Takeiri, R.~Kumazawa,
  S.~Kubo, K.~Ida, S.~Morita, K.~Narihara, et~al.}, {\em Overview of progress
  in {LHD} experiments}, Fusion Science and Technology, 50 (2006),
  pp.~136--145.

\bibitem{Komori2003}
{\sc A.~Komori, N.~Ohyabu, H.~Yamada, O.~Kaneko, K.~Kawahata, K.~Ida,
  Y.~Nakamura, T.~Akiyama, N.~Ashikawa, M.~Emoto, et~al.}, {\em Recent results
  from the {Large Helical Device}}, 45 (2003), p.~671.

\bibitem{kong2009}
{\sc M.~G. Kong, G.~Kroesen, G.~Morfill, T.~Nosenko, T.~Shimizu, J.~van Dijk,
  and J.~L. Zimmermann}, {\em Plasma medicine: an introductory review}, New
  Journal of Physics, 11 (2009), p.~115012.

\bibitem{Kozlov1983}
{\sc V.~V. Kozlov}, {\em Integrability and non-integrability in {Hamiltonian}
  mechanics}, Russian Mathematical Surveys, 38 (1983), p.~1.

\bibitem{Kress1986}
{\sc R.~Kress}, {\em On constant-alpha force-free fields in a torus}, Journal
  of Engineering Mathematics, 20 (1986), pp.~323--344.

\bibitem{Kress1989}
\leavevmode\vrule height 2pt depth -1.6pt width 23pt, {\em Linear integral
  equations}, vol.~82, Springer, 1989.

\bibitem{Krommes2009}
{\sc J.~A. Krommes and A.~H. Reiman}, {\em Plasma equilibrium in a magnetic
  field with stochastic regions}, Physics of Plasmas, 16 (2009), p.~072308.

\bibitem{Kruskal1958}
{\sc M.~D. Kruskal and R.~Kulsrud}, {\em Equilibrium of a magnetically confined
  plasma in a toroid}, The Physics of Fluids, 1 (1958), pp.~265--274.

\bibitem{Ku2010}
{\sc L.~Ku and A.~Boozer}, {\em New classes of quasi-helically symmetric
  stellarators}, Nuclear Fusion, 51 (2010), p.~013004.

\bibitem{Ku2008}
{\sc L.~Ku, P.~Garabedian, J.~Lyon, A.~Turnbull, A.~Grossman, T.~Mau,
  M.~Zarnstorff, and A.~Team}, {\em Physics design for {ARIES-CS}}, Fusion
  Science and Technology, 54 (2008), pp.~673--693.

\bibitem{kuberry2012}
{\sc P.~Kuberry, A.~Larios, L.~G. Rebholz, and N.~E. Wilson}, {\em Numerical
  approximation of the {Voigt} regularization for incompressible
  {Navier--Stokes} and magnetohydrodynamic flows}, Computers \& Mathematics
  with Applications, 64 (2012), pp.~2647--2662.

\bibitem{Landreman2011}
{\sc M.~Landreman}, {\em Electric Fields and Transport in Optimized
  Stellarators}, PhD thesis, MIT, 2011.

\bibitem{Landreman2017}
{\sc M.~Landreman}, {\em An improved current potential method for fast
  computation of stellarator coil shapes}, Nuclear Fusion, 57 (2017),
  p.~046003.

\bibitem{Landreman2019}
\leavevmode\vrule height 2pt depth -1.6pt width 23pt, {\em Optimized
  quasisymmetric stellarators are consistent with the {Garren--Boozer}
  construction}, Plasma Physics and Controlled Fusion, 61 (2019), p.~075001.

\bibitem{Landreman2019b}
\leavevmode\vrule height 2pt depth -1.6pt width 23pt, {\em Quasisymmetry: A
  hidden symmetry of magnetic fields},  (2019).

\bibitem{Landreman2016}
{\sc M.~Landreman and A.~H. Boozer}, {\em Efficient magnetic fields for
  supporting toroidal plasmas}, Physics of Plasmas, 23 (2016), p.~032506.

\bibitem{Landreman2012}
{\sc M.~Landreman and P.~J. Catto}, {\em Omnigenity as generalized
  quasisymmetry}, Physics of Plasmas, 19 (2012), p.~056103.

\bibitem{Landreman2018a}
{\sc M.~Landreman and W.~Sengupta}, {\em Direct construction of optimized
  stellarator shapes. {Part 1. Theory} in cylindrical coordinates}, Journal of
  Plasma Physics, 84 (2018).

\bibitem{Landreman2018b}
{\sc M.~Landreman, W.~Sengupta, and G.~G. Plunk}, {\em Direct construction of
  optimized stellarator shapes. {Part 2. Numerical} quasisymmetric solutions},
  Journal of Plasma Physics, 85 (2019).

\bibitem{larios2014}
{\sc A.~Larios and E.~S. Titi}, {\em Higher-order global regularity of an
  inviscid {Voigt}-regularization of the three-dimensional inviscid resistive
  magnetohydrodynamic equations}, Journal of Mathematical Fluid Mechanics, 16
  (2014), pp.~59--76.

\bibitem{lax2002functional}
{\sc P.~Lax}, {\em Functional analysis}, Pure and applied mathematics, Wiley,
  2002.

\bibitem{Lazerson2016}
{\sc S.~A. Lazerson, J.~Loizu, S.~Hirshman, and S.~R. Hudson}, {\em
  Verification of the ideal magnetohydrodynamic response at rational surfaces
  in the {VMEC} code}, Physics of Plasmas, 23 (2016), p.~012507.

\bibitem{Lee1990}
{\sc D.~Lee, J.~Harris, and G.~Lee}, {\em Magnetic island widths due to field
  perturbations in toroidal stellarators}, Nuclear Fusion, 30 (1990), p.~2177.

\bibitem{levchenko2018}
{\sc I.~Levchenko, S.~Xu, G.~Teel, D.~Mariotti, M.~Walker, and M.~Keidar}, {\em
  Recent progress and perspectives of space electric propulsion systems based
  on smart nanomaterials}, Nature Communications, 9 (2018), p.~879.

\bibitem{Lichtenberg2013}
{\sc A.~J. Lichtenberg and M.~A. Lieberman}, {\em Regular and stochastic
  motion}, vol.~38, Springer Science \& Business Media, 2013.

\bibitem{lieberman2005}
{\sc M.~A. Lieberman, A.~J. Lichtenberg, et~al.}, {\em Principles of Plasma
  Discharges and Materials Processing}, vol.~2, Wiley Online Library, 2005.

\bibitem{Littlejohn1983}
{\sc R.~G. Littlejohn}, {\em Variational principles of guiding centre motion},
  Journal of Plasma Physics, 29 (1983), pp.~111--125.

\bibitem{Lobsien2020}
{\sc J.-F. Lobsien, M.~Drevlak, T.~Kruger, S.~Lazerson, C.~Zhu, and T.~S.
  Pedersen}, {\em Improved performance of stellarator coil design
  optimization}, Journal of Plasma Physics, 86 (2020), p.~815860202.

\bibitem{Lobsien2018}
{\sc J.-F. Lobsien, M.~Drevlak, T.~S. Pedersen, et~al.}, {\em Stellarator coil
  optimization towards higher engineering tolerances}, Nuclear Fusion, 58
  (2018), p.~106013.

\bibitem{Loizu2015}
{\sc J.~Loizu, S.~Hudson, A.~Bhattacharjee, and P.~Helander}, {\em Magnetic
  islands and singular currents at rational surfaces in three-dimensional
  magnetohydrodynamic equilibria}, Physics of Plasmas, 22 (2015), p.~022501.

\bibitem{Lore2014}
{\sc J.~D. Lore, T.~Andreeva, J.~Boscary, S.~Bozhenkov, J.~Geiger, J.~H.
  Harris, H.~Hoelbe, A.~Lumsdaine, D.~McGinnis, A.~Peacock, et~al.}, {\em
  Design and analysis of divertor scraper elements for the {W7-X} stellarator},
  IEEE Transactions on Plasma Science, 42 (2014), pp.~539--544.

\bibitem{Malhotra2019}
{\sc D.~Malhotra, A.~Cerfon, L.-M. Imbert-G{\'e}rard, and M.~O'Neil}, {\em
  Taylor states in stellarators: A fast high-order boundary integral solver},
  Journal of Computational Physics, 397 (2019), p.~108791.

\bibitem{Mccormick2002}
{\sc K.~McCormick, P.~Grigull, R.~Burhenn, R.~Brakel, H.~Ehmler, Y.~Feng,
  F.~Gadelmeier, L.~Giannone, D.~Hildebrandt, M.~Hirsch, et~al.}, {\em New
  advanced operational regime on the {W7-AS} stellarator}, Physical Review
  Letters, 89 (2002), p.~015001.

\bibitem{Mcmillan2014}
{\sc M.~McMillan and S.~A. Lazerson}, {\em {BEAMS3D} neutral beam injection
  model}, Plasma Physics and Controlled Fusion, 56 (2014), p.~095019.

\bibitem{Mericer1964}
{\sc C.~Mercier}, {\em Equilibrium and stability of a toroidal
  magnetohydrodynamic system in the neighbourhood of a magnetic axis}, Nuclear
  Fusion, 4 (1964), p.~213.

\bibitem{Mercier1974}
{\sc C.~Mercier and H.~Luc}, {\em The {MHD} approach to the problem of plasma
  confinement in closed magnetic configurations}, Lectures in Plasma Physics,
  Commission of the European Communities, Luxembourg,  (1974).

\bibitem{Merkel1986}
{\sc P.~Merkel}, {\em An integral equation technique for the exterior and
  interior {Neumann} problem in toroidal regions}, Journal of Computational
  Physics, 66 (1986), pp.~83--98.

\bibitem{Merkel1987}
\leavevmode\vrule height 2pt depth -1.6pt width 23pt, {\em Solution of
  stellarator boundary value problems with external currents}, Nuclear Fusion,
  27 (1987), p.~867.

\bibitem{Mikhailov2019}
{\sc M.~Mikhailov, J.~N{\"u}hrenberg, and R.~Zille}, {\em Elimination of
  current sheets at resonances in three-dimensional toroidal
  ideal-magnetohydrodynamic equilibria}, Nuclear Fusion, 59 (2019), p.~066002.

\bibitem{Moffatt1985}
{\sc H.~Moffatt}, {\em Magnetostatic equilibria and analogous euler flows of
  arbitrarily complex topology. {Part 1. Fundamentals}}, Journal of Fluid
  Mechanics, 159 (1985), pp.~359--378.

\bibitem{More1978}
{\sc J.~J. Mor{\'e}}, {\em The {Levenberg-Marquardt} algorithm: implementation
  and theory}, in Numerical analysis, Springer, 1978, pp.~105--116.

\bibitem{Morrison1998}
{\sc P.~J. Morrison}, {\em Hamiltonian description of the ideal fluid}, Reviews
  of Modern Physics, 70 (1998), p.~467.

\bibitem{Moser1962}
{\sc J.~M{o}ser}, {\em On invariant curves of area-preserving mappings of an
  annulus}, Nachr. Akad. Wiss. G{\"o}ttingen, II,  (1962), pp.~1--20.

\bibitem{Murakami2002}
{\sc S.~Murakami, A.~Wakasa, H.~Maa{\ss}berg, C.~Beidler, H.~Yamada,
  K.~Watanabe, L.~E. Group, et~al.}, {\em Neoclassical transport optimization
  of {LHD}}, Nuclear Fusion, 42 (2002), p.~L19.

\bibitem{Murakami2004}
{\sc S.~Murakami, H.~Yamada, M.~Sasao, M.~Isobe, T.~Ozaki, T.~Saida,
  P.~Goncharov, J.~Lyon, M.~Osakabe, T.~Seki, et~al.}, {\em Effect of
  neoclassical transport optimization on energetic ion confinement in {LHD}},
  Fusion Science and Technology, 46 (2004), pp.~241--247.

\bibitem{Mynick2006}
{\sc H.~Mynick}, {\em Transport optimization in stellarators}, Physics of
  Plasmas, 13 (2006), p.~058102.

\bibitem{Mynick2002}
{\sc H.~E. Mynick, N.~Pomphrey, and S.~Ethier}, {\em Exploration of stellarator
  configuration space with global search methods}, Physics of Plasmas, 9
  (2002), pp.~869--876.

\bibitem{Najmabadi2008}
{\sc F.~Najmabadi, A.~Raffray, S.~Abdel-Khalik, L.~Bromberg, L.~Crosatti,
  L.~El-Guebaly, P.~Garabedian, A.~Grossman, D.~Henderson, A.~Ibrahim, et~al.},
  {\em The {ARIES-CS} compact stellarator fusion power plant}, Fusion Science
  and Technology, 54 (2008), pp.~655--672.

\bibitem{Nemov1999}
{\sc V.~Nemov, S.~Kasilov, W.~Kernbichler, and M.~Heyn}, {\em Evaluation of
  1/$\nu$ neoclassical transport in stellarators}, Physics of Plasmas, 6
  (1999), pp.~4622--4632.

\bibitem{Nemov2008}
{\sc V.~Nemov, S.~Kasilov, W.~Kernbichler, and G.~Leitold}, {\em Poloidal
  motion of trapped particle orbits in real-space coordinates}, Physics of
  Plasmas, 15 (2008), p.~052501.

\bibitem{Nicholson1983}
{\sc D.~R. Nicholson}, {\em Introduction to Plasma Theory}, Wiley New York,
  1983.

\bibitem{Nishimura1990}
{\sc K.~Nishimura, K.~Matsuoka, M.~Fujiwara, K.~Yamazaki, J.~Todoroki,
  T.~Kamimura, T.~Amano, H.~Sanuki, S.~Okamura, M.~Hosokawa, et~al.}, {\em
  {Compact Helical System} physics and engineering design}, Fusion Technology,
  17 (1990), pp.~86--100.

\bibitem{Nuhrenberg1994}
{\sc J.~N{\"u}hrenberg, W.~Lotz, and S.~Gori}, {\em Theory of fusion plasmas},
  in Proceedings of the Joint Varenna-Lausanne International Workshop, 1994,
  p.~3.

\bibitem{Nuhrenberg1988}
{\sc J.~N{\"u}hrenberg and R.~Zille}, {\em Quasi-helically symmetric toroidal
  stellarators}, Physics Letters A, 129 (1988), pp.~113--117.

\bibitem{Oneil2018}
{\sc M.~O'Neil and A.~J. Cerfon}, {\em An integral equation-based numerical
  solver for {Taylor} states in toroidal geometries}, Journal of Computational
  Physics, 359 (2018), pp.~263--282.

\bibitem{Ott2002}
{\sc E.~Ott}, {\em {Chaos in Dynamical Systems}}, Cambridge University Press,
  2002, ch.~7.

\bibitem{Parsopoulos2002}
{\sc K.~E. Parsopoulos and M.~N. Vrahatis}, {\em Recent approaches to global
  optimization problems through particle swarm optimization}, Natural
  Computing, 1 (2002), pp.~235--306.

\bibitem{Paul2018}
{\sc E.~Paul, M.~Landreman, A.~Bader, and W.~Dorland}, {\em An adjoint method
  for gradient-based optimization of stellarator coil shapes}, Nuclear Fusion,
  58 (2018), p.~076015.

\bibitem{Paul2019}
{\sc E.~J. Paul, I.~G. Abel, M.~Landreman, and W.~Dorland}, {\em An adjoint
  method for neoclassical stellarator optimization}, Journal of Plasma Physics,
  85 (2019).

\bibitem{Pedersen2016}
{\sc T.~S. Pedersen, M.~Otte, S.~Lazerson, P.~Helander, S.~Bozhenkov,
  C.~Biedermann, T.~Klinger, R.~C. Wolf, H.-S. Bosch, T.~Wendelstein, et~al.},
  {\em Confirmation of the topology of the {Wendelstein 7-X} magnetic field to
  better than 1: 100,000}, Nature Communications, 7 (2016), p.~13493.

\bibitem{Plunk2020}
{\sc G.~Plunk}, {\em Perturbing an axisymmetric magnetic equilibrium to obtain
  a quasi-axisymmetric stellarator}, arXiv preprint arXiv:2005.02981,  (2020).

\bibitem{Plunk2018}
{\sc G.~Plunk and P.~Helander}, {\em Quasi-axisymmetric magnetic fields: weakly
  non-axisymmetric case in a vacuum}, Journal of Plasma Physics, 84 (2018).

\bibitem{Plunk2019}
{\sc G.~G. Plunk, M.~Landreman, and P.~Helander}, {\em Direct construction of
  optimized stellarator shapes. {Part 3. Omnigenity} near the magnetic axis},
  Journal of Plasma Physics, 85 (2019).

\bibitem{Pomphrey2001}
{\sc N.~Pomphrey, L.~Berry, A.~Boozer, A.~Brooks, R.~Hatcher, S.~Hirshman,
  L.-P. Ku, W.~Miner, H.~Mynick, W.~Reiersen, et~al.}, {\em Innovations in
  compact stellarator coil design}, Nuclear Fusion, 41 (2001), p.~339.

\bibitem{BdryIntFlows}
{\sc C.~Pozrikidis}, {\em Boundary Integral and Singularity Methods for
  Linearized Viscous Flow}, Cambridge Texts in Applied Mathematics, Cambridge
  University Press, 1992.

\bibitem{Reiman1999}
{\sc A.~Reiman, G.~Fu, S.~Hirshman, L.~Ku, D.~Monticello, H.~Mynick, M.~Redi,
  D.~Spong, M.~Zarnstorff, B.~Blackwell, et~al.}, {\em Physics design of a
  high-quasi-axisymmetric stellarator}, Plasma Physics and Controlled Fusion,
  41 (1999), p.~B273.

\bibitem{1986Reiman}
{\sc A.~Reiman and H.~Greenside}, {\em Calculation of three-dimesnional {MHD}
  equilibria with islands and stochastic regions}, Computer Physics
  Communications, 43 (1986), p.~157.

\bibitem{Reiman2007}
{\sc A.~Reiman, M.~Zarnstorff, D.~Monticello, A.~Weller, J.~Geiger, et~al.},
  {\em Pressure-induced breaking of equilibrium flux surfaces in the {W7-AS}
  stellarator}, Nuclear Fusion, 47 (2007), p.~572.

\bibitem{Rodriguez2020}
{\sc E.~Rodriguez, P.~Helander, and A.~Bhattacharjee}, {\em Necessary and
  sufficient conditions for quasisymmetry}, Physics of Plasmas, 27 (2020),
  p.~062501.

\bibitem{Sakakibara2008}
{\sc S.~Sakakibara, K.~Watanabe, Y.~Suzuki, Y.~Narushima, S.~Ohdachi,
  N.~Nakajima, F.~Watanabe, L.~Garcia, A.~Weller, K.~Toi, et~al.}, {\em {MHD}
  study of the reactor-relevant high-beta regime in the {Large Helical
  Device}}, Plasma Physics and Controlled Fusion, 50 (2008), p.~124014.

\bibitem{Sanchez2001}
{\sc R.~Sanchez, M.~Y. Isaev, S.~Hirshman, W.~Cooper, G.~Fu, J.~Jimenez, L.~Ku,
  M.~Mikhailov, D.~Monticello, A.~Reiman, et~al.}, {\em Ideal {MHD} stability
  calculations for compact stellarators}, Computer Physics Communications, 141
  (2001), pp.~55--65.

\bibitem{Sano2000}
{\sc F.~Sano, T.~Obiki, M.~Wakatani, K.~KONDO, and T.~MIZUUCHI}, {\em
  Experimental program of {Heliotron J}}, J. Plasma and Fusion Res. SERIES, 3
  (2000), p.~26.

\bibitem{Schnack2009}
{\sc D.~D. Schnack}, {\em Lectures in magnetohydrodynamics: with an appendix on
  extended MHD}, vol.~780, Springer, 2009.

\bibitem{Sengupta2019}
{\sc W.~Sengupta}, {\em Stellarator equilibrium axis-expansion to all orders in
  distance from the axis for arbitrary plasma beta}.
\newblock In preparation.

\bibitem{sermange1983}
{\sc M.~Sermange and R.~Temam}, {\em Some mathematical questions related to the
  {MHD} equations},  (1983).

\bibitem{Shadid2016}
{\sc J.~N. Shadid, R.~P. Pawlowski, E.~C. Cyr, R.~S. Tuminaro, L.~Chac{\'o}n,
  and P.~Weber}, {\em Scalable implicit incompressible resistive {MHD} with
  stabilized {FE} and fully-coupled {Newton--Krylov-AMG}}, Computer Methods in
  Applied Mechanics and Engineering, 304 (2016), pp.~1--25.

\bibitem{Shaing1989}
{\sc K.-C. Shaing, E.~Crume~Jr, J.~Tolliver, S.~Hirshman, and W.~Van~Rij}, {\em
  Bootstrap current and parallel viscosity in the low collisionality regime in
  toroidal plasmas}, Physics of Fluids B: Plasma Physics, 1 (1989), p.~148.

\bibitem{Shimizu2018}
{\sc A.~Shimizu, H.~Liu, M.~Isobe, S.~Okamura, S.~Nishimura, C.~Suzuki, Y.~Xu,
  X.~Zhang, B.~Liu, J.~Huang, et~al.}, {\em Configuration property of the
  {Chinese First Quasi-Axisymmetric Stellarator}}, Plasma and Fusion Research,
  13 (2018), pp.~3403123--3403123.

\bibitem{Sinclair1970}
{\sc R.~Sinclair, J.~Hosea, and G.~Sheffield}, {\em Magnetic surface mappings
  by storage of phase-stabilized low-energy electron beams}, Applied Physics
  Letters, 17 (1970), p.~92.

\bibitem{Singh2019}
{\sc L.~Singh, T.~Kruger, C.~Zhu, S.~Hudson, D.~Anderson, and A.~Bader}, {\em A
  new method for the optimization of finite build stellarator coils}, APS, 2019
  (2019), pp.~JP10--037.

\bibitem{Solov1970}
{\sc L.~Solov\`{e}v and V.~Shafranov}, {\em Plasma confinement in closed
  magnetic systems}, in Reviews of Plasma Physics, Springer, 1970, pp.~1--247.

\bibitem{Sovinec2004}
{\sc C.~Sovinec, A.~Glasser, T.~Gianakon, D.~Barnes, R.~Nebel, S.~Kruger,
  S.~Plimpton, A.~Tarditi, M.~Chu, and the NIMROD~Team}, {\em Nonlinear
  magnetohydrodynamics with high-order finite elements}, J. Comp. Phys., 195
  (2004), p.~355.

\bibitem{Spitzer1958}
{\sc L.~Spitzer~Jr}, {\em The stellarator concept}, The Physics of Fluids, 1
  (1958), pp.~253--264.

\bibitem{Spitzer1953}
{\sc L.~Spitzer~Jr and R.~H{\"a}rm}, {\em Transport phenomena in a completely
  ionized gas}, Physical Review, 89 (1953), p.~977.

\bibitem{Spong1998}
{\sc D.~Spong, S.~Hirshman, J.~Whitson, D.~Batchelor, B.~Carreras, V.~Lynch,
  and J.~Rome}, {\em J* optimization of small aspect ratio stellarator/tokamak
  hybrid devices}, Physics of Plasmas, 5 (1998), pp.~1752--1758.

\bibitem{Spong2001}
{\sc D.~Spong, S.~P. Hirshman, L.~Berry, J.~Lyon, R.~Fowler, D.~Strickler,
  M.~Cole, B.~Nelson, D.~Williamson, A.~Ware, et~al.}, {\em Physics issues of
  compact drift optimized stellarators}, Nuclear Fusion, 41 (2001), p.~711.

\bibitem{Spong2011}
{\sc D.~A. Spong}, {\em Three-dimensional effects on energetic particle
  confinement and stability}, Physics of Plasmas, 18 (2011), p.~056109.

\bibitem{Spong2010}
{\sc D.~A. Spong and J.~H. Harris}, {\em {New QP / QI Symmetric Stellarator
  Configurations}}, Plasma and Fusion Research, 5 (2010), p.~S2039.

\bibitem{Stix1998}
{\sc T.~H. Stix}, {\em Highlights in early stellarator research at
  {Princeton}}, J. Plasma Fusion Res. Ser, 1 (1998), pp.~3--8.

\bibitem{Storn1997}
{\sc R.~Storn and K.~Price}, {\em Differential evolution--a simple and
  efficient heuristic for global optimization over continuous spaces}, Journal
  of Global Optimization, 11 (1997), pp.~341--359.

\bibitem{Stratton2006}
{\sc B.~Stratton, A.~Brooks, T.~Brown, D.~Johnson, G.~Labik, E.~Lazarus,
  N.~Pomphrey, S.~Raftopoulos, and M.~Zarnstorff}, {\em External magnetic
  diagnostics for the {National Compact Stellarator Experiment}}, Review of
  Scientific Instruments, 77 (2006), p.~10E314.

\bibitem{Strickler2002}
{\sc D.~J. Strickler, L.~A. Berry, and S.~P. Hirshman}, {\em Designing coils
  for compact stellarators}, Fusion Science and Technology, 41 (2002),
  pp.~107--115.

\bibitem{Strickler2003}
{\sc D.~J. Strickler, L.~A. Berry, and S.~P. Hirshman}, {\em Integrated plasma
  and coil optimization for compact stellarators}, tech. rep., 2003.

\bibitem{Strickler2004}
{\sc D.~J. Strickler, S.~P. Hirshman, D.~A. Spong, M.~J. Cole, J.~F. Lyon,
  B.~E. Nelson, D.~E. Williamson, and A.~S. Ware}, {\em Development of a robust
  quasi-poloidal compact stellarator}, Fusion Science and Technology, 45
  (2004), pp.~15--26.

\bibitem{Strykowsky2009}
{\sc R.~Strykowsky, T.~Brown, J.~Chrzanowski, M.~Cole, P.~Heitzenroeder,
  G.~Neilson, D.~Rej, and M.~Viol}, {\em Engineering cost \& schedule lessons
  learned on {NCSX}}, in 2009 23rd IEEE/NPSS Symposium on Fusion Engineering,
  IEEE, 2009, pp.~1--4.

\bibitem{Sudo1990}
{\sc S.~Sudo, Y.~Takeiri, H.~Zushi, F.~Sano, K.~Itoh, K.~Kondo, and
  A.~Iiyoshi}, {\em Scalings of energy confinement and density limit in
  stellarator/heliotron devices}, Nuclear Fusion, 30 (1990), p.~11.

\bibitem{Sugiyama2001}
{\sc L.~Sugiyama, W.~Park, H.~Strauss, S.~Hudson, D.~Stutman, and X.-Z. Tang},
  {\em Studies of spherical tori, stellarators and anisotropic pressure with
  the {M3D} code}, Nuclear Fusion, 41 (2001), p.~739.

\bibitem{Sunn2017}
{\sc T.~Sunn~Pedersen, A.~Dinklage, Y.~Turkin, R.~Wolf, S.~Bozhenkov,
  J.~Geiger, G.~Fuchert, H.-S. Bosch, K.~Rahbarnia, H.~Thomsen, et~al.}, {\em
  Key results from the first plasma operation phase and outlook for future
  performance in {Wendelstein 7-X}}, Physics of Plasmas, 24 (2017), p.~055503.

\bibitem{Pedersen2017}
\leavevmode\vrule height 2pt depth -1.6pt width 23pt, {\em Key results from the
  first plasma operation phase and outlook for future performance in
  {Wendelstein 7-X}}, Physics of Plasmas, 24 (2017), p.~055503.

\bibitem{Suzuki2013}
{\sc Y.~Suzuki, K.~Ida, K.~Kamiya, M.~Yoshinuma, S.~Sakakibara, K.~Watanabe,
  H.~Yamada, L.~E. Group, et~al.}, {\em {3D} plasma response to the magnetic
  field structure in the {Large Helical Device}}, Nuclear Fusion, 53 (2013),
  p.~073045.

\bibitem{Suzuki2006}
{\sc Y.~Suzuki, N.~Nakajima, K.~Watanabe, Y.~Nakamura, and T.~Hayashi}, {\em
  Development and application of {HINT2} to helical system plasmas}, Nuclear
  Fusion, 46 (2006), p.~L19.

\bibitem{Suzuki2009}
{\sc Y.~Suzuki, K.~Watanabe, H.~Funaba, S.~Sakakibara, N.~Nakajima, N.~Ohyabu,
  L.~E. Group, et~al.}, {\em Effects of the stochasticity on transport
  properties in high-$\beta$ {LHD}}, Plasma and Fusion Research, 4 (2009),
  pp.~036--036.

\bibitem{Taylor1974}
{\sc J.~B. Taylor}, {\em Relaxation of toroidal plasma and generation of
  reverse magnetic fields}, Physical Review Letters, 33 (1974), p.~1139.

\bibitem{Taylor1986}
{\sc J.~B. Taylor}, {\em Relaxation and magnetic reconnection in plasmas}, Rev.
  Mod. Phys., 58 (1986), pp.~741--763.

\bibitem{Tikhonov1963}
{\sc A.~N. Tikhonov}, {\em On the solution of ill-posed problems and the method
  of regularization}, in Doklady Akademii Nauk, vol.~151, Russian Academy of
  Sciences, 1963, pp.~501--504.

\bibitem{turek2D}
{\sc S.~Turek, O.~Mierka, and K.~B{\"a}umler}, {\em Numerical benchmarking for
  {3D} multiphase flow: New results for a rising bubble}, in Numerical
  Mathematics and Advanced Applications ENUMATH 2017, F.~A. Radu, K.~Kumar,
  I.~Berre, J.~M. Nordbotten, and I.~S. Pop, eds., Cham, 2019, Springer
  International Publishing, pp.~593--601.

\bibitem{Van2007}
{\sc J.~G. Van~Bladel}, {\em Electromagnetic fields}, vol.~19, John Wiley \&
  Sons, 2007.

\bibitem{Wagner1998}
{\sc F.~Wagner}, {\em Stellarators and optimised stellarators}, Fusion
  Technology, 33 (1998), pp.~67--83.

\bibitem{Wakatani1996}
{\sc M.~Wakatani and S.~Sudo}, {\em Overview of {Heliotron E} results}, Plasma
  Physics and Controlled Fusion, 38 (1996), p.~937.

\bibitem{Weitzner2014}
{\sc H.~Weitzner}, {\em Ideal magnetohydrodynamic equilibrium in a
  non-symmetric topological torus}, Physics of Plasmas, 21 (2014), p.~022515.

\bibitem{Weitzner2016}
{\sc H.~Weitzner}, {\em Expansions of non-symmetric toroidal
  magnetohydrodynamic equilibria}, Physics of Plasmas, 23 (2016), p.~062512.

\bibitem{Weller2006}
{\sc A.~Weller, S.~Sakakibara, K.~Watanabe, K.~Toi, J.~Geiger, M.~Zarnstorff,
  S.~Hudson, A.~Reiman, A.~Werner, C.~N{\"u}hrenberg, et~al.}, {\em
  Significance of {MHD} effects in stellarator confinement}, Fusion Science and
  Technology, 50 (2006), p.~158.

\bibitem{wesson2011}
{\sc J.~Wesson and D.~J. Campbell}, {\em Tokamaks}, vol.~149, Oxford University
  Press, 2011.

\bibitem{Willis1953}
{\sc C.~H. Willis}, {\em Design and construction of {Model A} stellarator},
  tech. rep., Princeton Univ., NJ Project Matterhorn, 1953.

\bibitem{Wing1991}
{\sc G.~M. Wing}, {\em A primer on integral equations of the first kind: the
  problem of deconvolution and unfolding}, vol.~27, SIAM, 1991.

\bibitem{Wolf2019}
{\sc R.~Wolf, A.~Alonso, S.~{\"A}k{\"a}slompolo, J.~Baldzuhn, M.~Beurskens,
  C.~Beidler, C.~Biedermann, H.-S. Bosch, S.~Bozhenkov, R.~Brakel, et~al.},
  {\em Performance of {Wendelstein 7-X} stellarator plasmas during the first
  divertor operation phase}, Physics of Plasmas, 26 (2019), p.~082504.

\bibitem{wu2003}
{\sc J.~Wu}, {\em Generalized {MHD} equations}, Journal of Differential
  Equations, 195 (2003), pp.~284--312.

\bibitem{Wurden2017}
{\sc G.~A. Wurden, C.~Biedermann, F.~Effenberg, M.~Jakubowski, H.~Niemann,
  L.~Stephey, S.~Bozhenkov, S.~Brezinsek, J.~Fellinger, B.~Cannas, et~al.},
  {\em Limiter observations during {W7-X} first plasmas}, Nuclear Fusion, 57
  (2017), p.~056036.

\bibitem{Xanthopoulos2014}
{\sc P.~Xanthopoulos, H.~Mynick, P.~Helander, Y.~Turkin, G.~Plunk, F.~Jenko,
  T.~G{\"o}rler, D.~Told, T.~Bird, and J.~Proll}, {\em Controlling turbulence
  in present and future stellarators}, Physical Review Letters, 113 (2014),
  p.~155001.

\bibitem{Yamada2001}
{\sc H.~Yamada, K.~Watanabe, K.~Yamazaki, S.~Murakami, S.~Sakakibara,
  K.~Narihara, K.~Tanaka, M.~Osakabe, K.~Ida, N.~Ashikawa, et~al.}, {\em Energy
  confinement and thermal transport characteristics of net current free plasmas
  in the {Large Helical Device}}, Nuclear Fusion, 41 (2001), p.~901.

\bibitem{Yamazaki1992}
{\sc K.~Yamazaki, O.~Motojima, and M.~Asao}, {\em Design scalings and
  optimization for the superconducting {Large Helical Device}}, Fusion
  Technology, 21 (1992), pp.~147--160.

\bibitem{Yoccoz1992}
{\sc J.-C. Yoccoz}, {\em {An Introduction To Small Divisors Problems}},
  Springer Berlin Heidelberg, Berlin, Heidelberg, 1992, pp.~659--679.

\bibitem{Yokoyama2001}
{\sc M.~Yokoyama, K.~Itoh, S.~Okamura, K.~Matsuoka, and S.-I. Itoh}, {\em
  Maximum-{J} capability in a quasiaxisymmetric stellarator}, Phys. Rev. E, 64
  (2001), p.~015401.

\bibitem{Yokoyama1997}
{\sc M.~Yokoyama, Y.~Nakamura, and M.~Wakatani}, {\em An optimized helical axis
  stellarator with modulated l=1 helical coil}, J. Plasma Fusion Res, 73
  (1997), pp.~723--731.

\bibitem{Yoshikawa1985}
{\sc S.~Yoshikawa and T.~Stix}, {\em Experiments on the {Model C} stellarator},
  Nuclear Fusion, 25 (1985), p.~1275.

\bibitem{Zakharov2015}
{\sc L.~E. Zakharov}, {\em Implementation of {Hamada} principle in calculations
  of nested {3-D} equilibria}, Journal of Plasma Physics, 81 (2015).

\bibitem{Zarnstorff2001}
{\sc M.~Zarnstorff, L.~Berry, A.~Brooks, E.~Fredrickson, G.~Fu, S.~Hirshman,
  S.~Hudson, L.~Ku, E.~Lazarus, D.~Mikkelsen, et~al.}, {\em Physics of the
  compact advanced stellarator {NCSX}}, Plasma Physics and Controlled Fusion,
  43 (2001), p.~A237.

\bibitem{Zarnstorff2004}
{\sc M.~Zarnstorff, A.~Weller, J.~Geiger, E.~Fredrickson, S.~Hudson, J.~Knauer,
  A.~Reiman, A.~Dinklage, G.~Fu, L.~Ku, et~al.}, {\em 20th {IAEA} fusion energy
  conference}, tech. rep., EX/3-4 (IAEA, Vilamoura, 2004), 2004.

\bibitem{Zhu2020b}
{\sc C.~Zhu, K.~Hammond, T.~G. Brown, D.~A. Gates, M.~C. Zarnstorff,
  K.~Corrigan, M.~Sibilia, and E.~Feibush}, {\em Topology optimization of
  permanent magnets for stellarators}, Nuclear Fusion,  (2020).

\bibitem{Zhu2018}
{\sc C.~Zhu, S.~R. Hudson, Y.~Song, and Y.~Wan}, {\em {New method to design
  stellarator coils without the winding surface}}, Nuclear Fusion, 58 (2018),
  p.~016008.

\bibitem{Zhu2020a}
{\sc C.~Zhu, M.~Zarnstorff, D.~Gates, and A.~Brooks}, {\em Designing
  stellarators using perpendicular permanent magnets}, Nuclear Fusion, 60
  (2020), p.~076016.

\end{thebibliography}

\end{document}